\pdfoutput = 1

\documentclass[reprint,prb,nofootinbib]{revtex4-2}

\usepackage{tikz}
\usetikzlibrary{decorations,decorations.pathmorphing,decorations.markings,calc,snakes,positioning,fixedpointarithmetic,shapes,shapes.geometric,patterns}

\usepackage{etoolbox}
\AtBeginEnvironment{tikzpicture}{\catcode`$=3 } % $ % was important for auctex preview??

\tikzset{alignmid/.style={baseline={([yshift=-.5ex]current bounding box.center)}}} % adjust pictures vertically
\tikzset{every picture/.append style=alignmid}

\tikzset{
bottomzigzag/.style={postaction={draw,decorate, decoration={zigzag,amplitude=1pt,segment length=3pt,raise=1pt}}},
zigzag/.style={draw,decorate, decoration={zigzag,amplitude=1pt,segment length=3pt}},
rc/.style=rounded corners,
}

\tikzset{
    -|/.style={to path={-| (\tikztotarget)}},
    |-/.style={to path={|- (\tikztotarget)}},
}

\usepackage{ifthen}
\usepackage{xparse,pgffor}

\tikzset{
mark/.code={
\tikzset{postaction={/network/mark/.cd,#1,/tikz/.cd,decorate},decoration={name=markings,mark=at position \netmarkpos with{%+\netmarkposoff} with{
\begin{scope}[netmarktrafo]
\netmarkcode
\end{scope}
}}}
\def\netmarkpos{0.5}%\pgfdecoratedpathlength}
},
}

\def\netmarkpos{0.5}%\pgfdecoratedpathlength}
\def\netmarkposoff{0cm}
\def\netmarkcode{}

\tikzset{
netmarktrafo/.style={},
netmarkstyle/.style={solid,semithick,sharp corners},
}

\pgfkeys{
/network/mark/.cd,
sty/.code={
\tikzset{netmarkstyle/.style={#1}}
},
asty/.code={
\tikzset{netmarkstyle/.append style={#1}}
},
f/.style={asty=fill},
p/.code={ % relative positioning on decorated line
\def\netmarkpos{#1}%\pgfdecoratedpathlength}
},
poff/.code={ % absolute shift on decorated line in cm
\def\netmarkposoff{#1cm}
},
e/.code={ % mark at the End of line
\def\netmarkpos{\pgfdecoratedpathlength-0.005cm-\netmarkposoff}

\tikzset{netmarktrafo/.append style={shift={(-\netmarkwidth,0)}}}
},
s/.code={ % mark at the Start of line
\def\netmarkpos{0.005cm+\netmarkposoff}

\tikzset{netmarktrafo/.append style={shift={(\netmarkwidth,0)},xscale=-1,yscale=-1}}
},
a/.code={ % mark After end of line
\def\netmarkpos{\pgfdecoratedpathlength-0.005cm}

\tikzset{netmarktrafo/.append style={xscale=-1,shift={(-\netmarkwidth,0)}}}
},
b/.code={ % mark Before start of line
\def\netmarkpos{0.005cm}

\tikzset{netmarktrafo/.append style={xscale=-1,shift={(\netmarkwidth,0),yscale=-1}}}
},
-/.code={ % flip horizontally (mark backwards)
\tikzset{netmarktrafo/.append style={xscale=-1}}
},
r/.code={ % flip vertically (mark on the Right instead of left)
\tikzset{netmarktrafo/.append style={yscale=-1}}
},
sideoff/.code={ % shift sideways
\tikzset{netmarktrafo/.append style={shift={(0,#1)}}}
},
slab/.code={ % label next to the line
\def\netmarkwidth{0}
\def\netmarkcode{
\node[inner sep=0.04cm,netmarkstyle,draw=none] (mylabelwidthtest) at (0,0){\phantom{#1}};
\path let \p1=(mylabelwidthtest.north east), \p2=(mylabelwidthtest.south east), \n1 = {max(abs(\y1),abs(\y2))} in node[inner sep=0.04cm,netmarkstyle] at (0,\n1) {#1};
}
},
lab/.code={ % label on the line
\def\netmarkwidth{0}
\def\netmarkcode{
\node[inner sep=0.04cm,anchor=\netmarkanchor] (mylabelwidthtest) at (0,0) {\phantom{#1}};
\draw[white] (mylabelwidthtest.\pgfdecoratedangle)--(mylabelwidthtest.\pgfdecoratedangle+180);
\node[inner sep=0.04cm,anchor=\netmarkanchor,netmarkstyle] at (0,0) {#1};
}
},
arr/.code={ % arrow
\def\netmarkwidth{0.04}
\def\netmarkcode{\draw[netmarkstyle] (-0.04,0.08)--(0.04,0)--(-0.04,-0.08);}
},
rarr/.code={ % round arrow
\def\netmarkwidth{0.04}
\def\netmarkcode{\draw[netmarkstyle] (-0.04,-0.08)arc(90-180:90:0.08);}
},
dot/.code={ % dot
\def\netmarkwidth{0.08}
\def\netmarkcode{\draw[netmarkstyle] (0,0)circle(0.08);}
},
flag/.code={ % flag to the left
\def\netmarkwidth{0.06}
\def\netmarkcode{\draw[netmarkstyle] (-0.06,0)--(0,0.09)--(0.06,0)--cycle;}
},
darr/.code={ % "double arrow": half-arrow on the left
\def\netmarkwidth{0.08}
\def\netmarkcode{\draw[netmarkstyle] (-0.04,0)--(0.04,0)--(-0.04,0.08)--cycle;}
},
hcirc/.code={ % half-circle on the left
\def\netmarkwidth{0.1}
\def\netmarkcode{\draw[netmarkstyle] (-0.1,0) arc (180:0:0.1);}
},
bar/.code={ % tick, short line perpendicular to line
\def\netmarkwidth{0.05}
\def\netmarkcode{
\draw[netmarkstyle] (0,-0.08cm-0.5*\pgflinewidth)--(0,0.08cm+0.5*\pgflinewidth);
}
},
hbar/.code={ % tick, short perpendicular line towards the left
\def\netmarkwidth{0.05}
\def\netmarkcode{
\draw[netmarkstyle] (0, 0.5*\pgflinewidth)--++(0,0.12);
}
},
mill/.code={
\def\netmarkwidth{0.16}
\def\netmarkcode{
\draw[netmarkstyle] (0,-0.5*\pgflinewidth)--++(-0.08,-0.08)--++(0,0.08);
\draw[netmarkstyle] (0,0.5*\pgflinewidth)--++(0.08,0.08)--++(0,-0.08);
}
},
three dots/.code={
\def\netmarkwidth{0.2}
\def\netmarkcode{
\fill (-0.12,0) circle (0.5*0.05) (0,0) circle (0.5*0.05) (0.12,0) circle (0.5*0.05);
}
},
}

\tikzset{wid/.style={minimum width=#1cm}}
\tikzset{hei/.style={minimum height=#1cm}}

\tikzset{sx/.style={xshift=#1cm}}
\tikzset{sy/.style={yshift=#1cm}}

\tikzset{box/.style={draw,rectangle}}
\tikzset{fbox/.style={draw,rectangle, line width=1.1}}
\tikzset{roundbox/.style={draw,rectangle,rounded corners}}
\tikzset{froundbox/.style={draw,rectangle, rounded corners, line width=1.1}}
\tikzset{rounddiamond/.style={draw,diamond,rounded corners}}
\tikzset{dot/.style={draw, shape=circle, fill=black, scale=0.5}}

\tikzset{
netbox/.code={
\node[draw,netbdstyle] (\atomname) at (0,0) {#1};
\coordinate (\atomname-r) at (\atomname.east);
\coordinate (\atomname-l) at (\atomname.west);
\coordinate (\atomname-t) at (\atomname.north);
\coordinate (\atomname-b) at (\atomname.south);
\coordinate (\atomname-tr) at (\atomname.north east);
\coordinate (\atomname-br) at (\atomname.south east);
\coordinate (\atomname-tl) at (\atomname.north west);
\coordinate (\atomname-bl) at (\atomname.south west);
},
}

\tikzset{bdlw/.code={\tikzset{mybdstyle/.style={draw, line width=#1}}}}
\tikzset{bdcol/.code={\tikzset{mybdstyle/.append style={#1}}}}

\newcommand\setelements[1]{
\pgfkeys{/network/atom/.cd,#1}
}

\newcommand\atoms[2]{
\foreach \name/\keys in {#2}{
\expandafter\atom\expandafter{\keys,#1}{\name}
}
}

\newcommand\atom[2]{
\def\atomname{#2}
\tikzset{
nettrafo/.style={},
netatompos/.style={},
netdeco/.style={},
netpostdeco/.style={},
}

\pgfkeys{/network/atom/.cd,#1}

\begin{scope}[netatompos] % shift to atom position
\begin{scope}[nettrafo] % rotate, flip and scale
\netshapecoords % set the anchor coordinates
\fill[netbackstyle] \netshapepath;
\clip \netshapepath;
\tikzset{netdeco}
\draw[netbdstyle] \netshapepath;
\end{scope}
\tikzset{netpostdeco} % draw post-decorations, not rotated, flipped, or scaled
\end{scope}

}

\pgfkeys{
/network/atom/.cd,
square/.code={

\def\netshapepath{(-\tempsize,-\tempsize)rectangle (\tempsize,\tempsize)}
\def\netshapecoords{
\node[rectangle,wid=2*\tempsize,hei=2*\tempsize,inner sep=0,transform shape](\atomname)at(0,0){};
\coordinate(\atomname-c) at (0,0);
\coordinate(\atomname-r) at (\tempsize,0);
\coordinate(\atomname-l) at (-\tempsize,0);
\coordinate(\atomname-t) at (0,\tempsize);
\coordinate(\atomname-b) at (0,-\tempsize);
\coordinate(\atomname-br) at (\tempsize,-\tempsize);
\coordinate(\atomname-tr) at (\tempsize,\tempsize);
\coordinate(\atomname-bl) at (-\tempsize,-\tempsize);
\coordinate(\atomname-tl) at (-\tempsize,\tempsize);
}},
circ/.code={

\def\netshapepath{(0,0)circle(\tempsize)}
\def\netshapecoords{
\node[circle,wid=2*\tempsize,hei=2*\tempsize,inner sep=0,transform shape](\atomname)at(0,0){};
\coordinate(\atomname-c) at (0,0);
\coordinate(\atomname-r) at (\tempsize,0);
\coordinate(\atomname-l) at (-\tempsize,0);
\coordinate(\atomname-t) at (0,\tempsize);
\coordinate(\atomname-b) at (0,-\tempsize);
}},
triang/.code={

\def\netshapepath{(-30:\tempsize)--(90:\tempsize)--(-150:\tempsize)--cycle}
\def\netshapecoords{
\node[regular polygon,regular polygon sides=3,wid=2*\tempsize,inner sep=0,transform shape](\atomname)at(0,0){};
\coordinate(\atomname-c) at (0,0);
\coordinate(\atomname-cr) at (-30:\tempsize);
\coordinate(\atomname-cl) at (-150:\tempsize);
\coordinate(\atomname-ct) at (90:\tempsize);
\coordinate(\atomname-mb) at (-90:0.5*\tempsize);
\coordinate(\atomname-mr) at (30:0.5*\tempsize);
\coordinate(\atomname-ml) at (150:0.5*\tempsize);
}},
diamond/.code={

\def\netshapepath{(0,-\tempsize)--(\tempsize,0)--(0,\tempsize)--(-\tempsize,0)--cycle}
\def\netshapecoords{
\node[rotate=45,rectangle,wid=sqrt(2)*\tempsize,hei=sqrt(2)*\tempsize,inner sep=0,transform shape](\atomname)at(0,0){};
\coordinate(\atomname-c) at (0,0);
\coordinate(\atomname-r) at (\tempsize,0);
\coordinate(\atomname-l) at (-\tempsize,0);
\coordinate(\atomname-t) at (0,\tempsize);
\coordinate(\atomname-b) at (0,-\tempsize);
}},
pentagon/.code={

\def\netshapepath{(-126:\tempsize)--(-54:\tempsize)--(18:\tempsize)--(90:\tempsize)--(162:\tempsize)--cycle}
\def\netshapecoords{
\node[regular polygon,regular polygon sides=5,wid=2*\tempsize,inner sep=0,transform shape](\atomname)at(0,0){};
\coordinate(\atomname-c) at (0,0);
\coordinate (\atomname-mb)at(-90:{\tempsize*cos(36)});
\coordinate (\atomname-mbr)at(-18:{\tempsize*cos(36)});
\coordinate (\atomname-mtr)at(54:{\tempsize*cos(36)});
\coordinate (\atomname-mtl)at(126:{\tempsize*cos(36)});
\coordinate (\atomname-mbl)at(-162:{\tempsize*cos(36)});
\coordinate (\atomname-cbr)at(-54:\tempsize);
\coordinate (\atomname-cr)at(18:\tempsize);
\coordinate (\atomname-ct)at(90:\tempsize);
\coordinate (\atomname-cl)at(162:\tempsize);
\coordinate (\atomname-cbl)at(-126:\tempsize);
}},
halfcirc/.code={

\def\netshapepath{(\tempsize,0)arc(0:180:\tempsize)--++(0,-0.04)-|cycle}
\def\netshapecoords{
\node[circle,wid=2*\tempsize,hei=2*\tempsize,inner sep=0,transform shape](\atomname)at(0,0){};
\coordinate(\atomname-c) at (0,0);
\coordinate(\atomname-r) at (\tempsize,0);
\coordinate(\atomname-l) at (-\tempsize,0);
\coordinate(\atomname-t) at (0,\tempsize);
\coordinate(\atomname-b) at (0,0);
}},
void/.code={
\def\netshapepath{}
\def\netshapecoords{
\coordinate(\atomname) at (0,0);
\coordinate(\atomname-c) at (0,0);
}},
labbox/.code={
\def\netshapepath{(0,0)}
\def\netshapecoords{}
\tikzset{netpostdeco/.append style={netbox=#1}}
},
}

\pgfkeys{
/network/atom/.cd,
p/.code={\tikzset{netatompos/.append style={shift={(#1)}}}},
xscale/.code={\tikzset{nettrafo/.append style={xscale=#1}}},
yscale/.code={\tikzset{nettrafo/.append style={yscale=#1}}},
scale/.code={\tikzset{nettrafo/.append style={scale=#1}}},
rot/.code={\tikzset{nettrafo/.append style={rotate=#1}}},
hflip/.style={xscale=-1},
vflip/.style={yscale=-1},
big/.style={scale=3/2},
small/.style={scale=2/3},
huge/.style={scale=9/4},
tiny/.style={scale=4/9},
flat/.style={yscale=2/3},
fflat/.style={yscale=4/9},
}

\tikzset{
netbdstyle/.style={line width=0.15em}, % changed from pt(default)
netdecstyle/.style={},
netpostdecstyle/.style={},
netbackstyle/.style={white},
}
\pgfkeys{
/network/atom/.cd,
bdstyle/.code={\tikzset{netbdstyle/.style={#1}}},
bdastyle/.code={\tikzset{netbdstyle/.append style={#1}}},
decstyle/.code={\tikzset{netdecstyle/.style={#1}}},
postdecstyle/.code={\tikzset{netpostdecstyle/.style={#1}}},
backstyle/.code={\tikzset{netbackstyle/.style={#1}}},
whiteblack/.style={decstyle=white,backstyle=black},
nobd/.code={bdstyle={line width=0}},
style/.style={bdastyle=#1, decstyle=#1, postdecstyle=#1},
}

\tikzset{
netbscope/.code={\begin{scope}[#1]},
netescope/.code={\end{scope}},
}

\pgfkeys{
/network/atom/.cd,
bscope/.code={\tikzset{netdeco/.append style={netbscope={#1}}}},
escope/.code={\tikzset{netdeco/.append style={netescope}}},
dec/.style 2 args={bscope={#1},#2,escope}, % for applying transformation keys to certain decorations only
vbar/.style={dec={rotate=90}{hbar}},
dcross/.style={dec={rotate=45}{hbar},dec={rotate=-45}{hbar}},
cross/.style={hbar,vbar},
sect6/.style={sect=60},
sect8/.style={sect=45},
sect10/.style={sect=36},
}

\def\regdec#1{\pgfkeys{/network/atom/.cd,#1/.code={\tikzset{netdeco/.append style={net#1}}}}}
\regdec{all}\regdec{rhalf}\regdec{rquart}\regdec{brquart}\regdec{dot}\regdec{spiral}\regdec{swirl}\regdec{hstripe}\regdec{hbar}\regdec{rrey}
\pgfkeys{/network/atom/.cd,sect/.code={\tikzset{netdeco/.append style={netsect=#1}}}}

\tikzset{
netall/.code={\fill[netdecstyle] (-0.3,-0.3)rectangle (0.3,0.3);}, % fill all
netrhalf/.code={\fill[netdecstyle] (0,-0.3)rectangle (0.3,0.3);}, % right half
netrquart/.code={\fill[netdecstyle] (0.075,-0.3)rectangle (0.3,0.3);}, % right quarter
netbrquart/.code={\fill[netdecstyle] (0,0)rectangle (0.3,-0.3);}, % bottom right quarter
netsect/.code={\fill[netdecstyle] (0,0)--(0,-0.3)arc(-90:-90+#1:0.3)--cycle;}, % section of angle #1 starting from -90
netdot/.code={\fill[netdecstyle] (0,0)circle(0.07);}, % dot in the middle
netspiral/.code={\draw[netdecstyle] plot [variable=\t,domain=0:4] ({0.075*\t*cos(pi*(\t-0.5) r)},{0.075*\t*sin(pi*(\t-0.5) r)});}, % spiral
netswirl/.code={\fill[netdecstyle] plot [variable=\t,domain=0:2] ({0.15*\t*cos(pi*(\t-0.5) r)},{0.15*\t*sin(pi*(\t-0.5) r)}) arc(-90:-450:0.3)--cycle;}, % filled swirl
nethstripe/.code={\fill[netdecstyle] (-0.3,-0.05)rectangle(0.3,0.05);}, % horizontal stripe
nethbar/.code={\draw[netdecstyle] (-0.3,0)--(0.3,0);}, % horizontal line
netrrey/.code={\draw[netdecstyle] (0,0)--(0.3,0);} % line from the middle to the right
}

\pgfkeys{
/network/atom/.cd,
postbscope/.code={\tikzset{netpostdeco/.append style={netbscope={#1}}}},
postescope/.code={\tikzset{netpostdeco/.append style={netescope}}},
postdec/.style 2 args={postbscope={#1},#2,postescope}, % for applying transformation keys to certain decorations only
arc/.code={\tikzset{netpostdeco/.append style={netarc=#1}}},
lab/.code={\tikzset{netpostdeco/.append style={netlab={#1}}}},
shadecirc/.code={\tikzset{netpostdeco/.append style=netshadecirc}},
shaderect/.code={\tikzset{netpostdeco/.append style={netshaderect={#1}}}},
debug/.code={\tikzset{netpostdeco/.append style=netdebug}},
markline/.code 2 args={\tikzset{netpostdeco/.append style={netmarkline={#1}{#2}}}},
postcirc/.code={\tikzset{netpostdeco/.append style=netpostcirc}},
}

\tikzset{
netlab/.code={
\pgfkeys{/network/atom/lab/.cd,#1}
\node[netpostdecstyle] at (\ifdefined\netlabpos\netlabpos\else\netlabang:\netlabdist\fi) {\netlabwrap{\netlabtext}};
},
netarc/.code args={#1:#2:#3}{
\draw[netpostdecstyle] (#1:#3) arc (#1:#2:#3);
},
netshadecirc/.code= {
\fill[opacity=0.4,netpostdecstyle] (0,0)circle(0.4);
},
netpostcirc/.code= {
\draw[netpostdecstyle] (0,0)circle(0.15);
},
netshaderect/.code= {
\fill[rc,opacity=0.4,netpostdecstyle] ($-1*(#1)$) rectangle (#1);
},
netdebug/.code= {
\node[red] at (0,0){\atomname};
},
netmarkline/.code 2 args= {
\draw (\atomname)edge[mark={#2}]++(#1);
},
}

\def\netlabwrap#1{#1}
\pgfkeys{
/network/atom/lab/.cd,
t/.code={\def\netlabtext{#1}}, % label text
p/.code={\def\netlabpos{#1}}, % position of the label
ang/.code={\def\netlabang{#1}},
dist/.code={\def\netlabdist{#1}},
wrap/.code={\def\netlabwrap##1{#1}}, % store in doesn't work
}

\usepackage{xcolor}
\usepackage{amsmath}
\usepackage{amssymb}
\usepackage{bbold}
\usepackage{mathtools}
\usepackage{hyperref}
\usepackage{enumitem}

\setlist[description]{
itemsep=0pt,
font={\normalfont\itshape},
}
\def\deshead#1{
\vspace{0.2cm}
\noindent
\textbf{#1}
}

\def\zz{\mathbb{Z}}
\def\cc{\mathbb{C}}

\def\cone{\operatorname{Cone}}
\def\calb{\mathcal{B}}
\def\calm{\mathcal{M}}
\def\calf{\mathcal{F}}
\def\calt{\mathcal{T}}
\def\calx{\mathcal{X}}
\def\lagr{\mathcal{L}}

\newcommand{\breakcell}[2][t]{\begin{tabular}[#1]{@{}l@{}}#2\end{tabular}}
\def\smallint{\begingroup\textstyle \int\endgroup}

\def\manifoldcol{black}
\def\bdcol{cyan}
\def\gluecol{red}
\def\secondmanifoldcol{brown}
\def\spincol{purple}
\def\omegacol{purple}
\def\cohomologyAcol{green}
\def\cohomologyBcol{red}

\tikzset{
ind/.style={mark={lab=$#1$,a}}, % normal open index label
startind/.style={mark={lab=$#1$,b}}, % normal open index label
irrep/.style={line width=1.7},
multip/.style={zigzag},
front/.style={preaction={draw,white,line width=3}},
bdcol/.style={\bdcol},
spin structure/.style={line width=0.2cm, opacity=0.3, \spincol},
fav/.style={mark={bar,poff=0.0,#1}},
glue/.style={\gluecol, densely dotted, line width=0.15cm, opacity=0.3},
bdcellulation/.style={\bdcol, dotted,line width=1.6},
bdribbon/.style={\bdcol,line width=1.6},
cellulation/.style={\manifoldcol, dotted},
cohomologyA/.style={postaction={draw,\cohomologyAcol,dash pattern=on 1pt off 2pt,line width=3}},
cohomologyB/.style={postaction={draw,\cohomologyBcol,dash pattern=on 1pt off 2pt,line width=3}},
cohomologyAfull/.style={postaction={pattern=north west lines, pattern color=\cohomologyAcol}},
cohomologyBfull/.style={postaction={pattern=north west lines, pattern color=\cohomologyBcol}},
}

\usetikzlibrary{patterns.meta}
\tikzset{
manifold/.style={fill=\manifoldcol,fill opacity=.4},
doublemanifold/.style={manifold,postaction=manifold},
manifoldboundary/.style={bdcol,line width=1.6},
manifoldbdfull/.style={pattern={Dots[radius=0.025cm]},pattern color=\bdcol},
}

\setelements{
vertex/.style={circ,tiny,all},
bdvertex/.style={vertex,bdastyle=bdcol,decstyle=bdcol},
regionvertex/.style={circ,tiny},
bdvertexodd/.style={vertex,bdastyle=bdcol,decstyle=bdcol, postcirc, postdecstyle=\spincol},
cellvertex/.style={circ,tiny,bdstyle=\manifoldcol},
bdcellvertex/.style={circ,tiny,bdstyle=\bdcol},
corner/.style={postdecstyle={line width=1.2,\bdcol},arc=#1:0.2},
omega/.style={postdecstyle=\omegacol,postdec={scale=0.5}{shadecirc}},
glueedge/.style={void,postdecstyle=\gluecol,postdec={scale=0.5}{shadecirc}},
}

\begin{document}
\title{Disentangling modular Walker-Wang models via fermionic invertible boundaries}
\author{Andreas Bauer}
\email{andibauer@zedat.fu-berlin.de}
\affiliation{Freie Universit{\"a}t Berlin, Arnimallee 14, 14195 Berlin, Germany}
\date{\today}

\begin{abstract}
Walker-Wang models are fixed-point models of topological order in $3+1$ dimensions constructed from a braided fusion category. For a modular input category $\mathcal M$, the model itself is invertible and is believed to be in a trivial topological phase, whereas its standard boundary is supposed to represent a $2+1$-dimensional chiral phase. In this work we explicitly show triviality of the model by constructing an invertible domain wall to vacuum as well as a disentangling generalized local unitary circuit in the case where $\mathcal M$ is a Drinfeld center. Moreover, we show that if we allow for fermionic (auxiliary) degrees of freedom inside the disentangling domain wall or circuit, the model becomes trivial for a larger class of modular fusion categories, namely those in the Witt classes generated by the Ising UMTC. In the appendices, we also discuss general (non-invertible) boundaries of general Walker-Wang models and describe a simple axiomatization of extended TQFT in terms of tensors.
\end{abstract}

\maketitle
\tableofcontents

\section{Introduction}
Exactly solvable fixed-point models of topological order are a highly successful approach to the study of topologically ordered phases. They provide a way to classify and study the properties of phases in an exact algebraic way while still retaining an explicit microscopic description.

Unfortunately, to date, fixed-point models fail to describe one important class of phases, namely so-called \emph{chiral} topological order in $2+1$ dimensions, or more precisely, phases that do not possess a gapped/topological boundary. Those phases include the integer and fractional quantum Hall states, one of the few topological phases that have been observed experimentally. Chiral intrinsic bosonic topological phases in $2+1$ dimensions are those for which the \emph{unitary modular tensor category} (UMTC) describing the anyon content is not a \emph{Drinfeld center} of a fusion category. It is argued in Ref.~\cite{Kapustin2019} that those chiral phases do not possess any (fixed-point) commuting-projector Hamiltonians due to their non-zero thermal Hall conductance. However, this does not rule out more general discrete fixed-path path integrals as argued in Ref.~\cite{universal_liquid}.

There is a way to resolve the issue of chiral phases, if we are willing to pay a high price, namely the introduction of an additional auxiliary dimension. This auxiliary dimension contains a $3+1$-dimensional state-sum construction introduced by Crane and Yetter \cite{Crane1993}, which was later studied as a Hamiltonian model in the context of topological phases by Walker and Wang \cite{Walker2011}, and which we refer to as the \emph{CYWW model}. The CYWW model takes as input a \emph{unitary braided fusion category}, but for our purposes the interesting case is when that braided fusion category is actually \emph{modular}, and hence a UMTC. Then the standard ``smooth'', or ``cone'' $2+1$-dimensional boundary \cite{Keyserlingk2012} of the CYWW model is said to be a model for the chiral phase whose anyon content is described by the input UMTC.

Saying that a boundary of a $3+1$-dimensional model represents a standalone $2+1$-dimensional phase only makes sense if the $3+1$-dimensional bulk itself is in a trivial phase. Unfortunately, it is still unknown in how far exactly this is the case for the modular CYWW model. One can easily show that the modular CYWW model is invertible and has no non-trivial $0+1$-dimensional or $1+1$-dimensional defects. However, this does not necessarily mean that the model is in a trivial microscopic phase. Indeed, there are many examples of invertible but non-trivial phases such as the fermionic Kitaev chain in $1+1$ dimensions, SPT phases, or the bosonic $E_8$ phase in $2+1$ dimensions \cite{Kitaev2005, Lu2012}
\footnote{Note that the Kitaev chain and SPT phases do have non-trivial codimension-2 defects at the termination of fermion-parity- or symmetry defects. This is not the case for the $E_8$ phase.}
. In order to show that the model is actually trivial, we would have to use our first-principle microscopic definition of choice for a phase, such as via paths not closing the spectral gap, disentangling (``fuzzy'') local unitary circuits \cite{Chen2010, Kapustin2020}, or invertible domain walls to vacuum.

While the general question of whether and for which definition the modular CYWW model is indeed trivial is open, there is evidence that there cannot be an exact local unitary circuit disentangling the CYWW model if the input UMTC is chiral. Otherwise, as has been argued in Ref.~\cite{Haah2018}, one could conjugate the CYWW Hamiltonian with the circuit in the bulk and terminate it in a light-cone manner near the boundary. This would yield a commuting-projector Hamiltonian for the chiral boundary alone, contradicting Ref.~\cite{Kapustin2019}. In the first part of this paper, we show that the converse is true, that is, any modular CYWW model whose UMTC is a Drinfeld center can be disentangled by an exact generalized local unitary circuit. In a second part we allow for fermionic auxiliary degrees of freedom and find that this way we can disentangle a larger class of CYWW models beyond Drinfeld centers, namely those generated by the Ising UMTC, stacking, and adding/removing Drinfeld centers.

A rough outline of the remainder of this work is as follows. In Section~\ref{sec:state_sums_boundaries}, we revisit state-sum models and their (invertible) boundaries. In Section~\ref{sec:cyww_bulk}, we define the modular CYWW model in the bulk. In Section~\ref{sec:drinfeld_center_boundary}, we define invertible boundaries of CYWW models from \emph{Lagrangian modules}, which can be found for UMTCs that are a Drinfeld center. We also show how those invertible boundaries give rise to disentangling generalized local unitary circuits. In Section~\ref{sec:fermionic_boundaries} we find \emph{fermionic} invertible boundaries for a larger class of UMTCs, namely the Witt classes of the Kitaev 16-fold way.

In Appendix~\ref{sec:general_braided} we describe how small adaptions of the structures in the main text yield general topological boundaries for the CYWW model for braided fusion categories, constructed from \emph{braided modules} thereof. Appendix~\ref{sec:tensorial_tqft} puts the methods used in the main text in a much more general context, showing that all the structures are examples of \emph{tensorial TQFT}. In Appendix~\ref{sec:cohomology_boundaries}, we revisit the obtained invertible boundaries in a totally different language, namely that of lattice gauge theory based on simplicial (super-) cohomology.

\section{State-sum models and invertible boundaries}
\label{sec:state_sums_boundaries}
State-sum models are partition functions in discrete spacetime representing microscopic fixed-point models for topological phases. To this end, we assign discrete variables to different sorts of places (edges/vertices/corners/etc.) in a triangulation. The partition function is a sum over all configurations of those local variables, and each summand consists of a product of local weights depending on the values of the variables in a constant-size neighborhood. Equivalently, one can consider tensor-network path integrals on arbitrary triangulations \cite{liquid_intro}. The central property of the models is their invariance under local \emph{moves} of the triangulation representing discrete homeomorphisms, such as \emph{Pachner moves}, which makes them exactly solvable. While state-sum models such as those constructed in Refs.~\cite{Levin2004, Xi2021, Aasen2017} are conjectured to cover all phases that possess a topological (i.e., gapped) boundary, the situation is less clear for phases without topological boundary \cite{universal_liquid}.

State-sum models exist not only for topological phases of the bulk, but also for ``higher order phases'' of arbitrary types of boundaries, anyons, defects, fusion events, etc. E.g., for boundaries we define the partition function on triangulated manifolds with boundary, and show that it is invariant under attachment/removal of $n$-simplices to the boundary (in $n$ spacetime dimensions).

Sometimes it is interesting to restrict state-sum models further by demanding their invariance under topology-changing moves in addition to the topology-preserving ones. An example for this are models of \emph{invertible} phases, which have additional invariance under \emph{surgery operations}. If we write $B_i$ for the $i$-ball and $S_i$ for the $i$-sphere, those moves are given by
\begin{equation}
S_i \times B_{n-i} = B_{i+1}\times S_{n-i-1}
\end{equation}
in $n$ spacetime dimensions, for all $0\leq i\leq n$. This means we cut out a patch of triangulation with the topology of the left and paste a patch with the topology of the right, using that the boundary of both sides is identified with
\begin{equation}
S_i\times S_{n-i-1}\;.
\end{equation}
Note that for each topology-changing move, it suffices to impose them for only one particular pair of triangulations/cellulations representing the left- and right-hand side, sharing the same boundary. Due to the Pachner move/re-cellulation invariance, pairs of moves with the same topologies on the left- and on the right-hand side are equivalent.

Invertible topological phases are sometimes said to lack ``long-range entanglement'' since they do not possess non-trivial lower-dimensional defects such as anyons. However, invertible models do not have to be in an actually trivial phase, and important examples are given by the fermionic Kitaev chain in $1+1$ dimensions, or the bosonic $E_8$ phase in $2+1$ dimensions (which does not have a fixed-point description to date though). The common way to prove triviality of the phase for state-sum models is to construct an \emph{invertible domain wall} to the trivial model (vacuum), or in other words, an \emph{invertible boundary}. Invertibility of the boundary is a stronger condition than invertibility of the bulk, but it also corresponds to invariance of the path integral under topology-changing moves. 

When we deal with partition functions on topological manifolds with boundaries, it will be necessary to distinguish between those \emph{physical boundaries} and the \emph{space boundaries} \footnote{At the space boundary, we get an open configuration of state-sum variables, i.e., a ``state'' on a ``spatial surface''. It should be noted though that we are dealing with a Euclidean spacetime where there is actually no distinction between space-like and time-like directions as such. We use the ``$d+1$'' notation only to make clear what is the spacetime and what the space dimension, but not to indicate a Lorentzian signature of a metric.} we might get from cutting out patches. The space boundary itself has a physical boundary which is one dimension lower than that of the spacetime bulk. To denote the topology-changing moves, we introduce names for a few manifolds with space and/or physical boundary. Namely, we will write $B_i^s$ for an $i$-ball with space boundary, $B_i^p$ for an $i$-ball with physical boundary, $S_i$ an $i$-sphere, and $B^{ps}_i$ for the $i$-ball whose boundary $i-1$-sphere is divided into two $i-1$-balls, one of which is space boundary and the other is physical boundary.

Now, the moves for the invertibility of a boundary in $n$ spacetime dimensions are given by $M_i$ for each $0\leq i\leq n$, which consists in attaching/removing an $i$-handle to the physical boundary,
\begin{equation}
\label{eq:boundary_invertibility_moves}
M_i: S_{i-1} \times B^{ps}_{n-i+1} =  B^s_i \times B^p_{n-i}\;.
\end{equation}
The space boundary shared by both sides is
\begin{equation}
 S_{i-1} \times B_{n-i}^p\;.
\end{equation}
For example, in $n=3$ spacetime dimensions with $i=2$, we get
\begin{equation}
M_2:\qquad
\begin{tikzpicture}
\path[manifold] (0,0)to[bend left=90,looseness=0.5](1.5,0)to[bend right=30](1.5,1.5)to[bend right=90,looseness=0.5](0,1.5)to[bend right=30]cycle;
\path[manifoldbdfull] (0,0)to[bend right=90,looseness=0.5](1.5,0)to[bend left=40](1.5,1.5)to[bend left=90,looseness=0.5](0,1.5)to[bend left=40]cycle;
\path[pattern={Dots[radius=0.025cm,xshift=0.05cm,yshift=0.05cm]},pattern color=cyan] (0,0)to[bend left=90,looseness=0.5](1.5,0)to[bend left=40](1.5,1.5)to[bend right=90,looseness=0.5](0,1.5)to[bend left=40]cycle;
\path[manifold] (0,0)to[bend right=90,looseness=0.5](1.5,0)to[bend right=30](1.5,1.5)to[bend left=90,looseness=0.5](0,1.5)to[bend right=30]cycle;
\end{tikzpicture}=
\begin{tikzpicture}
\path[manifold] (0,0)to[bend left=90,looseness=0.5](1.5,0)to[bend right=30](1.5,1.5)to[bend right=90,looseness=0.5](0,1.5)to[bend right=30]cycle;
\path[manifoldbdfull] (0,1.5)to[bend right=90,looseness=0.5](1.5,1.5)to[bend right=90,looseness=0.5]cycle;
\path[manifoldbdfull] (0,0)to[bend right=90,looseness=0.5](1.5,0)to[bend right=90,looseness=0.5]cycle;
\path[manifold] (0,0)to[bend right=90,looseness=0.5](1.5,0)to[bend right=30](1.5,1.5)to[bend left=90,looseness=0.5](0,1.5)to[bend right=30]cycle;
\end{tikzpicture}\;,
\end{equation}
where the physical boundary is dotted in blue and the space boundary in gray. The left-hand side is a solid torus where the ``inner'' half of the boundary is physical, and the outer half is space boundary. The right-hand side is a solid cylinder with physical boundary on the top and bottom and space boundary on the side. The space boundary itself on both sides is an annulus.

\section{The modular CYWW model}
\label{sec:cyww_bulk}
In this section we will introduce the modular CYWW model focusing on the state-sum description. We will give a TQFT-style description/definition of UMTCs which are the input of the construction, such that the topological invariance and invertibility of the $3+1$-dimensional model follow from purely geometric/topological/combinatorial considerations.
\subsection{UMTCs}
\label{sec:tensorial_umtc}
\emph{Unitary modular tensor categories} (UMTCs) are known to describe the anyon statistics of models of topological order in $2+1$ dimensions \cite{Kitaev2005}. UMTCs come with a calculus of string diagrams, and those string diagrams describe histories of anyon worldlines and their fusion events in a $2+1$-dimensional Euclidean spacetime, to which the UMTC assigns amplitudes. We will work with a definition of UMTCs that is directly based on this interpretation. To make the main text more digestible, we will here omit one subtle feature of the full definition, which is discussed in Appendix~\ref{sec:umtc_tensorial_tqft}, related to the \emph{chiral anomaly}. If we would really follow through with the simplified definition in this section, this would correspond to restricting to UMTCs with zero \emph{chiral central charge}, $c=0$. We will discuss the relation of our TQFT-style definition to common categorical definitions in Appendix~\ref{sec:umtc_appendix}, and derive our definition from a more general framework of \emph{tensorial TQFT} in Appendix~\ref{sec:umtc_tensorial_tqft}.

Formally, a UMTC will be defined as map assigning a collection of amplitudes, i.e., an array with multiple labels/indices, or \emph{tensor}, to geometric/combinatorial/topological objects representing spacetime histories. Let us start by defining those topological objects.

A \emph{ribbon manifold} is a compact oriented $3$-manifold with a network of embedded ribbon segments. A ribbon segment is a line embedded into the $3$-manifold with a \emph{normal framing} and an orientation. \footnote{Note that contrary to boundaries, codimension-2 submanifolds cannot be equipped with a standard orientation using the bulk orientation.} A normal framing is a non-zero section of the normal bundle, or in other words, a continuous choice of identification of the infinitesimal normal circle around each point of the line with a standard circle. The ends of those ribbon segments meet at \emph{fusion vertices}. The fusion vertices are \emph{framed} as well, meaning that the infinitesimal sphere around them is identified with a standard sphere-with-points which we call the \emph{link} of the fusion vertex. The points inside the link themselves carry an orientation, meaning they point either ``inwards'' or ``outwards''. Those points also carry a framing, i.e., an identification of an infinitesimal circle around them with the standard circle.

In this document we will draw ribbon manifolds by projecting them onto the drawing plane, with the framing being inside the plane as well. The orientation is indicated by equipping the lines with arrow directions, and the framing is assumed to be pointing to the right when looking along the arrow direction. Vertex links just differing in the positions of their points can emulate each other, so we can assume without loss of generality that all links have their points only along the equator, and this equator is inside the drawing plane. If the identification with the link at a fusion vertices would be ambiguous due to symmetries of the adjacent ribbons and their directions, we will remove this ambiguity by marking a ``favorite'' adjacent ribbon with a tick. An example is given by
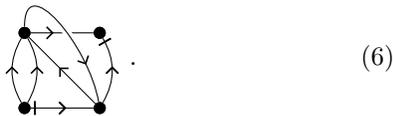
\begin{equation}
\begin{tikzpicture}
\atoms{vertex}{0/p={-0.5,-0.5}, 1/p={0.5,-0.5}, 2/p={-0.5,0.5}, 3/p={0.5,0.5}}
\draw (0)edge[bend left, mark=arr](2) (0)edge[bend right, mark=arr](2) (1)edge[mark=arr](2) (1)edge[bend right, mark=arr, fav=e](3) (0)edge[mark=arr,fav=s](1) (2)edge[mark={arr,p=0.3}](3) (2)edge[mark={arr,p=0.7},out=90,in=100,looseness=1.4, front](1);
\pgfresetboundingbox
\path[use as bounding box] (-0.8,-0.6) rectangle(0.8,0.9);
\end{tikzpicture}\;.
\end{equation}
The overall 3-manifold will usually be a 3-sphere, otherwise we will describe its topology and the homotopy classes of cycles of ribbon segments in words.

A ($c=0$) UMTC $\mathcal{M}$ is a map that associates a tensor (i.e., a complex multi-index array) to every ribbon manifold, with one \emph{ribbon label} at every ribbon segment, and one \emph{fusion index} at every fusion vertex. More precisely, ribbon labels are drawn from a fixed finite set, and the values of the fusion indices correspond to basis vectors in a finite-dimensional fusion vector space. The (dimension of the) latter depends on the link of the fusion vertex and on the values of the labels at the adjacent ribbons. A UMTC also comes with a \emph{quantum dimension} $d_i$ for each ribbon label $i$, and
\begin{equation}
D=(\sum_i d_i^2)^{1/2}
\end{equation}
is called the \emph{total quantum dimension}.

A UMTC $\mathcal{M}$ has to obey a collection of \emph{gluing axioms}, which are commutative diagrams involving twice the map $\mathcal{M}$, one \emph{gluing operation} $G$ of ribbon manifolds (such as a surgery), and one summation/contraction $C$ of tensors,
\begin{equation}
\label{eq:gluing_axiom_general}
\begin{tikzpicture}
\node (0) at (0,0){Ribbon manifold};
\node (1) at (4,0){Ribbon manifold};
\node (2) at (0,-1.5){Tensor};
\node (3) at (4,-1.5){Tensor};
\draw (0)edge[mark={arr,e},mark={slab=$G$}](1) (0)edge[mark={arr,e},mark={slab=$\calm$}](2) (1)edge[mark={arr,e},mark={slab=$\calm$}](3) (2)edge[mark={arr,e},mark={slab=$C$}](3);
\end{tikzpicture}\;.
\end{equation}
Concretely, we have the following gluing axioms.
\begin{itemize}
\item $G$ is the disjoint union of two ribbon manifolds, and $C$ is the tensor product of tensors. This axiom does not exactly fit into Eq.~\eqref{eq:gluing_axiom_general} since both $G$ and $C$ are 2-valent, so the left vertical arrow in Eq.~\eqref{eq:gluing_axiom_general} should be $\calm \otimes \calm$ instead of $\calm$.
\item For every link $L$ of a fusion vertex, let $G$ be a \emph{fusion 0-surgery} by gluing two fusion vertices whose links are $L$ and $\bar L$ (the orientation-reversed copy of $L$),
\begin{equation}
\label{eq:fusion_0surgery}
\cone(L)\times S_0
\rightarrow
L\times B_1\;.
\end{equation}
Here, $\cone(L)$ is $L\times [0,1]$ with $L\times 0$ identified with a single point, i.e., $\cone(L)$ consists of a small 3-ball around the fusion vertex. The according $C$ consists of 1) projecting onto the space where pairs of adjacent ribbons carry the same label, and 2) contracting the two fusion indices. Considering only the ribbon networks and not the 3-manifold topology, the gluing axiom is given by, e.g., for $L$ a 2-sphere with four points,
\begin{equation}
\sum_\alpha
\begin{tikzpicture}
\atoms{vertex}{0/lab={t=$\alpha$, p=0:0.25}, {1/p={1,0}, lab={t=$\alpha$,p=180:0.25}}}
\draw (0)edge[mark=arr, mark={slab=$a$,r,p=0.7}, mark={a, three dots}]++(110:0.8);
\draw (0)edge[mark=arr, mark={slab=$b$,r,p=0.7}, mark={a, three dots}]++(150:0.8);
\draw (0)edge[mark={arr,-}, mark={slab=$c$,p=0.7}, mark={a, three dots}]++(-150:0.8);
\draw (0)edge[mark=arr, mark={slab=$d$,p=0.7}, mark={a, three dots}]++(-110:0.8);
\draw (1)edge[mark={arr,-}, mark={slab=$a$,p=0.7}, mark={a, three dots}]++(70:0.8);
\draw (1)edge[mark={arr,-}, mark={slab=$b$,p=0.7}, mark={a, three dots}]++(30:0.8);
\draw (1)edge[mark=arr, mark={slab=$c$,r,p=0.7}, mark={a, three dots}]++(-30:0.8);
\draw (1)edge[mark={arr,-}, mark={slab=$d$,r,p=0.7}, mark={a, three dots}]++(-70:0.8);
\end{tikzpicture}
=
\begin{tikzpicture}
\draw (-1,0.8)edge[mark={a,three dots}, mark={b,three dots}, mark={slab=$a$}, mark={arr,-}, bend right=30](1,0.8);
\draw (-1,0.3)edge[mark={a,three dots}, mark={b,three dots}, mark={slab=$b$}, mark={arr,-}, bend right=20](1,0.3);
\draw (-1,-0.8)edge[mark={a,three dots}, mark={b,three dots}, mark={slab=$d$,r}, mark={arr,-}, bend left=30](1,-0.8);
\draw (-1,-0.3)edge[mark={a,three dots}, mark={b,three dots}, mark={slab=$c$,r}, mark={arr}, bend left=20](1,-0.3);
\end{tikzpicture}\;.
\end{equation}
\item $G$ can be a \emph{loop 1-surgery},
\begin{equation}
\label{eq:loop_1surgery}
\cone(S_1)\times S_1\rightarrow S_1\times B_2\;.
\end{equation}
Note that $\cone(S_1)$ is a $2$-ball with a single oriented, framed point in the center, which becomes a ribbon loop through the product with $S_1$. In other words, we remove the solid 3-torus neighborhood of a ribbon loop and replace it by its complementary solid 3-torus inside a 3-sphere, or yet in other words, we cut out the solid torus and paste it again after applying an $S$ modular transformation to the boundary. The according $C$ is given by summing over all ribbon labels $a$ weighted by $d_a/D$. Looking at the ribbon networks only, we have,
\begin{equation}
\sum_a \frac{d_a}{D}
\begin{tikzpicture}
\draw[mark=arr,mark={slab=$a$,p=0.25,r}] (0,0)arc(0:360:0.4);
\end{tikzpicture}
=\hspace{1cm}\;.
\end{equation}
\item $G$ is a \emph{plain 0-surgery} without fusion vertices,
\begin{equation}
\label{eq:0surgery_move}
B_3\times S_0 \rightarrow S_2\times B_1\;.
\end{equation}
$C$ consists in multiplying the overall tensor with the total quantum dimension $D$. Since $C$ is invertible, the axiom might also be applied backwards, in which case $G^{-1}$ is a 2-surgery. It might be viewed as a special case of fusion 0-surgery, if we are allowed to insert trivial fusion vertices with no adjacent ribbons at a cost of $D^{1/2}$. Physically, this axiom imposes that the ground state degeneracy on the $n-1$-sphere is $1$, which explicitly rules out symmetry-breaking phases.
\item Since we are doing quantum physics, the UMTC has to obey the following \emph{unitarity} condition.\footnote{A better name for this might be \emph{Hermiticity}, since our spacetimes have an imaginary and not a real time component.} $G$ is given by orientation reversal, and $C$ is given by complex conjugation.
\end{itemize}

Note that changing all UMTC tensors by applying a unitary map $(U^{ab}_c)_\alpha^\beta$ to every fusion index with adjacent ribbon labels $a$, $b$, and $c$,
\begin{equation}
\begin{tikzpicture}
\atoms{vertex}{0/lab={t=$\beta$,p=90:0.25}}
\draw (0)edge[mark={arr,-},mark={slab=$a$},mark={a,three dots}]++(150:0.8) (0)edge[mark={arr,-},mark={slab=$b$,r},mark={a,three dots}]++(30:0.8) (0)edge[mark={arr},mark={slab=$c$},mark={a,three dots}]++(-90:0.8);
\end{tikzpicture}
\coloneqq
\sum_\alpha
(U^{ab}_c)_\alpha^\beta
\begin{tikzpicture}
\atoms{vertex}{0/lab={t=$\alpha$,p=90:0.25}}
\draw (0)edge[mark={arr,-},mark={slab=$a$},mark={a,three dots}]++(150:0.8) (0)edge[mark={arr,-},mark={slab=$b$,r},mark={a,three dots}]++(30:0.8) (0)edge[mark={arr},mark={slab=$c$},mark={a,three dots}]++(-90:0.8);
\end{tikzpicture}\;.
\end{equation}
Such a transformation is sometimes called a \emph{gauge transformation}, and \emph{gauge equivalent} UMTCs describe the same topological phases.

Let us quickly sketch the relation of this definition to how UMTCs are usually presented in the physics literature. For more details we refer the reader to Appendix~\ref{sec:umtc_appendix}. In the common presentation of UMTCs, we assign linear operators to deformations of string diagrams, such as the \emph{$F$-matrix} to
\begin{equation}
\label{eq:umtc_fmove}
\begin{tikzpicture}
\atoms{vertex}{{0/lab={t=$\gamma$,p=-150:0.25}}, {1/p={-0.5,0.5},lab={t=$\delta$,p=-150:0.25}}}
\draw (1)edge[mark=arr,mark={slab=$f$}](0) (0)edge[mark=arr, mark={lab=$d$, a}]++(0,-0.5) (1)edge[mark={arr,-}, mark={lab=$a$, a}]++(135:0.5) (1)edge[mark={arr,-}, mark={lab=$b$, a}]++(45:0.5) (0)edge[mark={arr,-}, mark={lab=$c$, a}]++(45:0.5);
\end{tikzpicture}
=
\sum_{e,\alpha,\beta}
(F^{abc}_d)_{e\alpha\beta}^{f\gamma\delta}\quad
\begin{tikzpicture}
\atoms{vertex}{{0/lab={t=$\alpha$,p=-30:0.25}}, {1/p={0.5,0.5},lab={t=$\beta$,p=-30:0.25}}}
\draw (1)edge[mark=arr,mark={slab=$e$}](0) (0)edge[mark=arr, mark={lab=$d$, a}]++(0,-0.5) (1)edge[mark={arr,-}, mark={lab=$b$, a}]++(135:0.5) (1)edge[mark={arr,-}, mark={lab=$c$, a}]++(45:0.5) (0)edge[mark={arr,-}, mark={lab=$a$, a}]++(135:0.5);
\end{tikzpicture}\;.
\end{equation}
In our language, we connect open ribbons on the left and right side (one of them orientation-reversed), and obtain a single ribbon diagram to which we assign $F$ (or other linear operators) as a tensor,
\begin{equation}
\label{eq:umtc_ftensor}
(F^{abc}_d)_{e\alpha\beta}^{f\gamma\delta}
=
\calm(
\begin{tikzpicture}
\atoms{vertex}{{0/lab={t=$\alpha$,p=-150:0.25}}, {1/p={1,1},lab={t=$\beta$,p=-90:0.25}}, {2/p={0,1}, lab={t=$\delta$,p=-90:0.25}}, {3/p={1,0},lab={t=$\gamma$,p=-30:0.25}}}
\draw (1)edge[mark={arr,p=0.7,-},mark={slab=$e$,p=0.7}](0) (2)edge[front,mark={arr,p=0.3},mark={slab=$f$,p=0.3}](3) (1)edge[looseness=2,in=45,out=135,mark={arr},mark={slab=$b$,r}](2) (1)edge[looseness=2,in=45,out=45,mark={arr,-},mark={slab=$c$}](3) (0)edge[looseness=2,in=135,out=135,mark={arr},mark={slab=$a$}](2) (0)edge[looseness=2,in=-90,out=-90,mark={arr,-},mark={slab=$d$,r}](3);
\end{tikzpicture}
)\;.
\end{equation}
Vice versa, the $F$-move can be performed by simply gluing some vertices of the $F$ ribbon manifold to the corresponding section of another ribbon manifold,
\begin{equation}
\begin{tikzpicture}
\atoms{vertex}{{0/}, {1/p={-0.5,0.5}}}
\draw (1)edge[mark=arr](0) (0)edge[mark=arr]++(0,-0.5) (1)edge[mark={arr,-}]++(135:0.5) (1)edge[mark={arr,-}]++(45:0.5) (0)edge[mark={arr,-}]++(45:0.5);
\end{tikzpicture}
=
\begin{tikzpicture}
\atoms{vertex}{{0/}, {1/p={1,1}}, {2/p={0,1}}, {3/p={1,0}}}
\draw (1)edge[mark={arr,p=0.7,-}](0) (2)edge[front,mark={arr,p=0.3}](3) (1)edge[looseness=2,in=45,out=135,mark={arr}](2) (1)edge[looseness=2,in=45,out=45,mark={arr,-}](3) (0)edge[looseness=2,in=135,out=135,mark={arr}](2) (0)edge[looseness=2,in=-90,out=-90,mark={arr,-}](3);
\atoms{vertex}{{0x/p={1.7,0.3}}, {1x/p={2.2,0.8}}}
\draw (1x)edge[mark=arr](0x) (0x)edge[mark=arr]++(0,-0.5) (1x)edge[mark={arr,-}]++(135:0.5) (1x)edge[mark={arr,-}]++(45:0.5) (0x)edge[mark={arr,-}]++(135:0.5);
\draw[glue] (0)to[out=0,in=180](0x) (1)to[out=0,in=180](1x);
\atoms{glueedge}{x/p={0.7,0.7}}
\end{tikzpicture}
\;.
\end{equation}
The right-hand side is the disjoint union of the 3-sphere ribbon manifold in Eq.~\eqref{eq:umtc_ftensor}, together with an arbitrary ribbon manifold $X$ which contains a section that looks like the right-hand side of Eq.~\eqref{eq:umtc_fmove}. We then perform the fusion 0-surgery in Eq.~\eqref{eq:fusion_0surgery} to two pairs of vertices as indicated by the red semi-transparent thick dashed lines, and then the loop 1-surgery in Eq.~\eqref{eq:loop_1surgery} as indicated by the semi-transparent red circle on an edge that is part of the loop. This results in changing the ribbon network of $X$ as in Eq.~\eqref{eq:umtc_fmove} without changing the 3-manifold topology of $X$.

Note that there can be different ``moves'' of string diagrams like Eq.~\eqref{eq:umtc_fmove} that yield the same ribbon manifold after connecting the open ends of the two sides. An advantage of our presentation is that those moves automatically correspond to the same amplitudes, whereas this would have to be ensured by additional axioms in the conventional formulation. In general, a tensor-based formulation seems to be more natural than an operator-based formulation since we are in Euclidean spacetime where there is no notion of time direction. Our closed-up language also makes it much easier to see the precise topological/geometric interpretation to UMTCs, and in particular the important role of the embedding 3-manifold topology, which is at the heart of what \emph{Reshetikhin-Turaev TQFT} \cite{Reshetikhin1991} is about.

\subsection{The state-sum path integral}
\label{sec:CYWW_path_integral}
We will now define the modular CYWW model for a UMTC $M$ as a state-sum/tensor-network path integral in $3+1$ dimensions.

Consider a $3$-manifold cellulation $C$. Let $R(C)$ be the 3-manifold equipped with the Poincar\'e dual 1-skeleton
\footnote{By this we mean the 1-skeleton of the cellulation Poincar\'e dual to $C$. Also, by ``dual $i$-cell'', we will refer to an $i$-cell in the dual cellulation, not an $n-i$-cell dual to an $i$-cell of the cellulation.}
of $C$ as a ribbon network. That is, $R(C)$ has one fusion vertex in the center of every 3-cell of $C$, and the fusion vertices of neighboring 3-cells are connected by a ribbon intersecting the 2-cell separating them. We assume that each 2-cell and 3-cell is identified with a canonical representative, which we will call a \emph{branching}. The 2-cell identification is needed to fix the framing of the ribbon perpendicular to it, and the 3-cell identification is needed for the framing of the contained fusion vertex. E.g., for a triangulation, the branching can be determined by what is commonly known as a \emph{branching structure} and consists of a choice of edge directions acyclic at every triangle. Depending on the branching (structure) we can pick a convention to ``flatten'' the fusion vertex in the center of a 3-cell, such as for a tetrahedron,
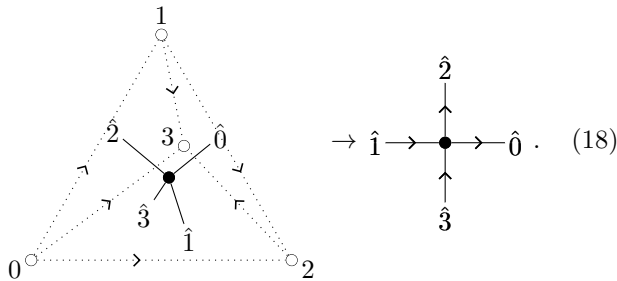
\begin{equation}
\label{eq:bulk_fusion_vertex_ordering}
\begin{tikzpicture}
\atoms{cellvertex}{{0/p={-150:2},lab={t=$0$,p=-150:0.25}}, {2/p={-30:2},lab={t=$2$,p=-30:0.25}}, {1/p={90:2},lab={t=$1$,p=90:0.25}}, {3/p={60:0.6},lab={t=$3$,p=150:0.25}}}
\draw[cellulation] (0)edge[mark={arr,p=0.4}](1) (0)edge[mark={arr,p=0.4}](2) (0)edge[mark=arr](3) (1)edge[mark={arr,p=.6}](2) (2)edge[mark=arr](3) (1)edge[mark=arr](3);
\atoms{vertex}{x/p={0.1,0.1}}
\draw (x)edge[mark={lab=$\hat 1$,a}](-60:0.6) (x)edge[mark={lab=$\hat 0$,a}](40:0.85) (x)edge[mark={lab=$\hat 2$,a}](130:0.8) (x)edge[mark={lab=$\hat 3$,a}](-0.1,-0.2);
\end{tikzpicture}
\rightarrow
\begin{tikzpicture}
\atoms{vertex}{0/}
\draw (0)edge[mark={lab=$\hat 0$,a},mark=arr]++(0:0.8) (0)edge[mark={lab=$\hat 2$,a},mark=arr]++(90:0.8) (0)edge[mark={lab=$\hat 1$,a},mark={arr,-}]++(180:0.8) (0)edge[mark={lab=$\hat 3$,a},mark={arr,-}]++(-90:0.8);
\end{tikzpicture}
\;.
\end{equation}
Here, and in the following, we will draw an underlying cellulation using dotted lines and ``hollow'' vertices. If the branching is orientation reversed, then so is the link of the fusion vertex. E.g., the boundary of a 4-simplex is a simple cellulation of a 3-sphere by branching-structure tetrahedra, to which we associate the following ribbon manifold,
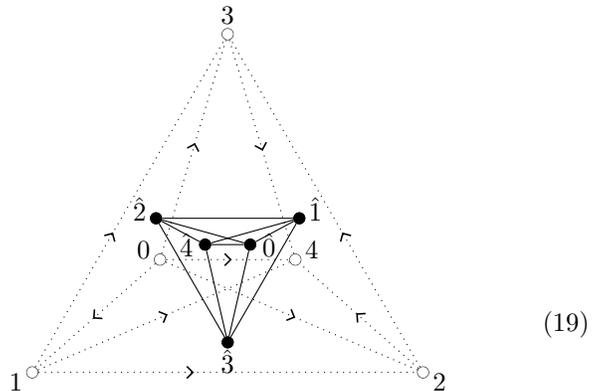
\begin{equation}
\label{eq:4simplex_ribbon_manifold}
\begin{multlined}
\begin{tikzpicture}
\atoms{cellvertex}{{0/p={-0.9,0},lab={t=$0$,p=150:0.25}}, {1/p={-150:3},lab={t=$1$,p=-150:0.25}}, {2/p={-30:3},lab={t=$2$,p=-30:0.25}}, {3/p={90:3},lab={t=$3$,p=90:0.25}}, {4/p={0.9,0},lab={t=$4$,p=30:0.25}}}
\atoms{vertex}{{0x/p={0.3,0.2},lab={t=$\hat 0$,p=0:0.25}}, {1x/p={30:1.1},lab={t=$\hat 1$,p=30:0.25}}, {2x/p={150:1.1},lab={t=$\hat 2$,p=150:0.25}}, {3x/p={-90:1.1},lab={t=$\hat 3$,p=-90:0.25}}, {4x/p={-0.3,0.2},lab={t=$\hat 4$,p=180:0.25}}}
\draw[cellulation] (0)edge[mark=arr](1) (0)edge[mark=arr](2) (0)edge[mark=arr](3) (1)edge[mark={arr,p=0.4}](2) (2)edge[mark={arr,p=0.4}](3) (1)edge[mark={arr,p=0.4}](3) (0)edge[mark=arr](4) (1)edge[mark=arr](4) (2)edge[mark=arr](4) (3)edge[mark=arr](4);
\draw (0x)--(1x)--(2x)--(3x)--(4x)--(0x) (0x)--(2x)--(4x)--(1x)--(3x)--(0x);
\end{tikzpicture}\\
\rightarrow
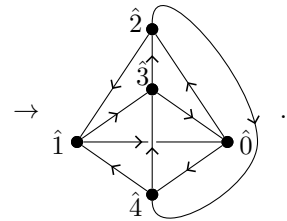
\begin{tikzpicture}
\clip (-0.4,-1.2) rectangle (2.6,2);
\atoms{vertex}{{0/p={2,0},lab={t=$\hat 0$,p=0:0.25}}, 1/lab={t=$\hat 1$,p=180:0.25}, {2/p={1,1.5},lab={t=$\hat 2$,p=150:0.25}}, {3/p={1,0.7},lab={t=$\hat 3$,p=120:0.25}}, {4/p={1,-0.7},lab={t=$\hat 4$,p=-150:0.25}}}
\draw (0)edge[mark={arr,-,p=0.6}](1) (1)edge[mark={arr,-}](2) (0)edge[mark=arr](2) (0)edge[mark={arr,-}](3) (1)edge[mark=arr](3) (2)edge[mark={arr,-}](3) (0)edge[mark=arr](4) (1)edge[mark={arr,-}](4) (3)edge[mark={arr,-,p=0.6},front](4);
\draw[mark=arr] (2)to[out=90,in=90](2.4,0)to[out=-90,in=-90](4);
\end{tikzpicture}\;.
\end{multlined}
\end{equation}

Now, the CYWW model is a state-sum that associates to each 4-cell $X$ the tensor
\begin{equation}
\calm(C(\partial X))\;,
\end{equation}
i.e., the tensor that $\calm$ assigns to the boundary 3-cellulation of $X$ with the Poincar\'e dual ribbon network. E.g., for a triangulation of a 4-manifold, we would assign the tensor for Eq.~\eqref{eq:4simplex_ribbon_manifold} to every 4-simplex.

The indices of those tensors are contracted as follows. At every 3-cell, there is a pair of fusion indices with opposite-orientation links from the tensors at the adjacent 4-cells, which are contracted. At each 2-cell, there are ribbon labels $a_1,a_2,\ldots$ from the tensors at the adjacent 4-cells. Those are contracted using a $\delta$-tensor (i.e., first projected onto the subspace where they all take the same value, and then summed over), weighted the quantum dimension,
\begin{equation}
\frac{d_{a_1}}{D} \delta_{a_1,a_2,\ldots}=
\frac{d_{a_1}}{D} \cdot
\begin{cases}
1 & \text{if}\quad a_1=a_2=\ldots\\
0 & \text{otherwise}
\end{cases}\;.
\end{equation}
In addition, there is a normalization of $\frac1D$ at every edge, and $D$ at every vertex.

\subsection{Evaluation of the path integral}
\label{sec:CYWW_evaluation}
Let us evaluate the state-sum within a cellulation $Y$ of a 4-manifold with (space) boundary. We will find that the evaluation is given by $\calm(R(\partial Y))$, where $\partial Y$ is the 3-cellulation of the (space) boundary of $Y$. This can be shown very directly using the gluing axioms, which imply that performing the contractions of the state-sum tensors is equivalent to gluing operations of the corresponding ribbon manifolds. The evaluation can be divided into the following steps.
\begin{enumerate}
\item We start with one tensor $\calm(R(\partial X))$ at every 4-cell $X$ of $Y$. Using the disjoint union axiom, we have
\begin{equation}
\bigotimes_{X\in C_4[Y]}\calm(R(\partial X)) = \calm(R(\bigcup_{X\in C_4[Y]} \partial X))\eqqcolon \calm(R_0)\;,
\end{equation}
writing $C_i[Y]$ for the set of $i$-cells of $Y$. We notice that as a 3-manifold, $R_0$ is just the boundary of the neighborhood of the interior dual 0-skeleton of $Y$, a collection of 3-spheres.
\item At each interior 3-cell of $Y$, we contract a pair of fusion indices. Calling this contraction $C$, we have
\begin{equation}
C(\calm(R_0))=\calm(G(R_0))\eqqcolon \calm(R_1)\;,
\end{equation}
where $G$ is the fusion $0$-surgery in Eq.~\eqref{eq:fusion_0surgery} between every pair of fusion vertices. As a 3-manifold, $R_1$ is the boundary of the neighborhood of the dual 1-skeleton of $Y$. At $\partial Y$, the ribbon network of $R_1$ is the dual 1-skeleton of the 3-cellulation $\partial Y$. At each interior 2-cell of $Y$, $R_1$ contains a ribbon around the corresponding non-contractible loop.
\item At every interior 2-cell of $Y$, we sum over the surrounding ribbon labels $a$ weighted by $d_a/D$. Calling this summation $C$, we have
\begin{equation}
C(\calm(R_1))=\calm(G(R_1))\eqqcolon \calm(R_2)\;,
\end{equation}
where $G$ is the loop 1-surgery in Eq.~\eqref{eq:loop_1surgery} around every ribbon loop of $R_1$. Thus, $R_2$ as a 3-manifold which is the boundary of the the neighborhood of the dual 2-skeleton of $Y$. There are no ribbons in the bulk of $Y$, and the ribbon network at $\partial Y$ is the same as for $R_1$.
\item At every interior 1-cell, there is a normalization of $1/D$. Writing $C$ for this normalization, we have
\begin{equation}
C(\calm(R_2))=\calm(G(R_2))\eqqcolon \calm(R_3)\;,
\end{equation}
where $G$ is the \emph{backwards} 0-surgery (which is a 2-surgery) in Eq.~\eqref{eq:0surgery_move} at all 1-cells. $R_3$ contains one disconnected ribbonless 3-sphere for every interior vertex of $Y$, which can be canceled by the weight of $D$ at every vertex, yielding $R_4$. $R_4$ is the same as $R(\partial Y)$, and thus we find that the evaluation of the CYWW state-sum on $Y$ is given by $\calm(R(\partial Y))$.
\end{enumerate}

Since the evaluation of the CYWW model is independent of the bulk cellulation, the CYWW model is re-cellulation invariant, i.e., a topological state-sum fixed-point model. Since it also does not depend on the topology of the bulk, it further follows that the (modular) CYWW model is invertible.

\subsection{Ground state and Hamiltonian}
Since the modular CYWW model is invertible, it has a unique ground state on any space manifold/cellulation. This ground state on a 3-cellulation $X$ is simply given by $\calm(R(X))$. More precisely, we need to multiply by a factor of $(d_i/D)^{1/2}$ at each 2-cell, where $i$ is the label of the ribbon perpendicular to the 2-cell, if we want the state to be the ground state of a Hermitian Hamiltonian.

Any state-sum construction of a simple enough type
\footnote{This includes all established state-sum constructions, but not those suggested in Ref.~\cite{universal_liquid}.}
yields a commuting-projector Hamiltonian by considering the state-sum on diamond-shaped cells. On a 3-cellulation, the standard prescription would result in one Hamiltonian term at every vertex involving the degrees of freedom on the 3-cells containing that vertex. However, the CYWW model has a very specific structure that allows for a simpler set of local projectors, one at every edge/dual face. Namely, we take the ribbon loop around the dual face together with ``loose'' ribbon segments connected to it, and combine those loose ribbon ends with an orientation-reversed copy, e.g.,
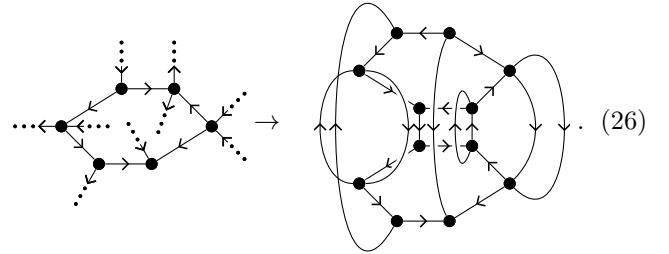
\begin{equation}
\begin{tikzpicture}
\atoms{vertex}{0/, 1/p={0.5,-0.5}, 2/p={1.2,-0.5}, 3/p={2,0}, 4/p={1.5,0.5}, 5/p={0.8,0.5}}
\draw (0)edge[mark=arr](1) (1)edge[mark=arr](2) (2)edge[mark={arr,-}](3) (3)edge[mark=arr](4) (4)edge[mark={arr,-}](5) (5)edge[mark=arr](0);
\draw (0)edge[mark={three dots,a},mark={arr,e}]++(180:0.3) (0)edge[mark={three dots,a},mark={arr,e,-}]++(0:0.3) (1)edge[mark={three dots,a},mark={arr,e}]++(-120:0.3) (2)edge[mark={three dots,a},mark={arr,e,-}]++(120:0.3) (3)edge[mark={three dots,a},mark={arr,e,-}]++(-45:0.3) (3)edge[mark={three dots,a},mark={arr,e,-}]++(45:0.3) (4)edge[mark={three dots,a},mark={arr,e}]++(90:0.3) (4)edge[mark={three dots,a},mark={arr,e}]++(-110:0.3) (5)edge[mark={three dots,a},mark={arr,e,-}]++(90:0.3);
\end{tikzpicture}
\rightarrow\quad
\begin{tikzpicture}
\clip (-0.6,-1) rectangle(2.8,2.6);
\atoms{vertex}{0/, 1/p={0.5,-0.5}, 2/p={1.2,-0.5}, 3/p={2,0}, 4/p={1.5,0.5}, 5/p={0.8,0.5}}
\atoms{vertex}{0x/p={0,1.5}, 1x/p={0.5,2}, 2x/p={1.2,2}, 3x/p={2,1.5}, 4x/p={1.5,1}, 5x/p={0.8,1}}
\draw (0)edge[mark=arr](1) (1)edge[mark=arr](2) (2)edge[mark={arr,-}](3) (3)edge[mark=arr](4) (4)edge[mark={arr,-}](5) (5)edge[mark=arr](0);
\draw (0x)edge[mark={arr,-}](1x) (1x)edge[mark={arr,-}](2x) (2x)edge[mark={arr}](3x) (3x)edge[mark={arr,-}](4x) (4x)edge[mark={arr}](5x) (5x)edge[mark={arr,-}](0x);
\draw (0)edge[mark={arr},out=180,in=180](0x) (0)edge[mark={arr,-},out=0,in=0,looseness=1.3,front](0x) (1)edge[mark={arr},out=-120,in=120,looseness=2](1x) (2)edge[mark={arr,-},out=120,in=-120,looseness=0.5,front](2x) (3)edge[mark={arr,-},out=-45,in=45,looseness=2](3x) (3)edge[mark={arr,-},out=45,in=-45](3x) (4)edge[mark={arr},out=90,in=-90](4x) (4)edge[mark={arr},out=-110,in=110,looseness=3,front](4x) (5)edge[mark={arr,-},out=90,in=-90](5x);
\end{tikzpicture}\;.
\end{equation}
The according tensor can be interpreted as a linear operator from the ribbon labels and fusion indices at the bottom to those at the top, parametrically depending on the labels of the ribbons connecting bottom and top. If we normalize each bottom and top ribbon by $(d_i/D)^{1/2}$, then it follows directly from the gluing axioms that this linear operator is a projector. It can also be easily seen that the gluing axioms imply that projectors at different dual 2-cells commute. Furthermore, the unitarity condition implies that the projector is Hermitian. The Hamiltonian is just the sum of the inverses of all such local projectors.

\section{Invertible boundaries for Drinfeld centers}
\label{sec:drinfeld_center_boundary}
In this section we will define an invertible boundary for modular CYWW models whose input UMTC is a Drinfeld center. In this (and only this) case (c.f.~Appendix~\ref{sec:tqft_umtc_module}), there exists an extension of the UMTC to a categorical structure that also associates amplitudes to manifolds with boundary, and which we will refer to as a \emph{Lagrangian module}. We will first give a TQFT-style definition of those Lagrangian modules, and then use them to define the invertible boundary. We will formulate the latter first as a state-sum, then in terms of ground states/Hamiltonians, and finally as a generalized local unitary circuit disentangling the modular CYWW model.
\subsection{Lagrangian modules}
\label{sec:tensorial_umtc_module}
If a topological model in $2+1$ dimensions has a topological boundary, then there can also be anyons traveling inside the boundary. We can consider spacetime histories containing both bulk and boundary anyons as well as their fusion events. An assignment of amplitudes to all those histories yields what is known as a \emph{module category} of the UMTC. Imposing all the gluing axioms restricts that module category further to one that arises from condensing a \emph{Lagrangian commutative Frobenius algebra object inside the UMTC}, and thus we will refer to it as a \emph{Lagrangian module}. We will here again resort to a TQFT-style definition via gluing axioms, and comment more on the relation to existing categorical notions in Appendix~\ref{sec:tqft_umtc_module}.

We define a \emph{boundary ribbon manifold} as a 3-manifold with boundary, and a network of ribbons in the bulk as well as a network of \emph{boundary ribbons} within the boundary. In addition to the fusion vertices in the bulk, there are \emph{boundary fusion vertices} where bulk- as well as boundary ribbons meet. Like any other submanifold, the boundary fusion vertices are framed, i.e., the infinitesimal ``sphere'' around each of them is identified with its \emph{link}, a disk with points in both bulk and boundary. When we draw a boundary ribbon manifold, we again project it to 2 dimensions. We will draw boundary ribbons and vertices in blue, and think of the bulk as being ``behind'' the boundary, e.g.,
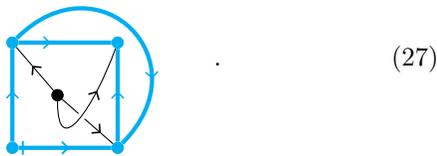
\begin{equation}
\begin{tikzpicture}
\atoms{bdvertex}{0/p={-0.7,-0.7}, 1/p={0.7,-0.7}, 2/p={-0.7,0.7}, 3/p={0.7,0.7}}
\atoms{vertex}{4/p={-0.1,0}}
\draw (4)edge[mark={arr,p=0.7}](1) (4)edge[mark={arr}](2) (4)edge[out=-90,in=-110,front,looseness=2,mark={arr,p=0.6}](3);
\draw[bdribbon] (0)edge[mark=arr](2) (1)edge[mark=arr](3) (0)edge[mark=arr,fav=s](1) (2)edge[mark={arr,p=0.3}](3) (2)edge[mark={arr,p=0.7},out=45,in=45,looseness=2](1);
\end{tikzpicture}\;.
\end{equation}

A Lagrangian module $\calb$ (containing a UMTC $\calm$) is a map that associates to every boundary ribbon manifold a tensor, with one label for each boundary ribbon and one index for each boundary fusion vertex (in addition to the UMTC labeling in the bulk). Again, the labels are from a fixed finite set (which is different from the set of bulk labels), and the dimension of the boundary fusion indices depends on their link and labeling of the adjacent ribbons. This contains $\calm$ by restricting to manifolds without boundary. A Lagrangian module also contains a quantum dimension $d^{\calb}_i$ for each boundary ribbon label $i$, and like in the bulk, we define
\begin{equation}
D^\calb=(\sum_i {d^\calb_i}^2)^{1/2}\;.
\end{equation}
There are the following gluing axioms (cf.~Eq.~\eqref{eq:gluing_axiom_general}) between gluing operations of the boundary ribbon manifolds and contractions of the tensors, in addition to those for the UMTC.
\begin{itemize}
\item $G$ is disjoint union and $C$ is the tensor product.
\item For each link $L$ of a boundary fusion vertex, $G$ is the \emph{fusion 1-handle attachment} given by gluing two boundary fusion vertices with links $L$ and $\bar L$,
\begin{equation}
\label{eq:fusion_1handle_attachment}
\cone(L)\times S_0 \rightarrow L\times B^s_1\;.
\end{equation}
The according $C$ consists in projecting onto the subspace where the adjacent ribbons carry pairwise equal labels, and contracting the two fusion indices themselves. Looking at the ribbon networks, we have, e.g., for $L$ a disk with two bulk- and two boundary points,
\begin{equation}
\label{eq:1handle_attachment}
\sum_\alpha
\begin{tikzpicture}
\atoms{bdvertex}{0/lab={t=$\alpha$, p=0:0.25}, {1/p={1,0}, lab={t=$\alpha$,p=180:0.25}}}
\draw[bdribbon] (0)edge[mark=arr, mark={slab=$a$,r,p=0.7}, mark={a, three dots}]++(110:0.8);
\draw (0)edge[mark=arr, mark={slab=$b$,r,p=0.7}, mark={a, three dots}]++(150:0.8);
\draw[bdribbon] (0)edge[mark={arr,-}, mark={slab=$c$,p=0.7}, mark={a, three dots}]++(-150:0.8);
\draw (0)edge[mark=arr, mark={slab=$d$,p=0.7}, mark={a, three dots}]++(-110:0.8);
\draw[bdribbon] (1)edge[mark={arr,-}, mark={slab=$a$,p=0.7}, mark={a, three dots}]++(70:0.8);
\draw (1)edge[mark={arr,-}, mark={slab=$b$,p=0.7}, mark={a, three dots}]++(30:0.8);
\draw[bdribbon] (1)edge[mark=arr, mark={slab=$c$,r,p=0.7}, mark={a, three dots}]++(-30:0.8);
\draw (1)edge[mark={arr,-}, mark={slab=$d$,r,p=0.7}, mark={a, three dots}]++(-70:0.8);
\end{tikzpicture}
=
\begin{tikzpicture}
\draw[bdribbon] (-1,0.8)edge[mark={a,three dots}, mark={b,three dots}, mark={slab=$a$}, mark={arr,-}, bend right=30](1,0.8);
\draw (-1,0.3)edge[mark={a,three dots}, mark={b,three dots}, mark={slab=$b$}, mark={arr,-}, bend right=20](1,0.3);
\draw (-1,-0.8)edge[mark={a,three dots}, mark={b,three dots}, mark={slab=$d$,r}, mark={arr}, bend left=30](1,-0.8);
\draw[bdribbon] (-1,-0.3)edge[mark={a,three dots}, mark={b,three dots}, mark={slab=$c$,r}, mark={arr,-}, bend left=20](1,-0.3);
\end{tikzpicture}\;.
\end{equation}
\item $G$ is a \emph{loop 2-handle attachment} which can be applied to the tubular neighborhood of a boundary ribbon loop,
\begin{equation}
\label{eq:loop_2handle_attachment}
\cone(B_1^p) \times S_1
\rightarrow
B^s_2\times B^p_1\;.
\end{equation}
Note that $\cone(B_1^p)$ is $B^{ps}_2$ with a single point inside the physical-boundary 1-ball, generating a boundary ribbon in the product with $S_1$. The according $C$ is given by summing over all labels $a$ for this boundary ribbon, weighted by $d_a^\calb/D^\calb$. On the level of ribbon networks, this looks like
\begin{equation}
\label{eq:2handle_attachment}
\sum_a \frac{d^\calb_a}{D^\calb}
\begin{tikzpicture}
\draw[bdribbon,mark=arr,mark={slab=$a$,p=0.25,r}] (0,0)arc(0:360:0.4);
\end{tikzpicture}
=\hspace{1cm}\;.
\end{equation}
\item $G$ is a \emph{plain $1$-handle attachment} without fusion vertices,
\begin{equation}
\label{eq:boundary_1surgery}
B^{ps}_3\times S_0 \rightarrow B^p_2\times B^s_1\;.
\end{equation}
The according $C$ is given by multiplying the tensor with an overall factor of $D^\calb$. This is a special case of the fusion 1-handle attachment if we can add trivial boundary fusion vertices at a cost of $(D^\calb)^{1/2}$. Physically, this imposes that the ground state degeneracy on the 3-ball is 1, again explicitly ruling out symmetry-breaking at the boundary.
\end{itemize}

\subsection{The invertible-boundary state-sum}
\label{sec:boundary_state_sum}
Let us now describe the state-sum formulation of the invertible boundary for a UMTC $\calm$ and its Lagrangian module $\calb$. Consider a cellulation $X$ of a 3-manifold with (physical) boundary. Let $R(X)$ denote the manifold $X$ with a ribbon network that is the dual 1-skeleton in the bulk, and a boundary ribbon network that is the dual 1-skeleton of the 2-cellulation $\partial X$. I.e., there is a boundary fusion vertex at the center of every boundary 2-cell, where one bulk ribbon and a number of boundary ribbons meet. A very simple 3-cellulation with boundary is given by a single 3-cell. E.g., for a cubic 3-cell $X$ (with dotted lines and hollow vertices), $R(X)$ would look as follows,
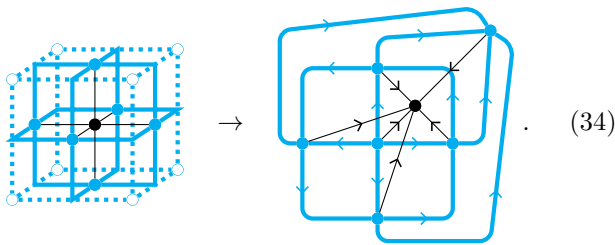
\begin{equation}
\begin{tikzpicture}
\atoms{bdcellvertex}{0/, 1/p={1.6,0}, 2/p={0,1.6}, 3/p={1.6,1.6}, 4/p={0.6,0.4}, 5/p={2.2,0.4}, 6/p={0.6,2}, 7/p={2.2,2}}
\draw[bdcellulation] (0)--(1)--(3)--(2)--(0) (4)--(5)--(7)--(6)--(4) (0)--(4) (1)--(5) (2)--(6) (3)--(7);
\atoms{bdvertex}{x0/p={0.8,0.8}, x1/p={1.1,0.2}, x2/p={1.9,1}, x3/p={1.1,1.8}, x4/p={0.3,1}, x5/p={1.4,1.2}}
\atoms{vertex}{m/p={1.1,1}}
\draw[bdribbon] (x0)--++(0,-0.8)--(x1) (x0)--++(0.8,0)--(x2) (x0)--++(0,0.8)--(x3) (x0)--++(-0.8,0)--(x4) (x1)--++(0.8,0)--(x2) (x2)--++(0,0.8)--(x3) (x3)--++(-0.8,0)--(x4) (x4)--++(0,-0.8)--(x1) (x5)--++(0,-0.8)--(x1) (x5)--++(0.8,0)--(x2) (x5)--++(0,0.8)--(x3) (x5)--++(-0.8,0)--(x4);
\draw (m)--(x0) (m)--(x1) (m)--(x2) (m)--(x3) (m)--(x4) (m)--(x5);
\end{tikzpicture}
\quad\rightarrow\quad
\begin{tikzpicture}
\atoms{bdvertex}{0/, 1/p=-90:1, 2/p=0:1, 3/p=90:1, 4/p=180:1, 5/p={1.5,1.5}}
\atoms{vertex}{m/p={0.5,0.5}}
\draw[bdribbon] (0)edge[mark=arr](1) (0)edge[mark=arr](2) (0)edge[mark=arr](3) (0)edge[mark=arr](4) (1)edge[rc, -|, mark={arr,p=0.3}](2) (2)edge[rc, |-, mark={arr,p=0.3}](3) (3)edge[rc, -|, mark={arr,p=0.3}](4) (4)edge[rc, |-, mark={arr,p=0.3}](1);
\draw[bdribbon,rc,mark=arr] (2)--++(0.4,0)--(5);
\draw[bdribbon,rc,mark=arr] (3)--++(0,0.4)--(5);
\draw[bdribbon,rc,mark=arr] (1)--++(0,-0.3)--++(1.5,0)--(1.8,1.3)--(5);
\draw[bdribbon,rc,mark=arr] (4)--++(-0.3,0)--++(0,1.5)--(1.3,1.8)--(5);
\draw (0)edge[mark=arr](m) (1)edge[mark=arr](m) (2)edge[mark=arr](m) (3)edge[mark=arr](m) (4)edge[mark=arr](m) (5)edge[mark=arr](m);
\end{tikzpicture}\;.
\end{equation}
In order to determine the ordering and arrow directions of the strings adjacent to a boundary fusion vertex (here chosen at random), we need to use the branching, i.e., the identification of the boundary 2-cells with a canonical representative. For a branching-structure triangulation, a possible convention is
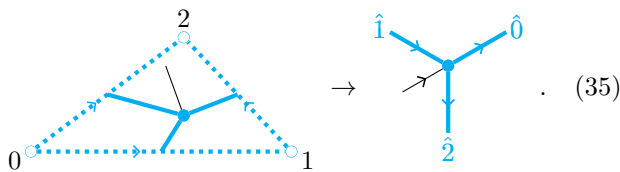
\begin{equation}
\begin{tikzpicture}
\atoms{bdcellvertex}{{0/p={-150:2},lab={t=$0$,p=-150:0.25}}, {2/p={-30:2},lab={t=$1$,p=-30:0.25}}, {3/p={60:0.6},lab={t=$2$,p=90:0.25}}}
\draw[bdcellulation] (0)edge[mark={arr,p=0.4}](2) (0)edge[mark={arr,p=0.4}](3) (2)edge[mark={arr,p=0.4}](3);
\atoms{bdvertex}{x1/p=-60:0.6}
\draw (x1)--++(110:0.7);
\draw[bdribbon] (x1)--($(0)!0.5!(2)$) (x1)--($(0)!0.5!(3)$) (x1)--($(2)!0.5!(3)$);
\end{tikzpicture}
\rightarrow
\begin{tikzpicture}
\atoms{bdvertex}{0/}
\draw (0)edge[mark={arr,-}]++(-150:0.7);
\draw[bdribbon] (0)edge[mark={lab=$\hat 2$,a},mark=arr]++(-90:0.9) (0)edge[mark={lab=$\hat 0$,a},mark=arr]++(30:0.9) (0)edge[mark={lab=$\hat 1$,a},mark={arr,-}]++(150:0.9);
\end{tikzpicture}\;.
\end{equation}
That is, we complete the triangle to a tetrahedron with a ``$-1$-vertex'' and then use the same convention as for the bulk fusion vertices of this tetrahedron in Eq.~\eqref{eq:bulk_fusion_vertex_ordering}. If we do the same for the ribbon diagram associated to the 4-simplex, we get a copy of the bulk ribbon network in Eq.~\eqref{eq:4simplex_ribbon_manifold} where we replace some of the bulk ribbons by boundary ribbons,
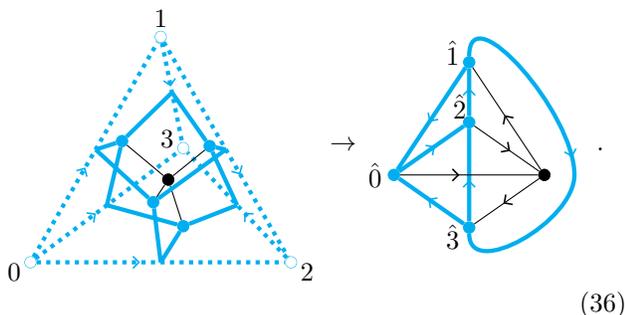
\begin{equation}
\label{eq:boundary_tetrahedron_tensor}
\begin{tikzpicture}
\atoms{bdcellvertex}{{0/p={-150:2},lab={t=$0$,p=-150:0.25}}, {2/p={-30:2},lab={t=$2$,p=-30:0.25}}, {1/p={90:2},lab={t=$1$,p=90:0.25}}, {3/p={60:0.6},lab={t=$3$,p=150:0.25}}}
\draw[bdcellulation] (0)edge[mark={arr,p=0.4}](1) (0)edge[mark={arr,p=0.4}](2) (0)edge[mark={arr,p=0.4}](3) (1)edge[mark={arr,p=.6}](2) (2)edge[mark={arr,p=0.4}](3) (1)edge[mark={arr,p=0.4}](3);
\atoms{vertex}{x/p={0.1,0.1}}
\atoms{bdvertex}{x1/p=-60:0.6, x0/p=40:0.85, x2/p=130:0.8, x3/p={-0.1,-0.2}}
\draw (x)--(x1) (x)--(x2) (x)--(x3) (x)--(x0);
\draw[bdribbon] (x1)--($(0)!0.5!(2)$) (x1)--($(0)!0.5!(3)$) (x1)--($(2)!0.5!(3)$) (x2)--($(0)!0.5!(1)$) (x2)--($(0)!0.5!(3)$) (x2)--($(1)!0.5!(3)$) (x0)--($(1)!0.5!(2)$) (x0)--($(1)!0.5!(3)$) (x0)--($(2)!0.5!(3)$) (x3)--($(0)!0.5!(1)$) (x3)--($(0)!0.5!(2)$) (x3)--($(1)!0.5!(2)$);
\end{tikzpicture}
\rightarrow
\begin{tikzpicture}
\clip (-0.4,-1.2) rectangle (2.6,2);
\atoms{bdvertex}{1/lab={t=$\hat 0$,p=180:0.25}, {2/p={1,1.5},lab={t=$\hat 1$,p=150:0.25}}, {3/p={1,0.7},lab={t=$\hat 2$,p=120:0.25}}, {4/p={1,-0.7},lab={t=$\hat 3$,p=-150:0.25}}}
\atoms{vertex}{{0/p={2,0}}}
\draw[bdribbon] (1)edge[mark={arr,-}](2) (1)edge[mark=arr](3) (2)edge[mark={arr,-}](3) (1)edge[mark={arr,-}](4) (3)edge[mark={arr,-,p=0.6},front](4);
\draw[mark=arr,bdribbon] (2)to[out=90,in=90](2.4,0)to[out=-90,in=-90](4);
\draw (0)edge[mark=arr](2) (0)edge[mark={arr,-}](3) (0)edge[mark={arr,-,p=0.6}](1) (0)edge[mark=arr](4);
\end{tikzpicture}\;.
\end{equation}

The invertible-boundary state-sum simply associates to each boundary 3-cell $X$ the tensor $\calb(R(X))$. E.g., in a branching-structure triangulation, we would have the tensor in Eq.~\eqref{eq:boundary_tetrahedron_tensor} at every boundary tetrahedron.
For each boundary $2$-cell, the two adjacent boundary fusion indices are contracted. At each boundary 3-cell, the central fusion index is contracted with a fusion index of the tensor at the adjacent bulk 4-cell. At each boundary 1-cell, the boundary ribbon labels of the surrounding boundary 3-cell tensors are contracted with a $\delta$-tensor, weighted by $d_a^\calb/D^\calb$. At each boundary 0-cell, we multiply by $D^\calb$.

\subsection{Evaluation of the path integral}
\label{sec:cyww_boundary_evaluation}
Let us now evaluate the state-sum on an arbitrary 4-manifold cellulation $Y$ with physical boundary $X$. The space boundaries of $Y$ and $X$ will be denoted $\partial Y$ and $\partial X$, such that $\partial X$ is the physical boundary of $\partial Y$. We will find that the evaluation is given by $\calb(R(\partial Y))$. To show this, we perform the contractions of the state-sum and use the fact that they correspond to gluing operations via the gluing axioms, completely analogous to the procedure for the bulk in Section~\ref{sec:CYWW_path_integral}. Concretely, the steps are as follows.
\begin{itemize}
\item To begin with, we evaluate the state-sum on all bulk 4-cells as in Section~\ref{sec:CYWW_evaluation}, obtaining
\begin{equation}
\calm(R_4)=\calm(R(\partial Y\cup X))\;,
\end{equation}
noting that $\partial Y\cup X$ is the full boundary of $Y$ consisting of the physical and space part.
\item Then at every 3-cell of $X$, take the tensor product with the corresponding state-sum tensor and contract its bulk fusion index with that at the adjacent bulk 4-cell of $Y$. Calling this $C$, we have
\begin{equation}
C(\calb(R_4))=\calb(G(R_4))\eqqcolon \calb(R_5)\;,
\end{equation}
where $G$ is the disjoint union with 3-ball ribbon manifolds followed by gluing them to $R_4$ via Eq.~\eqref{eq:fusion_0surgery}. The resulting $R_5$ is $R_4$ where the neighborhood of the fusion vertex at the center of every 3-cell of $X$ has been removed, and replaced by a physical-boundary 2-sphere whose boundary ribbon network is the dual 1-skeleton of the 3-cell boundary. $R_5$ also has one bulk ribbon through every 2-cell of $X$, connecting the boundary fusion vertices at the two holes on both sides.
\item Next, at each interior 2-cell of $X$, we contract two boundary fusion indices of the neighboring boundary 3-cells, and we sum over the bulk ribbon labels at this 2-cell. Denoting this contraction by $C$, we have
\begin{equation}
C(\calb(R_5))=\calb(G(R_5))\eqqcolon \calb(R_6)\;,
\end{equation}
where $G$ is given by the 1-handle attachment in Eq.~\eqref{eq:fusion_1handle_attachment} followed by the loop 1-surgery in Eq.~\eqref{eq:loop_1surgery} at every 2-cell. Both operations together will result in removing the 1-handle that is the neighborhood of every bulk ribbon connecting two holes. So $R_6$ as a manifold is $X\cup \partial Y$ with the neighborhood of the dual 1-skeleton of $X$ removed. At each dual 2-cell of $X$, there is one boundary ribbon loop winding around the corresponding boundary loop that is contractible through the bulk. Near $\partial X$ we end up with a ``tunnel system'' containing a boundary ribbon network that is the dual 1-skeleton of the 2-cellulation $\partial X$.
\item Next, at each 1-cell of $X$, we contract the surrounding boundary ribbon labels with a $\delta$-tensor, and we include the normalization of $1/D$ from the bulk. Denoting this contraction by $C$, we have
\begin{equation}
C(\calb(R_6))=\calb(G(R_6))\eqqcolon \calb(R_7)\;,
\end{equation}
where $G$ corresponds to the 2-handle attachment Eq.~\eqref{eq:loop_2handle_attachment} followed by the backwards plain 0-surgery in Eq.~\eqref{eq:0surgery_move}. Both gluing operations together will have the effect of removing the 2-handle given by the neighborhood of the 2-cell dual to the 1-cell. Thus, $R_7$ as a manifold is $X\cup \partial Y$ with the dual 2-skeleton of $X$ removed, i.e., $\partial Y$ together with one disjoint ribbon-less $3$-ball for every vertex of $X$. Those 3-balls can be removed canceling the normalization of $D^\calb$, yielding a ribbon manifold $R_8$. We find $R_8=R(\partial Y)$, and thus the evaluation is given by $\calb(R(\partial Y))$ as claimed.
\end{itemize}

So, we have found that evaluating the state-sum on an arbitrary 4-cellulation $Y$ with physical and space boundary depends only on the space boundary $\partial Y$ (whose physical boundary is $\partial X$), and is independent of the spacetime bulk. As a consequence, the defined boundary for $\calb$ is recellulation invariant. Since the evaluation does not depend on the topology either, it follows that this boundary is also invertible since it is invariant under the moves in Eq.~\eqref{eq:boundary_invertibility_moves}.

\subsection{Ground state and Hamiltonian}
Since both the modular CYWW and its Lagrangian-module boundary are invertible, they have a unique ground state on any space manifold/cellulation with boundary. Like for the bulk CYWW, this ground state on a 3-cellulation $Y$ (with physical boundary $X$) is simply given by $\calb(R(Y))$, multiplied by $(d^\calb_i/D^\calb)^{1/2}$ for all boundary ribbon segments with label $i$ and $(d_i/D)^{1/2}$ for all bulk segments.

The local projectors of a commuting-projector Hamiltonian are again defined for every dual 2-cell of the 3-cellulation either in the bulk or at the boundary, i.e., for every plaquette of the ribbon network. E.g., we have
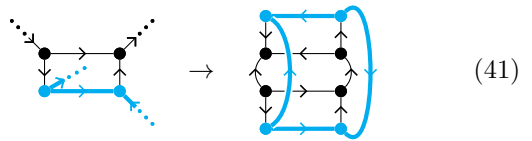
\begin{equation}
\begin{tikzpicture}
\atoms{bdvertex}{0/, 1/p={1,0}}
\atoms{vertex}{2/p={1,0.5}, 3/p={0,0.5}}
\draw (0)edge[bdribbon,mark=arr](1) (1)edge[mark=arr](2) (2)edge[mark={arr,-}](3) (3)edge[mark=arr](0);
\draw (0)edge[bdribbon,mark={three dots,a},mark={arr,e}]++(30:0.3) (1)edge[bdribbon,mark={three dots,a},mark={arr,e,-}]++(-45:0.3) (2)edge[mark={three dots,a},mark={arr,e}]++(45:0.3) (3)edge[mark={three dots,a},mark={arr,e,-}]++(135:0.3);
\end{tikzpicture}
\quad\rightarrow\quad
\begin{tikzpicture}
\atoms{bdvertex}{0/, 1/p={1,0}, 0x/p={0,1.5}, 1x/p={1,1.5}}
\atoms{vertex}{2/p={1,0.5}, 3/p={0,0.5}, 2x/p={1,1}, 3x/p={0,1}}
\draw (0)edge[bdribbon,mark=arr](1) (1)edge[mark=arr](2) (2)edge[mark={arr,-}](3) (3)edge[mark=arr](0);
\draw (0x)edge[bdribbon,mark={arr,-}](1x) (1x)edge[mark={arr,-}](2x) (2x)edge[mark={arr}](3x) (3x)edge[mark={arr,-}](0x);
\draw (0)edge[bdribbon,mark={arr},out=30,in=-30,front,looseness=0.7](0x) (1)edge[bdribbon,mark={arr,-},out=-45,in=45](1x) (2)edge[mark={arr},out=45,in=-45,front](2x) (3)edge[mark={arr},out=135,in=-135](3x);
%\draw (0)edge[bdribbon,mark={three dots,a},mark={arr,e}]++(30:0.3) (1)edge[bdribbon,mark={three dots,a},mark={arr,e,-}]++(-45:0.3) (2)edge[mark={three dots,a},mark={arr,e}]++(45:0.3) (3)edge[mark={three dots,a},mark={arr,e,-}]++(135:0.3);
\end{tikzpicture}
\end{equation}
for a bulk plaquette involving a single boundary ribbon, and similarly for plaquettes involving more or only boundary ribbons.
\subsection{The generalized local unitary circuit}
\label{sec:gLU_circuit}
We have shown the triviality of the Drinfeld-center CYWW model by constructing an invertible boundary of the corresponding state-sum path integral constructed from a Lagrangian module $\calb$. Another way of showing that two fixed-point models are in the same phase, which might be more familiar to the physically inclined reader, is via a so-called \emph{generalized local unitary circuit} (\emph{gLU circuit}) \cite{Chen2010}. Such a gLU circuit is a constant-depth homogeneous circuit of gLU operators $I_x$ of constant-size spatial support, mapping a family of ground states $A$ to another family of ground states $B$. Each operator $I_x$ does not need to be unitary but only needs to behave like an isometry when applied to the current state after applying all previous gLU operators to $A$,
\begin{equation}
\label{eq:glu_condition}
I_x^\dagger I_x (\prod_{y<x} I_y) A=  (\prod_{y<x} I_y) A\;.
\end{equation}
Physically speaking, gLU circuits allow both the addition and removal of disentangled (auxiliary) qubits, as opposed to local (strictly) unitary circuits.

Any state-sum invertible domain wall of a simple type \footnote{This includes all established state-sum constructions, e.g., ones with one tensor per highest-dimensional simplex/cell depending only on that cell. This is opposed to so-called \emph{vertex-liquid models} \cite{universal_liquid}.} automatically gives rise to a generalized local unitary circuit. The gLU conditions follow directly from retriangulation/recellulation invariance. Intuitively, we can glue patches of state-sum manifolds onto the state to create islands (0-handles) of $B$ inside $A$, then connect those 0-handles by $B$ 1-handles, then fill loops of $B$ 1-handles by $B$ 2-handles, and so on, till we obtain the $B$ ground state. E.g., in $1+1$ dimensions, a ``horizontal'' invertible domain wall state-sum (in blue) between $A$ (in gray) and $B$ (in brown) can be cut into pieces to yield a gLU circuit as follows,
\begin{equation}
\begin{tikzpicture}
\path[manifold] (0,0)rectangle(3,-1);
\fill[\secondmanifoldcol,opacity=0.4] (0,0)rectangle(3,1);
\draw[manifoldboundary] (0,0)--(3,0);
\end{tikzpicture}
=
\begin{tikzpicture}
\clip (0,-0.5)rectangle(4,1.7);
\path[manifold] (0,0)rectangle(4,-0.5);
\path[rc,manifold] (0.1,0.1)rectangle++(0.8,0.5) (1.1,0.1)rectangle++(0.8,0.5) (2.1,0.1)rectangle++(0.8,0.5) (3.1,0.1)rectangle++(0.8,0.5);
\fill[\secondmanifoldcol!40] (0.25,0.6){[rc]|-++(0.5,-0.3)}--++(0,0.3)--cycle (1.25,0.6){to[|-,rc]++(0.5,-0.3)}--++(0,0.3)--cycle (2.25,0.6){to[|-,rc]++(0.5,-0.3)}--++(0,0.3)--cycle (3.25,0.6){to[|-,rc]++(0.5,-0.3)}--++(0,0.3)--cycle;
\draw[manifoldboundary] (0.25,0.6)to[|-,rc]++(0.5,-0.3)to[rc]++(0,0.3) (1.25,0.6)to[|-,rc]++(0.5,-0.3)to[rc]++(0,0.3) (2.25,0.6)to[|-,rc]++(0.5,-0.3)to[rc]++(0,0.3) (3.25,0.6)to[|-,rc]++(0.5,-0.3)to[rc]++(0,0.3);
\fill[rc,\secondmanifoldcol,opacity=0.4] (-0.4,0.7)rectangle++(0.8,0.5) (0.6,0.7)rectangle++(0.8,0.5) (1.6,0.7)rectangle++(0.8,0.5) (2.6,0.7)rectangle++(0.8,0.5) (3.6,0.7)rectangle++(0.8,0.5);
\fill[\manifoldcol!40] (-0.25,0.7){to[|-,rc]++(0.5,0.3)}--++(0,-0.3)--cycle (0.75,0.7){to[|-,rc]++(0.5,0.3)}--++(0,-0.3)--cycle (1.75,0.7){to[|-,rc]++(0.5,0.3)}--++(0,-0.3)--cycle (2.75,0.7){to[|-,rc]++(0.5,0.3)}--++(0,-0.3)--cycle (3.75,0.7){to[|-,rc]++(0.5,0.3)}--++(0,-0.3)--cycle;
\draw[manifoldboundary] (-0.25,0.7){to[|-,rc]++(0.5,0.3)}--++(0,-0.3) (0.75,0.7){to[|-,rc]++(0.5,0.3)}--++(0,-0.3) (1.75,0.7){to[|-,rc]++(0.5,0.3)}--++(0,-0.3) (2.75,0.7){to[|-,rc]++(0.5,0.3)}--++(0,-0.3) (3.75,0.7){to[|-,rc]++(0.5,0.3)}--++(0,-0.3);
\fill[\secondmanifoldcol,opacity=0.3] (0,1.3)rectangle++(4,0.5);
\end{tikzpicture}\;.
\end{equation}
The gLU operators consist of the state-sum evaluated on the small patches in between, interpreted as operators from the bottom to the top. They fulfill the gLU property, e.g.,
\begin{equation}
II^\dagger = 
\begin{tikzpicture}
\fill[rc,\secondmanifoldcol, opacity=0.4] (-0.4,0.7)rectangle++(0.8,0.5);
\fill[\manifoldcol!40] (-0.25,0.7){to[|-,rc]++(0.5,0.3)}--++(0,-0.3)--cycle;
\draw[manifoldboundary] (-0.25,0.7){to[|-,rc]++(0.5,0.3)}--++(0,-0.3);
\fill[rc,\secondmanifoldcol, opacity=0.4] (-0.4,1.3)rectangle++(0.8,0.5);
\fill[\manifoldcol!40] (-0.25,1.8){to[|-,rc]++(0.5,-0.3)}--++(0,0.3)--cycle;
\draw[manifoldboundary] (-0.25,1.8){to[|-,rc]++(0.5,-0.3)}--++(0,0.3);
\end{tikzpicture}
=
\begin{tikzpicture}
\fill[rc,\secondmanifoldcol, opacity=0.4] (-0.4,0.7)rectangle++(0.8,1.1);
\fill[\manifoldcol!40] (-0.25,0.7){to[|-,rc]++(0.5,0.3)}--++(0,-0.3)--cycle;
\draw[manifoldboundary] (-0.25,0.7){to[|-,rc]++(0.5,0.3)}--++(0,-0.3);
\fill[\manifoldcol!40] (-0.25,1.8){to[|-,rc]++(0.5,-0.3)}--++(0,0.3)--cycle;
\draw[manifoldboundary] (-0.25,1.8){to[|-,rc]++(0.5,-0.3)}--++(0,0.3);
\end{tikzpicture}
=
\begin{tikzpicture}
\fill[rc,\secondmanifoldcol, opacity=0.4] (-0.4,0.7)rectangle++(0.8,1.1);
\fill[\manifoldcol!40] (-0.25,0.7){to[|-,rc]++(0.5,0.3)}--++(0,-0.3)--cycle;
\fill[\manifoldcol!40] (-0.25,0.7)--++(0,1.1)--++(0.5,0)--++(0,-1.1)--cycle;
\draw[manifoldboundary] (-0.25,0.7)--++(0,1.1) (0.25,0.7)--++(0,1.1);
\end{tikzpicture}\;.
\end{equation}
The second equality holds due to the invertibility axioms of the domain wall. The first equality is only true if the geometry of the state-sum construction is simple enough, which is the case for any established construction including the CYWW model. In this case, the right-hand side is a local projector whose support is a state with two domain walls, i.e., $II^\dagger$ acts trivially on the intermediate state of the circuit.

In our case, $A$ is the $3+1$-dimensional modular CYWW model, $B$ is the trivial model, and the domain wall is the invertible Lagrangian-module boundary. The disentangling circuit consists of the following four steps, in each of which we apply gLU operators in parallel with non-overlapping spatial support.
We start with the CYWW ground state on a 3-cellulation $Y$, which is $\calm(R(Y))\eqqcolon \calm(R_0)$, normalized by $(d_a/D)^{1/2}$ at every dual 2-cell.

In the first step, for each 3-cell $C$, consider $\calb(R(C))$, such as for a tetrahedron 3-cell,
\begin{equation}
\begin{multlined}
I_{\alpha}^{abcdef\beta\gamma\delta\epsilon}(wxyz)
\\
=
(\frac{d^\calb_ad^\calb_bd^\calb_cd^\calb_dd^\calb_ed^\calb_f}{(D^\calb)^6})^{1/2}
(\frac{D^\calb}{D})^{4/2}
\hspace{-0.5cm}
\begin{tikzpicture}
\atoms{bdcellvertex}{{0/p={-150:2}}, {2/p={-30:2}}, {1/p={90:2}}, {3/p={60:0.6}}}
\draw[bdcellulation] (0)--(1) (0)--(2) (0)--(3) (1)--(2) (2)--(3) (1)--(3);
\atoms{vertex}{{x/p={0.1,0.1},lab={t=$\alpha$,p=90:0.25}}}
\atoms{bdvertex}{{x1/p=-60:0.6,lab={t=$\beta$,p=-60:0.25}}, {x0/p=40:0.85,lab={t=$\gamma$, p=30:0.25}}, {x2/p=130:0.8,lab={t=$\delta$,p=130:0.25}}, {x3/p={-0.1,-0.2},lab={t=$\epsilon$,p=-130:0.25}}}
\draw (x)edge[mark={slab=$x$}](x1) (x)edge[mark={slab=$y$,r}](x2) (x)edge[mark={slab=$z$,r}](x3) (x)edge[mark={slab=$w$,r}](x0);
\draw[bdribbon] (x1)edge[mark={lab=$a$,a}]($(0)!0.5!(2)$) (x1)edge[mark={lab=$b$,a}]($(0)!0.5!(3)$) (x1)edge[mark={lab=$c$,a}]($(2)!0.5!(3)$) (x2)edge[mark={lab=$d$,a}]($(0)!0.5!(1)$) (x2)--($(0)!0.5!(3)$) (x2)edge[mark={lab=$e$,a}]($(1)!0.5!(3)$) (x0)edge[mark={lab=$f$,a}]($(1)!0.5!(2)$) (x0)--($(1)!0.5!(3)$) (x0)--($(2)!0.5!(3)$) (x3)--($(0)!0.5!(1)$) (x3)--($(0)!0.5!(2)$) (x3)--($(1)!0.5!(2)$);
\end{tikzpicture}\;.
\end{multlined}
\end{equation}
As depicted, we interpret this tensor as an operator $I$ from the single bulk fusion index to the indices of all boundary ribbons and boundary vertices. The operator parametrically depends on the bulk ribbon labels, that is, those are both input and output but their value does not change. The normalization consists of $(d^\calb_a/D^\calb)^{1/2}$ for every boundary ribbon label $a$, and $(D^\calb/D)^{1/2}$ for each vertex of the 3-cell. This operator is an isometry, which follows from performing the gluing operations corresponding to $I^\dagger I$. Now we apply the operator $I$ to the fusion index in the center of each bulk 3-cell of $Y$. When we perform the corresponding gluing operations between $R_0$ and the different $R(C)$, we obtain another ribbon manifold $R_1$. $R_1$ as a manifold is $Y$ with the dual 0-skeleton removed, that is, there is a hole in the center of every 3-cell. On the boundary of each hole we have a ribbon network dual to the boundary of the corresponding 3-cell, and for every 2-cell there is a bulk ribbon connecting two boundary fusion vertices on the holes of the adjacent 3-cells.

In the second step, for each 2-cell, consider the following ``globe'' 3-ball boundary ribbon manifold. Along the equator, we have the ribbon network dual to the boundary of the 2-cell, that is, for an $n$-gon we divide the equator into $n$ boundary ribbon segments joined at $n$ fusion vertices. There are two fusion vertices at the poles, and ``meridian'' boundary ribbons connecting all of the equatorial fusion vertices with the pole vertices. Finally, there is a bulk ribbon connecting the two pole vertices. E.g., for a 4-gon 2-cell we get
\begin{equation}
\begin{multlined}
I_{x\mu\nu}^{abcd\alpha\beta\gamma\delta}(efghijkl)
\\=
(\frac{d^\calb_ad^\calb_bd^\calb_cd^\calb_d}{(D^\calb)^4})^{1/2} (\frac{d_x}{D})^{1/2}
\begin{tikzpicture}
\atoms{bdcellvertex}{{0/p={0,0}}, {1/p={2.6,0}}, {2/p={1,0.8}}, {3/p={3.6,0.8}}}
\draw[bdcellulation] (0)--(1)--(3)--(2)--(0);
\atoms{bdvertex}{{x0/p={$(0)!0.5!(1)$},lab={t=$\alpha$,p=110:0.25}}, {x1/p={$(1)!0.5!(3)$},lab={t=$\beta$, p=0:0.25}}, {x2/p={$(3)!0.5!(2)$},lab={t=$\gamma$,p=-130:0.25}}, {x3/p={$(2)!0.5!(0)$},lab={t=$\delta$,p=180:0.25}}, {x4/p={1.8,1.6},lab={t=$\mu$,p=90:0.25}}, {x5/p={1.8,-0.8},lab={t=$\nu$,p=-90:0.25}}}
\draw (x4)edge[mark={slab=$x$}](x5);
\draw[bdribbon] (x0)edge[mark={slab=$a$,p=0.6}](x1) (x1)edge[mark={slab=$b$}](x2) (x2)edge[mark={slab=$c$,p=0.4}](x3) (x3)edge[mark={slab=$d$}](x0) (x0)edge[mark={slab=$e$,p=0.6}](x4) (x1)edge[mark={slab=$f$,r}](x4) (x3)edge[mark={slab=$g$}](x4) (x2)edge[mark={slab=$h$}](x4) (x0)edge[mark={slab=$i$}](x5) (x1)edge[mark={slab=$j$}](x5) (x3)edge[mark={slab=$k$,r}](x5) (x2)edge[mark={slab=$l$,p=0.65}](x5);
\end{tikzpicture}\;.
\end{multlined}
\end{equation}
As depicted, we interpret the corresponding Lagrangian-module tensor as an operator $I$ from the bulk ribbon and pole vertex labels to the equatorial ribbon and fusion vertex labels, parametrically depending on the meridian ribbon labels. The normalization consists of $(d_a^\calb/D^\calb)^{1/2}$ for all equatorial boundary ribbon labels $a$, and $(d_x/D)^{1/2}$ for the bulk ribbon label $x$. $I$ is an isometry, which follows from performing the gluing operations corresponding to contracting $I^\dagger I$. 
At every dual 2-cell, $R_1$ contains one bulk ribbon segment connecting two boundary fusion vertices through the bulk, e.g., for a 4-gon,
\begin{equation}
\begin{tikzpicture}
\atoms{bdcellvertex}{{0/p={0,0}}, {1/p={1.5,0}}, {2/p={0.7,0.5}}, {3/p={2.2,0.5}}}
\draw[bdcellulation] (0)--(1)--(3)--(2)--(0);
\atoms{bdvertex}{{x4/p={1.1,1},lab={t=$\mu$,p=120:0.25}}, {x5/p={1.1,-0.5},lab={t=$\nu$,p=-60:0.25}}}
\draw (x4)edge[mark={slab=$x$}](x5);
\draw[bdribbon] (x4)edge[mark={slab=$e$,p=0.6},mark={three dots,a}]++(-140:0.6) (x4)edge[mark={slab=$f$,p=0.6,r},mark={three dots,a}]++(0:0.6) (x4)edge[mark={slab=$h$,p=0.6},mark={three dots,a}]++(40:0.6) (x4)edge[mark={slab=$g$,p=0.6,r},mark={three dots,a}]++(180:0.6);
\draw[bdribbon] (x5)edge[mark={slab=$i$,p=0.6},mark={three dots,a}]++(-140:0.6) (x5)edge[mark={slab=$j$,p=0.6,r},mark={three dots,a}]++(0:0.6) (x5)edge[mark={slab=$l$,p=0.6},mark={three dots,a}]++(40:0.6) (x5)edge[mark={slab=$k$,p=0.6,r},mark={three dots,a}]++(180:0.6);
\end{tikzpicture}\;.
\end{equation}
At each such bulk segment, we apply the operator $I$ as indicated by the choice of labels.
Performing the corresponding gluing operations between $R_1$ and the ``globe'' ribbon manifolds, we get a new ribbon manifold $R_2$. Specifically, we first apply the 1-handle attachment of Eq.~\eqref{eq:1handle_attachment} to both $\mu$ and $\nu$, and then the 2-handle attachment of Eq.~\eqref{eq:2handle_attachment} to the resulting $x$ loop. The corresponding $R_2$ as a manifold is $Y$ with the neighborhood of the dual 1-skeleton removed. At each dual edge of $Y$ we have a boundary ribbon winding around the corresponding non-contractible loop. At each dual face, we have a boundary ribbon winding around the corresponding loop that is non-contractible inside the boundary, but contractible through the bulk. The loops of each adjacent pair of dual edge and face intersect once at a 4-valent fusion vertex.

In the third step, for each 1-cell, consider the ribbon manifold that looks like the ``globe'' from the previous step for the dual 2-cell, just without the bulk ribbon connecting the poles. E.g., for a 4-valent 1-cell, we have
\begin{equation}
\label{eq:unitary_step3}
\begin{multlined}
I_{abcd\alpha\beta\gamma\delta}^{\mu\nu}(efghijkl)\\
=
(\frac{d_ad_bd_cd_d}{(D^\calb)^4})^{1/2}
\begin{tikzpicture}
\atoms{void}{{0/p={0,0}}, {1/p={2.6,0}}, {2/p={1,0.8}}, {3/p={3.6,0.8}}}
\atoms{bdcellvertex}{e0/p={1.8,1.7}, e1/p={1.8,-0.9}}
\draw[bdcellulation] (e0)--(e1);
\atoms{bdvertex}{{x0/p={$(0)!0.5!(1)$},lab={t=$\alpha$,p=110:0.25}}, {x1/p={$(1)!0.5!(3)$},lab={t=$\beta$, p=0:0.25}}, {x2/p={$(3)!0.5!(2)$},lab={t=$\gamma$,p=-130:0.25}}, {x3/p={$(2)!0.5!(0)$},lab={t=$\delta$,p=180:0.25}}, {x4/p={1.8,1.4},lab={t=$\mu$,p=140:0.25}}, {x5/p={1.8,-0.6},lab={t=$\nu$,p=-140:0.25}}}
\draw[bdribbon] (x0)edge[mark={slab=$a$,p=0.6}](x1) (x1)edge[mark={slab=$b$}](x2) (x2)edge[mark={slab=$c$,p=0.4}](x3) (x3)edge[mark={slab=$d$}](x0) (x0)edge[mark={slab=$e$,p=0.6}](x4) (x1)edge[mark={slab=$f$,r}](x4) (x3)edge[mark={slab=$g$}](x4) (x2)edge[mark={slab=$h$,p=0.4}](x4) (x0)edge[mark={slab=$i$}](x5) (x1)edge[mark={slab=$j$}](x5) (x3)edge[mark={slab=$k$,r}](x5) (x2)edge[mark={slab=$l$,p=0.65}](x5);
\end{tikzpicture}\;.
\end{multlined}
\end{equation}
As depicted, we interpret the associated Lagrangian-module tensor as an operator $I$ from the equatorial ribbon and vertex labels to the pole vertex labels, parametrically depending on the meridian ribbon labels, normalized as shown. This tensor is \emph{not} an isometry, but still defines a gLU operator as we will see in a moment.
After the previous step, the ribbon configuration in $R_2$ around each edge looks like, e.g., for a 4-valent edge,
\begin{equation}
\begin{tikzpicture}
\atoms{void}{{0/p={0,0}}, {1/p={2.6,0}}, {2/p={1,0.8}}, {3/p={3.6,0.8}}}
\atoms{bdcellvertex}{e0/p={1.8,1.7}, e1/p={1.8,-0.9}}
\draw[bdcellulation] (e0)--(e1);
\atoms{bdvertex}{{x0/p={$(0)!0.5!(1)$},lab={t=$\alpha$,p=110:0.25}}, {x1/p={$(1)!0.5!(3)$},lab={t=$\beta$, p=0:0.25}}, {x2/p={$(3)!0.5!(2)$},lab={t=$\gamma$,p=-130:0.25}}, {x3/p={$(2)!0.5!(0)$},lab={t=$\delta$,p=180:0.25}}}
\draw[bdribbon] (x0)edge[mark={slab=$a$,r}](x1) (x1)edge[mark={slab=$b$,r}](x2) (x2)edge[mark={slab=$c$,r}](x3) (x3)edge[mark={slab=$d$,r}](x0) (x0)edge[mark={slab=$e$,p=0.8},mark={three dots,a}]++(100:1.1) (x0)edge[mark={slab=$i$,p=0.8},mark={three dots,a}]++(-100:0.9) (x1)edge[mark={slab=$f$,p=0.8},mark={three dots,a}]++(60:0.9) (x1)edge[mark={slab=$j$,p=0.8},mark={three dots,a}]++(-60:0.9) (x2)edge[mark={slab=$h$,p=0.8},mark={three dots,a}]++(80:0.9) (x2)edge[mark={slab=$l$,p=0.8},mark={three dots,a}]++(-80:1.1) (x3)edge[mark={slab=$g$,p=0.8},mark={three dots,a}]++(120:0.9) (x3)edge[mark={slab=$k$,p=0.8},mark={three dots,a}]++(-120:0.9);
\end{tikzpicture}\;,
\end{equation}
with bulk 3-manifold on the inside. We apply the operator $I$ to this configuration as indicated by the labeling. Performing the corresponding gluing operations between $R_2$ and all the ``globe'' ribbon manifolds, we get another ribbon manifold $R_3$. Specifically, we first apply the 1-handle attachment of Eq.~\eqref{eq:1handle_attachment} to the labels $\alpha$ $\beta$, $\gamma$, and $\delta$, and then the 2-handle attachment of Eq.~\eqref{eq:2handle_attachment} to the resulting loops $a$, $b$, $c$, and $d$. Finally, we apply the backwards plain 0-surgery in Eq.~\eqref{eq:0surgery_move}.
$R_3$ as a 3-manifold is $Y$ with the dual 2-skeleton removed, i.e., one disconnected 3-ball at every dual 3-cell. On the boundary of each 3-ball we have the ribbon network dual to the boundary 2-cellulation of the dual 3-cell.

Now imagine applying $I^\dagger$ to this resulting state. This corresponds to gluing the ribbon manifolds in Eq.~\eqref{eq:unitary_step3} at the fusion vertices $\mu$ and $\nu$ in between the disconnected 3-balls of $R_3$ at the endpoints of every edge, by 1-handle attachments as in Eq.~\eqref{eq:1handle_attachment}. This will restore the ribbon manifold $R_2$, and thus $I$ fulfills the gLU condition in Eq.~\eqref{eq:glu_condition}.

After the third step, $\calb(R_3)$ is product state, which usually is taken as the definition for having ``disentangled'' a state. If we want, we can apply a last layer of gLU operators to ``remove'' the product state and obtain complete vacuum without any degrees of freedom. The corresponding operator $I$ is the Lagrangian-module tensor on the dual 3-cell with dual boundary ribbon network, as an operator from all indices to nothing, i.e., the projector onto each factor of the product state.

In this work we use the mathematically natural formulation where the total Hilbert space of the model is as a direct sum indexed by the different ribbon label configurations. Each summand is the tensor product of fusion vector spaces at the different fusion vertices, which only depend locally on the adjacent ribbon labels. In physics, people might be more used to Hilbert spaces that are plain tensor products.

A common ad hoc way of arriving at a plain tensor-product Hilbert space is to simply enlarge the dimensions of all the vertex fusion spaces to $N_{\text{max}}\coloneqq\max_{a,b,c}N^{ab}_c$, independent of $a$, $b$, and $c$. That is, the enlarged UMTC tensors equal the original UMTC tensors when all fusion vertex labels $\alpha$ take values $0\leq \alpha<N^{ab}_c$, and are equal to $0$ if any fusion vertex label takes values $N^{ab}_c\leq \alpha<N_{\text{max}}$. All the UMTC tensors, as well as the CYWW ground state, do then live in a tensor-product Hilbert space with one qu-$d$-it with $d=\#\text{anyons}$ at every face dual to a ribbon, and a qu-$d$-it with $d=N_{\max}$ at every volume dual to a fusion vertex. Note that the enlarged UMTC tensors still fulfils all the axioms in Section~\ref{sec:tensorial_umtc} and Section~\ref{sec:tensorial_umtc_module}, just not the convention
\begin{equation}
\begin{tikzpicture}
\atoms{vertex}{0/lab={t=$\alpha$,p=180:0.25}, {1/p={1.5,0},lab={t=$\beta$,p=0:0.25}}}
\draw (0)edge[mark=arr,mark={slab=$a$}](1) (0)edge[mark=arr,bend left=40,mark={slab=$b$}](1) (0)edge[mark={arr,-},bend right=40,mark={slab=$c$,r}](1);
\end{tikzpicture}
=
\delta_{\alpha,\beta}\qquad \forall a, b, c\;,
\end{equation}
which we would usually impose. We thus see that the gLU defined in this section is even a gLU on the enlarged tensor-product Hilbert space, not only on the mathematical direct-sum space constrained by $N^{ab}_c$. Note that in the physics literature, $N_{\text{max}}=1$ is usually assumed such that we do not need any degrees of freedom at the fusion vertices, and going to the enlarged Hilbert space simply means also including ribbon configurations violating the fusion rules. It can also be directly seen that the circuit proposed in this section remains gLU after enlarging the Hilbert space: While the circuit acts as zero on configurations where any vertex index $\alpha$ takes values $N^{ab}_c\leq \alpha<N_{\text{max}}$, the circuit is still gLU since the ground state amplitudes on those values are zero as well.

A mathematically more natural way is to use a tensor-product Hilbert space with one qu-$d$-it with $d=\sum_{a,b,c}N^{ab}_c$ at every fusion vertex only. This qu-$d$-it can be thought of as a composite of three ribbon labels $a$, $b$, and $c$ and fusion index of dimension $N^{ab}_c$. We can associate each of $a$, $b$, and $c$ to one adjacent ribbon, such that every ribbon has two associated labels, say $a$ and $a'$, coming from its two end vertices. Now, the direct-sum Hilbert space can be identified with the tensor-product Hilbert space where $a=a'$ at every ribbon. This leads to an equivalent tensorial definition of UMTCs and Lagrangian modules with indices only at the vertices. In this case, fusion 0-surgery simply corresponds to index contraction. For a fusion 0-surgery producing a loop, we need to insert a weight matrix into the contraction (related to the factor $d_a/D$ in the direct-sum definition). Loop 1-surgery corresponds to doing nothing. This is equivalent to the bare tensorial TQFT described in Appendix~\ref{sec:tensorial_tqft} before dimension-reduction block-diagonalization. All constructions in this paper can be carried out using those alternative tensorial definitions, in which case all the occurring Hilbert spaces are again plain tensor-product spaces.

Note that a disentangling gLU circuit is equivalent to a proper strict local unitary circuit if we allow for the addition of auxiliary qu-$d$-its. To this end, we arbitrarily extend each generalized/partial isometry $I$ to outside the local support of the ground state, yielding a full isometry. If the output dimension of a partial isometry is smaller than the input dimension, we need to add an auxiliary qu-$d$-it for this to be possible. It is unclear whether CYWW models can be disentangled using an exact strict local unitary circuit without auxiliary degrees of freedom. A plausible scenario would be that this works for all abelian UMTCs, but not the non-abelian ones. Note, however, that in the context of phases of matter, gLUs are the natural notion to consider and not strict local unitary circuits.

\subsection{For Drinfeld centers of UMTCs}
\label{sec:braided_drinfeld_center}
If a fusion category $\calf$ can be extended with a modular braiding to an UMTC $\calm$, then the Drinfeld center of $F$ is simply given by the tensor/Deligne product \cite{Mueger2001},
\begin{equation}
Z(\calf)=\calm \otimes \overline{\calm}\;.
\end{equation}
Then, also the CYWW model for $Z(\calf)$ is simply given by a non-interacting stack of the CYWW model for $\calm$ and an orientation-reversed copy. The invertible boundary for this stack is given by the \emph{folding boundary}, schematically in 2 dimensions lower,
\begin{equation}
\begin{tikzpicture}
\fill[opacity=0.3] (0,0)--++(2,0)--++(60:1.5)--++(-2,0)--cycle;
\fill[opacity=0.3] (2,0)arc(-90:-30:0.3)--++(60:1.5)arc(-30:-90:0.3)--cycle;
\fill[opacity=0.3] (2,0.6)arc(90:-30:0.3)--++(60:1.5)arc(-30:90:0.3)--cycle;
\fill[opacity=0.3] (0,0.6)--++(2,0)--++(60:1.5)--++(-2,0)--cycle;
\draw[decorate,decoration=brace] (-0.1,0)--++(0,0.6)node[midway,left]{$\scriptstyle\text{CYWW}(Z(\calf))$};
\draw (2.6,0.6)edge[<-,out=0,in=-120,mark={lab={\breakcell{Folding\\boundary}},a}] (4,1);
\draw (1,-0.1)edge[<-,out=-70,in=180,mark={lab={$\scriptstyle\text{CYWW}(\calm)$},a}]++(1,-0.8);
\draw (1,0.5)edge[<-,out=-70,in=180,mark={lab={$\scriptstyle\overline{\text{CYWW}(\calm)}$},a}]++(1,-0.8);
\end{tikzpicture}\;.
\end{equation}
The invertibility of the folding boundary follows directly from the invertibility of the bulk. The same folding boundary construction in one dimension lower can be used to obtain a Lagrangian module for $Z(\calf)$ as follows. Given a boundary ribbon manifold, we double the bulk, replace the boundary with a fold, and replace the boundary ribbons by bulk ribbons within the fold. Precomposing/pulling back $\calm$ with this geometric operation will yield the Lagrangian module for $Z(\calf)=\calm\otimes \overline{\calm}$. The folding boundary for the CYWW model of $Z(\calf)$ above is the same as the Lagrangian-module boundary for this Lagrangian module.

\subsection{Example: Toric code}
\label{sec:toric_code_boundary}
The \emph{toric code} UMTC is the Drinfeld center of the arguably simplest fusion category, namely the $\zz_2$ group category. At the same time, the (untwisted) $\zz_2$ category, unlike many other simple fusion categories, cannot be equipped with a modular braiding, and thus does not belong to the somewhat trivial case described in Section~\ref{sec:braided_drinfeld_center}. The toric code UMTC has $4$ ribbon labels $1$, $e$, $m$, $\epsilon$, with $\zz_2\times \zz_2$ fusion rules, i.e., the dimension of a fusion vertex is $1$ if there is an even number of adjacent $e$ or $\epsilon$ ribbons and an even number of $m$ or $\epsilon$ ribbons, and $0$ otherwise.

The amplitude for a labeled ribbon manifold can be obtained by reducing it to the empty manifold in steps. Every connected 3-manifold can be obtained from applying 1-surgeries to a set of framed loops inside the 3-sphere \cite{Lickorish1997}. Applying the inverse 1-surgeries to the ribbon manifold, we obtain a ribbon 3-sphere with a set of additional loops. The amplitude is thus obtained by summing the ribbon 3-sphere amplitudes for all configurations of ribbon labels $i$ on the additional loops weighted by $d_i/D$. The amplitude for a ribbon 3-sphere can be further reduced as follows. We first remove all $1$-labeled ribbon segments, resolve every $\epsilon$-labeled ribbon into a pair of $e$ and $m$ ribbon,
\begin{equation}
\begin{tikzpicture}
\draw (0,0)edge[mark={slab=$\epsilon$},mark=arr](1,0);
\end{tikzpicture}
\rightarrow
\begin{tikzpicture}
\draw (0,0)edge[mark={slab=$e$},mark=arr]++(1,0) (0,-0.3)edge[mark={slab=$m$,r},mark=arr]++(1,0);
\end{tikzpicture}\;,
\end{equation}
and also define a way to resolve every fusion vertex into uninterrupted ribbons, e.g.,
\begin{equation}
\begin{tikzpicture}
\atoms{vertex}{0/}
\draw (0)edge[mark={arr,-},mark={slab=$e$}]++(135:0.7) (0)edge[mark={arr,-},mark={slab=$m$,r}]++(45:0.7) (0)edge[mark={arr},mark={slab=$\epsilon$,r}]++(-90:0.7);
\end{tikzpicture}
\quad\rightarrow\quad
(2)
\begin{tikzpicture}
\draw[mark={arr,-,p=0.7},mark={slab=$m$,p=0.3},rc] (-0.15,0)--++(0,0.7)--++(45:1);
\draw[mark={arr,-,p=0.7},mark={slab=$e$,p=0.3,r},rc,front] (0.15,0)--++(0,0.7)--++(135:1);
\end{tikzpicture}\;.
\end{equation}
Different choices of resolution yield different but gauge equivalent UMTCs. We also drop the direction and the framing of the $e$ and $m$ ribbons, such that we obtain a collection of $e$ and $m$ loops which are disjoint but might be linked with another. Now we unlink them via
\begin{equation}
\label{eq:toric_code_unlinking}
\begin{tikzpicture}
\draw (0,0)edge[mark={slab=$m$,p=0.7,r,sideoff=-0.1}](1,1.5);
\draw (1,0)edge[front, mark={slab=$e$,p=0.7,sideoff=-0.1}](0,1.5);
\end{tikzpicture}
=
(-1)
\begin{tikzpicture}
\draw (1,0)edge[mark={slab=$e$,p=0.7,sideoff=-0.1}](0,1.5);
\draw (0,0)edge[front,mark={slab=$m$,p=0.7,r,sideoff=-0.1}](1,1.5);
\end{tikzpicture}\;.
\end{equation}
This corresponds to the fact that the full braiding between $e$ and $m$ is $-1$. In the end, we can remove unlinked loops, and the amplitude for a 3-sphere without ribbons is $1/D=1/2$.

The Lagrangian module is given as follows. There are two boundary ribbon labels, $1$ and $e'$ with quantum dimension $1$, so $D=\sqrt2$. The fusion rules inside the boundary are $\zz_2$, i.e., the dimension of the fusion indices is $1$ if there is an even number of adjacent $e'$ ribbons and $0$ otherwise. So the fusion category formed by the boundary ribbons alone is just the $\zz_2$ group category whose Drinfeld center is the toric code UMTC in the bulk.
The non-zero dimensions of the boundary fusion vertices with one bulk and one boundary ribbon are given by
\begin{equation}
\begin{multlined}
\begin{tikzpicture}
\atoms{bdvertex}{0/lab={p=90:0.25, t=$1$}}
\draw (0)edge[bdribbon,mark={arr},mark={slab=$1$}]++(0.8,0) (0)edge[mark={arr,-},mark={slab=$1$,r}]++(-0.8,0);
\end{tikzpicture}
\;,
\begin{tikzpicture}
\atoms{bdvertex}{0/lab={p=90:0.25, t=$1$}}
\draw (0)edge[bdribbon,mark={arr},mark={slab=$e'$}]++(0.8,0) (0)edge[mark={arr,-},mark={slab=$e$,r}]++(-0.8,0);
\end{tikzpicture}
\;,\\
\begin{tikzpicture}
\atoms{bdvertex}{0/lab={p=90:0.25, t=$1$}}
\draw (0)edge[bdribbon,mark={arr},mark={slab=$1$}]++(0.8,0) (0)edge[mark={arr,-},mark={slab=$m$,r}]++(-0.8,0);
\end{tikzpicture}
\;,
\begin{tikzpicture}
\atoms{bdvertex}{0/lab={p=90:0.25, t=$1$}}
\draw (0)edge[bdribbon,mark={arr},mark={slab=$e'$}]++(0.8,0) (0)edge[mark={arr,-},mark={slab=$\epsilon$,r}]++(-0.8,0);
\end{tikzpicture}\;.
\end{multlined}
\end{equation}
So physically speaking, the $m$ anyon ``condenses'' at the boundary, whereas the $e$ anyon ``confines'' and turns into a boundary anyon.

The amplitude for a ribbon network on a 3-ball is again obtained by reducing it to a bulk network. Namely, we neglect $1$ ribbons, and transform $e'$ ribbons into $e$ ribbons pushing them slightly into the bulk,
\begin{equation}
\begin{tikzpicture}
\draw[bdribbon] (0,0)edge[mark={slab=$e'$}](1,0);
\end{tikzpicture}
\rightarrow
\begin{tikzpicture}
\draw (0,0)edge[mark={slab=$e$}](1,0);
\end{tikzpicture}\;,
\end{equation}
and accordingly the condensation vertices,
\begin{equation}
\begin{tikzpicture}
\atoms{bdvertex}{0/}
\draw (0)edge[bdribbon,mark={slab=$e'$}]++(0.8,0) (0)edge[mark={slab=$e$,r}]++(-0.8,0);
\end{tikzpicture}
\rightarrow
(2^{1/2})
\begin{tikzpicture}
\draw (0,0)edge[mark={slab=$e$}]++(1,0);
\end{tikzpicture}\;.
\end{equation}
After this, we end up with a collection of $e$ and $m$ loops, as well as $m$ segments ending at two boundary vertices. We unlink them via Eq.~\eqref{eq:toric_code_unlinking}, and in the end remove contractible loops or $m$ segments.

The corresponding CYWW model and its invertible boundary is a state-sum with one $\zz_2\times \zz_2$-valued variable at each 2-cell, and a $\zz_2$-valued variable at each boundary 1-cell. We can learn more about the concrete model by expressing it as a simplicial gauge theory, which we do in Appendix~\ref{sec:gauge_toric_code}.
\section{Fermionic invertible boundaries for Kitaev 16-fold way Witt classes}
\label{sec:fermionic_boundaries}
In this section we will find fermionic invertible boundaries for certain CYWW models beyond those for Drinfeld center UMTCs. That is, even though the bulk CYWW model has only purely bosonic degrees of freedom, there are fermionic degrees of freedom within the boundary. As a consequence, the gLU circuit disentangling the bosonic model, which we can construct from the invertible boundary, uses fermionic auxiliary degrees of freedom. Concretely, we find that using auxiliary fermionic degrees of freedom, we can show the triviality of the CYWW model not only for Drinfeld centers but additionally for UMTCs in the \emph{Witt classes} of the \emph{Kitaev 16-fold way}.

\subsection{Fermionic Lagrangian modules}
\label{sec:fermionic_umtc_module}
A \emph{fermionic Lagrangian module} is the fermionic analog of a Lagrangian module, and like any fermionic analog differs from the latter by three points. First, the associated tensors are \emph{fermionic tensors}. Second, the ribbon manifolds carry a spin structure (only inside the boundary). Third, the spin structure interacts with the tensors according to a \emph{spin-statistics relation}.

Starting with the first point, a fermionic tensor (c.f.~Ref.~\cite{liquid_intro}) is a tensor whose indices live in super vector spaces with a \emph{super dimension} $a|b$ consisting of an even dimension $a$ and an odd dimension $b$, such that every configuration $x$ has an even (0) or odd (1) \emph{parity}
\begin{equation}
|x|=
\begin{cases}
0\in\zz_2 & \text{if}\quad 0\leq x<a\\
1\in\zz_2 & \text{if}\quad a\leq x<a+b
\end{cases}\;.
\end{equation}
A fermionic tensor also includes an ordering of its indices, and a distinction between \emph{input} indices and \emph{output} indices. So it is a pair, which we can denote like
\begin{equation}
(T_{abcxyz},bzx\bar yc\bar a)\;,
\end{equation}
where $y$ and $a$ are input indices, as marked by the bar. An equivalence of two such pairs is given by transposing two consecutive indices $x$ and $y$ in the ordering and at the same time multiplying the tensor by $(-1)^{|x||y|}$, e.g.,
\begin{equation}
(T_{abcxyz},bzx\bar yc\bar a)\sim ((-1)^{|x||y|} T_{abcxyz}, bz\bar yxc\bar a)\;.
\end{equation}
Furthermore, the tensor entry is $0$ or undefined for index configurations whose total parity is not $0$. To contract two indices $x$ and $\bar y$ of a fermionic tensor, we first permute the index ordering such that they appear consecutively as $\bar y x$, and then contract tensor normally, removing the indices from the ordering. Last, in order to block two output indices $x$ and $y$ into a single index $(xy)$ we first permute the ordering such that they appear consecutively as $xy$, but for blocking two input indices $\bar x$ and $\bar y$ into $\overline{ (xy)}$, they should appear as $\bar y\bar x$. An equivalent way to denote a fermionic tensor is via \emph{Grassmann variables} $\theta$,
\begin{equation}
\begin{multlined}
(T_{abcxyz},bzx\bar yc\bar a)\\
\leftrightarrow
T_{abcxyz} (\theta_b)^{|b|} (\theta_z)^{|z|} (\theta_x)^{|x|} (\overline{\theta_y})^{|y|} (\theta_c)^{|c|} (\overline{\theta_a})^{|a|}\;.
\end{multlined}
\end{equation}
The basic rules for manipulating expressions with tensors and Grassmann variables are then
\begin{align}
\label{eq:grassmann1}
\theta_a \theta_b &= - \theta_b\theta_a\;,\\
\label{eq:grassmann2}
\overline{\theta_a\theta_b}&= \overline{\theta_b}\quad\overline{\theta_a}\;,\\
\label{eq:grassmann3}
\sum_a T_{\ldots a\ldots a\ldots} (\overline{\theta_a})^{|a|}(\theta_a)^{|a|} &= \sum_a T_{\ldots a\ldots a\ldots}\;.
\end{align}
Note that for our purpose, the UMTC itself is still non-fermionic, and accordingly only the boundary fusion indices in the Lagrangian module can carry a non-zero odd dimension.

Coming to the second point, let us now discuss how to define a spin structure inside the 2-dimensional boundary of our ribbon manifolds. To this end we work with a concrete triangulation/cellulation of the boundary 2-manifold such that the boundary ribbons coincide with a subset of the Poincar\'e dual 1-skeleton of that cellulation. Then a combinatorial spin structure can be implemented using obstruction theory and simplicial/cellular cohomology as described in Ref.~\cite{Gaiotto2015}. A spin structure is a $\zz_2$-valued simplicial (or rather, cellular) 1-chain $\eta_2$ (c.f.~Appendix~\ref{sec:simplicial cohomology}) such that
\begin{equation}
\label{eq:spin_structure_definition}
\delta \eta_2=\omega_2\;,
\end{equation}
where $\omega_2$ is a fixed 0-cycle representing the \emph{second Stiefel-Whitney class}. There exist simple formulas computing the value of $\omega_2$ on a vertex depending only on the combinatorics of the cellulation in a small neighborhood, such as for a triangulation in Ref.~\cite{Goldstein1976}. In 2 dimensions, a simple formula can be obtained by choosing a ``favorite'' vertex $v_0(f)$ of every face $f$, a favorite vertex $v_0(e)$ of every edge $e$ and we automatically have $v_0(v)=v$ for every vertex $v$. Then the $\zz_2$ weight at a vertex $v$ is given by
\begin{equation}
\omega_2(v) = \sum_{f\in C_2} \delta_{v,v_0(f)} + \sum_{e\in C_1} \delta_{e,v_0(e)} + \sum_{v\in C_0} \delta_{v,v_0(v)}\;,
\end{equation}
where $C_i$ is the set of $i$-cells, and the formula is $\operatorname{mod} 2$.
E.g., consider the following cellulation with favorite vertices of edges corresponding to outgoing arrows and favorite vertices of faces marked by putting little angles into the corresponding corners. The vertices $v$ with $\omega_2(v)=1$ and edges $e$ with $\eta_2(e)=1$ are marked in green as follows,
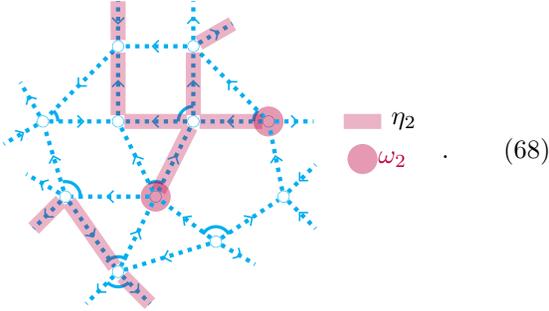
\begin{equation}
\begin{tikzpicture}
\atoms{bdcellvertex}{{0/p={-1.5,0},corner=0:45}, 1/p={-0.5,0}, {2/p={0.5,0},corner=90:180}, {3/p={1.5,0},corner=135:180}, 4/p={-0.5,1}, 5/p={0.5,1}, {6/p={-1.2,-1},corner=0:100}, {7/p={0,-1},corner=60:120}, {8/p={1.7,-1}}, {9/p={-0.5,-2},corner=55:140,corner=-45:-135}, {10/p={0.8,-1.6},corner=30:150}}
\atoms{void}{{x/p=3,omega}, {y/p=7, omega}}
\draw[bdcellulation] (0)edge[mark=arr](1) (2)edge[mark=arr](1) (3)edge[mark=arr](2) (1)edge[mark=arr](4) (2)edge[mark=arr](5) (7)edge[mark=arr](1) (7)edge[mark=arr](2) (7)edge[mark=arr](6) (9)edge[mark=arr](7) (10)edge[mark=arr](7) (6)edge[mark=arr](0) (4)edge[mark=arr](0) (5)edge[mark=arr](4) (3)edge[mark=arr](5) (8)edge[mark=arr](3) (10)edge[mark=arr](8) (10)edge[mark=arr](9) (6)edge[mark=arr](9);
\draw[bdcellulation] (0)edge[mark={arr,-}]++(-150:0.6) (0)edge[mark=arr]++(120:0.6) (4)edge[mark={arr,-}]++(90:0.6) (5)edge[mark=arr]++(90:0.6) (5)edge[mark=arr]++(30:0.6) (3)edge[mark=arr]++(0:0.6) (8)edge[mark={arr,-}]++(45:0.6) (8)edge[mark={arr,-}]++(-45:0.6) (10)edge[mark=arr]++(-30:0.6) (9)edge[mark=arr]++(-135:0.6) (9)edge[mark=arr]++(-45:0.6) (6)edge[mark=arr]++(150:0.6) (6)edge[mark=arr]++(-135:0.6);
\draw[spin structure] (7)--(2)--(1)--(4)--++(90:0.6) (3)--(2)--(5)--++(30:0.6) (9)--(6)--++(-135:0.6) (9)--++(-45:0.6);
\draw[spin structure] (2.5,0)edge[mark={lab=$\eta_2$,a,sty={text=black,opacity=1}}]++(0.5,0);
\atoms{void,omega}{{x/p={2.75,-0.5},lab={t=$\omega_2$,p=0:0.4}}}
\end{tikzpicture}\;.
\end{equation}
Note that $\omega_2$ is fixed by the cellulation, whereas $\eta_2$ can be chosen freely subject to the constraint $\delta\eta_2=\omega_2$.

We also need to define how the spin structure behaves under the gluing operations. To be concrete, we assign to each possible link for a boundary fusion vertex (or its orientation-reversed partner) a fixed 2-cell representative $f$ (or its orientation-reversed partner) in which each such vertex is contained.
At a 1-handle attachment as in Eq.~\eqref{eq:fusion_1handle_attachment}, a cellulation of the new boundary is obtained by simply gluing the two containing boundary 2-cells $f_1$ and $f_2$ identified with $\bar f$ and $f$. When doing so, every pair of edges $(e_1,e_2)$ of $(f_1,f_2)$ becomes a single edge $e$. The spin structure after the gluing on $e$ is given by
\begin{equation}
\label{eq:spin_structure_1handle}
\eta_2(e) = \eta_2(e_1) + \eta_2(e_2) + \eta_1^f(e)\;,
\end{equation}
where $\eta_1^f$ is the $1$-chain within the boundary of $f$ representing the orientation of $f$, consisting of all ``clockwise'' edges. That is, we have $\delta\eta_1^f=\omega_1^f$, where
\begin{equation}
\omega_1^f(v) =  \sum_{e\in C_1[f]} \delta_{e,v_0(e)} + \sum_{v\in C_0[f]} \delta_{v,v_0(v)}
\end{equation}
is the first Stiefel-Whitney class, a 0-cycle inside the boundary of $f$. Accordingly, we find that for a vertex $v$ originating from gluing the pair of vertices $(v_1,v_2)$ of $(f_1,f_2)$, $\omega_2$ is given by
\begin{equation}
\omega_2(v) = \omega_2(v_1) + \omega_2(v_2) + \omega_1^f(v)
\;,
\end{equation}
such that Eq.~\eqref{eq:spin_structure_1handle} is consistent with Eq.~\eqref{eq:spin_structure_definition}.

At a 2-handle attachment as in Eq.~\eqref{eq:loop_2handle_attachment}, the physical boundary is transformed like
\begin{equation}
\label{eq:2handle_attachment_spinstructure}
B_1\times S_1\rightarrow S_0\times B_2\;,
\end{equation}
with common boundary $S_0\times S_1=S_1\sqcup S_1$. This boundary has $4$ different spin structures, as each circle can be bounding or non-bounding. However, only the bounding-bounding spin structure is possible on the right-hand side of Eq.~\eqref{eq:2handle_attachment_spinstructure} since both $S_1$ are the boundary of a $B_2$. So we can only apply this gluing operation if $S_1$ is bounding on the left-hand side.

Coming to the third point, we now discuss how the associated tensors depend on $\eta_2$ via the spin-statistics relation. Namely, if we change $\eta_2$ by $\eta_2'=\eta_2+df$ for a boundary 2-cell $f$, then the assigned tensor gets multiplied by $(-1)^{|a_f|}$, where $a_f$ is the configuration of the boundary fusion index at the center of $f$ (if there is one), e.g.,
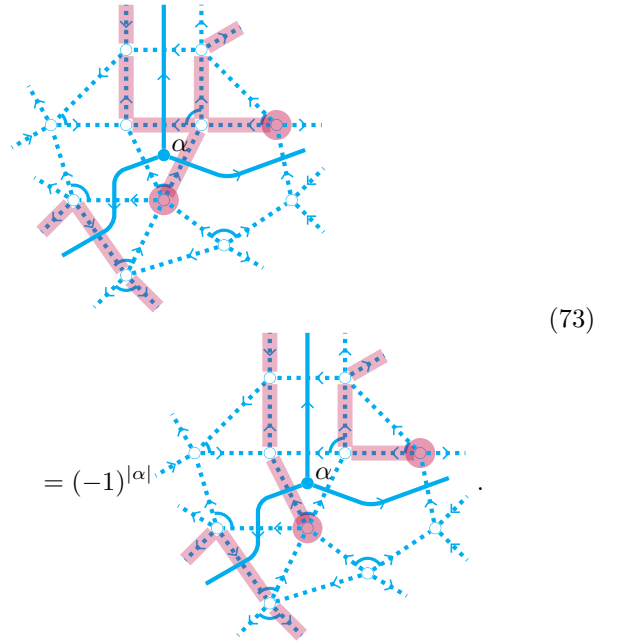
\begin{equation}
\begin{multlined}
\begin{tikzpicture}
\atoms{bdcellvertex}{{0/p={-1.5,0},corner=0:45}, 1/p={-0.5,0}, {2/p={0.5,0},corner=90:180}, {3/p={1.5,0},corner=135:180}, 4/p={-0.5,1}, 5/p={0.5,1}, {6/p={-1.2,-1},corner=0:100}, {7/p={0,-1},corner=60:120}, {8/p={1.7,-1}}, {9/p={-0.5,-2},corner=55:140,corner=-45:-135}, {10/p={0.8,-1.6},corner=30:150}}
\atoms{void}{{x/p=3,omega}, {y/p=7, omega}}
\draw[bdcellulation] (0)edge[mark={arr,p=0.3}](1) (2)edge[mark={arr,p=0.3}](1) (3)edge[mark={arr,p=0.3}](2) (1)edge[mark={arr,p=0.3}](4) (2)edge[mark={arr,p=0.3}](5) (7)edge[mark={arr,p=0.3}](1) (7)edge[mark={arr,p=0.3}](2) (7)edge[mark={arr,p=0.3}](6) (9)edge[mark={arr,p=0.3}](7) (10)edge[mark={arr,p=0.3}](7) (6)edge[mark={arr,p=0.3}](0) (4)edge[mark={arr,p=0.3}](0) (5)edge[mark={arr,p=0.3}](4) (3)edge[mark={arr,p=0.3}](5) (8)edge[mark={arr,p=0.3}](3) (10)edge[mark={arr,p=0.3}](8) (10)edge[mark={arr,p=0.3}](9) (6)edge[mark={arr,p=0.3}](9);
\draw[bdcellulation] (0)edge[mark={arr,-}]++(-150:0.6) (0)edge[mark=arr]++(120:0.6) (4)edge[mark={arr,-}]++(90:0.6) (5)edge[mark=arr]++(90:0.6) (5)edge[mark=arr]++(30:0.6) (3)edge[mark=arr]++(0:0.6) (8)edge[mark={arr,-}]++(45:0.6) (8)edge[mark={arr,-}]++(-45:0.6) (10)edge[mark=arr]++(-30:0.6) (9)edge[mark=arr]++(-135:0.6) (9)edge[mark=arr]++(-45:0.6) (6)edge[mark=arr]++(150:0.6) (6)edge[mark=arr]++(-135:0.6);
\draw[spin structure] (7)--(2)--(1)--(4)--++(90:0.6) (3)--(2)--(5)--++(30:0.6) (9)--(6)--++(-135:0.6) (9)--++(-45:0.6);
\atoms{bdvertex}{{0/p={0,-0.4},lab={t=$\alpha$,p=30:0.25}}}
\draw[bdribbon,mark=arr,rc] (0)--++(-160:0.7)--++(-90:0.7)--++(-150:0.8);
\draw[bdribbon,mark=arr,rc] (0)--++(-20:0.9)--++(20:1.1);
\draw[bdribbon,mark=arr,rc] (0)--++(90:2);
\end{tikzpicture}
\\
=
(-1)^{|\alpha|}
\begin{tikzpicture}
\atoms{bdcellvertex}{{0/p={-1.5,0},corner=0:45}, 1/p={-0.5,0}, {2/p={0.5,0},corner=90:180}, {3/p={1.5,0},corner=135:180}, 4/p={-0.5,1}, 5/p={0.5,1}, {6/p={-1.2,-1},corner=0:100}, {7/p={0,-1},corner=60:120}, {8/p={1.7,-1}}, {9/p={-0.5,-2},corner=55:140,corner=-45:-135}, {10/p={0.8,-1.6},corner=30:150}}
\atoms{void}{{x/p=3,omega}, {y/p=7, omega}}
\draw[bdcellulation] (0)edge[mark={arr,p=0.3}](1) (2)edge[mark={arr,p=0.3}](1) (3)edge[mark={arr,p=0.3}](2) (1)edge[mark={arr,p=0.3}](4) (2)edge[mark={arr,p=0.3}](5) (7)edge[mark={arr,p=0.3}](1) (7)edge[mark={arr,p=0.3}](2) (7)edge[mark={arr,p=0.3}](6) (9)edge[mark={arr,p=0.3}](7) (10)edge[mark={arr,p=0.3}](7) (6)edge[mark={arr,p=0.3}](0) (4)edge[mark={arr,p=0.3}](0) (5)edge[mark={arr,p=0.3}](4) (3)edge[mark={arr,p=0.3}](5) (8)edge[mark={arr,p=0.3}](3) (10)edge[mark={arr,p=0.3}](8) (10)edge[mark={arr,p=0.3}](9) (6)edge[mark={arr,p=0.3}](9);
\draw[bdcellulation] (0)edge[mark={arr,-}]++(-150:0.6) (0)edge[mark=arr]++(120:0.6) (4)edge[mark={arr,-}]++(90:0.6) (5)edge[mark=arr]++(90:0.6) (5)edge[mark=arr]++(30:0.6) (3)edge[mark=arr]++(0:0.6) (8)edge[mark={arr,-}]++(45:0.6) (8)edge[mark={arr,-}]++(-45:0.6) (10)edge[mark=arr]++(-30:0.6) (9)edge[mark=arr]++(-135:0.6) (9)edge[mark=arr]++(-45:0.6) (6)edge[mark=arr]++(150:0.6) (6)edge[mark=arr]++(-135:0.6);
\draw[spin structure] (7)--(1)--(4)--++(90:0.6) (3)--(2)--(5)--++(30:0.6) (9)--(6)--++(-135:0.6) (9)--++(-45:0.6);
\atoms{bdvertex}{{0/p={0,-0.4},lab={t=$\alpha$,p=30:0.25}}}
\draw[bdribbon,mark=arr,rc] (0)--++(-160:0.7)--++(-90:0.7)--++(-150:0.8);
\draw[bdribbon,mark=arr,rc] (0)--++(-20:0.9)--++(20:1.1);
\draw[bdribbon,mark=arr,rc] (0)--++(90:2);
\end{tikzpicture}\;.
\end{multlined}
\end{equation}
As in the bosonic case, the tensors also depend on the orientation via Hermiticity/unitarity. The fusion index inside of a boundary 2-cell $f$ is an output index if $f$ is identified with its standard representative, and an input index if $f$ is identified with the orientation-reversed partner thereof. As a consequence, instead of just complex conjugation, we also need to invert the index ordering in the Hermiticity condition.

Having defined fermionic tensors and their contraction, as well as manifolds with spin structures and their gluing operations, the definition of a fermionic Lagrangian module is a straight-forward modification of the bosonic case in Section~\ref{sec:tensorial_umtc_module}. First, instead of ribbon manifolds, we take ribbon manifolds with spin structure inside the boundary, second, instead of tensors we take fermionic tensors, and third, we impose the spin-statistics relation.

Despite the direct analogy between the bosonic and fermionic definitions, the instances of fermionic UMTCs show a new qualitative phenomenon which is not present in the bosonic case. There, the dimension of a ``trivial'' boundary fusion vertex,
\begin{equation}
\begin{tikzpicture}
\atoms{bdvertex}{0/}
\draw[bdribbon] (0)edge[mark=arr, mark={slab=$a$}]++(0.7,0) (0)edge[mark={arr,-}, mark={slab=$b$,r}]++(-0.7,0);
\end{tikzpicture}\;,
\end{equation}
is $\delta_{a,b}$, and we can add such a vertex to an $a$ ribbon at a cost of $(d_a/D)^{1/2}$. In the fermionic case, the fusion super-dimension is still $0|0$ if $a\neq b$, but it can be either $1|0$ or $1|1$ if $a=b$. In Ref.~\cite{Aasen2017}, $a$ is referred to as \emph{$m$-type} in the former, and \emph{$q$-type} in the latter case. As we argue at the end of Appendix~\ref{sec:umtc_tensorial_tqft}, this is due to the fact that there are two irreducible super-algebras, the trivial one and the Clifford algebra $Cl_1$, as opposed to only the trivial algebra in the bosonic case. Physically speaking, it is due to the existence of a non-trivial invertible fermionic phase in $1+1$ dimensions, the Kitaev chain.

\subsection{The invertible-boundary state-sum and gLU circuit}
\label{sec:fermionic_state_sum}
The invertible fermionic-Lagrangian-algebra CYWW boundary is a straight-forward fermionic analog of the bosonic case in Section~\ref{sec:boundary_state_sum}, taking into account the three differences, fermionic tensors, spin structure, and spin-statistics relation. For the fermionic tensors, the CYWW boundary 3-cell tensors are fermionic tensors, and the fusion indices shared among them can have a non-zero odd-parity dimension. All indices contracted inside the bulk or between bulk and boundary are purely even-parity. Note that in every discrete fermionic partition function, the fermion parity of a fixed configuration of indices/state-sum variables yields a $\zz_2$-valued simplicial $n-1$-cocycle (or 1-cycle) $a$, and the permutation of Grassmann variables in the evaluation of the fermionic tensor-network gives rise to a \emph{reordering sign} $\sigma[a]$ \cite{Gaiotto2015}, which we discuss in Appendix~\ref{sec:reordering_selflinking}.

The spin structure needs to be defined within the 3-dimensional physical-boundary cellulation, which again can be done using simplicial/cellular cohomology. This time, $\omega_2$ is a 1-cycle and $\eta_2$ a 2-chain with $\delta \eta_2=\omega_2$.
The formula for $\omega_2$ on an edge depends on a choice of a spin structure $\eta_2^v$ for the boundary of every volume (representative) $v$ and an orientation $\eta_1^f$ for the boundary of every face (representative) $f$,
\begin{equation}
\label{eq:3d_spin_structure}
\omega_2(e) = \sum_{e\subset v\in C_3} \eta_2^v(e) + \sum_{e\subset f\in C_2} \eta_1^f(e) + 1\;.
\end{equation}

For the spin-statistics relation, the spin structure and the state-sum have to interact in the following way. For every boundary 2-cell $f$ in configuration $a_f$, we get a sign
\begin{equation}
(-1)^{|a_f|\eta_2(f)}
\end{equation}
in addition to the contraction of the fermionic tensor network.

Let us discuss how the spin-statistics relation for Lagrangian modules is consistent with the spin-statistics relation for the fermionic CYWW boundary state-sum. The state-sum tensor $\calb(R(v))$ on a boundary volume $v$ depends on a spin structure on $\partial v$, and we choose this spin structure to be the same as $\eta_2^v$ in Eq.~\eqref{eq:3d_spin_structure}. Furthermore, whether a fermionic index of $
\calb(R(v))$ at a face $f$ of $v$ is input or output depends on an orientation of $f$, and we choose this orientation to be the same as $\eta_1^f$ in Eq.~\eqref{eq:3d_spin_structure}. Now, imagine changing $\eta_2^v$ on a boundary 3-cell $v$ by $\eta_2^v = \eta_2^v+df$, for a boundary 2-cell $f$. Then $\omega_2$ of the state-sum changes by $\omega_2'=\omega_2+df$, which can be fixed by changing $\eta_2'=\eta_2+f$. So both the state-sum and the Lagrangian-module tensor associated to $v$ change by $(-1)^{|a_f|}$. In this sense, the state-sum is independent of the choice of $\eta_2^v$.

When evaluating the CYWW boundary state-sum on a 4-manifold with physical and space boundary as in Section~\ref{sec:cyww_boundary_evaluation}, we apply the 2-handle attachment in Eq.~\eqref{eq:loop_2handle_attachment} to a non-contractible ribbon loop around each boundary edge $e$, in order to get from the ribbon manifold $R_6$ to $R_7$. For this to be possible, we must have a bounding spin structure on the ribbon loop around $e$, which is guaranteed by the spin-statistics relation of the state-sum. Whether the ribbon loop is bounding or not is given by the value of $\omega_2$ on that edge, and having $\eta_2(f)=1$ for $f$ a face containing $e$ effectively changes the spin structure from bounding to non-bounding. Since $d\eta_2=\omega_2$, the spin structure at the ribbon loop around $e$ is always bounding. Thus, the evaluation on a 4-manifold cellulation $Y$ with space boundary $\partial Y$ is simply $\calb(R(\partial Y))$ as in the bosonic case.

Note that also the definition of what it means for a boundary to be invertible has to change in the fermionic case, by equipping the physical boundary in the moves in Eq.~\eqref{eq:boundary_invertibility_moves} with a spin structure. For $M_2$ in Eq.~\eqref{eq:boundary_invertibility_moves}, the physical boundary on both sides is
\begin{equation}
\label{eq:invertibility_m2_boundary}
S_1\times B_2\leftrightarrow B_2\times S_1\;.
\end{equation}
The common space boundary $S_1\times S_1$ of this physical boundary can have four spin structures since each circle can be bounding or non-bounding. Eq.~\eqref{eq:invertibility_m2_boundary} can only hold for the bounding times bounding spin structure as each circle is the boundary of $B_2$ on either side. The other moves $M_i$ only allow for a unique spin structure of the space boundary.

The gLU circuit disentangling the CYWW model resulting from the invertible boundary is a direct fermionic analog of the bosonic case in Section~\ref{sec:gLU_circuit} as well. The initial ground state is a completely bosonic state. However, the gLU circuit locally creates pairs of odd-parity configurations in the beginning, which are then acted on in intermediate steps and annihilated in the end. It should be noted though, that the disentangling circuit depends on a microscopic choice of combinatorial spin structure, and is in particular only defined on 3-manifolds that admit a spin structure.

\subsection{Examples: Three-fermion and Ising UMTCs}
\label{sec:umtc_module_examples}
In this section, we will discuss two simple examples for fermionic Lagrangian modules, namely for the three-fermion UMTC as well as the Ising UMTC, giving rise to invertible boundaries of the corresponding CYWW models.

\paragraph{Three-fermion UMTC}
Let us start with the three-fermion UMTC, one of the simplest UMTCs that are not a Drinfeld center. It has four bulk ribbon labels, $1$, $f_1$, $f_2$, $f_3$ whose quantum dimensions are all $1$, so $D=2$. The fusion rules are $\zz_2\times \zz_2$, i.e., the dimension at a (3-valent) fusion vertex is $1$ if there is an even number of adjacent $f_1$ or $f_3$ labels and an even number of $f_2$ or $f_3$ labels, and $0$ otherwise.

As in Section~\ref{sec:toric_code_boundary}, the amplitude for a labeled ribbon manifold is determined by the amplitudes of ribbon 3-spheres. The latter can be reduced to a trivial network as follows. We first remove all $1$-labeled ribbon segments, resolve every $f_3$-labeled ribbon into a pair of $f_1$ and $f_2$ ribbon,
\begin{equation}
\begin{tikzpicture}
\draw (0,0)edge[mark={slab=$f_3$},mark=arr](1,0);
\end{tikzpicture}
\rightarrow
\begin{tikzpicture}
\draw (0,0)edge[mark={slab=$f_1$},mark=arr]++(1,0) (0,-0.3)edge[mark={slab=$f_2$,r},mark=arr]++(1,0);
\end{tikzpicture}\;,
\end{equation}
and also resolve every fusion vertex into uninterrupted ribbons, e.g.,
\begin{equation}
\begin{tikzpicture}
\atoms{vertex}{0/}
\draw (0)edge[mark={arr,-},mark={slab=$f_1$}]++(135:0.7) (0)edge[mark={arr,-},mark={slab=$f_2$,r}]++(45:0.7) (0)edge[mark={arr},mark={slab=$f_3$,r}]++(-90:0.7);
\end{tikzpicture}
\quad\rightarrow\quad
(2^{1/2})
\begin{tikzpicture}
\draw[mark={arr,-,p=0.7},mark={slab=$f_2$,p=0.3},rc] (-0.15,0)--++(0,0.7)--++(45:1);
\draw[mark={arr,-,p=0.7},mark={slab=$f_1$,p=0.3,r},rc,front] (0.15,0)--++(0,0.7)--++(135:1);
\end{tikzpicture}\;.
\end{equation}
Different choices of resolution yield different but gauge equivalent UMTCs. We obtain a collection of $f_1$ and $f_2$ ribbon loops which are disjoint but might be linked with another and twisted. We first drop the ribbon directions, and then unlink the loops and remove the twists by
\begin{equation}
\begin{gathered}
\begin{tikzpicture}
\draw (0,0)edge[mark={slab=$f_1$,p=0.7,r,sideoff=-0.1}](1,1.5);
\draw (1,0)edge[front, mark={slab=$f_1$,p=0.7,sideoff=-0.1}](0,1.5);
\end{tikzpicture}
=
\begin{tikzpicture}
\draw (1,0)edge[mark={slab=$f_1$,p=0.7,sideoff=-0.1}](0,1.5);
\draw (0,0)edge[front,mark={slab=$f_1$,p=0.7,r,sideoff=-0.1}](1,1.5);
\end{tikzpicture}
\;,
\\
\begin{tikzpicture}
\draw[looseness=2] (0,0)edge[out=0,in=0](0.8,1) (0.8,1)edge[out=180,in=180,mark={slab=$f_1$,s},front](1.6,0);
\end{tikzpicture}
=
(-1)
\begin{tikzpicture}
\draw (0,0)edge[mark={slab=$f_1$}](1,0);
\end{tikzpicture}\;,
\end{gathered}
\end{equation}
the same for $f_2$, as well as
\begin{equation}
\begin{tikzpicture}
\draw (0,0)edge[mark={slab=$f_2$,p=0.7,r,sideoff=-0.1}](1,1.5);
\draw (1,0)edge[front, mark={slab=$f_1$,p=0.7,sideoff=-0.1}](0,1.5);
\end{tikzpicture}
=
(-1)
\begin{tikzpicture}
\draw (1,0)edge[mark={slab=$f_1$,p=0.7,sideoff=-0.1}](0,1.5);
\draw (0,0)edge[front,mark={slab=$f_2$,p=0.7,r,sideoff=-0.1}](1,1.5);
\end{tikzpicture}\;.
\end{equation}
This corresponds to the fact that the full braiding between $f_1$ and $f_2$ as well as the twists of $f_1$ and $f_2$ are all $-1$. In the end, we can remove unlinked and untwisted loops, and the empty 3-sphere has amplitude $1/D$.

The Lagrangian module is given as follows. There are two boundary ribbon labels, $1$ and $\beta$ with quantum dimension $1$, so $D^\calb=\sqrt2$. The super-dimension of the boundary fusion vertices is $1|0$ if an even number of adjacent boundary ribbon labels are $\beta$, and $0|0$ otherwise. 
The non-zero super-dimensions of the boundary fusion vertices with one bulk and one boundary ribbon are given by
\begin{equation}
\begin{multlined}
\begin{tikzpicture}
\atoms{bdvertex}{0/lab={t=$1|0$,p=90:0.25}}
\draw (0)edge[bdribbon,mark={arr},mark={slab=$1$}]++(0.8,0) (0)edge[mark={arr,-},mark={slab=$1$,r}]++(-0.8,0);
\end{tikzpicture}
\;,
\begin{tikzpicture}
\atoms{bdvertex}{0/lab={t=$1|0$,p=90:0.25}}
\draw (0)edge[bdribbon,mark={arr},mark={slab=$\beta$}]++(0.8,0) (0)edge[mark={arr,-},mark={slab=$f_1$,r}]++(-0.8,0);
\end{tikzpicture}
\;,\\
\begin{tikzpicture}
\atoms{bdvertex}{0/lab={t=$0|1$,p=90:0.25}}
\draw (0)edge[bdribbon,mark={arr},mark={slab=$1$}]++(0.8,0) (0)edge[mark={arr,-},mark={slab=$f_2$,r}]++(-0.8,0);
\end{tikzpicture}
\;,
\begin{tikzpicture}
\atoms{bdvertex}{0/lab={t=$0|1$,p=90:0.25}}
\draw (0)edge[bdribbon,mark={arr},mark={slab=$\beta$}]++(0.8,0) (0)edge[mark={arr,-},mark={slab=$f_3$,r}]++(-0.8,0);
\end{tikzpicture}\;.
\end{multlined}
\end{equation}
So physically speaking, we ``condense'' the $f_2$ anyon, but we do this with an odd fermion parity, in other words, condensing $f_2$ yields a fundamental fermion. This is equivalent to what is known as \emph{fermion condensation} \cite{Aasen2017}.

The amplitude for a ribbon network inside a 3-ball is again obtained by reducing it to a trivial network, which in the bulk is done via the procedure described above. At the boundary, we neglect $1$ ribbons, ribbon directions, and push $\beta$ ribbons slightly into the bulk,
\begin{equation}
\begin{tikzpicture}
\draw[bdribbon] (0,0)edge[mark={slab=$\beta$}](1,0);
\end{tikzpicture}
\rightarrow
\begin{tikzpicture}
\draw (0,0)edge[mark={slab=$f_1$}](1,0);
\end{tikzpicture}\;,
\end{equation}
and accordingly the condensation vertices,
\begin{equation}
\begin{tikzpicture}
\atoms{bdvertex}{0/}
\draw (0)edge[bdribbon]++(0.8,0) (0)edge[mark={slab=$f_1$,r}]++(-0.8,0);
\end{tikzpicture}
\rightarrow
(2^{1/2})
\begin{tikzpicture}
\draw (0,0)edge[mark={slab=$f_1$}]++(1,0);
\end{tikzpicture}\;.
\end{equation}
Applying then the procedure in the bulk, we end up with only a collection of $f_2$ segments ending at two odd-parity boundary vertices. Those segments can be removed, which however depends on the fermionic index ordering and the spin structure within the boundary. To be concrete, we choose a cellulation of the boundary where each odd boundary vertex is contained inside a 2-gon cell with acyclic edge orientations. A ribbon segment can be removed if the 2-gons containing its boundary vertices share an edge that is not part of $\eta$,
\begin{equation}
\label{eq:fermion_segment_removal}
\begin{tikzpicture}
\atoms{bdvertex}{0/lab={t=$\textcolor{\spincol}{0}$,p=90:0.25}, {1/p={1,0},lab={t=$\textcolor{\spincol}{1}$,p=90:0.25}}}
\atoms{bdcellvertex}{a/p={0.5,-0.6}, b/p={0.5,0.6}}
\draw (0)edge[bend right=20,mark={slab=$f_2$,p=0.3}](1);
\draw[bdcellulation] (a)edge[bend left=80,mark=arr,looseness=2](b) (a)edge[bend right=90,mark=arr,looseness=2](b) (a)edge[mark={arr,p=0.3}](b);
\end{tikzpicture}
=
\hspace{2cm}\;,
\end{equation}
where the fermionic index ordering was indicated by the green colored labels.
Such a configuration can be achieved by local recellulations on the boundary, and local changes of $\eta$ taking into account the spin-statistics relation,
\begin{equation}
\begin{tikzpicture}
\atoms{bdvertex}{0/}
\atoms{bdcellvertex}{a/p={0,-0.6}, b/p={0,0.6}}
\draw (0)edge[mark={slab=$f_2$,p=0.8}]++(180:1);
\draw[bdcellulation] (a)edge[bend left=60,mark={arr,p=0.4}](b) (a)edge[bend right=60,mark=arr](b);
\end{tikzpicture}
=
(-1)
\begin{tikzpicture}
\atoms{bdvertex}{0/}
\atoms{bdcellvertex}{a/p={0,-0.6}, b/p={0,0.6}}
\draw (0)edge[mark={slab=$f_2$,p=0.8}]++(180:1);
\draw[bdcellulation] (a)edge[bend left=60,mark={arr,p=0.4}](b) (a)edge[bend right=60,mark=arr](b);
\draw[spin structure] (a)edge[bend left=60,mark={arr,p=0.4}](b) (a)edge[bend right=60,mark=arr](b);
\end{tikzpicture}
\;.
\end{equation}

The odd fermion parity together with the spin structure is vital for the Lagrangian module to obey the gluing axioms. E.g., consider the following consistency condition,
\begin{equation}
\label{eq:three_fermion_boundary_consistency}
\begin{multlined}
\begin{tikzpicture}
\atoms{bdvertex,postdecstyle=\spincol}{0/lab={t=$0$,p=180:0.25}, {1/p={1,0},lab={t=$1$,p=0:0.25}}}
\draw (0)edge[mark={slab=$f_2$}](1);
\end{tikzpicture}
=
(-1)
\begin{tikzpicture}
\atoms{bdvertex,postdecstyle=\spincol}{0/lab={t=$0$,p=180:0.25}, {1/p={2,0},lab={t=$1$,p=0:0.25}}}
\draw[looseness=2,mark={slab=$f_2$}] (0)--++(0.3,0)to[out=0,in=0](1,1)to[out=180,in=180] ($(1)+(-0.3,0)$)--(1);
\end{tikzpicture}
\\
=
(-1)
\begin{tikzpicture}
\atoms{bdvertex,postdecstyle=\spincol}{0/lab={t=$0$,p=180:0.25}, {1/p={2,0},lab={t=$1$,p=0:0.25}}}
\draw (0)edge[mark={slab=$f_2$}](1);
\draw[spin structure] (1)circle(0.5);
\end{tikzpicture}
=
\begin{tikzpicture}
\atoms{bdvertex,postdecstyle=\spincol}{0/lab={t=$0$,p=180:0.25}, {1/p={1,0},lab={t=$1$,p=0:0.25}}}
\draw (0)edge[mark={slab=$f_2$}](1);
\end{tikzpicture}\;.
\end{multlined}
\end{equation}
We start with an $f_2$ segment, which is then twisted in the bulk yielding a factor of $(-1)$. Since the $f_2$ line terminates on the boundary, the twist can be undone by a topology-preserving deformation. However, when the boundary is cellulated, this deformation has to be implemented by local recellulation moves that effectively rotate the 2-gons containing the two fusion vertices against each other. No matter how we perform those moves, this will create a pair of $\omega_2$ points, move one around the ball between the fusion vertices, and annihilate them again (or something equivalent in $\zz_2$ cohomology). That is, by untwisting we create an $\eta$ loop separating the two fusion vertices, and pulling this $\eta$ loop through either the fusion vertex on the left or right yields another factor of $-1$. Without spin structure and odd fermion parity this would yield an inconsistency, corresponding to the fact that we can only condense bosons, i.e., anyons with trivial spin.

Furthermore, the spin structure is immediately necessary to cancel reordering signs in the equations coming from symmetries of the ribbon manifolds. E.g., a 3-ball with $f_2$ ribbon going from north to south pole (which are right and left in the diagrams below) is symmetric under a $\pi$ rotation around an axis going through the equator. This rotation changes the ordering of the two odd-parity boundary fusion vertices on the poles, but also introduces an $\eta$ line around the equator. The two contributions cancel each other,
\begin{equation}
\begin{multlined}
\begin{tikzpicture}
\atoms{bdvertex,postdecstyle=\spincol}{0/lab={t=$0$,p=180:0.25}, {1/p={1.5,0},lab={t=$1$,p=0:0.25}}}
\draw (0)edge[mark={slab=$f_2$}](1);
\end{tikzpicture}
=
\begin{tikzpicture}
\atoms{bdvertex,postdecstyle=\spincol}{0/lab={t=$1$,p=180:0.25}, {1/p={1.5,0},lab={t=$0$,p=0:0.25}}}
\draw (0)edge[mark={slab=$f_2$}](1);
\draw[spin structure] (1)circle(0.5);
\end{tikzpicture}
\\
=
(-1)
\begin{tikzpicture}
\atoms{bdvertex,postdecstyle=\spincol}{0/lab={t=$0$,p=180:0.25}, {1/p={1.5,0},lab={t=$1$,p=0:0.25}}}
\draw (0)edge[mark={slab=$f_2$}](1);
\draw[spin structure] (1)circle(0.5);
\end{tikzpicture}
=
\begin{tikzpicture}
\atoms{bdvertex,postdecstyle=\spincol}{0/lab={t=$0$,p=180:0.25}, {1/p={1.5,0},lab={t=$1$,p=0:0.25}}}
\draw (0)edge[mark={slab=$f_2$}](1);
\end{tikzpicture}\;.
\end{multlined}
\end{equation}
That $\eta$ behaves this way can be easily seen by choosing a triangulation/cellulation of the sphere. E.g., if we choose a cellulation that is symmetric under the $\pi$ rotation, then we get two $\omega_2=1$ vertices at the two points where the rotation axis ends at the equator, and $\eta$ has to connect those two vertices via one of two sides. Then when we rotate, this side changes such that we add an $\eta$ loop around the equator.
%Another option would be to choose a pole-axis rotation-symmetric cellulation of the upper hemisphere and then glue it with an equator-axis rotated copy at the lower hemisphere. When doing this the edge directions of the equatorial edges do not match, and we have to choose the directions of, say, the north hemisphere. When we rotate by $\pi$ around the equatorial axis, the direction of the equatorial edges changes, and when changing it back we move a $\omega_2=1$ vertex from one side to the other, adding $\eta$ on every equatorial edge.

Note that the spin structure does \emph{not} enter in the consistency under mirroring at the equatorial plane. When doing so, the Hermiticity/unitarity condition tells us that we have to invert the index ordering (in addition to complex conjugation).

We will revisit the same invertible boundary of the three-fermion CYWW model from a totally different perspective in Appendix~\ref{sec:3fermion_cohomology}, namely as a (super-)cohomology simplicial gauge theory.

\paragraph{Ising UMTC}
Now let us consider the Ising UMTC, with three ribbon labels, $1$, $\sigma$, $\psi$ with quantum dimensions $d_1=d_\psi=1$ and $d_\sigma=\sqrt2$, so $D=2$. In fact, ``Ising UMTC'' refers to a family of $8$ different but very similar UMTCs with the same fusion rules, parametrized by an odd element $\nu\in \zz_{16}$ \cite{Kitaev2005}. The fusion dimensions of 3-valent vertices are $1$ if the adjacent ribbons have labels $\psi,\psi,1$, or $\sigma,\sigma,1$, or $\sigma,\sigma,\psi$, and $0$ otherwise.

The amplitude for a labeled ribbon 3-sphere (determining any other ribbon manifold amplitude) is obtained as follows. First we remove all $1$-labeled ribbons, drop the direction of $\psi$ ribbons, and resolve fusion vertices, e.g.,
\begin{equation}
\begin{tikzpicture}
\atoms{vertex}{0/}
\draw (0)edge[mark={arr},mark={slab=$\sigma$}]++(0.8,0) (0)edge[mark={arr},mark={slab=$\sigma$,r}]++(-0.8,0) (0)edge[mark={slab=$\psi$,r}]++(0,0.8);
\end{tikzpicture}
\rightarrow
(2^{-1/4})
\begin{tikzpicture}
\atoms{vertex}{0/,1/p={0.6,0}}
\draw (0)edge[mark={arr,-},mark={slab=$\sigma$}](1) (0)edge[mark={arr},mark={slab=$\sigma$,r}]++(-0.8,0) (1)edge[fav=s,mark={arr},mark={slab=$\sigma$}]++(0.6,0) (0)edge[mark={slab=$\psi$,r}]++(0,0.8);
\end{tikzpicture}\;,
\end{equation}
such that they are all of the following links,
\begin{equation}
\begin{tikzpicture}
\atoms{vertex}{0/}
\draw (0)edge[mark={arr,-},mark={slab=$\sigma$}]++(0.8,0) (0)edge[mark={arr},mark={slab=$\sigma$,r}]++(-0.8,0) (0)edge[mark={slab=$\psi$,r}]++(0,0.8);
\end{tikzpicture}\;,
\begin{tikzpicture}
\atoms{vertex}{0/}
\draw (0)edge[fav=s,mark={arr},mark={slab=$\sigma$}]++(0.8,0) (0)edge[mark={arr},mark={slab=$\sigma$,r}]++(-0.8,0);
\end{tikzpicture}
\end{equation}
and their orientation-reversed partners.

The ribbon network will then be a collection of $\sigma$ loops, $\psi$ loops, and $\psi$ segments connecting different vertices on $\sigma$ loops. We can reduce the labeled network to the trivial network by the following steps. First, we can fuse pairs of $\sigma$ loops to a single loop if they are connected by any $\psi$ segments using the following move,
\begin{equation}
\begin{tikzpicture}
\atoms{vertex}{0/, 1/p={0,0.8}}
\draw (0)edge[mark={arr,-},mark={slab=$\sigma$}]++(0.8,0) (0)edge[mark={arr},mark={slab=$\sigma$,r}]++(-0.8,0) (1)edge[mark={arr},mark={slab=$\sigma$}]++(0.8,0) (1)edge[mark={arr,-},mark={slab=$\sigma$,r}]++(-0.8,0) (0)edge[mark={slab=$\psi$,r}](1);
\end{tikzpicture}
=
\begin{tikzpicture}
\atoms{vertex}{0/, 1/p={0.8,0}}
\draw (0)edge[mark={arr,-},mark={slab=$\sigma$}]++(0,0.8) (0)edge[mark={arr},mark={slab=$\sigma$,r}]++(0,-0.8) (1)edge[mark={arr},mark={slab=$\sigma$}]++(0,0.8) (1)edge[mark={arr,-},mark={slab=$\sigma$,r}]++(0,-0.8) (0)edge[mark={slab=$\psi$,r}](1);
\end{tikzpicture}
+
(2^{1/2})
\begin{tikzpicture}
\draw (0,0)edge[mark={arr,-},mark={slab=$\sigma$}]++(0,0.8) (0.8,0)edge[mark={arr},mark={slab=$\sigma$}]++(0,0.8);
\end{tikzpicture}\;.
\end{equation}
We do this until for every $\psi$ segment its two endpoints are on the same $\sigma$ loop.
Pairs of direction-changing vertices on the $\sigma$ loops can be removed by
\begin{equation}
\label{eq:ising_pair_creation}
\begin{tikzpicture}
\atoms{vertex}{0/,1/p={0.6,0}}
\draw (0)edge[fav=e,fav=s,mark={arr,-},mark={slab=$\sigma$}](1) (0)edge[mark={arr,-},mark={slab=$\sigma$,r}]++(-0.6,0) (1)edge[mark={arr},mark={slab=$\sigma$}]++(0.6,0);
\end{tikzpicture}
=
(2^{1/2})
\begin{tikzpicture}
\draw (0,0)edge[mark=arr,mark={slab=$\sigma$}](0.8,0);
\end{tikzpicture}\;.
\end{equation}
For this we might first need to change the favorite ribbon of one vertex using
\begin{equation}
\begin{tikzpicture}
\atoms{vertex}{0/}
\draw (0)edge[fav=s,mark={arr},mark={slab=$\sigma$}]++(0.8,0) (0)edge[mark={arr},mark={slab=$\sigma$,r}]++(-0.8,0);
\end{tikzpicture}
=
(\kappa_\sigma)
\begin{tikzpicture}
\atoms{vertex}{0/}
\draw (0)edge[mark={arr},mark={slab=$\sigma$}]++(0.8,0) (0)edge[fav=s,mark={arr},mark={slab=$\sigma$,r}]++(-0.8,0);
\end{tikzpicture}\;,
\end{equation}
where
\begin{equation}
\kappa_\sigma=(-1)^{\nu^2-1}=\pm 1
\end{equation}
is the \emph{Frobenius-Schur indicator} of $\sigma$. Furthermore, $\psi$ segments can be moved across direction-changing lines,
\begin{equation}
\begin{tikzpicture}
\atoms{vertex}{0/,1/p={0.6,0}}
\draw (0)edge[fav=e,mark={arr,-},mark={slab=$\sigma$}](1) (0)edge[mark={arr},mark={slab=$\sigma$,r}]++(-0.8,0) (1)edge[mark={arr},mark={slab=$\sigma$}]++(0.6,0) (0)edge[mark={slab=$\psi$,r}]++(0,0.8);
\end{tikzpicture}
=
\begin{tikzpicture}
\atoms{vertex}{0/,1/p={0.6,0}}
\draw (0)edge[mark={arr},mark={slab=$\sigma$}](1) (0)edge[fav=s,mark={arr},mark={slab=$\sigma$,r}]++(-0.8,0) (1)edge[mark={arr},mark={slab=$\sigma$}]++(0.6,0) (1)edge[mark={slab=$\psi$,r}]++(0,0.8);
\end{tikzpicture}\;.
\end{equation}
We can change the side of a $\sigma$ ribbon where $\psi$ is attached,
\begin{equation}
\begin{tikzpicture}
\atoms{vertex}{0/}
\draw (0)edge[mark={arr,-},mark={slab=$\sigma$,r}]++(-0.8,0) (0)edge[mark={arr},mark={slab=$\sigma$}]++(0.8,0) (0)edge[mark={slab=$\psi$,r}]++(0,0.8);
\end{tikzpicture}
=
(-i^\nu)
\begin{tikzpicture}
\atoms{vertex}{0/}
\draw (0)edge[mark={arr,-},mark={slab=$\sigma$,r}]++(-0.8,0) (0)edge[mark={arr,p=0.3},mark={slab=$\sigma$,p=0.3}]++(0.8,0);
\draw[front,mark={slab=$\psi$,r,p=0.8}] (0)to[out=-90,in=-90,looseness=2]++(0.5,0)--++(0,0.8);
\end{tikzpicture}\;.
\end{equation}

Finally we can detach pairs of $\psi$ vertices from the $\sigma$ loops via
\begin{equation}
\begin{tikzpicture}
\atoms{vertex}{0/, 1/p={0.8,0}}
\draw (0)edge[mark={arr,-},mark={slab=$\sigma$}]++(1) (1)edge[mark={arr,-},mark={slab=$\sigma$}]++(0.8,0) (0)edge[mark={arr},mark={slab=$\sigma$,r}]++(-0.8,0) (0)edge[mark={slab=$\psi$,r}]++(0,0.8) (1)edge[mark={slab=$\psi$,r}]++(0,0.8);
\end{tikzpicture}
=
(2^{1/2})
\begin{tikzpicture}
\draw (0,0)edge[mark={arr,-},mark={slab=$\sigma$}]++(1,0);
\draw (0,0.8)edge[mark={slab=$\psi$}]++(1,0);
\end{tikzpicture}
\;.
\end{equation}

After the previous steps, we obtain a collection of disjoint but possibly linked and twisted $\sigma$- and $\psi$ loops. The twists can be removed via
\begin{equation}
\begin{gathered}
\begin{tikzpicture}
\draw[looseness=2] (0,0)edge[out=0,in=0](0.8,1) (0.8,1)edge[out=180,in=180,mark={slab=$\psi$,s},front](1.6,0);
\end{tikzpicture}
=
(-1)
\begin{tikzpicture}
\draw (0,0)edge[mark={slab=$\psi$}](1,0);
\end{tikzpicture}
\;,
\\
\begin{tikzpicture}
\draw[looseness=2] (0,0)edge[out=0,in=0](0.8,1) (0.8,1)edge[mark={arr,p=0.1},out=180,in=180,mark={slab=$\sigma$,s},front](1.6,0);
\end{tikzpicture}
=
(\theta_\sigma)
\begin{tikzpicture}
\draw (0,0)edge[mark={arr},mark={slab=$\sigma$}](1,0);
\end{tikzpicture}\;,
\end{gathered}
\end{equation}
where
\begin{equation}
\theta_\sigma= e^{\frac{2\pi i}{16}\nu}
\end{equation}
is the \emph{topological twist} of $\sigma$.
Furthermore, any configuration with two linked $\sigma$ loops has amplitude $0$,
\begin{equation}
\begin{tikzpicture}
\draw[mark={arr,s},mark={slab=$\sigma$,e,r}] (0,0)arc(0:180:0.6);
\draw[front, mark=arr,mark={slab=$\sigma$,e,r}] (0.4,0)arc(0:360:0.6);
\draw[front] (0,0)arc(0:-180:0.6);
\end{tikzpicture}
=0\;.
\end{equation}
Finally, $\sigma$ and $\psi$ loops can be unlinked yielding a factor of $-1$,
\begin{equation}
\begin{tikzpicture}
\draw (0,0)edge[mark={arr,p=0.4},mark={slab=$\sigma$,p=0.7,r}](1,1.5);
\draw (1,0)edge[front, mark={slab=$\psi$,p=0.7}](0,1.5);
\end{tikzpicture}
=
(-1)
\begin{tikzpicture}
\draw (1,0)edge[mark={slab=$\psi$,p=0.7}](0,1.5);
\draw (0,0)edge[front,mark={arr,p=0.4},mark={slab=$\sigma$,p=0.7,r}](1,1.5);
\end{tikzpicture}\;.
\end{equation}
Unlinked $\psi$ loops can be removed, $\sigma$ loops can be removed yielding a factor of $\sqrt2$, and the amplitude of the empty $3$-sphere is $1/D$.

We can define a fermionic Lagrangian module for the Ising UMTC as follows. The boundary fermionic fusion category has been referred to as $C_2$ in Ref.~\cite{Aasen2017} and has two boundary ribbon labels $1$ and $\beta$, with quantum dimension $d^\calb_1=d^\calb_\beta=1$, and total quantum dimension $D^\calb=2^{1/2}$. \footnote{In some conventions, the quantum dimension of $\beta$ (and any other anyon with $Cl_1$ automorphism algebra) might include an extra factor of $2^{1/2}$. However, for Eq.~\eqref{eq:2handle_attachment} to hold, we do not include this factor.} The super-dimension at a 3-valent boundary fusion vertex is $1|1$ if two of the adjacent ribbon labels are $\beta$, $1|0$ if all are $1$, and $0|0$ otherwise. Accordingly, $\beta$ has a $Cl_1$ automorphism algebra. The non-zero super-dimensions of the boundary fusion indices with one bulk and one boundary anyon are given by
\begin{equation}
\begin{tikzpicture}
\atoms{bdvertex}{0/lab={t=$1|0$,p=90:0.25}}
\draw (0)edge[bdribbon,mark={arr},mark={slab=$1$}]++(0.8,0) (0)edge[mark={arr,-},mark={slab=$1$,r}]++(-0.8,0);
\end{tikzpicture}
\;,
\begin{tikzpicture}
\atoms{bdvertex}{0/lab={t=$0|1$,p=90:0.25}}
\draw (0)edge[bdribbon,mark={arr},mark={slab=$1$}]++(0.8,0) (0)edge[mark={arr,-},mark={slab=$\psi$,r}]++(-0.8,0);
\end{tikzpicture}
\;,
\begin{tikzpicture}
\atoms{bdvertex}{0/lab={t=$1|1$,p=90:0.25}}
\draw (0)edge[bdribbon,mark={arr},mark={slab=$\beta$}]++(0.8,0) (0)edge[mark={arr,-},mark={slab=$\sigma$,r}]++(-0.8,0);
\end{tikzpicture}\;.
\end{equation}

The amplitude for a ribbon graph inside a 3-ball can be obtained by reducing it to the trivial network. First, at the boundary, we neglect all $1$-labeled ribbons, denote odd-parity boundary fusion vertices by
\begin{equation}
\begin{tikzpicture}
\atoms{bdvertexodd}{0/}
\draw[bdribbon] (0)edge[mark={arr}]++(0.8,0) (0)edge[mark={arr,-}]++(-0.8,0);
\end{tikzpicture}\;,
\end{equation}
neglect such even-parity vertices at a cost of $2^{1/4}$, and resolve odd direction-changing vertices as
\begin{equation}
\begin{tikzpicture}
\atoms{bdvertexodd}{0/}
\draw[bdribbon] (0)edge[mark={arr}]++(0.8,0) (0)edge[mark={arr}]++(-0.8,0);
\end{tikzpicture}
\rightarrow
(2^{1/4})
\begin{tikzpicture}
\atoms{bdvertexodd}{0/}
\atoms{bdvertex}{0x/p={0.6,0}}
\draw[bdribbon] (0)edge[mark={arr,-}](0x) (0x)edge[mark={arr}]++(0.6,0) (0)edge[mark={arr}]++(-0.6,0);
\end{tikzpicture}\;,
\end{equation}
and odd condensation vertices as
\begin{equation}
\begin{tikzpicture}
\atoms{bdvertexodd}{0/}
\draw (0)edge[bdribbon,mark={arr}]++(0.8,0) (0)edge[mark={arr,-}]++(-0.8,0);
\end{tikzpicture}
\rightarrow
(2^{1/4})
\begin{tikzpicture}
\atoms{bdvertex}{0/}
\atoms{bdvertexodd}{0x/p={0.6,0}}
\draw (0)edge[bdribbon,mark={arr}](0x) (0x)edge[bdribbon,mark={arr}]++(0.6,0) (0)edge[mark={arr,-}]++(-0.6,0);
\end{tikzpicture}\;.
\end{equation}

Then we move $\beta$ ribbons from the boundary slightly into the bulk,
\begin{equation}
\begin{tikzpicture}
\draw[bdribbon] (0,0)edge[mark={arr},mark={slab=$\beta$}](1,0);
\end{tikzpicture}
\rightarrow
\begin{tikzpicture}
\draw (0,0)edge[mark={arr},mark={slab=$\sigma$}](1,0);
\end{tikzpicture}\;.
\end{equation}
Accordingly, also odd-parity-, direction-changing-, and condensation vertices will be pushed into the bulk,
\begin{equation}
\begin{gathered}
\begin{tikzpicture}
\atoms{bdvertexodd}{0/}
\draw[bdribbon] (0)edge[mark={arr}]++(0.8,0) (0)edge[mark={arr,-}]++(-0.8,0);
\end{tikzpicture}
\rightarrow
(2^{-1/2})
\begin{tikzpicture}
\atoms{vertex}{0/}
\atoms{bdvertex}{a/p={0,0.6}}
\draw (0)edge[mark={arr},mark={slab=$\sigma$}]++(0.8,0) (0)edge[mark={arr,-},mark={slab=$\sigma$,r}]++(-0.8,0) (0)edge[mark={slab=$\psi$,r}](a);
\end{tikzpicture}
\;,\\
\begin{tikzpicture}
\atoms{bdvertex}{0/}
\draw[bdribbon] (0)edge[fav=s,mark={arr}]++(0.8,0) (0)edge[mark={arr}]++(-0.8,0);
\end{tikzpicture}
\rightarrow
\begin{tikzpicture}
\atoms{vertex}{0/}
\draw (0)edge[fav=s,mark={arr},mark={slab=$\sigma$}]++(0.8,0) (0)edge[mark={arr},mark={slab=$\sigma$,r}]++(-0.8,0);
\end{tikzpicture}
\;,\\
\begin{tikzpicture}
\atoms{bdvertex}{0/}
\draw (0)edge[bdribbon,mark={arr}]++(0.8,0) (0)edge[mark={arr,-},mark={slab=$\sigma$,r}]++(-0.8,0);
\end{tikzpicture}
\rightarrow
(2^{1/4})
\begin{tikzpicture}
\draw (0,0)edge[mark={arr},mark={slab=$\sigma$}](1,0);
\end{tikzpicture}
\;.
\end{gathered}
\end{equation}
For the first replacement, we glue two pairs of vertices and apply one 1-surgery to the $\psi$ ribbon on the right instead of gluing two odd-parity vertices on the left. For the 1-surgery we need to include the normalization of $(d_\psi/D)^{1/2}=2^{-1/2}$ on the right.

If we now apply the procedure for the bulk, we will end up with a collection of $\psi$ segments each ending at two odd boundary vertices. Those $\psi$ segments can be removed depending on the spin structure as shown in Eq.~\eqref{eq:fermion_segment_removal} with $\psi$ instead of $f_2$.

Just as for the three-fermion Lagrangian module, the odd fermion parity for $\psi$ terminating at the boundary is needed for a consistency condition like Eq.~\eqref{eq:three_fermion_boundary_consistency}. Note that the fact that a 2-valent $\beta,\beta$ fusion vertex can have either even or odd parity is directly related to the fact that $\sigma$ and $\sigma$ can either fuse to $1$ or $\psi$.

\subsection{Invertible boundaries for a \texorpdfstring{$\zz_{16}$}{Z16} Witt subgroup}
\label{sec:z16_witt_boundaries}
In this section we argue that the fermionic invertible boundary of the Ising CYWW model gives rise to fermionic invertible boundaries for a large family of modular CYWW models, namely the Witt classes of Ising and products thereof. Two UMTCs $A$ and $B$ are said to be in the same \emph{Witt class} \cite{Davydov2010} if the UMTC $A\otimes \bar B$ has a Lagrangian module, or equivalently is a Drinfeld center. Two UMTCs in the same Witt class must have the same chiral central charge. Vice versa, many known examples of pairs of UMTCs with the same chiral central charge belong to the same Witt class, but there exist counter examples such as the semion and the $U(1)_4$ UMTCs. The Witt classes form a group under stacking, where the inverse of $A$ is the Witt class of the complex conjugate $\bar A$.

Now assume the UMTC $B$ has a Lagrangian module and $A$ is another UMTC in the same Witt class. Then we can combine the Lagrangian module of $A\otimes \bar B$ with the Lagrangian module of $B$, yielding a Lagrangian module for $A$. The same holds on the level of CYWW models, namely combining the invertible boundary for $A\otimes \bar B$ with the invertible boundary for $B$ yields an invertible boundary for $A$. More formally, consider a geometric/topological/combinatorial mapping replacing the boundary of $A$ by a thin strip of $\bar B$ bulk which on the one side has a $B$-boundary and on the other side the $A\otimes \bar B$-boundary together with $A$, for illustration in 2 dimensions lower,
\begin{equation}
\begin{multlined}
\begin{tikzpicture}
\fill[opacity=0.3] (0,0)--++(2,0)--++(60:1.5)--++(-2,0)--cycle;
\draw[manifoldboundary] (2,0)--++(60:1.5);
\draw[bdcol] (2.5,0.7)edge[<-,out=0,in=90,mark={lab={\breakcell{Boundary of\\$\text{CYWW}(A)$}},a}]++(0.6,-0.5);
\node[below] at (0.8,0){$\text{CYWW}(A)$};
\end{tikzpicture}
\\
\coloneqq
\begin{tikzpicture}
\fill[opacity=0.3] (0,0)--++(2,0)--++(60:1.5)--++(-2,0)--cycle;
\fill[opacity=0.3] (2,0)arc(-90:-45:0.3)coordinate(x)--++(60:1.5)arc(-45:-90:0.3)--cycle;
\fill[opacity=0.3] (x)arc(45:90:0.3)coordinate(y)--++(60:1.5)arc(90:45:0.3)--cycle;
\fill[opacity=0.3] (y)--++(-0.5,0)coordinate(z)--++(60:1.5)--++(0.5,0)--cycle;
\draw[manifoldboundary] (z)--++(60:1.5);
\draw[manifoldboundary] (x)--++(60:1.5);
\draw[bdcol] (2.6,0.6)edge[<-,out=0,in=-120,mark={lab={\breakcell{Boundary of\\$\text{CYWW}(A\otimes \bar B)$}},a}] (4,1);
\node[below] at (0.8,0){$\text{CYWW}(A)$};
\draw (1.8,0.15)edge[<-,out=-70,in=180,mark={lab={$\overline{\text{CYWW}(B)}$},a}]++(0.8,-0.3);
\draw[bdcol] (1.8,0.9)edge[<-,out=150,in=-90,mark={lab={\breakcell{Boundary of\\$\text{CYWW}(B)$}},a}]++(-0.6,0.5);
\end{tikzpicture}\;.
\end{multlined}
\end{equation}
Pulling back this mapping onto the level of state-sums/tensors, we obtain the invertible boundary of the $A$ CYWW model. This holds for ordinary Lagrangian modules as well as for fermionic ones. Note that this construction can be used both directly in $3+1$ dimensions for the CYWW models, but also in $2+1$ dimensions to get a Lagrangian module of $A$. All in all, we have found fermionic invertible boundaries of CYWW models for any UMTC in the same Witt class as a stack of copies of the Ising UMTC.

We find that $16$ copies of the Ising UMTC are again in a trivial Witt class, as it can be trivialized by condensing $15$ pairs of fermions $\psi_i\psi_{i+1}$ for $0\leq i<14$ between different copies, as well as the composite of all $\sigma$ particles $\sigma_0\sigma_1\cdots\sigma_{15}$. From the fact that the Ising UMTC has chiral central charge $c=\frac12$ we can infer that the Witt class cannot be trivial for any smaller number of copies. So the Witt classes generated by copies of Ising form a $\zz_{16}$ subgroup of the general Witt group. A set of simple representative UMTCs for those Witt classes are known under the name \emph{Kitaev $16$-fold way} \cite{Kitaev2005}. We discuss those UMTCs as well as the relation to \emph{fermion condensation} in more detail in Appendix~\ref{sec:condensation_appendix}.

\section{Conclusion}
In this work, we have derived two main results. First, we have shown that CYWW models whose input category is a Drinfeld center possess an invertible domain wall to vacuum, which can be reshaped into a disentangling gLU circuit. While we do not expect this result to be surprising to many experts in the field, an explicit proof seems to be unknown or unpublished so far. As a second, perhaps more surprising result, we have shown that if we allow for fermionic degrees of freedom, we can find invertible domain walls to vacuum not only for the trivial Witt class (i.e., Drinfeld centers), but for the Witt classes containing the Kitaev 16-fold way UMTCs. Those invertible domain walls yield disentangling gLU circuits with fermionic auxiliary degrees of freedom, which are however not uniform on arbitrary triangulations, but depend on a microscopic choice of spin structure.

Throughout this work and in the appendices, we have also established a general formalism of ``tensorial extended TQFT'', which allows us to give precise and simple geometric/topological interpretations/definitions to categorical structures related to the classification of phases of matter. This approach allows us to derive all results of this work from purely geometric/topological considerations, and makes it quite easy to derive similar results. For example, as we show in Appendix~\eqref{sec:general_braided}, adding an ``auxiliary'' $4$-dimensional bulk to our ribbon manifolds yields CYWW models for general braided fusion categories, and general boundaries thereof for \emph{braided modules}.

The main motivation for this work was the study of topological phases of matter in $3+1$ dimensions. It is believed that modular CYWW models are in some sense trivial as phases of matter in $3+1$ dimensions, such that their $2+1$-dimensional standard/cone boundary could be viewed as a standalone microscopic realization of a chiral phase. However, it is unclear to which extent and under which definition of a phase this is actually true. For example, if we were to take exact disentangling gLU circuits as a definition of a phase (which only makes sense for fixed-point models), then non-Drinfeld-center modular CYWW models would actually be in non-trivial (but invertible) phases, since such a disentangling circuit would yield a commuting-projector Hamiltonian for a chiral phase contradicting Ref.~\cite{Kapustin2019}, as argued in Ref.~\cite{Haah2018}. In this case, our fermionic gLU circuits lead to an interesting new notion. Namely, the 16-fold way Witt class CYWW models would represent non-trivial phases ``protected by the absence of fermions''. That is, they can be disentangled if we have access to fundamental fermions which can be used as auxiliary degrees of freedom, but not otherwise.

While the absence of gLU circuits makes it hard to prove, it still seems likely that general modular CYWW models are trivial if we pick the proper definition of a phase. One obvious way to go beyond exact gLU circuits is to allow for some ``exponential tails'' in the locality of the gLU operators, or consider ``fuzzy'' gLU circuits which are time evolutions under local Hamiltonians \cite{Kapustin2020}. Those are roughly equivalent to the ``standard'' definition via gapped paths through \emph{quasi-adiabatic evolution} \cite{Hastings2005, Chen2010}. While it is possible that going from strict to approximate locality will trivialize modular CYWW models, this does not strike as a very elegant and natural solution. A fixed-point algebraic proof of triviality could in contrast be given by invertible domain walls which are state-sums of a more general type as described in Ref.~\cite{universal_liquid}, but their existence is still speculative to date. As a last phase definition one might consider disentangling \emph{quantum cellular automata (QCAs)}, which have been shown to exist for many abelian modular CYWW models \cite{Haah2018, Haah2019, Shirley2022} and presumably exist for all such models. However, while a QCA is by definition a locality-preserving automorphism of the operator algebra of a many-body model, it is not itself a microscopic local object and does in particular not directly give rise to an invertible domain wall or a gLU circuit, so it is unclear how well it captures the notion of a phase. Also, it is questionable whether such QCAs also exist in the non-abelian case, since they disentangle the full spectrum of the Hamiltonian and not only its ground state. Certainly, non-abelian CYWW models do not have a stabilizer description which is central in Refs.~\cite{Haah2018, Haah2019, Shirley2022}.

To summarize, the question of whether general modular CYWW models are in a trivial phase is still open. In this work, we have settled the case of UMTCs that are Drinfeld centers, as well as those in a Witt class of the Kitaev 16-fold way if we allow for fermionic auxiliaries.

\section*{Acknowledgments}
Special thanks goes to David Aasen for hours of discussions and answering many of my questions. I would like to thank the DFG (EI 519/15-1, and CRC 183 project B01) for support.
\appendix

\bibliography{walker_wang_refs_v2}{}
\newpage

\section{\texorpdfstring{$c\neq 0$}{c!=0} UMTCs, braided fusion categories, and braided modules}
\label{sec:general_braided}
In this appendix, we show how to deal with a non-zero chiral central charge of UMTCs. We also introduce general braided fusion categories and the corresponding (non-invertible) CYWW model, as well as general braided modules and the corresponding (non-invertible) CYWW boundaries.
\subsection{Braided fusion categories and \texorpdfstring{$c\neq 0$}{c!=0} UMTCs}
\label{sec:umtc_appendix}
As indicated, the definition we used in Section~\ref{sec:tensorial_umtc} is not equivalent to UMTCs in general, but only to ones for with $c=0$. Physically, this is due to the \emph{chiral anomaly}, or \emph{framing anomaly} of the $c\neq 0$ topological phases. As a consequence, the amplitude assigned to some spacetime history does not only depend on the topology, but also on an additional \emph{$p_1$-structure}, or equivalently an \emph{Atyiah 2-framing} of the $3$-manifolds. This dependence can be explicitly seen in \emph{Chern-Simons theory}, a microscopic field-theory model for many of those phases.

There are three ways to incorporate $c\neq 0$ UMTCs into the definition. The first is to simply equip the ribbon manifolds, to which we assign tensors, with a $p_1$-structure. The second, but probably least elegant, is to change the domain of the map to \emph{projective tensors}, i.e., equivalence classes of tensors up to global phase factors. The third, and probably most insightful, is to add a fourth dimension to our ribbon manifolds. That is, the ribbon manifolds to which we associate tensors are now 4-manifolds with boundary with a ribbon network embedded inside the 3-dimensional boundary. $0$-surgery when gluing two fusion vertices in Eq.~\eqref{eq:fusion_0surgery} and in Eq.~\eqref{eq:0surgery_move} is replaced by attaching a 1-handle to the 4-manifolds, and 1-surgery in Eq.~\eqref{eq:loop_1surgery} by attaching a 2-handle. In addition, we add as gluing axiom the invariance under arbitrary surgery operations of the $4$-dimensional bulk, in other words, the tensor only depends on the cobordism class of the bulk.

Note that cobordism classes of \emph{closed} 4-manifolds form a $\zz$ group under disjoint union, generated by $\cc P(2)$ and its orientation-reversed copy $\overline{\cc P(2)}$. Thus, any two 4-manifolds $X$ and $Y$ with $\partial X=\partial Y=M$ are related by 4-manifold surgery together with disjoint union with $k$ of copies of $\cc P(2)$ (or $|k|$ copies of $\overline{\cc P(2)}$ if $k<0$). Thus, the tensors for different bulks of the same 3-manifolds only differ by global scalars,
\begin{equation}
\calm(X)=\calm(Y)\cdot \calm(\cc P(2))^k\;.
\end{equation}
We will see in Section~\ref{sec:umtc_generators} that $\calm(\cc P(2))$ is very directly related to the chiral central charge $c$ of $\calm$,
\begin{equation}
\label{eq:cp2_umtc}
\calm(\cc P(2)) = e^{2\pi i \frac{c}{8}}\;.
\end{equation}

A topological invariant that exactly determines the cobordism class of a closed 4-manifold $X$ is the \emph{signature} $\sigma(X)$. According to the \emph{Hirzebruch signature theorem}, we have
\begin{equation}
3 \sigma(X) = \smallint_X p_1\;,
\end{equation}
where $p_1$ is the first \emph{Pontryagin characteristic class}, a $\zz$-valued 4-cocycle that is integrated over the 4-manifold, i.e., evaluated against the fundamental class. So in general, the amplitudes of two $4$-dimensional bulks $X$ and $Y$ with the same boundary $\partial X=\partial Y=M$ are related by a prefactor
\begin{equation}
\calm(X) = \calm(Y) \cdot e^{2\pi i \frac{c}{24}\int_{(X\sqcup \bar{Y})/(\partial X=\partial Y)} p_1}\;.
\end{equation}
This makes it clear why the 4-dimensional bulk can be replaced with a $p_1$-structure.

Another type of tensorial TQFT (c.f.~Appendix~\ref{sec:tensorial_general_definition}) that is interesting in this context is when we do \emph{not} impose invariance of the tensors under surgery operations of 4-manifolds. Then, the tensors might not only depend on the cobordism class of the bulk, and not only up to global prefactors. What we get in this case is equivalent to \emph{unitary braided fusion categories} which are not necessarily modular.

The CYWW model for a braided fusion category, or a UMTC with $c\neq 0$, is completely analogous to Section~\eqref{sec:cyww_bulk}. For every 4-manifold $X$ whose boundary $\partial X$ is cellulated, define $R(X)$ as the ribbon manifold $R(\partial X)$ whose 4-dimensional bulk is that of $X$. Every 4-cell $C$ can be interpreted as a 4-sphere with a 3-cellulation as boundary, and the associated tensor is simply given by $\calm(R(C))$. Analogous to Section~\ref{sec:CYWW_evaluation}, the evaluation of the CYWW model on a 4-cellulation $Y$ is $\calm(R(\partial^{\text{top}}Y))$, where $\partial^{\text{top}}Y$ is the (space) boundary cellulation together with the bulk topology of $Y$ (but without the bulk cellulation). Thus, the model is recellulation invariant, but the evaluation \emph{does} depend on the bulk topology of $Y$. However, in the modular case, it is invariant under surgery operations on the bulk of $Y$, and therefore the CYWW model is invertible. This is no longer true in the general braided case.

\subsection{Braided modules}
We can also add a bulk to the ribbon manifolds in the definition of Lagrangian modules. That is, we add a 4-dimensional bulk to the ribbon 3-manifold, but also a 3-dimensional bulk to its 2-dimensional boundary. The boundary of the bulk 4-manifold is then divided into the ribbon 3-manifold, and the bulk of the boundary 2-manifold. A map that assigns tensors to such ribbon manifolds is equivalent to what we will call a \emph{braided module}.

One could also think about imposing invariance under attaching handles to the 3-dimensional bulk, analogous to the invariance under surgery of the 4-dimensional bulk for UMTCs. However, this is equivalent to not having a bulk at all, since all 4-manifolds with boundary are related by handle attachments/removals and accordingly the 3-dimensional oriented cobordism group is trivial. In more physical terms, the 2-dimensional boundary of a 3-dimensional model itself cannot have any chiral anomaly.

Let us now adapt the CYWW boundary from Section~\ref{sec:drinfeld_center_boundary} to braided modules. Consider a 4-manifold $Y_{\text{top}}$ whose boundary is divided into a physical boundary $X_{\text{top}}$ and a space boundary $\partial Y$. Add a 3-cellulation of $\partial Y$ that restricts to a 2-cellulation of $\partial X$ at its physical boundary. Define $R(Y_{\text{top}})$ as the boundary ribbon manifold $R(\partial Y)$ where we add $Y$ and $X$ as bulk. Now, for any 3-cell $C$, we can set $Y_{\text{top}}$ to be a 4-ball, $X_{\text{top}}$ to be a 3-ball, $\partial Y$ to be the 3-cellulation with a single 3-cell $C$, and $\partial X$ to be $\partial C$. With this identification of $C$ with $Y_{\text{top}}$, the tensor associated to a boundary 3-cell $C$ is simply $R(C)$. Now, consider the evaluation of the so-defined CYWW boundary for a braided module $\calb$ on a 4-cellulation $Y$ with physical boundary $X$ and space boundary $\partial Y$. Analogous to Section~\ref{sec:cyww_boundary_evaluation}, we get $\calb(R(Y_{\text{top}}))$, where $Y_{\text{top}}$ is $Y$ without the bulk cellulation, that is, keeping only the topology of $Y$ and the cellulation of $\partial Y$. Thus, the evaluation is recellulation invariant, but it does depend on the bulk topology and is not invertible unless the braided module is Lagrangian.

A special case of this is when the braided fusion category and hence the CYWW bulk is trivial. Then braided modules associate tensors to ribbon 2-manifolds with 3-dimensional bulk, and are just fusion categories. The CYWW boundary is then a $2+1$-dimensional state-sum without any (non-trivial) bulk, namely the Turaev-Viro-Barrett-Westbury state-sum \cite{Turaev1992, Barrett1993} for the fusion category. For any braided fusion category $\calb$, a particularly simple braided module is the one resulting from condensing the trivial algebra, where the fusion category equals $\calb$ without braiding. This braided module is the pullback of $\calb$ along the following simple geometric mapping. The bulk of the boundary is mapped to boundary with no ribbons, and boundary ribbons are mapped to bulk ribbons. This braided module yields the ``smooth'' or ``cone'' boundary of the CYWW state-sum. For a UMTC, this is the boundary that is supposed to contain the (chiral) phase described by the UMTC.

Note that a different characterization of CYWW boundaries has been proposed in Ref.~\cite{Bridgeman2020}, based on a \emph{commutative Frobenius algebra object} inside the unitary braided fusion category. Such algebra objects are not equivalent to braided modules. As also stated in Appendix~\ref{sec:tqft_collection}, they correspond to a tensorial TQFT that is almost-fully extended at both bulk and boundary. Thus, the corresponding ribbon manifolds have points in the boundary where bulk ribbons can end, but no boundary ribbons. In order to yield a topologically invariant CYWW boundary, those tensorial algebra objects should also be invariant under plain 2-handle attachments. This is necessary to make up for the loop 2-handle attachments in Section~\ref{sec:tensorial_umtc_module}, which are missing due to the lack of boundary ribbons. Such algebra objects yield braided modules via \emph{condensation}, namely pulling back along a geometric mapping where boundary ribbons and fusion vertices are replaced by slightly immersed into the bulk. Braided modules are more general, since they also allow to stack a standalone $2+1$-dimensional Turaev-Viro state-sum on top of the boundary, as discussed in the previous paragraph.

The contribution of the bulk 4-manifold $X$,
\begin{equation}
\label{eq:chiral_anomaly}
e^{2\pi i \frac{c}{24} \int_X p_1}\;,
\end{equation}
to the amplitude of the UMTC, i.e., the chiral anomaly, can be represented by a state-sum in 4 dimensions that is physically trivial in the following sense (c.f.~Ref.~\cite{universal_liquid}). It associates phase factors to the vertices of a triangulation that only depend on the combinatorics of the surrounding triangulation. There are no state-sum variables that are summed over and therefore no degrees of freedom.
So, if the cone boundary of the CYWW model is a microscopic model for the input UMTC, the modular CYWW model itself can be viewed as a physically non-trivial way of representing the chiral anomaly. Indeed, it is known that the CYWW model on a 4-manifold $X$ evaluates to Eq.~\eqref{eq:chiral_anomaly} \cite{Walker2011}.

Let us now interpret the CYWW model as a trivially-fermionic model, i.e., restrict its definition to spin 4-manifolds. It is known that for spin 4-manifolds we have
\begin{equation}
\label{eq:pontryagin_spin_manifold}
p_1 \mod 16=0\;.
\end{equation}
That is, there is a characteristic 1-chain $\gamma$ such that $d\gamma=p_1\operatorname{mod} 16$. Furthermore, on any 4-manifold $X$, we have
\begin{equation}
\int_X p_1\bmod 3= 3\sigma(X)\bmod 3=0\;,
\end{equation}
which only holds globally for the cohomology class, or after integration. Thus, if $c=n/2$ for an integer $n$, evaluating the CYWW partition function yields $1$ on all spin $4$-manifolds. So as trivially-fermionic UMTCs, half-integer central-charge UMTCs such as Ising do not have a chiral anomaly. It is therefore possible to find fermionic invertible domain walls to (non-anomalous) vacuum.

A UMTC does not fully determine the phase of a model. Instead, there is an infinite stack of different microscopic models with the same anyon statistics but in different phases, described by the chiral central charge $c$ which is only determined $\operatorname{mod} 8$ by the UMTC. The different models differ by stacking copies of the invertible $E_8$ phase with $c=8$. The bosonic Ising UMTC has central charge $\frac12$, so the the possible chiral anomaly is given by
\begin{equation}
e^{\frac{2\pi i}{24} (\frac12+8n) p_1}\;.
\end{equation}
for different $n\in \zz$. If we choose $n=2$, then we get
\begin{equation}
e^{\frac{2\pi i}{24} (\frac12+16) p_1}=
e^{2\pi i 11 \frac{p_1}{16}} = 1
\end{equation}
using Eq.~\eqref{eq:pontryagin_spin_manifold}. Thus, the Ising UMTC can have a fermionic realization without chiral anomaly.

\subsection{Generators for UMTCs}
\label{sec:umtc_generators}
In this section we will, without detailed proof, give a generating set of fusion vertices, ribbon manifolds, and gluing axioms for UMTCs. More precisely, we will do this simultaneously for UMTCs with $c=0$, $c\neq 0$, and braided fusion categories. We will also describe how our definitions relate to the common categorical definitions.

We first observe that we can restrict ourselves to a single link for the fusion vertices, namely a sphere with three points, two with positive and one with negative orientation,
\begin{equation}
\begin{tikzpicture}
\atoms{vertex}{0/}
\draw (0)edge[mark={arr,-}]++(150:0.6) (0)edge[mark={arr,-}]++(30:0.6) (0)edge[mark={arr}]++(-90:0.6);
\end{tikzpicture}\;,
\end{equation}
together with its orientation-reversed partner. 
All other types of fusion vertices can be constructed from this 3-valent one, e.g., a 4-valent one via
\begin{equation}
\begin{tikzpicture}
\atoms{vertex}{0/lab={t=$x\alpha\beta$,p=0:0.45}}
\draw (0)edge[mark={arr,-}]++(135:0.6) (0)edge[mark={arr,-}]++(45:0.6) (0)edge[mark={arr}]++(-45:0.6) (0)edge[mark={arr}]++(-135:0.6);
\end{tikzpicture}
\coloneqq
(d_x^{1/2}D^{-1/2})
\begin{tikzpicture}
\atoms{vertex}{0/lab={t=$\alpha$,p=-60:0.25}, {1/p={0.8,0}, lab={t=$\beta$,p=-120:0.25}}}
\draw (0)edge[mark={arr,-}]++(135:0.6) (1)edge[mark={arr,-}]++(45:0.6) (1)edge[mark={arr}]++(-45:0.6) (0)edge[mark={arr}]++(-135:0.6) (0)edge[mark=arr,mark={slab=$x$}](1);
\end{tikzpicture}\;.
\end{equation}
Gluing the 4-valent vertex is equivalent to gluing at $\alpha$, $\beta$, and applying the loop 1-surgery in Eq.~\eqref{eq:loop_1surgery} to the resulting $x$ loop.
Also 1-valent vertices can be obtained via
\begin{equation}
\begin{tikzpicture}
\atoms{vertex}{0/lab={t=$x\alpha$,p=0:0.35}}
\draw (0)edge[mark={arr,-},mark={slab=$a$}]++(180:0.6);
\end{tikzpicture}
\coloneqq
(d_x^{1/2})
\begin{tikzpicture}
\atoms{vertex}{0/lab={t=$\alpha$,p=0:0.25}}
\draw (0)edge[mark={arr,-},mark={slab=$a$}]++(180:0.6);
\draw[mark={arr,p=0.4}, mark={slab=$x$},looseness=2] (0)to[out=60,in=90]++(0.6,0)to[out=-90,in=-60](0);
\end{tikzpicture}\;.
\end{equation}
Gluing the 4-valent vertex is equivalent to gluing at $\alpha$, applying the loop 1-surgery in Eq.~\eqref{eq:loop_1surgery} to the resulting $x$ loop, and then applying plain 0-surgery in Eq.~\eqref{eq:0surgery_move} backwards. We notice that the generating links of fusion vertices are the same as the generating extended manifolds of ordinary 2-1-extended TQFT in Eq.~\eqref{eq:2dtqft_generating_manifolds}, apart from the fact that we do not need the 1-point sphere as generator since we have the additional gluing operation in Eq.~\eqref{eq:0surgery_move}.

Note that we could even define ``singular'' fusion vertices, whose link is not a 2-sphere, but, e.g., a torus,
\begin{equation}
\begin{tikzpicture}
\atoms{vertex}{{0/lab={t=$x\alpha$,p=90:0.25},lab={t=$(S_1\times S_1)$, p=-90:0.25}}}
\end{tikzpicture}
\coloneqq
(d_x^{1/2}D^{-1/2})
\begin{tikzpicture}
\atoms{vertex}{0/lab={t=$\alpha$,p=180:0.25}}
\draw[mark={arr,-},mark={slab=$x$,r}] (0)arc(-180:180:0.4);
\end{tikzpicture}\;,
\end{equation}
such that instead of gluing a pair of singular vertices, we glue a pair of 2-valent vertices and then apply the loop 1-surgery.

Furthermore, any ribbon manifold can be obtained via gluing operations from the following two generating ribbon manifolds.
\begin{itemize}
\item The \emph{tetrahedron},
\begin{equation}
\label{eq:tetrahedron_tensor}
\begin{tikzpicture}
\atoms{vertex}{0/lab={t=$\alpha$,p=-90:0.25}, {1/p={-150:1},lab={t=$\alpha$,p=-150:0.25}}, {2/p={-30:1},lab={t=$\gamma$,p=-30:0.25}}, {3/p={90:1},lab={t=$\delta$,p=90:0.25}}}
\draw (0)edge[mark=arr, mark={slab=$a$,r,p=0.4}](1) (0)edge[mark=arr, mark={slab=$b$,p=0.4}](2) (0)edge[mark={arr,-}, mark={slab=$c$,p=0.4}](3) (1)edge[mark=arr, mark={slab=$d$,r}](2) (2)edge[mark=arr, mark={slab=$e$,r}](3) (1)edge[mark=arr, mark={slab=$f$}](3);
\end{tikzpicture}\;,
\end{equation}
is a ribbon network inside a 3-sphere, which is at the boundary of a 4-ball in the $c\neq 0$ or braided non-modular case.
The associated tensor equals what is conventionally known as $F$-tensor of the UMTC/braided fusion category, up to a prefactor,
\begin{equation}
(\frac{d_a}{D})^{-1/2} (\frac{d_e}{D})^{-1/2} {F^{fdb}_c}^{a\alpha\beta}_{e\delta\gamma}\;.
\end{equation}
We already have explained towards the end of Section~\ref{sec:tensorial_umtc} how this ribbon manifold corresponds to a ``move'' of string diagrams.
\item The \emph{braiding}, consisting of two fusion vertices connected by three ribbons such that two of them are exchanged,
\begin{equation}
\label{eq:braiding_tensor}
\begin{tikzpicture}
\atoms{vertex}{0/lab={t=$\alpha$,p=-90:0.25}, {1/p={0,1},lab={t=$\beta$,p=90:0.25}}}
\draw[mark={arr,p=0.3},mark={slab=$a$}] (0)to[bend left=60](1);
\draw[mark={arr,-,p=0.25},mark={slab=$b$,r,p=0.6}] (0)to[out=90,in=-90]++(0.5,0.8)to[out=90,in=0](1);
\draw[front,mark={arr,p=0.25},mark={slab=$c$,p=0.6}] (1)to[out=-90,in=90]++(0.5,-0.8)to[out=-90,in=0](0);
\end{tikzpicture}
\;,
\end{equation}
again inside the 3-sphere, potentially at the boundary of the 4-ball. The associated tensor is known as the $R$-tensor
\begin{equation}
{R^{bc}_a}_\alpha^\beta
\end{equation}
for a UMTC/braided fusion category. The well-known $R$-move exchanging two ribbons adjacent to a fusion vertex can be performed by taking the disjoint union with Eq.~\eqref{eq:braiding_tensor} and then gluing with one of the vertices.
\end{itemize}
We notice that the generating ribbon manifolds correspond to the generating axioms of 2-1-extended TQFT in Appendix~\ref{sec:11_tensorial_tqft}. To this end, we combine the two sides of the axiom. Every vertex of the ribbon manifold corresponds to a generating 2-1 manifold of the axiom with the same link, and every ribbon corresponds to a gluing of two vertices of the 2-1 manifold. Specifically, the tetrahedron tensor in Eq.~\eqref{eq:tetrahedron_tensor} corresponds to the associativity axiom in Eq.~\eqref{eq:21_extended_associativity}, and the braiding in Eq.~\eqref{eq:braiding_tensor} corresponds to the commutativity axiom in Eq.~\eqref{eq:21_commutativity}. So in a sense, UMTCs or braided fusion categories can be viewed as a categorification of commutative Frobenius algebras.

One can also show that all the gluing axioms from Section~\ref{sec:tensorial_umtc} are generated from a finite set of axioms. In a more conventional extended-TQFT language, a result along these lines has been derived in Ref.~\cite{Bartlett2014}. The most notable generating axioms are as follows.
\begin{itemize}
\item What is known as \emph{pentagon equation} of fusion categories yields a gluing axiom with three tetrahedra on one side and two on the other side. It holds both with or without the $4$-dimensional bulk.
\item The \emph{hexagon equation} involves the tetrahedron and the braiding, and is the central axiom that makes a fusion category into a braided one,
\begin{equation}
\begin{multlined}
\begin{tikzpicture}
\atoms{vertex}{0/, 1/p={0,1.6}, 2/p={0.5,0.8}, 3/p={1,0.8}}
\draw (0)edge[mark=arr](1) (0)edge[mark=arr](2) (0)edge[mark={arr,-}](3) (1)edge[mark=arr](2) (2)edge[mark=arr](3) (1)edge[mark=arr](3);
\atoms{vertex}{0x/p={1.5,0.6}, 1x/p={2.5,0.6}}
\draw[mark={arr,-,p=0.3}] (0x)to[bend right=60]coordinate[pos=0.7](gl)(1x);
\draw[mark={arr,p=0.25}] (0x)to[out=0,in=180]++(0.6,0.5)to[out=0,in=90](1x);
\draw[front,mark={arr,-,p=0.25}] (1x)to[out=180,in=0]++(-0.6,0.5)to[out=180,in=90](0x);
\atoms{vertex}{0y/p={3,0}, 1y/p={3,1.6}, 2y/p={3.5,0.8}, 3y/p={4,0.8}}
\draw (0y)edge[mark=arr](1y) (0y)edge[mark=arr](2y) (0y)edge[mark={arr,-}](3y) (1y)edge[mark=arr](2y) (2y)edge[mark=arr](3y) (1y)edge[mark=arr](3y);
\draw[glue] (2)to[bend left=20](1y) (3)--(0x) (1x)--(0y);
\atoms{glueedge}{glx/p={gl}}
\end{tikzpicture}
=
\begin{tikzpicture}
\atoms{vertex}{0/, 1/p={0,1.6}, 2/p={0.5,0.8}}
\draw (0)edge[mark=arr](1) (0)edge[mark=arr](2) (1)edge[mark=arr](2);
\atoms{vertex}{1y/p={1.6,1.6}, 2y/p={2.1,0.8}, 3y/p={2.8,0.8}}
\draw (1y)edge[mark=arr](2y) (2y)edge[mark=arr](3y) (1y)edge[mark=arr](3y);
\draw (0)edge[mark={arr,-},out=0,in=-135](3y) (1)edge[mark={arr,p=0.7},out=0,in=-150](2y) (2)edge[front,mark={arr,p=0.7},out=0,in=-120]coordinate[pos=0.3](gl)(1y);
\draw[glue] (2)to[bend left=20](1y);
\atoms{glueedge}{glx/p={gl}}
\end{tikzpicture}\\
=
\begin{tikzpicture}
\atoms{vertex}{0/, 1/p={0,1.6}, 2/p={1.2,0.8}, 3/p={2,0.8}}
\draw (0)edge[mark=arr](1) (2)edge[mark=arr](3) (0)edge[mark={arr,-},bend right=20](3) (1)edge[looseness=2,out=0,in=180,mark={arr,p=0.3}](2) (0)edge[front,out=60,in=90,mark={arr,p=0.3}](2) (1)edge[front,out=-45,in=90,mark={arr,p=0.7}](3);
\end{tikzpicture}
=
\begin{tikzpicture}
\atoms{vertex}{0/, 1/p={0,1.6}, 2/p={0.5,0.8}, 3/p={1,0.8}}
\draw (0)edge[mark=arr](1) (0)edge[mark=arr](2) (0)edge[mark={arr,-}](3) (1)edge[mark=arr](2) (2)edge[mark=arr](3) (1)edge[mark=arr](3);
\atoms{vertex}{0x/p={-1.5,1}, 1x/p={-0.5,1}}
\draw[mark={arr,-,p=0.3}] (0x)to[bend right=60](1x);
\draw[mark={arr,p=0.25}] (0x)to[out=0,in=180]++(0.6,0.5)to[out=0,in=90](1x);
\draw[front,mark={arr,-,p=0.25}] (1x)to[out=180,in=0]++(-0.6,0.5)to[out=180,in=90](0x);
\atoms{vertex}{0y/p={1.5,1}, 1y/p={2.5,1}}
\draw[mark={arr,-,p=0.3}] (0y)to[bend right=60](1y);
\draw[mark={arr,p=0.25}] (0y)to[out=0,in=180]++(0.6,0.5)to[out=0,in=90](1y);
\draw[front,mark={arr,-,p=0.25}] (1y)to[out=180,in=0]++(-0.6,0.5)to[out=180,in=90](0y);
\draw[glue] (1x)--(1) (2)edge[bend left=20](0y);
\end{tikzpicture}
\;.
\end{multlined}
\end{equation}
It holds both with and without the $4$-dimensional bulk.
\item The \emph{modularity condition} is what conventionally makes a braided fusion category into a modular one. Without a 4-dimensional bulk, the gluing axiom is
\begin{equation}
\label{eq:umtc_modularity}
\begin{tikzpicture}
\draw (0,0)arc(-180:0:0.5);
\draw[front] (0.3,0)arc(-180:0:0.5);
\draw[mark={arr,s,-}] (0.3,0)arc(180:0:0.5);
\draw[front,mark={arr,s,-}] (0,0)arc(180:0:0.5);
\begin{scope}[shift={(3,0)},xscale=-1]
\draw (0,0)arc(-180:0:0.5);
\draw[front] (0.3,0)arc(-180:0:0.5);
\draw[mark={arr,s}] (0.3,0)arc(180:0:0.5);
\draw[front,mark={arr,s}] (0,0)arc(180:0:0.5);
\end{scope}
\draw[glue] (1.3,0)--(1.7,0);
\end{tikzpicture}
=
\begin{gathered}
\begin{tikzpicture}
\draw[mark={arr,s},mark={slab=$a$}] (0,0)arc(-180:180:0.5);
\draw[mark={arr,s},mark={slab=$b$}] (1.3,0)arc(-180:180:0.5);
\end{tikzpicture}\\
\in S_2\times S_1,\\
a\sim x\times S_1,b\sim y\times S_1
\end{gathered}
\;.
\end{equation}
Here ``gluing two loops'' without any fusion vertices is a shortcut for first inserting trivial fusion vertices on the glued loops via Eq.~\eqref{eq:trivial_vertex_introduction}, then gluing those via Eq.~\eqref{eq:fusion_0surgery}, and finally applying the 1-surgery in Eq.~\eqref{eq:loop_1surgery} to the resulting single loop. $x$ and $y$ are any two points of $S_2$. The tensor assigned to the right-hand side is $\delta_{a,b}$ which follows from gluing axioms and a normalization condition. Thus, the matrix
\begin{equation}
S_{ab}\coloneqq
\begin{tikzpicture}
\draw[mark={slab=$a$,s}] (0,0)arc(-180:0:0.5);
\draw[front] (0.3,0)arc(-180:0:0.5);
\draw[mark={arr,s,-},mark={slab=$b$,e}] (0.3,0)arc(180:0:0.5);
\draw[front,mark={arr,s,-}] (0,0)arc(180:0:0.5);
\end{tikzpicture}\;,
\end{equation}
commonly known as \emph{$S$-matrix} is unitary, $SS^\dagger=S^\dagger S=\mathbb{1}$. The ribbon manifold of $S$ can be glued from two braidings,
\begin{equation}
\begin{tikzpicture}
\atoms{vertex}{0/, 1/p={1,0}}
\draw[mark={arr,-,p=0.3}] (0)to[bend right=60]coordinate[pos=0.7](g1)(1);
\draw[mark={arr,p=0.25}] (0)to[out=0,in=180]++(0.6,0.5)to[out=0,in=90](1);
\draw[front,mark={arr,-,p=0.25}] (1)to[out=180,in=0]++(-0.6,0.5)to[out=180,in=90](0);
\begin{scope}[,shift={(1,-1)}, rotate=180]
\atoms{vertex}{0x/, 1x/p={1,0}}
\draw[mark={arr,-,p=0.3}] (0x)to[bend right=60](1x);
\draw[mark={arr,p=0.25}] (0x)to[out=0,in=180]++(0.6,0.5)to[out=0,in=90](1x);
\draw[front,mark={arr,-,p=0.25}] (1x)to[out=180,in=0]++(-0.6,0.5)to[out=180,in=90](0x);
\end{scope}
\draw[glue] (1)to[out=0,in=0](0x) (0)to[out=180,in=180](1x);
\atoms{glueedge}{glx/p={g1}}
\end{tikzpicture}
=
\begin{tikzpicture}
\draw (0,0)arc(-180:0:0.5);
\draw[front] (0.3,0)arc(-180:0:0.5);
\draw[mark={arr,s,-}] (0.3,0)arc(180:0:0.5);
\draw[front,mark={arr,s,-}] (0,0)arc(180:0:0.5);
\end{tikzpicture}\;.
\end{equation}
Now, let us add a 4-dimensional bulk. Gluing the left-hand side of the modularity condition in Eq.~\eqref{eq:umtc_modularity} yields $S_2\times B_2$ as bulk, whereas on the right-hand side, the bulk that yields the $\delta$-tensor is $B_3\times S_1$. However, those two bulks are still related by a surgery operation and in the same cobordism class. So we indeed find that this gluing axiom holds for UMTCs with both $c=0$ and $c\neq 0$, but not for more general braided fusion categories.
\item The \emph{anomaly-freeness condition} is a gluing axiom only for ribbon manifolds without a 4-dimensional bulk. Consequently, it only holds for UMTCs with $c=0$,
\begin{equation}
\label{eq:anomaly_freeness}
\begin{tikzpicture}
\draw (0,0)edge[mark={arr,p=0.3},out=90,in=-90,looseness=3] (1,0) (0,0)edge[front,out=-90,in=90,looseness=3] (1,0);
\atoms{glueedge}{x/p={0,0}}
\end{tikzpicture}=
S_3\;.
\end{equation}
I.e., we apply the 1-surgery in Eq.~\eqref{eq:loop_1surgery} to a once-twisted loop inside the 3-sphere and obtain another 3-sphere. Now, if the left- and right-hand side was the boundary of a 4-ball and the 1-surgery was a 2-handle attachment, this would result in
\begin{equation}
\cc P(2)-B_4\;,
\end{equation}
i.e., the complex projective plane with the neighborhood of a point removed, c.f.~page 47 of Ref.~\cite{Mandelbaum1980}. $B_4$ and the above 4-manifold differ by connected sum with $\cc P(2)$, so this gluing axiom does not hold for general braided fusion categories or for UMTCs with $c\neq 0$. Indeed, the tensor assigned to a once-twisted loop as in Eq.~\eqref{eq:anomaly_freeness} with label $a$ is given by
\begin{equation}
\frac{d_a\theta_a}{D}\;,
\end{equation}
where $\theta_a$ is known as the \emph{topological twist} of $a$. With this, Eq.~\eqref{eq:anomaly_freeness} becomes
\begin{equation}
\sum_a \frac{d_a}{D} \frac{d_a\theta_a}{D} = \frac1D\;,
\end{equation}
and thus $c=0$ since $c$ is defined as
\begin{equation}
 e^{2\pi i \frac{c}8} = \sum_a \frac{d_a^2\theta_a}{D}\;.
\end{equation}
From this, we can also read off Eq.~\eqref{eq:cp2_umtc}.
\end{itemize}
The axioms above do not strictly generate all gluing axioms in Section~\ref{sec:tensorial_umtc}. For example, it does not suffice to only impose one single hexagon equation as above, but we need one version of it for any choice of ribbon directions (and similar for the pentagon equation). In order to derive those other versions from a single version, we can introduce one ``auxiliary'' fusion vertex,
\begin{equation}
\begin{tikzpicture}
\atoms{vertex}{0/}
\draw (0)edge[mark=arr,fav=s]++(180:0.6) (0)edge[mark=arr]++(180:-0.6);
\end{tikzpicture}\;,
\end{equation}
its orientation-reversed partner, three auxiliary tensors,
\begin{equation}
\begin{tikzpicture}
\atoms{vertex}{0/, 1/p=0:1, 2/p=60:1}
\draw (0)edge[mark=arr,bend right=20](1) (1)edge[mark=arr,bend right=20](0) (0)edge[mark=arr](2) (1)edge[mark=arr,fav=e](2);
\end{tikzpicture}\;,\qquad
\begin{tikzpicture}
\atoms{vertex}{0/, 1/p=0:1, 2/p=60:1}
\draw (0)edge[mark={arr,-},bend right=20](1) (1)edge[mark={arr,-},bend right=20](0) (0)edge[mark={arr}](2) (1)edge[mark={arr},fav=e](2);
\end{tikzpicture}\;,\qquad
\begin{tikzpicture}
\atoms{vertex}{0/, 1/p=0:1}
\draw (0)edge[mark=arr,bend right=45,fav=s](1) (0)edge[mark=arr,bend left=45,fav=e](1);
\end{tikzpicture}\;,
\end{equation}
and a couple of axioms reversing the direction of certain ribbons, e.g.,
\begin{equation}
\label{eq:auxiliary_umtc_axiom1}
\begin{multlined}
\begin{tikzpicture}
\atoms{vertex}{0/, 1/p={-150:1}, 2/p={-30:1}, 3/p={90:1}}
\draw (0)edge[mark=arr](1) (0)edge[mark=arr](2) (0)edge[mark={arr,-}](3) (1)edge[mark=arr](2) (2)edge[mark=arr](3) (1)edge[mark=arr](3);
\atoms{vertex}{0x/p={-1.2,-0.2}, 1x/p={-2.2,-0.2}, 2x/p={-1.7,-1}}
\draw (0x)edge[mark={arr,-},bend right=20](1x) (1x)edge[mark={arr,-},bend right=20](0x) (0x)edge[mark={arr,-},fav=e](2x) (1x)edge[mark={arr,-}](2x);
\draw[glue] (1)--(0x);
\end{tikzpicture}
=
\begin{tikzpicture}
\atoms{vertex}{0/, 1/p={-150:1}, 2/p={-30:1}, 3/p={90:1}, m/p={$(1)!0.5!(2)$}}
\draw (0)edge[mark=arr](1) (0)edge[mark=arr](2) (0)edge[mark={arr,-}](3) (1)edge[mark={arr,-}](m) (m)edge[mark=arr,fav=s](2) (2)edge[mark=arr](3) (1)edge[mark=arr](3);
\end{tikzpicture}
\\=
\begin{tikzpicture}
\atoms{vertex}{0/, 1/p={-150:1}, 2/p={-30:1}, 3/p={90:1}}
\draw (0)edge[mark=arr](1) (0)edge[mark=arr](2) (0)edge[mark={arr,-}](3) (1)edge[mark={arr,-}](2) (2)edge[mark=arr](3) (1)edge[mark=arr](3);
\atoms{vertex}{0x/p={1.2,-0.2}, 1x/p={2.2,-0.2}, 2x/p={1.7,-1}}
\draw (0x)edge[mark=arr,bend right=20](1x) (1x)edge[mark=arr,bend right=20](0x) (0x)edge[mark={arr,-}](2x) (1x)edge[mark={arr,-},fav=e](2x);
\draw[glue] (2)--(0x);
\end{tikzpicture}\;,
\end{multlined}
\end{equation}
or,
\begin{equation}
\begin{tikzpicture}
\atoms{vertex}{0/p={-1.2,0.5}, 1/p={-1.2,-0.5}, 2/p={-0.4,0}}
\draw (0)edge[mark=arr,bend right=20](1) (1)edge[mark=arr,bend right=20](0) (0)edge[mark={arr}](2) (1)edge[mark={arr},fav=e](2);
\atoms{vertex}{0x/, 1x/p=0:1}
\draw (0x)edge[mark=arr,bend right=45,fav=s](1x) (0x)edge[mark=arr,bend left=45,fav=e](1x);
\draw[glue] (2)--(0x);
\end{tikzpicture}
=
\begin{tikzpicture}
\atoms{vertex}{0/p={-1.2,0.5}, 1/p={-1.2,-0.5}, 2/p={-0.4,0}}
\draw (0)edge[mark=arr,bend right=20](1) (1)edge[mark=arr,bend right=20](0) (0)edge[mark={arr},fav=e](2) (1)edge[mark={arr}](2);
\end{tikzpicture}
\;.
\end{equation}

Note that UMTCs in the literature have a few extra constraints that are not physically necessary, but also do not obstruct any interesting examples. Most importantly, the fusion dimension of a trivial vertex,
\begin{equation}
\begin{tikzpicture}
\atoms{vertex}{0/}
\draw (0)edge[mark={arr,-},mark={slab=$a$,r}]++(180:0.6) (0)edge[mark=arr,mark={slab=$b$}]++(0:0.6);
\end{tikzpicture}\;,
\end{equation}
is assumed to be $\delta_{a,b}$, and we can add such a fusion vertex at a cost of a prefactor,
\begin{equation}
\label{eq:trivial_vertex_introduction}
\begin{tikzpicture}
\atoms{vertex}{0/}
\draw (0)edge[mark={arr,-},mark={slab=$a$,r}]++(180:0.6) (0)edge[mark=arr,mark={slab=$a$}]++(0:0.6);
\end{tikzpicture}
=
(\frac{d_a}{D})^{-1/2}
\begin{tikzpicture}
\draw (0,0)edge[mark={arr},mark={slab=$a$}](1,0);
\end{tikzpicture}\;.
\end{equation}

\subsection{Generators for Lagrangian modules}
\label{sec:tqft_umtc_module}
Analogous to the previous section, we will now describe sets of generating fusion vertex links, ribbon manifolds, and gluing axioms simultaneously for Lagrangian and braided modules. At the same time we will discuss how they are related to known categorical structures.

First, every fusion vertex can be generated from those with two different links, namely \emph{trivalent boundary vertices} where three boundary ribbons meet, and \emph{condensation vertices} where one bulk and one boundary ribbon meet,
\begin{equation}
\label{eq:module_generating_vertices}
\begin{tikzpicture}
\atoms{bdvertex}{0/}
\draw[bdribbon] (0)edge[mark={arr,-}]++(150:0.6) (0)edge[mark={arr,-}]++(30:0.6) (0)edge[mark={arr}]++(-90:0.6);
\end{tikzpicture}\;,
\quad
\begin{tikzpicture}
\atoms{bdvertex}{0/}
\draw (0)edge[bdribbon,mark={arr,-}]++(180:0.6) (0)edge[mark={arr}]++(0:0.6);
\end{tikzpicture}\;.
\end{equation}
Again, we notice that generating boundary fusion vertex links are the same as the generating extended manifolds for 2-dimensional open-closed TQFTs in Eq.~\eqref{eq:open_closed_generators}. The link of the trivalent vertices is the 3-boundary-point disk representing the non-commutative Frobenius algebra in Eq.~\eqref{eq:open_closed_generators}, and the link of the condensation vertices is the same as the bulk-boundary-point disk representing the central homomorphism in Eq.~\eqref{eq:open_closed_generators}. Note that again, we do not need the one-point disk since it can be obtained from the three-point disk through the additional gluing operation in Eq.~\eqref{eq:boundary_1surgery}.

Next, a set of generating boundary ribbon manifolds is as follows.
\begin{itemize}
\item The \emph{boundary tetrahedron}, a 3-ball with the 1-skeleton of a tetrahedron as boundary ribbon network,
\begin{equation}
\label{eq:boundary_tetrahedron}
\begin{tikzpicture}
\atoms{bdvertex}{0/, {1/p={-150:1}}, {2/p={-30:1}}, {3/p={90:1}}}
\draw[bdribbon] (0)edge[mark=arr](1) (0)edge[mark=arr](2) (0)edge[mark={arr,-}](3) (1)edge[mark=arr](2) (2)edge[mark=arr](3) (1)edge[mark=arr](3);
\end{tikzpicture}\;.
\end{equation}
For a braided module, the 3-ball has a 4-ball as bulk, and the boundary 2-sphere has a 3-ball as bulk.
The associated tensor is the $F$-symbol of a fusion category (without braiding).
\item The \emph{triple condensation}, a 3-ball with one 3-valent bulk fusion vertex and one 3-valent boundary fusion vertex, such that pairs of adjacent bulk- and boundary ribbons meet at a condensation fusion vertex each,
\begin{equation}
\label{eq:triple_condensation}
\begin{tikzpicture}
\atoms{bdvertex}{0/, {1/p={30:1}}, {2/p={150:1}}, {3/p={-90:1}}}
\atoms{vertex}{4/p=-30:0.3}
\draw[bdribbon] (0)edge[mark=arr](1) (0)edge[mark=arr](2) (0)edge[mark={arr,-}](3);
\draw (4)edge[bend right=10, mark={arr,-}](1) (4)edge[bend right,mark={arr,-}](2) (4)edge[bend left=10,mark={arr}](3);
\end{tikzpicture}\;.
\end{equation}
For a braided module, we again have a bulk 4-ball.
\item The \emph{boundary braiding}, a 3-ball with two boundary fusion vertices sharing two boundary ribbons dividing the boundary into two 2-balls. The third adjacent ribbons of both fusion vertices are attached at different $2$-balls so they cannot be connected with a boundary ribbon. Instead they are connected by a bulk ribbon via two condensation vertices,
\begin{equation}
\label{eq:boundary_braiding}
\begin{tikzpicture}
\atoms{bdvertex}{0/, 1/p={0.5,0}, 2/p={1.2,0}, 3/p={1.7,0}}
\draw[bdribbon] (0)edge[bend left=90,mark={arr}](2) (0)edge[bend right=90,mark={arr,-,p=0.3}](2) (0)edge[mark={arr,-}](1) (2)edge[mark={arr}](3);
\draw (1)edge[bend right=90, mark={arr,-,p=0.7}](3);
\end{tikzpicture}\;,
\end{equation}
again with a bulk 4-ball in the braided case.
\end{itemize}
Analogous to UMTCs, the generating ribbon manifolds correspond the generating axioms of 2-dimensional open-closed TQFT described in Appendix~\ref{sec:11_tensorial_tqft}. So the proof that those ribbon manifolds generate all ribbon manifolds follows directly from the proof that the knowledgeable-Frobenius-algebra axioms generate all axioms of 2-dimensional open-closed TQFT, presented in Ref.~\cite{Lauda2005}. Specifically, the boundary tetrahedron in Eq.~\eqref{eq:boundary_tetrahedron} corresponds to the associativity axiom for the non-commutative algebra in open-closed TQFT, involving two three-point disks on both sides. The triple condensation Eq.~\eqref{eq:triple_condensation} corresponds to the homeomorphism axiom in Eq.~\eqref{eq:open_closed_homomorphism_axiom}. The boundary braiding Eq.~\eqref{eq:boundary_braiding} corresponds to the knowledgeably axiom in Eq.~\eqref{eq:open_closed_knowledgeability_axiom}, but also the \emph{Cardy condition} in Ref.~\cite{Lauda2005}, which is a different axiom but yields the same when we combine the left- and right-hand side.

We further conjecture that the full set of axioms can be derived from the following axioms, in addition to the pentagon and hexagon equation of the UMTC/braided fusion category.
\begin{itemize}
\item The pentagon equation for the boundary fusion category.
\item An axiom consisting of four triple condensations, one bulk tetrahedron, and one boundary tetrahedron,
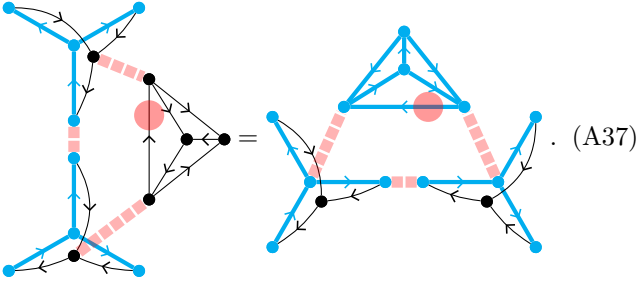
\begin{equation}
\begin{tikzpicture}
\atoms{bdvertex}{0/, {1/p={30:1}}, {2/p={150:1}}, {3/p={-90:1}}}
\atoms{vertex}{4/p=-30:0.3}
\draw[bdribbon] (0)edge[mark=arr](1) (0)edge[mark=arr](2) (0)edge[mark={arr,-}](3);
\draw (4)edge[bend right=10, mark={arr,-}](1) (4)edge[bend right,mark={arr,-}](2) (4)edge[bend left=10,mark={arr}](3);
\begin{scope}[shift={(0,-2.5)}]
\atoms{bdvertex}{0x/, {1x/p={-30:1}}, {2x/p={90:1}}, {3x/p={-150:1}}}
\atoms{vertex}{4x/p=-90:0.3}
\draw[bdribbon] (0x)edge[mark=arr](1x) (0x)edge[mark=arr](2x) (0x)edge[mark={arr,-}](3x);
\draw (4x)edge[bend right=10, mark={arr,-}](1x) (4x)edge[bend right,mark={arr,-}](2x) (4x)edge[bend left=10,mark={arr}](3x);
\end{scope}
\begin{scope}[shift={(-2,-2.05)}]
\atoms{vertex}{0y/p={3,0}, 1y/p={3,1.6}, 2y/p={3.5,0.8}, 3y/p={4,0.8}}
\draw (0y)edge[mark=arr](1y) (0y)edge[mark={arr,-}](2y) (0y)edge[mark={arr}](3y) (1y)edge[mark=arr](2y) (2y)edge[mark={arr,-}](3y) (1y)edge[mark=arr](3y);
\end{scope}
\draw[glue] (3)--(2x) (4x)--(0y) (4)--(1y);
\atoms{glueedge}{x/p={$(0y)!0.7!(1y)$}}
\end{tikzpicture}
=
\begin{tikzpicture}
\begin{scope}[rotate=-30]
\atoms{bdvertex}{0/, {1/p={30:1}}, {2/p={150:1}}, {3/p={-90:1}}}
\atoms{vertex}{4/p=-30:0.3}
\draw[bdribbon] (0)edge[mark=arr](1) (0)edge[mark=arr](2) (0)edge[mark={arr,-}](3);
\draw (4)edge[bend right=10, mark={arr,-}](1) (4)edge[bend right,mark={arr,-}](2) (4)edge[bend left=10,mark={arr}](3);
\end{scope}
\begin{scope}[shift={(2.5,0)},rotate=-30]
\atoms{bdvertex}{0x/, {1x/p={-30:1}}, {2x/p={90:1}}, {3x/p={-150:1}}}
\atoms{vertex}{4x/p=-90:0.3}
\draw[bdribbon] (0x)edge[mark=arr](1x) (0x)edge[mark=arr](2x) (0x)edge[mark={arr,-}](3x);
\draw (4x)edge[bend right=10, mark={arr,-}](1x) (4x)edge[bend right,mark={arr,-}](2x) (4x)edge[bend left=10,mark={arr}](3x);
\end{scope}
\atoms{bdvertex}{0y/p={0.45,1}, 1y/p={2.05,1}, 2y/p={1.25,1.5}, 3y/p={1.25,2}}
\draw[bdribbon] (0y)edge[mark={arr,-}](1y) (0y)edge[mark={arr}](2y) (0y)edge[mark={arr,-}](3y) (1y)edge[mark={arr,-}](2y) (2y)edge[mark={arr}](3y) (1y)edge[mark={arr,-}](3y);
\draw[glue] (1)--(3x) (0x)--(1y) (0)--(0y);
\atoms{glueedge}{x/p={$(0y)!0.7!(1y)$}}
\end{tikzpicture}\;.
\end{equation}
This makes the triple condensation a Frobenius algebra object in the tensor (i.e., Deligne) product of the boundary fusion category and the UMTC (without braiding).
\item An axiom consisting of two triple condensations, one boundary braiding, and one bulk braiding,
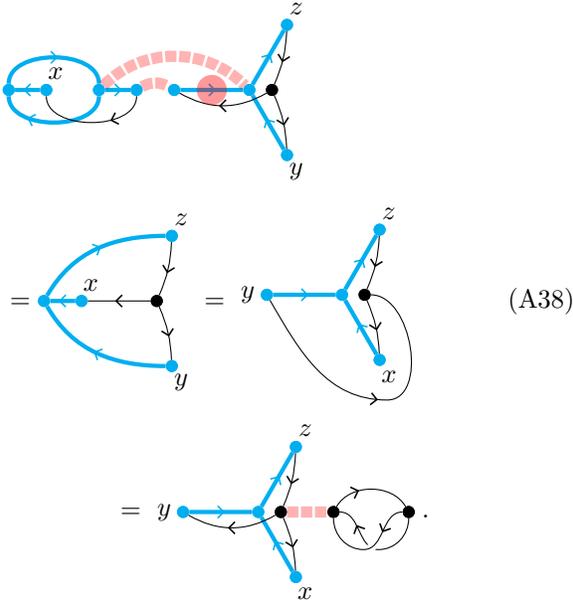
\begin{equation}
\begin{multlined}
\begin{tikzpicture}
\atoms{bdvertex}{0/, {1/p={0.5,0},lab={t=$x$,p=60:0.25}}, 2/p={1.2,0}, 3/p={1.7,0}}
\draw[bdribbon] (0)edge[bend left=90,mark={arr}](2) (0)edge[bend right=90,mark={arr,-,p=0.3}](2) (0)edge[mark={arr,-}](1) (2)edge[mark={arr}](3);
\draw (1)edge[bend right=90, mark={arr,-,p=0.7}](3);
\begin{scope}[shift={(3.2,0)}]
\atoms{bdvertex}{0x/, {1x/p={60:1},lab={t=$z$,p=60:0.25}}, {2x/p={180:1}}, {3x/p={-60:1},lab={t=$y$,p=-60:0.25}}}
\atoms{vertex}{4x/p=0:0.3}
\draw[bdribbon] (0x)edge[mark=arr](1x) (0x)edge[mark={arr,-}](2x) (0x)edge[mark={arr,-}](3x);
\draw (4x)edge[bend right=10, mark={arr,-}](1x) (4x)edge[bend left,mark={arr}](2x) (4x)edge[bend left=10,mark={arr}](3x);
\end{scope}
\draw[glue] (3)to[bend left](2x) (2)to[bend left=45](0x);
\atoms{glueedge}{x/p={$(0x)!0.5!(2x)$}}
\end{tikzpicture}
\\
=
\begin{tikzpicture}
\atoms{bdvertex}{0/p={-1.2,0}, {1/p={-0.7,0},lab={t=$x$,p=60:0.25}}}
\draw[bdribbon] (0)edge[mark={arr,-}](1);
\atoms{bdvertex}{{1x/p={60:1},lab={t=$z$,p=60:0.25}}, {3x/p={-60:1},lab={t=$y$,p=-60:0.25}}}
\atoms{vertex}{4x/p=0:0.3}
\draw[bdribbon] (0)edge[out=60,in=180,mark=arr](1x) (0)edge[out=-60,in=180,mark={arr,-}](3x);
\draw (4x)edge[bend right=10, mark={arr,-}](1x) (4x)edge[mark={arr}](1) (4x)edge[bend left=10,mark={arr}](3x);
\end{tikzpicture}
=
\begin{tikzpicture}
\atoms{bdvertex}{0/, {1/p={60:1},lab={t=$z$,p=60:0.25}}, {2/p={180:1},lab={t=$y$,p=180:0.25}}, {3/p={-60:1},lab={t=$x$,p=-60:0.25}}}
\atoms{vertex}{4/p=0:0.3}
\draw[bdribbon] (0)edge[mark=arr](1) (0)edge[mark={arr,-}](2) (0)edge[mark={arr,-}](3);
\draw (4)edge[bend right=10, mark={arr,-}](1) (4)edge[bend left=10,mark={arr}](3);
\draw[mark={arr}] (2)to[out=-60,in=180](0.6,-1.4)to[out=0,in=0](4);
\end{tikzpicture}
\\
=
\begin{tikzpicture}
\atoms{bdvertex}{0/, {1/p={60:1},lab={t=$z$,p=60:0.25}}, {2/p={180:1},lab={t=$y$,p=180:0.25}}, {3/p={-60:1},lab={t=$x$,p=-60:0.25}}}
\atoms{vertex}{4/p=0:0.3}
\draw[bdribbon] (0)edge[mark=arr](1) (0)edge[mark={arr,-}](2) (0)edge[mark={arr,-}](3);
\draw (4)edge[bend right=10, mark={arr,-}](1) (4)edge[bend left,mark={arr}](2) (4)edge[bend left=10,mark={arr}](3);
\atoms{vertex}{0x/p={1,0}, 1x/p={2,0}}
\draw[mark={arr,p=0.3}] (0x)to[bend right=-60](1x);
\draw[mark={arr,-,p=0.25}] (0x)to[out=0,in=180]++(0.6,-0.5)to[out=0,in=-90](1x);
\draw[front,mark={arr,p=0.25}] (1x)to[out=180,in=0]++(-0.6,-0.5)to[out=180,in=-90](0x);
\draw[glue] (4)--(0x);
\end{tikzpicture}\;.
\end{multlined}
\end{equation}
\item All the above gluing axioms hold with and without a 4-dimensional/3-dimensional bulk, and thus for braided as well as Lagrangian modules. The following \emph{Lagrangian condition} is analogous to the modularity condition for UMTCs in Eq.~\eqref{eq:umtc_modularity}. It only holds without the bulk,
\begin{equation}
\label{eq:lagrangian_condition1}
\begin{tikzpicture}
\atoms{bdvertex}{0/, 1/p={1,0}}
\draw (0)edge[mark={arr,p=0.7}](1);
\draw[mark={arr,-},bdribbon] (0)circle(0.5);
\begin{scope}[shift={(2.5,0)},xscale=-1]
\atoms{bdvertex}{0x/, 1x/p={1,0}}
\draw (0x)edge[mark={arr,-,p=0.7}](1x);
\draw[mark={arr},bdribbon] (0x)circle(0.5);
\end{scope}
\draw[glue] (1)--(1x) (0)to[bend right=50](0x);
\atoms{glueedge}{x/p={0.7,0}}
\end{tikzpicture}
=
\begin{gathered}
\begin{tikzpicture}
\draw[bdribbon,mark={arr,s},mark={slab=$a$}] (0,0)arc(-180:180:0.5);
\draw[bdribbon,mark={arr,s},mark={slab=$b$}] (1.3,0)arc(-180:180:0.5);
\end{tikzpicture}\\
\in B_2\times S_1,\\
a\sim x\times S_1,
b\sim y\times S_1
\end{gathered}
\;,
\end{equation}
where $x$ and $y$ are any two points on the boundary of $B_2$. It follows from the gluing axioms and a normalization condition that the tensor assigned to the right-hand side is $\delta_{a,b}$. On the other hand, we have
\begin{equation}
\label{eq:lagrangian_condition2}
\begin{multlined}
\begin{tikzpicture}
\atoms{bdvertex}{0/, 1/p={1,0}}
\draw (0)edge[mark={arr,p=0.7}](1);
\draw[mark={arr,-,p=0.25},bdribbon] (0)circle(0.5);
\begin{scope}[shift={(-1.5,0)},xscale=-1]
\atoms{bdvertex}{0x/, 1x/p={1,0}}
\draw (0x)edge[mark={arr,-,p=0.7}](1x);
\draw[mark={arr,p=0.25},bdribbon] (0x)circle(0.5);
\end{scope}
\draw[glue] (-0.5,0)--(-1,0);
\end{tikzpicture}
\\
=
\begin{gathered}
\begin{tikzpicture}
\atoms{bdvertex}{{0/lab={t=$\alpha$,p=180:0.25}}, {1/p={1,0},lab={t=$\alpha'$,p=0:0.25}}}
\draw (0)edge[mark={arr,p=0.7},mark={slab=$a$}](1);
\atoms{bdvertex}{{0x/p={2,0},lab={t=$\beta$,p=180:0.25}}, {1x/p={3,0},lab={t=$\beta'$,p=0:0.25}}}
\draw (0x)edge[mark={arr,p=0.7},mark={slab=$b$}](1x);
\end{tikzpicture}\\
\in B_1\times S_2,\\
a\sim B_1\times x, b\sim B_1\times y
\end{gathered}
\;,
\end{multlined}
\end{equation}
where $x$ and $y$ are any two points of $S_2$. Again, the tensor on the right-hand side is $\delta_{(a,\alpha,\alpha'), (b,\beta,\beta')}$. Thus, the matrix
\begin{equation}
\label{eq:lagrangian_condition}
L_{a\alpha\alpha'}^b\coloneqq
\begin{tikzpicture}
\atoms{bdvertex}{{0/lab={t=$\alpha$,p=180:0.25}}, {1/p={1,0},lab={t=$\alpha'$,p=0:0.25}}}
\draw (0)edge[mark={arr,p=0.7},mark={slab=$a$,p=0.7}](1);
\draw[mark={arr,-},bdribbon,mark={slab=$b$,r}] (0)circle(0.5);
\end{tikzpicture}
\end{equation}
is unitary, $LL^\dagger=L^\dagger L=\mathbb{1}$. Now, consider this ribbon manifold with a bulk $4$-ball $B_4^{br}$ whose boundary is half bulk, half ribbon manifold, using notation analogous to $B^{ps}_i$ in Section~\ref{sec:state_sums_boundaries}. Then the gluing operation on the left-hand side of Eq.~\eqref{eq:lagrangian_condition1} would yield $B^r_2\times B^b_2$, whereas the bulk for which we get the tensor $\delta_{a,b}$ on the right-hand side is $B^{br}_3\times S_1$. 
\end{itemize}
To actually prove that all gluing axioms follow from the above we would have to carry out an analog of Ref.~\cite{Lauda2005} in one dimension higher, or an analog of Ref.~\cite{Bartlett2014} with boundary, which would clearly exceed the scope of this paper. Note that to make this fully work, we again need to add a few small auxiliary fusion vertex links, ribbon manifolds, and gluing axioms similar to Eq.~\eqref{eq:auxiliary_umtc_axiom1}.

Let us now discuss the connection between our tensorial definition of Lagrangian/braided modules and known categorical structures. There are many different possible viewpoints giving rise to equivalent names for those structures. First, as we noticed, Lagrangian/braided modules can be viewed as a categorification of knowledgeable Frobenius algebras. That is, the axioms of the latter give rise to natural transformations, mapping between different string diagrams, like the $F$- or $R$-tensor. In our framework, we connect the open ends of the string diagrams on both sides to a single string diagram.

As another viewpoint, we have a bulk UMTC (or braided fusion category) $\calm$ together with a boundary fusion category $\calb$, interacting with each other. To study the interaction, let us define a new fusion vertex link,
\begin{equation}
\begin{tikzpicture}
\atoms{bdvertex}{0/}
\draw (0)edge[bdribbon,mark={arr,-}]++(180:0.6) (0)edge[bdribbon,mark={arr}]++(0:0.6) (0)edge[mark={arr,-}]++(90:0.6);
\end{tikzpicture}
\rightarrow
\begin{tikzpicture}
\atoms{bdvertex}{0/, 1/p={0,0.5}}
\draw (0)edge[bdribbon,mark={arr,-}]++(180:0.6) (0)edge[bdribbon,mark={arr,-}](1) (0)edge[bdribbon,mark={arr}]++(0:0.6) (1)edge[mark={arr,-}]++(90:0.6);
\end{tikzpicture}
\;.
\end{equation}
This corresponds to a bi-functor
\begin{equation}
\triangleright:\calm\times \calb\rightarrow\calb\;,
\end{equation}
where we do not need the braiding of $\calm$ and the monoidal (fusion-category) structure of $\calb$. Now, consider the following ribbon manifold,
\begin{equation}
\label{eq:bulk_boundary_tetrahedron}
\begin{tikzpicture}
\atoms{vertex}{0/}
\atoms{bdvertex}{{1/p={-150:1}}, {2/p={-30:1}}, {3/p={90:1}}}
\draw (0)edge[mark=arr](1) (0)edge[mark=arr](2) (0)edge[mark={arr,-}](3);
\draw[bdribbon] (1)edge[mark=arr](2) (2)edge[mark=arr](3) (1)edge[mark=arr](3);
\end{tikzpicture}\;.
\end{equation}
Cutting this into two pieces yields, e.g., the following move of string diagrams,
\begin{equation}
\begin{tikzpicture}
\atoms{bdvertex}{1x/, 2x/p={0.8,0}}
\draw[bdribbon] (1x)edge[mark={arr}](2x) (1x)edge[mark={three dots,a},mark={arr,-}]++(180:0.5) (2x)edge[mark={three dots,a},mark={arr}]++(0:0.5);
\draw (1x)edge[mark={three dots,a},mark={arr,-}]++(-90:0.5) (2x)edge[mark={three dots,a},mark={arr,-}]++(-90:0.5);
\end{tikzpicture}
\leftrightarrow
\begin{tikzpicture}
\atoms{vertex}{0/p={0,-0.7}}
\atoms{bdvertex}{1/}
\draw[bdribbon] (1)edge[mark={three dots,a},mark={arr,-}]++(180:0.5) (1)edge[mark={three dots,a},mark={arr}]++(0:0.5);
\draw (0)edge[mark={arr}](1) (0)edge[mark={three dots,a},mark={arr,-}]++(-135:0.5) (0)edge[mark={three dots,a},mark={arr,-}]++(-45:0.5);
\end{tikzpicture}
\;.
\end{equation}
This defines a natural isomorphism
\begin{equation}
(M \otimes N)\triangleright B
\rightarrow M\triangleright (N\triangleright B)\;,
\end{equation}
for all $M,N\in \calm$ and $B\in\calb$. Such a functor is the most important part of the definition of a \emph{module category} \cite{Ostrik2001}, which is sometimes also referred to as having $\calb$ as a $\calm$-module. The functor needs to obey a coherence condition that looks like the pentagon equation where four of the tetrahedron tensors are replaced by the functor. Note that the module category structure does not fully determine the braided module, since we have used neither the braiding of $\calm$ nor the tensor product (i.e., monoidal, or fusion-category structure) of $\calb$. For a module category, $\calm$ is just a fusion category and $\calb$ is a plain category.

In order to study the role of the braiding of $\calm$, let us introduce another boundary fusion vertex link with only a bulk ribbon attached,
\begin{equation}
\begin{tikzpicture}
\atoms{bdvertex}{0/}
\draw (0)edge[mark={arr,-}]++(90:0.6);
\end{tikzpicture}
\rightarrow
\begin{tikzpicture}
\atoms{bdvertex}{0/, 1/p={0,0.5}}
\draw (1)edge[mark={arr,-}]++(90:0.6);
\draw[bdribbon] (0)edge[mark={arr,-}](1);
\draw[bdribbon,mark={arr,p=0.4}, looseness=2] (0)to[out=0,in=0]++(0,-0.6)to[out=180,in=180](0);
\end{tikzpicture}\;.
\end{equation}
The dimension $A_i$ of this new fusion vertex with bulk ribbon with label $i$ determines a non-necessarily-simple (isomorphism class of) object(s)
\begin{equation}
A=\bigoplus_i A_i \cdot i
\end{equation}
of the UMTC.
Physically speaking, this object contains all the anyons that can condense at the boundary.
Then we consider the triple condensation of Eq.~\eqref{eq:bulk_boundary_tetrahedron} without boundary ribbons,
\begin{equation}
\begin{tikzpicture}
\atoms{vertex}{0/}
\atoms{bdvertex}{{1/p={-150:0.7}}, {2/p={-30:0.7}}, {3/p={90:0.7}}}
\draw (0)edge[mark=arr](1) (0)edge[mark=arr](2) (0)edge[mark={arr,-}](3);
\end{tikzpicture}\;,
\end{equation}
corresponding to an isomorphism
\begin{equation}
\begin{tikzpicture}
\atoms{bdvertex}{1x/, 2x/p={0.8,0}}
\draw (1x)edge[mark={three dots,a},mark={arr,-}]++(-90:0.5) (2x)edge[mark={three dots,a},mark={arr,-}]++(-90:0.5);
\end{tikzpicture}
\leftrightarrow
\begin{tikzpicture}
\atoms{vertex}{0/p={0,-0.7}}
\atoms{bdvertex}{1/}
\draw (0)edge[mark={arr}](1) (0)edge[mark={three dots,a},mark={arr,-}]++(-135:0.5) (0)edge[mark={three dots,a},mark={arr,-}]++(-45:0.5);
\end{tikzpicture}
\;.
\end{equation}
This isomorphism is an element of
\begin{equation}
\bigoplus_{i,j,k} A_i A_j A_k \operatorname{Hom}(i\otimes j, k) \simeq \operatorname{Hom}(A\otimes A, A)\;,
\end{equation}
and makes $A$ into an \emph{algebra object} in the UMTC/braided fusion category, or in other words, an algebra enriched by the UMTC,
The algebra has to be associative as well as commutative, which follows from the gluing axioms. It also has to obey the \emph{Frobenius property}, which is automatic in our language where there is no distinguished time direction. In a physics context, the commutativity ensures that all the condensing anyons are bosons (i.e., have topological twist 1) and braid trivially with each other,
\begin{equation}
\begin{tikzpicture}
\atoms{bdvertex}{1/}
\draw (1)edge[mark={three dots,a},mark={arr,-}]++(-90:0.8);
\end{tikzpicture}
=
\begin{tikzpicture}
\atoms{bdvertex}{1/}
%\draw[bdribbon, mark=arr,looseness=2] (1)to[out=0,in=0]++(0,0.5)to[out=180,in=180](1);
\draw[front,looseness=2,mark={three dots,a},mark={arr,-}] (1)to[out=-90,in=-90]++(0.6,-0.6)to[out=90,in=90](0,-1.2);
\end{tikzpicture}
\;,\qquad
\begin{tikzpicture}
\atoms{bdvertex}{1/, 2/p={0.8,0}}
\draw (1)edge[mark={three dots,a},mark={arr,-}]++(-90:0.8) (2)edge[mark={three dots,a},mark={arr,-}]++(-90:0.8);
\end{tikzpicture}
=
\begin{tikzpicture}
\atoms{bdvertex}{1/, 2/p={0.8,0}}
\draw (1)edge[out=-90,in=90](0.8,-0.7) (2)edge[front,out=-90,in=90](0,-0.7) (0,-0.7)edge[mark={arr,s},front,mark={three dots,a},out=-90,in=90](0.8,-1.4) (0.8,-0.7)edge[mark={arr,s},front,mark={three dots,a},out=-90,in=90](0,-1.4);
\end{tikzpicture}
\;.
\end{equation}

In order to study the role of the tensor product of $\calb$, we can consider the ribbon manifold
\begin{equation}
\begin{tikzpicture}
\atoms{vertex}{0/}
\atoms{bdvertex}{{1/p={60:0.8}}, {2/p={120:0.8}}, {3/p={-90:0.8}}, 4/p={-1.2,0}, 5/p={1.2,0}}
\draw (0)edge[mark=arr](1) (0)edge[mark=arr](2) (0)edge[mark={arr,-}](3);
\draw[bdribbon] (2)edge[mark={arr,-}](4) (3)edge[mark={arr,-}](4) (1)edge[mark=arr](5) (2)edge[mark=arr](5) (3)edge[mark=arr](5);
\draw[bdribbon,mark={arr,-}] (1)to[out=150,in=90](4);
\end{tikzpicture}
\end{equation}
corresponding to a natural isomorphism
\begin{equation}
\begin{tikzpicture}
\atoms{bdvertex}{0/}
\atoms{bdvertex}{1/p={-135:0.5}, 2/p={-45:0.5}}
\draw (1)edge[mark={three dots,a},mark={arr,-}]++(-45:0.5) (2)edge[mark={three dots,a},mark={arr,-}]++(45:0.5);
\draw[bdribbon] (0)edge[mark={arr,-}](1) (0)edge[mark={arr,-}](2) (1)edge[mark={three dots,a},mark={arr,-}]++(-135:0.5) (2)edge[mark={three dots,a},mark={arr,-}]++(-45:0.5) (0)edge[mark={three dots,a},mark={arr}]++(90:0.5);
\end{tikzpicture}
\leftrightarrow
\begin{tikzpicture}
\atoms{bdvertex}{0/p={0,-0.7}}
\atoms{bdvertex}{1/}
\atoms{vertex}{2/p={0.5,0}}
\draw (1)edge[mark={arr,-}](2) (2)edge[mark={three dots,a},mark={arr,-}]++(45:0.5) (2)edge[mark={three dots,a},mark={arr,-}]++(-45:0.5);
\draw[bdribbon] (0)edge[mark={arr}](1) (0)edge[mark={three dots,a},mark={arr,-}]++(-135:0.5) (0)edge[mark={three dots,a},mark={arr,-}]++(-45:0.5) (1)edge[mark={three dots,a},mark={arr}]++(90:0.5);
\end{tikzpicture}
\;.
\end{equation}
So a braided module corresponds to $\calb$ being module category of $\calm$, which is also consistent with the braiding of $\calm$ and the fusion of $\calb$.

A Lagrangian module has to obey the extra condition that $L$ defined in Eq.~\eqref{eq:lagrangian_condition} is a unitary matrix. A commutative Frobenius algebra is called Lagrangian if its category of local modules is trivial. In physics terms, this means that any anyon that does not condense, braids non-trivially with a condensing anyon and thus confines. The Lagrangian condition in Eq.~\eqref{eq:lagrangian_condition1} and Eq.~\eqref{eq:lagrangian_condition2} states that the braiding between condensing and boundary anyons is unitary. The boundary anyons are just confined bulk anyons, and the unitarity implies that only the trivial anyon (which always condenses) can braid trivially with all confined anyons.

A third approach is to view the UMTC/braided fusion category as a special case of a fusion 2-category. Fusion 2-categories are equivalent to a type of tensorial TQFT whose extended manifolds are 3-manifolds with an embedded network of membranes, lines where membranes meet, and vertices where lines meet, with a 4-dimensional bulk. That is, it is a TQFT that is 3-2-1-0-extended at the 3-dimensional boundary and trivial (4-extended) in the 4-dimensional bulk. From any such fusion 2-category we can define a 4-3-2-1-0-extended TQFT (i.e., a microscopic lattice model). Braided fusion categories are a special case of fusion 2-categories with only one membrane label, and so the CYWW model is a special case of the 4-3-2-1-0-extended fusion 2-category model \cite{Xi2021}. A 3-2-1-0-extended boundary of this 4-3-2-1-0-extended TQFT (i.e., a microscopic model for a boundary) corresponds to an \emph{enriched fusion category} in that fusion 2-category. A braided module is equivalent to a fusion category enriched in the UMTC.

As a fourth approach is via an operation known as the \emph{Drinfeld center} which maps a fusion category $\calb$ to a ($c=0$) UMTC $Z(\calb)$ by pulling back the following geometric mapping from 3-manifolds with bulk ribbons to 3-manifolds with boundary ribbons. Remove the tubular neighborhood of the bulk ribbons and fusion vertices, and embed a sufficient boundary ribbon network into the emerging boundary. Specifically, it suffices to take one boundary ribbon along every tube segment and one around the non-contractible loop, as well as one fusion vertex at every junction of tube segments. More precisely, this pullback yields a 3-2-1-extended tensorial TQFT, which has to be block-diagonalized as described in Appendix~\ref{sec:umtc_tensorial_tqft} to get a UMTC. The algebra that we need to block-diagonalize as in Eq.~\eqref{eq:321_210_mapping} is known as \emph{tube algebra}. Now, the condensation vertex in Eq.~\eqref{eq:module_generating_vertices} of a braided module provides a way to connect a bulk ribbon of $\calm$ with a tube of $Z(\calb)$. Categorically, this defines a functor from the braided fusion category $\calm$ to $Z(\calb)$.

For Lagrangian modules, the gluing axioms without bulk imply that the functor from $\calm$ to $Z(\calb)$ defines an equivalence of braided fusion categories. Note that this is analogous to the situation in one dimension lower: A knowledgeable Frobenius algebra is a commutative Frobenius algebra and another (not necessarily commutative) Frobenius algebra, together with an isomorphism between the former and the (commutative) center of the latter. The whole Lagrangian module can be constructed from only the boundary fusion category $\calb$ by pulling back a geometric mapping. A boundary ribbon manifold (with both bulk and boundary ribbons) is mapped to a ribbon manifold with ribbons only in the boundary as follows. Bulk ribbons are replaced by tubes just like in the Drinfeld center mapping. The boundary is left the same except that bulk ribbons adjacent to boundary fusion vertices are replaced by boundary ribbons coming out of holes in the boundary.

Another viewpoint is to think of the fusion category as a 3-2-1-0-extended TQFT together with a 2-1-0-extended boundary, namely a state sum known as the \emph{Turaev-Viro-Barrett-Westbury} model based on the fusion category together with its \emph{cone boundary} (like the ``smooth'' boundary of the toric code). It is a type of tensorial TQFT that associates tensors to 3-manifolds with physical and space boundary, with indices at the space boundary of the bulk as well as the space boundary of the physical boundary. The UMTC and its Lagrangian module can be obtained by pulling back the following geometric operation from boundary ribbon manifolds to manifolds with physical and space boundary. We remove a tubular neighborhood of both the bulk ribbons and the boundary ribbons yielding tubes (in the bulk) and stripes (at the boundary) of space-boundary. Restricted to the bulk, this mapping is again the Drinfeld center of the fusion category.

%  Now, a 3-2-1-0 extended TQFT should automatically contain a 3-2-1-extended TQFT, and this is also the case in our framework. The Turaev-Viro-Barrett-Westbury state sum associates tensors to extended manifolds with only an internal 3-region and a space 2-region. A ribbon manifold (to which the UMTC assigns tensors) can be made into such a 3-manifold with space boundary by simply cutting out the tubular neighborhood of the ribbon network. Pulling back this geometric operation yields a map from 3-2-1-0-extended TQFTs to 3-2-1-extended TQFTs, on the level of generators and relations. This is equivalent to what is known as \emph{Drinfeld center}.  We can perform an analogous construction at the boundary, cutting out the tubular neighborhood of the boundary ribbons to get a manifold with space boundary. Pulling back this operation, we get the Lagrangian algebra.

% So in summary, we find that a UMTC allows for a Lagrangian module iff it is the Drinfeld center of a fusion category. A general enriched fusion category in a braided fusion category corresponds to a \emph{functor} from the braided fusion category to the Drinfeld center of the fusion category. For a Lagrangian module this functor is an isomorphism between the UMTC and the Drinfeld center. 

\section{Tensorial TQFT}
\label{sec:tensorial_tqft}
\subsection{General framework}
\label{sec:tensorial_general_definition}
In this appendix, we propose a simple axiomatization of what appears to be equivalent to (a generalization of) extended TQFTs, and which we will refer to as \emph{tensorial (extended) TQFT}. Roughly, such a tensorial TQFT is a map from some type of geometric/topological objects to tensors, subject to \emph{gluing axioms}. The geometric objects are \emph{combinatorial extended manifolds} of a fixed \emph{type}, which roughly speaking are composites of manifolds of different dimensions being adjacent to or embedded into each other, and which can be defined inductively in their (maximal) dimension.

An \emph{extended manifold type} of (maximal) dimension $n$ consists of a set of \emph{regions}, and for each region $r$ a \emph{dimension} $0\leq d_r\leq n$ and a \emph{link} $L_r$ which is an extended manifold of maximal dimension $n-d_r-1$. A region of dimension $d_r$ will also be called a $d_r$-region. The type of $L_r$ can be obtained from the original type by making each region $r'$ of dimension $d_{r'}$ into a region of dimension $d_{r'}-d_r-1$ and then discarding regions with negative dimension. An extended manifold $M$ of a given type consists of 1) a (compact piece-wise linear) $d_r$-manifold $M_r$ with boundary $\partial M_r$ for every region $r$ and 2) for every pair of regions $r$, $r'$, a (piece-wise linear) map
\begin{equation}
\label{eq:extended_adjacency_map}
\psi^M_{r,r'}: M_{r}\times (L_r)_{r'}
\rightarrow
\partial M_{r'}
\end{equation}
subject to the relations
\begin{equation}
\label{eq:extended_manifold_relation}
\psi^M_{r_1,r_2}(\psi^M_{r_0,r_1}(x, p_0), p_1) = \psi^M_{r_0,r_2}(x, \psi^{L_{r_0}}_{r_1,r_2}(p_0, p_1))
\end{equation}
for all regions $r_0$, $r_1$, $r_2$, and $x\in r_0$, $p_0\in (L_{r_0})_{r_1}$, $p_1\in (L_{r_1})_{r_2}$. Additionally, we demand that
\begin{equation}
\begin{multlined}
(\bigcup_{r} \psi^M_{r, r'}) / [\text{Eq.~\eqref{eq:extended_manifold_relation}}]:
\\
(\bigcup_{r} M_r\times (L_r)_{r'}) / [\text{Eq.~\eqref{eq:extended_manifold_relation}}]\rightarrow \partial M_{r'}
\end{multlined}
\end{equation}
is a (piece-wise linear) homeomorphism for all regions $r'$. Two extended manifolds of the same type are considered equivalent if all of their regions are homeomorphic and the homeomorphisms commute with $\psi$.

While the above definition is precise, the following informal \emph{collapsed picture} might be simpler to understand or draw. We first extend $M_{r'}$ by gluing $M_r\times \cone((L_r)_{r'})$ to the image of the map $\psi_{r,r'}^{M}$ within $\partial M_{r'}$, recalling that $\cone(X)$ is $[0,1]\times X$ with $0\times X$ identified with a single point $c$. We then think of $M_r$ as being attached to the extended $M_{r'}$ along $M_r\times c$. We do this recursively for the manifold itself as well as all the links. That way we obtain a space directly consisting of the different $M_r$ that are attached to each other in different ways. Within this space, consider the normal space of maximal dimension $n-d_r$ at a point of $M_r$ for a region $r$. The distance-$\epsilon$ neighborhood within the extended manifold around the point restricted to this normal space is $L_r$. As an example, consider an extended manifold type of dimension 2 with a 2-region $a$ and a 1-region $b$ whose link consists of two points. Note that the link of a maximal-dimension region such as $a$ is a $-1$-manifold, and we choose the convention that there is one single (empty) $-1$-manifold. The collapsed picture for such an extended manifold is a 2-manifold ($a$) with embedded circles ($b$). The $\epsilon$-neighborhood around a point of $b$ restricted to a normal space consists of $2$ points, which is the link of $b$. As another example, consider an extended manifold type of dimension 3 with a 3-region $a$, a 2-region $b$ with point link, and a 0-region $c$ with disk link. The collapsed picture for such an extended manifold is a 3-manifold ($a$) with boundary ($b$) and embedded points within the boundary ($c$). The distance-$\epsilon$ neighborhood of a point of $b$ restricted to the normal space consists of one point, the link of $b$. The distance-$\epsilon$ neighborhood of a point in $c$ is a disk, and since everything is normal to a point, this is the link of $c$. Note that the collapsed picture loses one important aspect of the formal definition, namely a normal framing of the various submanifolds. E.g., in our first example the embedded circles have a favorite side, and in the second example, the points embedded into the boundary carry a unit vector inside the boundary (which, however, disappears when we consider equivalence classes).

A \emph{combinatorial extended manifold} is given by an extended manifold where $M_r$ and $(L_r)_{r'}$ for all regions $r$ and $r'$ are cellulated such that the map $\psi$ in Eq.~\eqref{eq:extended_adjacency_map} is a map of cellulations. Furthermore, in order to unambiguously define gluing below, we need a branching as defined in the main text, i.e., an identification of every $r$-simplex in $M_r$ with a standard representative. As the name suggests, a combinatorial extended manifold contains only the combinatorial data describing the cellulation and the cell map $\psi$, which suffices to specify $M$ up to equivalence/homeomorphism.
\footnote{We can restrict without loss of generality to triangulations, with only one standard representative, the $d_r$-simplex $\operatorname{Sx}_{d_r}$, such that the branching is a branching structure. In this case  $M_{r}\times (L_{r})_{r'}$ is not a triangulation but consists of cells of the form $\operatorname{Sx}_{d_r}\times \operatorname{Sx}_{d_{r'}-d_r-1}$, so in order to make $\psi$ a map of triangulations we have to choose a standard triangulation of $\operatorname{Sx}_{d_r}\times \operatorname{Sx}_{d_{r'}-d_r-1}$.}

A \emph{tensorial TQFT type} consists of 1) an extended manifold type and 2) a labeling of each region as either an \emph{internal} or a \emph{space} region. A \emph{tensorial TQFT} of given type is a map that associates a tensor to every combinatorial extended manifold, with one index associated to every $d_r$-cell of $M_r$ for every space region $r$. The vector-space dimension of the index can depend on $r$ and the standard representative of the $d_r$-cell. A tensorial TQFT also includes a \emph{weight matrix} $D$ for every $d_r-1$-cell of every standard representative of a $d_r$-cell for every space region $r$.

The map is subject to consistency conditions we will call \emph{gluing axioms}. A gluing axiom is a commutative diagram as in Eq.~\eqref{eq:gluing_axiom_general} where $G$ is a ``gluing'' operation, and $C$ is a ``contraction'' of the associated tensors. There are three types of gluing axioms. First, $G$ is topology-preserving recellulation (such as bistellar flips/Pachner moves) inside one of the internal regions, and $C$ is trivial. In other words, the tensors are independent of the cellulation of the internal regions, but only depend on their topology. Second, $G$ is the disjoint union of combinatorial extended manifolds and $C$ is the tensor product.

The third type of gluing axiom is when $G$ is a ``proper'' gluing operation, and $C$ is a ``proper'' index contraction. A proper gluing operation is defined as follows for any space region $r$ and any pair of $d_r$-cells $C_1$ and $C_2$ of $M_r$ with the same standard representative. First, we remove $C_1$ and $C_2$ from $M_r$ and identify/glue $\partial C_1$ with $\partial C_2$. Second, we identify/glue $\psi_{rr'}(C_1\times (L_r)_{r'})$ with $\psi_{rr'}(C_2\times (L_r)_{r'})$ for every region $r'$. If there are any $d_r-1$-cells of the standard representative that coincide for $\partial C_1$ and $\partial C_2$, then those are removed when gluing. If $\partial C_1$ and $\partial C_2$ share $d_r-1$-cells that do not correspond to the same cell of the standard representative, the identification/gluing of cells is extended transitively. The corresponding proper index contraction is the Einstein summation over two indices located at $C_1$ and $C_2$, apart from the following. For every $d_r-1$-cell of the standard representative that coincides for $\partial C_1$ and $\partial C_2$, we have to insert the according weight matrix $D$. That is, if the indices at $C_1$ and $C_2$ are labeled $a_1$ and $a_2$, then the tensor $T$ is contracted as
\begin{equation}
\sum_{a_1,a_2} T_{a_1,a_2,\ldots} D_{a_1,a_2}\;.
\end{equation}
The different weight matrices for a standard representative commute among each other, and if $C_1$ and $C_2$ have a coinciding $d_r-1$-cell, then the weight matrix can be moved from $C_1$ to $C_2$. For example, if we glue two branching-structure triangles, there are three different weight matrices $D^{01}$, $D^{02}$, and $D^{12}$, which we need to include when they share either their $01$, their $02$, or their $12$ edge. For an example of how this works in the case of $2+1$-dimensional lattice TQFT we refer the reader to Ref.~\cite{liquid_intro}.

With this, the definition of a tensorial TQFT is complete. It is interesting to consider the topological operations that can be implemented by a proper gluing of two cells $C_1$ and $C_2$. For any $0\leq x\leq d_r$ and every embedding of $S_x\times B_{d_r-x}$ into $M_r$, we can perform an $x$-surgery on $M_r$,
\begin{equation}
\label{eq:general_gluing}
S_x\times B_{d_r-x} \rightarrow B_{x+1}\times S_{d_r-x-1}\;,
\end{equation}
and attach
\begin{equation}
B_{x+1}\times B_{d_r-x} \times (L_r)_{r'}
\end{equation}
to $M_{r'}$ along $\psi_{r,r'}(S_x \times B_{d_r-x} \times (L_r)_{r'})$, where $S_x$ is identified with $\partial B_{x+1}$. To this end, the cells of the standard representative that coincide for $\partial C_1$ and $\partial C_2$ must be topologically equal to $S_{x-1}\times B_{r-x+1}$. In particular, we have the case $x=0$ if $\partial C_1$ and $\partial C_2$ are disjoint. If the set of coinciding cells has a different topology or $\partial C_1$ and $\partial C_2$ share any other $d_r-1$-cells, then gluing can correspond to other topological operations. However, all those operations are generated by the surgery operations above.

Depending on the type, tensorial TQFT as defined above can describe many different things. First of all, tensorial TQFT with a single internal $n$-region and space $d$-regions for $d\leq n-i-1$ with arbitrary links seems to be equivalent to \emph{extended TQFT}, extended down to $i$-manifolds. Let us motivate this claim by demonstrating equivalence in three cases. First, TQFT extended to $n-1$-manifolds is just ordinary axiomatic TQFT. This corresponds to a tensorial TQFT type whose extended manifolds have one internal $n$-region with empty link, and a spatial $0$-region for every compact (connected) $n-1$-manifold as link. An extended manifold of that type is an $n$-manifold whose boundary is identified with a disjoint union of $n-1$-manifolds, i.e., a cobordism without input/output distinction. Note that in the collapsed picture, every $n-1$-manifold is shrinked to a point, which is ``singular''  if the boundary component is not an $n-1$-sphere. Second, fully extended TQFT is supposed to describe microscopic fixed-point models, or lattice TQFT. This is a tensorial TQFT with one internal $n$-region with empty link and one space $n-1$-region whose link is a single point. That is, we associate tensors to cellulated $n$-manifolds, with indices at the boundary $n-1$-cells. An index configuration inside the space boundary corresponds to an open boundary condition for the $n$-dimensional microscopic model, and the corresponding tensor entry is just the partition function of the model on the $n$-dimensional space(-time) with this boundary condition. Third, UMTCs give rise to Reshetikhin-Turaev TQFTs which are known to be 3-2-1 extended TQFTs. As we argue in Appendix~\ref{sec:umtc_tensorial_tqft}, UMTCs are equivalent to a tensorial TQFT type.

Note that in order to fully determine an extended TQFT, we do not necessarily need all possible links, but some can be generated from others. It seems to be the case that a \emph{finite} generating set of links exists precisely for fully-extended and almost-fully-extended ($i=1$) TQFTs. In those two cases, it seems like we can also define a finite set of generating combinatorial extended manifolds and generating gluing axioms. Those two cases are thus useful for the classification of topological phases, whereas less extended TQFTs are much more difficult to handle explicitly mathematically. Moreover, it seems unlikely that they would be enough to fully specify a microscopic topological phase. In particular, as already implied in the previous paragraph, for fully-extended tensorial TQFTs, arbitrary lower-dimensional regions $M_x$ with arbitrary link $L_x$ can be generated from only a single space $n-1$-region with point link. To this end, we pull back a simple geometric mapping, namely transforming $M_x$ into the space $n-1$-region by a cartesian product with $L_x$. This seems to be the tensorial-TQFT analog of the \emph{cobordism hypothesis}. For example, consider a $2+1$-dimensional lattice TQFT associating tensors to the boundary of 3-manifolds. Now add a 1-region with circle link to the extended manifold type. The associated tensor can be obtained by taking the cartesian product of this 1-region with the circle to obtain a 2-region, and taking the associated tensor. This is equivalent to what is known as the \emph{tube algebra} or \emph{Drinfeld center} of a $2+1$-dimensional lattice TQFT. For almost-fully-extended TQFT, the generating set of space regions is more involved, consisting of an $n-2$-region with circle link but also a finite set of lower-dimensional space regions.

More general structures are obtained by also choosing more than one internal region. For example, if we choose an extended manifold type with one space ``boundary'' region for every internal region, we get microscopic state-sum models/lattice TQFT including boundaries, anyons, domain walls, or arbitrary sorts of defects or interfaces. We can also mix fully extended and almost-fully extended descriptions for different regions of the model.

The above formalism can be generalized in several ways, out of which we will sketch two. First, tensorial TQFTs as described above describe phases of quantum spin/qu-$d$-it systems. We can add fermionic degrees of freedom or impose symmetries, in three steps. First, we need to equip all or some regions $r$ of an extended manifold type with an extra geometric/topological structure. That is, we equip $M_r$ of the manifold itself but also of the different links with the extra structure, and let $\psi$ be a map preserving the extra structure. Concretely, an orientation is standard for tensorial TQFTs describing quantum topological phases, and its absence would correspond to a $\zz_2$ time-reversal symmetry. A $1$-cohomology class (valued in the symmetry group) is needed for symmetries, and a spin structure is needed if we have fermionic degrees of freedom. All of these can be represented combinatorially in terms of simplicial cohomology. Second, we need a different ``target category'' of the tensorial TQFT, i.e., instead of ``array'' tensors we associate tensors of a different \emph{tensor type} \cite{tensor_type}, such as fermionic tensors or tensors whose indices are equipped with group representations. Third, there is some standard interplay between the extra geometric structure and the tensors. \emph{Hermiticity} equates orientation reversal and complex conjugation, a \emph{spin-statistics relation} equates spin-structure defects with the fermion parity, and for models with symmetry, the symmetry defects have to correspond to the symmetry action.

The second generalization is to consider different types of combinatorial representations of extended manifolds and gluing, and different locations for the indices for the same extended manifold type. Often, the resulting mathematical structures for different combinatorial representations will be equivalent to the \emph{standard simplicial type} of tensorial TQFT defined above. For example, one can define fully-extended 3-dimensional tensorial TQFT with indices on the edges of a cellulation, such that the resulting algebraic structure is very similar to weak Hopf algebras \cite{liquid_intro}. The latter are equivalent to spherical fusion categories which are the algebraic structure corresponding to the standard simplicial version of the TQFT (after dimension-reduction block-diagonalization, see Appendix~\ref{sec:umtc_tensorial_tqft}). Note that in general, gluing could be an arbitrarily complicated combinatorial operation acting on a finite-size patch of the combinatorial extended manifold, leading to potentially more general mathematical structures. Tensorial TQFTs with complicated gluing operations can however be simplified diagrammatically by coarse graining. For fully extended/lattice TQFT, this does not lead to the standard simplicial type, but to \emph{vertex-liquid models} \cite{universal_liquid}. Also for almost-fully extended TQFT, we need a more complicated combinatorial structure if we want to be diagramatically universal.
% However, for non-fully/almost-fully extended TQFT where the largest dimension of a space region is $n-2$, we can obtain a fully-extended $n-1$-dimensional TQFT by pulling backthe cartesian product of the internal region with the circle. This lattice TQFT is guaranteed to have a topological boundary as the circle bounds the disk. As argued in Ref.~\cite{universal_liquid}, lattice TQFTs with topological boundary can be coarse-grained to a standard simplicial type. So all in all, for almost-fully-extended TQFTs it suffices to take a standard simplicial type, whereas fully extended/lattice TQFTs of a vertex-liquid type can be potentially more general.

\subsection{\texorpdfstring{$1+1$}{1+1}-dimensional TQFTs as tensorial TQFTs}
\label{sec:11_tensorial_tqft}
Let us consider four simple examples of tensorial TQFTs. We start with the well-known 2-1-extended (ordinary) TQFT \cite{Atiyah1988} in $1+1$ dimensions \cite{Kock2004}. As tensorial TQFT, there is one internal $2$-region, and one space $0$-region with circle link. Extended manifolds of this type are nothing but 2-dimensional cobordisms without an input/output distinction. In the collapsed picture, we shrink the boundary circles and obtain 2-manifolds with embedded points. The circle link of the $0$-region can be cellulated with a single edge, such that a combinatorial extended manifold is given by a cellulated 2-manifold with 1-gon holes. However, modulo recellulation invariance of the internal 2-region, the full information of a combinatorial manifold is contained in the 2-manifold with embedded points. We can glue two points of the space 0-region by cutting out two small disks around them and replacing those by an annulus. This is the same as gluing boundary circles of cobordisms in ordinary TQFT. To every extended manifold, we associate a tensor, with one index per point. This is the same as the linear map associated to a cobordism in ordinary TQFT. Disjoint union and gluing at points have to be compatible with tensor product and contraction of index pairs for the associated tensors. A set of extended manifolds that generate all others via gluing operations is given by the pair of pants, or sphere with three points, and disk, or sphere with one point,
\begin{equation}
\label{eq:2dtqft_generating_manifolds}
\begin{tikzpicture}
\path[doublemanifold] (0,0) circle(0.8);
\atoms{vertex}{0/p=-150:0.5, 1/p=-30:0.5, 2/p=90:0.5}
\end{tikzpicture}
\;,\qquad
\begin{tikzpicture}
\path[doublemanifold] (0,0) circle(0.8);
\atoms{vertex}{0/}
\end{tikzpicture}\;.
\end{equation}
Thus, the whole (tensorial) TQFT is determined by the two associated tensors. One can also write down a finite set of generating gluing axioms. Those are such that the generating tensors define the multiplication and unit of a \emph{commutative Frobenius algebra}. E.g., associativity follows from
\begin{equation}
\label{eq:21_extended_associativity}
\begin{multlined}
\begin{tikzpicture}
\path[doublemanifold] (0,0) circle(0.8);
\atoms{vertex}{{0/p=-150:0.5,lab={t=$a$,p=-90:0.25}}, {1/p=-30:0.5,lab={t=$b$,p=-90:0.25}}, 2/p=90:0.5}
\begin{scope}[shift={(0,2)}, yscale=-1]
\path[doublemanifold] (0,0) circle(0.8);
\atoms{vertex}{{0x/p=-150:0.5,lab={t=$c$,p=-90:0.25}}, {1x/p=-30:0.5,lab={t=$d$,p=-90:0.25}}, 2x/p=90:0.5}
\end{scope}
\draw[glue] (2)--(2x);
\end{tikzpicture}
=
\begin{tikzpicture}
\path[doublemanifold] (0,0) circle(0.8);
\atoms{vertex}{{0/p=-135:0.5,lab={t=$a$,p=-135:0.25}}, {1/p=-45:0.5,lab={t=$b$,p=-45:0.25}}, {1/p=135:0.5,lab={t=$c$,p=135:0.25}}, {1/p=45:0.5,lab={t=$d$,p=45:0.25}}}
\end{tikzpicture}
\\
=
\begin{tikzpicture}
\begin{scope}[rotate=-90]
\path[doublemanifold] (0,0) circle(0.8);
\atoms{vertex}{{0/p=-150:0.5,lab={t=$c$,p=-90:0.25}}, {1/p=-30:0.5,lab={t=$a$,p=-90:0.25}}, 2/p=90:0.5}
\end{scope}
\begin{scope}[shift={(2,0)}, rotate=90]
\path[doublemanifold] (0,0) circle(0.8);
\atoms{vertex}{{0x/p=-150:0.5,lab={t=$b$,p=-90:0.25}}, {1x/p=-30:0.5,lab={t=$d$,p=-90:0.25}}, 2x/p=90:0.5}
\end{scope}
\draw[glue] (2)--(2x);
\end{tikzpicture}\;,
\end{multlined}
\end{equation}
and commutativity from
\begin{equation}
\label{eq:21_commutativity}
\begin{tikzpicture}
\path[doublemanifold] (0,0) circle(0.8);
\atoms{vertex}{{0/p=-150:0.5,lab={t=$a$,p=-90:0.25}}, {1/p=-30:0.5,lab={t=$b$,p=-90:0.25}}, {1/p=90:0.5,lab={t=$c$,p=90:0.25}}}
\end{tikzpicture}
=
\begin{tikzpicture}
\path[doublemanifold] (0,0) circle(0.8);
\atoms{vertex}{{0/p=-150:0.5,lab={t=$b$,p=-90:0.25}}, {1/p=-30:0.5,lab={t=$a$,p=-90:0.25}}, {1/p=90:0.5,lab={t=$c$,p=90:0.25}}}
\end{tikzpicture}\;.
\end{equation}
To be precise, for a quantum mechanical interpretation, we need to equip the manifolds with an orientation and label each point as ``inwards'' or ``outwards'' with respect to that orientation. We then only allow gluing between inward and outward points, and demand Hermiticity as a gluing axiom with $G$ orientation reversal and $C$ complex conjugation.

As a second example take 2-dimensional open-closed TQFT \cite{Lauda2005} that is 2-1-extended in the bulk, and 1-0-extended at the $1$-dimensional boundary. There is one internal 2-region with empty link (the bulk), one internal 1-region with point link (the physical/closed boundary), one space 0-region with circle link (point in the bulk), and one space 0-region with interval link (points on the boundary). Extended manifolds of this type are the same as open-closed cobordisms without the input/output distinction, namely 2-manifolds whose boundary consists of ``open'' circles, ``closed'' circles, or circles consisting of a sequence of open and closed intervals. In the collapsed picture, all open circles are shrinked to bulk points, and all open intervals to boundary points. As in the previous example, a combinatorial extended manifold modulo recellulation on the internal regions carries the same information as the extended manifold. A set of generating extended manifolds (together with those of ordinary TQFT above) is given by disks with one or three boundary points, or one bulk and one boundary point,
\begin{equation}
\label{eq:open_closed_generators}
\begin{tikzpicture}
\draw[manifoldboundary,manifold] (0,0) circle(0.8);
\atoms{bdvertex}{0/p=-150:0.8, 1/p=-30:0.8, 2/p=90:0.8}
\end{tikzpicture}
\;,\qquad
\begin{tikzpicture}
\draw[manifoldboundary,manifold] (0,0) circle(0.5);
\atoms{bdvertex}{2/p=90:0.5}
\end{tikzpicture}
\;,\qquad
\begin{tikzpicture}
\draw[manifoldboundary,manifold] (0,0) circle(0.8);
\atoms{bdvertex}{2/p=90:0.8}
\atoms{vertex}{0/}
\end{tikzpicture}\;.
\end{equation}
So the TQFT is determined by the associated tensors. There is also a finite generating set of gluing axioms, which make those tensors into what is called a \emph{knowledgeable Frobenius algebra} in Ref.~\cite{Lauda2005}. Those consist of a commutative Frobenius algebra (the bulk TQFT without boundary), a (non-commutative) Frobenius algebra (whose multiplication is the disk with three boundary points) and a homomorphism (the disk with one bulk and one boundary point) from the former to the latter whose image is the center of the non-commutative algebra. E.g., in addition to associativity (but not commutativity) for the 3-point disk, the defining property of the homomorphism is
\begin{equation}
\label{eq:open_closed_homomorphism_axiom}
\begin{multlined}
\begin{tikzpicture}
\draw[manifoldboundary,manifold] (0,0) circle(0.5);
\atoms{bdvertex}{{0/p=-150:0.5}, {1/p=-30:0.5}, {2/p=90:0.5,lab={t=$c$,p=90:0.25}}}
\begin{scope}[rotate=120]
\draw[manifoldboundary,manifold] (0,1.5) circle(0.5);
\atoms{bdvertex}{0x/p={0,1}}
\atoms{vertex}{{0y/p={0,1.5},lab={t=$a$,p=90:0.25}}}
\end{scope}
\begin{scope}[rotate=-120]
\draw[manifoldboundary,manifold] (0,1.5) circle(0.5);
\atoms{bdvertex}{1x/p={0,1}}
\atoms{vertex}{{1y/p={0,1.5},lab={t=$b$,p=90:0.25}}}
\end{scope}
\draw[glue] (0)--(0x);
\draw[glue] (1)--(1x);
\end{tikzpicture}
\\
=
\begin{tikzpicture}
\draw[manifoldboundary,manifold] (0,0) circle(0.8);
\atoms{vertex}{{0/p=180:0.5,lab={t=$a$,p=-90:0.25}}, {1/p=0:0.5,lab={t=$b$,p=-90:0.25}}}
\atoms{bdvertex}{{x/p=90:0.8,lab={t=$c$,p=90:0.25}}}
\end{tikzpicture}
=
\begin{tikzpicture}
\path[doublemanifold] (0,0) circle(0.8);
\atoms{vertex}{{0/p=-150:0.5,lab={t=$a$,p=-150:0.25}}, {1/p=-30:0.5,lab={t=$b$,p=-30:0.25}}, 2/p=90:0.5}
\draw[manifoldboundary,manifold] (0,1.5) circle(0.5);
\atoms{bdvertex}{{x/p={0,2},lab={t=$c$,p=90:0.25}}}
\atoms{vertex}{y/p={0,1.5}}
\draw[glue] (2)--(y);
\end{tikzpicture}\;,
\end{multlined}
\end{equation}
and the image being the center manifests itself via
\begin{equation}
\label{eq:open_closed_knowledgeability_axiom}
\begin{tikzpicture}
\draw[manifoldboundary,manifold] (0,0) circle(0.5);
\atoms{bdvertex}{{0/p=-150:0.5,lab={t=$a$,p=-150:0.25}}, {1/p=-30:0.5,lab={t=$b$,p=-30:0.25}}, 2/p=90:0.5}
\draw[manifoldboundary,manifold] (0,1.5) circle(0.5);
\atoms{bdvertex}{0x/p={0,1}}
\atoms{vertex}{{0y/p={0,1.5},lab={t=$c$,p=90:0.25}}}
\draw[glue] (2)--(0x);
\end{tikzpicture}
=
\begin{tikzpicture}
\draw[manifoldboundary,manifold] (0,0) circle(0.8);
\atoms{bdvertex}{{0/p=180:0.8,lab={t=$a$,p=180:0.25}}, {1/p=0:0.8,lab={t=$b$,p=0:0.25}}}
\atoms{vertex}{x/lab={t=$c$,p=90:0.25}}
\end{tikzpicture}
=
\begin{tikzpicture}
\draw[manifoldboundary,manifold] (0,0) circle(0.5);
\atoms{bdvertex}{{0/p=-150:0.5,lab={t=$b$,p=-150:0.25}}, {1/p=-30:0.5,lab={t=$a$,p=-30:0.25}}, 2/p=90:0.5}
\draw[manifoldboundary,manifold] (0,1.5) circle(0.5);
\atoms{bdvertex}{0x/p={0,1}}
\atoms{vertex}{{0y/p={0,1.5},lab={t=$c$,p=90:0.25}}}
\draw[glue] (2)--(0x);
\end{tikzpicture}
\;,
\end{equation}
which is called the \emph{knowledgeably property} in Ref.~\cite{Lauda2005}.

Third, let us consider lattice TQFT in $1+1$ dimensions \cite{Fukuma1992}, which is 2-1-0-extended. We have one internal 2-region with empty link and one space 1-region with point link. Extended manifolds of this type in the collapsed picture are 2-manifolds with boundary. The space 1-region is cellulated, that is, this 1-manifold is divided into edges. We can glue two edges, in which case they disappear from the boundary. The branching can be implemented by giving each edge a direction, which makes gluing unambiguous. The associated tensors have one index per edge, and gluing edges is compatible with Einstein summation over the corresponding index pair. If two glued edges coincide in their $0$ vertex or $1$ vertex \footnote{Here we number the vertices of a branching-structure simplex such that the branching structure points from lower to higher numbers.} , we need to contract via one of two weight matrices $D^{0/1}$, e.g.,
\begin{equation}
\label{eq:weight_gluing}
\begin{tikzpicture}
\atoms{regionvertex}{0/,1/p={0.6,0},2/p={1.2,0}}
\draw (0)edge[mark=arr](1) (1)edge[mark={arr,-}](2) (0)edge[mark={three dots,a}]++(-0.2,0) (2)edge[mark={three dots,a}]++(0.2,0);
\draw[glue] (0.3,0)to[out=90,in=90,looseness=1.5] (0.9,0);
\end{tikzpicture}
=
\sum_{a,b} D^1_{a,b}
\begin{tikzpicture}
\atoms{regionvertex}{0/,1/p={0.6,0},2/p={1.2,0}}
\draw (0)edge[mark=arr,mark={slab=$a$}](1) (1)edge[mark={arr,-},mark={slab=$b$}](2) (0)edge[mark={three dots,a}]++(-0.2,0) (2)edge[mark={three dots,a}]++(0.2,0);
\end{tikzpicture}\;.
\end{equation}

Every combinatorial extended manifold can be glued from a single one, namely a disk whose boundary is divided into three edges,
\begin{equation}
\label{eq:generating_triangle}
\begin{tikzpicture}
\atoms{regionvertex}{0/, 1/p=0:1, 2/p={60:1}}
\path[manifold] (0-c)--(1-c)--(2-c)--cycle;
\draw (0)edge[mark=arr](1) (0)edge[mark=arr](2) (2)edge[mark=arr](1);
\end{tikzpicture}\;.
\end{equation}
Again there is a finite set of generating axioms, the solutions of which can in fact be fully classified. To this end we combine the three-index tensor above and the weight matrix to obtain a \emph{special Frobenius algebra}. E.g., the 3-index tensor itself already defines an algebra, which is associative by
\begin{equation}
\begin{tikzpicture}
\atoms{regionvertex}{0/, 1/p=0:1, 2/p={60:1}}
\path[manifold] (0-c)--(1-c)--(2-c)--cycle;
\draw (0)edge[mark=arr](1) (0)edge[mark=arr](2) (2)edge[mark=arr](1);
\begin{scope}[shift={(0,-0.5)},yscale=-1]
\atoms{regionvertex}{0x/, 1x/p=0:1, 2x/p={60:1}}
\path[manifold] (0x-c)--(1x-c)--(2x-c)--cycle;
\draw (0x)edge[mark=arr](1x) (0x)edge[mark=arr](2x) (2x)edge[mark={arr,-}](1x);
\end{scope}
\draw[glue] ($(0)!0.5!(1)$)-- ($(0x)!0.5!(1x)$);
\end{tikzpicture}
=
\begin{tikzpicture}
\atoms{regionvertex}{0/, 1/p={0.8,-0.8}, 2/p={1.6,0}, 3/p={0.8,0.8}}
\path[manifold] (0-c)--(1-c)--(2-c)--(3-c)--cycle;
\draw (0)edge[mark=arr](1) (0)edge[mark=arr](3) (3)edge[mark=arr](2) (2)edge[mark=arr](1);
\end{tikzpicture}
=
\begin{tikzpicture}
\begin{scope}[rotate=90]
\atoms{regionvertex}{0/, 1/p=0:1, 2/p={60:1}}
\path[manifold] (0-c)--(1-c)--(2-c)--cycle;
\draw (0)edge[mark={arr,-}](1) (0)edge[mark={arr,-}](2) (2)edge[mark=arr](1);
\end{scope}
\begin{scope}[shift={(0.5,0)},xscale=-1,rotate=90]
\atoms{regionvertex}{0x/, 1x/p=0:1, 2x/p={60:1}}
\path[manifold] (0x-c)--(1x-c)--(2x-c)--cycle;
\draw (0x)edge[mark={arr,-}](1x) (0x)edge[mark={arr,-}](2x) (2x)edge[mark={arr,-}](1x);
\end{scope}
\draw[glue] ($(0)!0.5!(1)$)-- ($(0x)!0.5!(1x)$);
\end{tikzpicture}\;.
\end{equation}
It is known that special Frobenius algebras are always semisimple and can be block-diagonalized. That is, there is a change of basis where the algebra is given by a direct sum of full matrix algebras. So for any tensorial TQFT, there is an isomorphism, a set of dimensions $I_\alpha$, and a weight vector $\omega_\alpha$, such that after applying the isomorphism we have
\begin{equation}
\label{eq:block_diagonal_algebra}
\begin{tikzpicture}
\atoms{regionvertex}{0/, 1/p=0:1, 2/p={60:1}}
\path[manifold] (0-c)--(1-c)--(2-c)--cycle;
\draw (0)edge[mark=arr,mark={slab=$a\alpha a'$,r}](1) (0)edge[mark=arr,mark={slab=$b\beta b'$}](2) (2)edge[mark=arr,mark={slab=$c\gamma c'$}](1);
\end{tikzpicture}
=
\begin{tikzpicture}
\atoms{circ,tiny,all}{{0/}}
\atoms{circ,small}{{w/p={0.1,0.7},lab={t=$\omega$,p=90:0.25}}}
\draw[irrep] (0)edge[mark={lab=$\alpha$,a}]++(-90:0.8) (0)edge[mark={lab=$\gamma$,a}]++(0:0.8) (0)edge[mark={lab=$\beta$,a}]++(180:0.8);
\draw[irrep] (0)--(w);
\draw[rc,mark={lab=$a$,a},mark={lab=$b$,b}] (-0.8,-0.3)--(-0.3,-0.3)--(-0.3,-0.8);
\draw[rc,mark={lab=$a'$,a},mark={lab=$c'$,b}] (0.8,-0.3)--(0.3,-0.3)--(0.3,-0.8);
\draw[mark={lab=$c$,a},mark={lab=$b'$,b}] (-0.8,0.3)--(0.8,0.3);
\end{tikzpicture}
= \delta_{\alpha, \beta,\gamma} \delta_{a,b}\delta_{b',c} \delta_{a',c} \omega_\alpha
\;,
\end{equation}
and
\begin{equation}
D^{0/1}_{a\alpha a', b\beta b'}
=
\begin{tikzpicture}
\atoms{circ,tiny,all}{{0/}}
\atoms{circ,small}{{w/p={0.1,0.7},lab={t=$I^{-1}\omega^{-2}$,p=90:0.3}}}
\draw[irrep] (-0.8,0)edge[mark={lab=$\alpha$,b}, mark={lab=$\beta$,a}](0.8,0);
\draw[irrep] (0)--(w);
\draw[mark={lab=$b$,a},mark={lab=$a$,b}] (-0.8,-0.3)--(0.8,-0.3);
\draw[mark={lab=$b'$,a},mark={lab=$a'$,b}] (-0.8,0.3)--(0.8,0.3);
\end{tikzpicture}
= \delta_{\alpha, \beta} \delta_{a,b}\delta_{a',b'} \omega_\alpha
\;.
\end{equation}
The diagrams in the middle are tensor-network diagrams including a $\delta$-tensor (the small black dot), two or three identity matrices (the line segments), and a vector $\omega_\alpha$ or $I_\alpha^{-1} \omega_\alpha^{-2}$. Each index on the left is a composite of one \emph{irrep index} (drawn with fat line style) and two \emph{block indices}. The $\delta$-tensor is defined to be $1$ if all of its irrep index labels are equal, and $0$ otherwise. In fact, the tensor-network notation is slightly generalized since the dimension $I_\alpha$ of the block indices depends on the value $\alpha$ of the irrep indices. Note that the total dimension of each index on the left is $\sum_\alpha I_\alpha^2$.

As a last example let us consider lattice TQFT in $1+1$ dimensions with boundary. The regions contain those of ordinary lattice TQFT, plus additionally an internal 1-region with point link (the bulk of the boundary) and a space 0-region whose link is an interval. The interval has two boundary points, one corresponding to the internal 1-region, and the other corresponding to the space 1-region. There is a single generating combinatorial extended manifold, in addition to Eq.~\eqref{eq:generating_triangle},
\begin{equation}
\begin{tikzpicture}
\atoms{bdvertex}{0/, 1/p={1,0}}
\path[manifold] (0-c)--(1-c)to[bend left=40]cycle;
\draw (0)edge[mark=arr](1);
\draw[manifoldboundary] (1)to[bend left=40](0);
\end{tikzpicture}\;,
\end{equation}
where the internal and space region of the boundary are drawn in blue. The gluing axioms imply that the corresponding 3-index tensor is a representation of the special Frobenius algebra above. Such representations can be block-diagonalized along with their Frobenius algebras, yielding
\begin{equation}
\label{eq:representation_blockdiagonalization}
\begin{tikzpicture}
\atoms{bdvertex}{0/lab={t=$b\beta\sigma$,p=180:0.45}, {1/p={1,0},lab={t=$c\gamma\sigma'$,p=0:0.5}}}
\path[manifold] (0-c)--(1-c)to[bend left=40]cycle;
\draw (0)edge[mark=arr,mark={slab=$a\alpha a'$}](1);
\draw[manifoldboundary] (1)to[bend left=40](0);
\end{tikzpicture}
=
\begin{tikzpicture}
\atoms{circ,all,tiny}{{0/}}
\atoms{circ,small}{{w/p={0.1,-0.7},lab={t=$\omega$,p=-90:0.25}}}
\draw[irrep] (0)edge[mark={lab=$\alpha$,a}]++(90:0.8) (0)edge[mark={lab=$\gamma$,a}]++(0:0.8) (0)edge[mark={lab=$\beta$,a}]++(180:0.8);
\draw[irrep] (0)--(w);
\draw[rc,mark={lab=$a$,a},mark={lab=$b$,b}] (-0.8,0.3)--(-0.3,0.3)--(-0.3,0.8);
\draw[rc,mark={lab=$a'$,a},mark={lab=$c$,b}] (0.8,0.3)--(0.3,0.3)--(0.3,0.8);
%\draw[multip] (-0.8,-0.3)--(0.8,-0.3);
\draw[zigzag] (-0.8,-0.3)--(0.8,-0.3);
\path[mark={lab=$\sigma'$,a},mark={lab=$\sigma$,b}] (-0.8,-0.3)--(0.8,-0.3);
\end{tikzpicture}\;.
\end{equation}
Each index on the left is a triple consisting of one irrep index, one block index, and one \emph{multiplicity index} drawn as a zigzag line. Again, the dimension $M_\alpha$ of the multiplicity index depends on the irrep label $\alpha$, and is different from the dimension $I_\alpha$ of the block indices.

\subsection{UMTCs and Lagrangian modules as tensorial TQFTs}
\label{sec:umtc_tensorial_tqft}
In this section we discuss how our definitions of UMTCs and Lagrangian modules in Section~\ref{sec:tensorial_umtc} and Section~\ref{sec:tensorial_umtc_module} are equivalent to tensorial TQFT types. We will first discuss UMTCs, and the case of Lagrangian modules will follow by analogy. For UMTCs, the equivalent tensorial TQFT type is 3-2-1-extended. The corresponding extended manifolds have an internal 3-region $a$ with empty link, a space 1-region $b$ with circle link, and space 0-regions $c_x$ whose link is a sphere with $x$ embedded points. This makes sense, since the ribbon manifolds in the definition of a UMTC are precisely extended manifolds of this type: $a$ is the bulk, $b$ are the ribbons, and $c_x$ are the $x$-valent fusion vertices. The framings of ribbons and fusion vertices are automatically included by our notion of extended manifold. Note that this is also in accordance with Reshetikhin-Turaev TQFT \cite{Reshetikhin1991} constructed from a UMTC, which is known to be 3-2-1-extended.

For a tensorial TQFT we need combinatorial extended manifolds of this type, i.e., we need to cellulate all regions and links. The circle $L_b$ can be cellulated as a 1-gon with a single edge, so $M_b\times L_b$ consists of sequences of ``tube segments'' which are the cartesion product of a 1-gon and a single edge. For each link $L_{c_x}$ we choose a cellulated 2-manifold with $x$ 1-gon holes, such that $M_{c_x}\times L_{c_x}$ is a collection of spheres-with-holes. Those 1-gon holes are joined with the ends of sequences of tube segments of $M_b\times L_b$. The so-obtained ``tube system'' is identified with the boundary of the 3-cellulation of $M_a$. Note that by representing $M_b$ as a sequence of tube segments instead of just edges embedded into $M_a$, we combinatorially remember the normal framing of $M_b$. The same is true for representing $M_{c_x}$ as a collection of spheres-with-holes instead of just vertices embedded into $M_a$. Modulo the recellulation invariance inside $M_a$, it is easier to draw $M_b$ and $M_{c_x}$ in the collapsed picture as sequences of edges meeting at vertices, and to describe the topology of $M_a$ and the normal framing in words.

To each such combinatorial extended manifold we associate a tensor with one index at each edge/tube segment of $M_b$, as well as one index at each vertex/sphere-with-holes of $M_{c_x}$. Two edges/tube segments or two vertices/spheres-with-holes can be glued by removing them from the combinatorial extended manifold and gluing the cells of adjacent regions. This is compatible with contracting the corresponding index pairs. If two glued edges share their $0$ or $1$ vertex, we need to contract the indices via a weight matrix $D^{0/1}$ as shown in Eq.~\eqref{eq:weight_gluing}.

We can already see that the definitions of UMTCs and tensorial TQFTs are very similar. The major difference is that a UMTC only has one label assigned to a whole ribbon, whereas a tensorial TQFT has one index for every edge/tube segment into which the ribbon is decomposed. In the following, we will show that the two algebraic structures are equivalent by defining a map $\calx$ that transforms every UMTC in a tensorial TQFT, and an opposite map $\overline \calx$. Very roughly, loops of edges of the tensorial TQFT define a special Frobobenius algebra, and the ribbon labels of the UMTC are the irreps of that algebra.

Let us start by defining the tensorial TQFT $\calx(\calm)$ for a UMTC $\calm$. The $\calx(\calm)$ vector space of the edge indices is spanned by the ribbon labels of $\calm$, and the vector space of fusion vertex indices has dimension
\begin{equation}
\sum_{a,b,\ldots} N_{a,b,\ldots}\;,
\end{equation}
where $N$ is the dimension of the fusion vertex of a given link where the adjacent ribbons are labeled $a$, $b$, $\ldots$. To obtain $\calx(\calm)(R)$ for a combinatorial extended manifold $R$ we consider the ribbon manifold $\widetilde R$ that is $R$ where we forget the cellulation of the ribbons. $\calx(\calm)(R)$ is then obtained from $\calm(\widetilde R)$ as follows. For every ribbon of $R$ with $n$ edges, we make $n+2$ copies of the ribbon label $a$ by contracting this index of $\calm(\widetilde R)$ with a $\delta$-tensor. Those copies are associated to the different edges as well as the two endpoints of the ribbon. If the ribbon is a loop, we only make $n$ copies. In addition, we multiply by $(d_a/D)^{-n/2}$, e.g.,
\begin{equation}
\begin{multlined}
\begin{tikzpicture}
\atoms{vertex}{0/lab={t=$\alpha x_0x_1x_2$,p=180:0.7}, {1/p={2,0},lab={t=$\beta y_0y_1y_2$,p=0:0.7}}, {2/p={1,1},lab={t=$\gamma z_0z_1$,p=90:0.3}}}
\atoms{regionvertex}{3/p={0.5,-0.2}, 4/p={1,-0.3}, 5/p={1.5,-0.2}, 6/p={0.5,0.6}, 8/p={1,0.4}}
\draw (0)edge[mark=arr,mark={slab=$a$,r}](3) (3)edge[mark={arr,-},mark={slab=$b$,r}](4) (4)edge[mark=arr,mark={slab=$c$,r}](5) (5)edge[mark=arr,mark={slab=$d$,r}](1) (0)edge[mark={arr,-},mark={slab=$e$}](6) (6)edge[mark=arr,mark={slab=$f$}](2) (2)edge[mark={arr,-},mark={slab=$g$}](1) (0)edge[mark={arr,-},mark={slab=$i$,r,p=0.6}](8) (8)edge[mark=arr,mark={slab=$j$,r,p=0.4}](1);
\end{tikzpicture}
\\=
\delta_{a,b,c,d,x_2,y_0} \delta_{e,f,x_0,z_0} \delta_{g,z_1,y_1} \delta_{i,j,x_1,y_2}\\
(\frac{d_a}{D})^{-2}(\frac{d_i}{D})^{-1}(\frac{d_g}{D})^{-\frac12}(\frac{d_e}{D})^{-1}
\begin{tikzpicture}
\atoms{vertex}{0/lab={t=$\alpha$,p=180:0.25}, {1/p={2,0},lab={t=$\beta$,p=0:0.25}}, {2/p={1,1},lab={t=$\gamma$,p=90:0.25}}}
%\atoms{vertex}{0/, 1/p={2,0}, 2/p={1,1}}
\draw (0)edge[mark=arr,mark={slab=$a$,r},bend right](1) (0)edge[mark={arr,-},mark={slab=$e$},bend left=10](2) (2)edge[mark=arr,mark={slab=$g$},bend left=10](1) (0)edge[mark=arr,mark={slab=$i$},bend left](1);
\end{tikzpicture}\;.
\end{multlined}
\end{equation}
We have drawn the vertices within a ribbon with a white filling to distinguish them from the black-filled fusion vertices. Both weight matrices of $\calx(\calm)$ are given via the quantum dimensions of $\calm$ as
\begin{equation}
D^{0/1}_{a,b}=\delta_{a,b} \frac{d_a}{D}\;.
\end{equation}
The gluing axiom of $\calx(\calm)$ for gluing two vertices follows directly from the fusion 0-surgery axiom of $\calm$ in Eq.~\eqref{eq:fusion_0surgery}. Gluing two edges/tube segments of $R$ corresponds to introducing two 2-valent vertices on the corresponding ribbons of $\widetilde R$ via Eq.~\eqref{eq:trivial_vertex_introduction} and then gluing those two vertices. Those two operations are compatible because both adding an extra edge to $R$ for $\calx(\calm)$ and adding an extra 2-valent fusion vertex to $\widetilde R$ for $\calm$ yields a weight of $(d_a/D)^{-1/2}$. If the two glued edges of $R$ share a vertex, then gluing the 2-valent fusion vertices of $\widetilde R$ yields an extra ribbon loop. This extra loop needs to be removed using the loop 1-surgery in Eq.~\eqref{eq:loop_1surgery}. Both the removal of the loop for $\calm$ and the inclusion of the weight matrix $D^{0/1}$ for $\calx(\calm)$ yield a factor of $d_a/D$.

Constructing the UMTC $\overline \calx(\calt)$ from a tensorial TQFT $\calt$ is slightly more involved. It is given by an operation that we will refer to as \emph{dimension-reduction block-diagonalization}. In the following, we will describe dimension-reduction block-diagonalization for the present case of 3-2-1-extended tensorial TQFT, but an analog operation can be applied to any tensorial TQFT type. Let $A$ be the extended manifold type with an internal 2-region and a point-link space 1-region, and let $B$ be the extended manifold type defining $\calt$. Consider the geometric/topological mapping from combinatorial extended manifolds of type $A$ to type $B$, which maps the internal 2-region to the internal 3-region by the cartesian product with the circle, leaving the space 1-region as it is. E.g., a 2-ball is mapped to a 3-sphere with an embedded ribbon loop,
\begin{equation}
\label{eq:321_210_mapping}
\begin{tikzpicture}
\atoms{regionvertex}{0/, 1/p=0:1, 2/p={60:1}}
\path[manifold] (0-c)--(1-c)--(2-c)--cycle;
\draw (0)edge[mark=arr](1) (0)edge[mark=arr](2) (2)edge[mark=arr](1);
\end{tikzpicture}
\quad\rightarrow\quad
\begin{tikzpicture}
\atoms{regionvertex}{0/, 1/p=0:1, 2/p={60:1}}
\draw (0)edge[mark=arr](1) (0)edge[mark=arr](2) (2)edge[mark=arr](1);
\end{tikzpicture}\in S_3\;.
\end{equation}
Composing this map with $\calt$, we obtain a 2-1-0-extended/lattice TQFT. Next, let $A$ be the extended manifold type describing $1+1$-dimensional lattice TQFT with boundary as in Appendix~\ref{sec:11_tensorial_tqft}. For every fusion vertex link (with $x$ points) we can define a mapping from $A$ to $B$ as follows. We take the internal 2-region times a collection of $x$ circles, the internal 1-region times the fusion vertex link, and the space 1-region times $x$ points. E.g., for $x=3$, we have
\begin{equation}
\label{eq:fusionvertex_reduction_mapping}
\begin{tikzpicture}
\atoms{bdvertex}{0/p={0,-0.2}, 3/p={1.8,-0.2}}
\atoms{regionvertex}{1/p={0.6,0}, 2/p={1.2,0}}
\path[manifold] (0-c)--(1-c)--(2-c)--(3-c)to[bend left=40]cycle;
\draw (0)edge[mark=arr](1) (1)edge[mark=arr](2) (2)edge[mark=arr](3);
\draw[manifoldboundary] (3)to[bend left=40](0);
\end{tikzpicture}
\quad\rightarrow\quad
\begin{tikzpicture}
\atoms{vertex}{0/p={0,-0.2}, 3/p={1.8,-0.2}}
\atoms{regionvertex}{1/p={0.6,0}, 2/p={1.2,0}}
\atoms{regionvertex}{1x/p={0.6,0.4}, 2x/p={1.2,0.4}}
\atoms{regionvertex}{1y/p={0.6,-0.4}, 2y/p={1.2,-0.4}}
\draw (0)edge[mark=arr](1) (1)edge[mark=arr](2) (2)edge[mark=arr](3);
\draw (0)edge[mark=arr](1x) (1x)edge[mark=arr](2x) (2x)edge[mark=arr](3);
\draw (0)edge[mark={arr,-}](1y) (1y)edge[mark={arr,-}](2y) (2y)edge[mark={arr,-}](3);
\end{tikzpicture}
\in S_3\;.
\end{equation}
Composing this map with $\calt$, we obtain a 2-dimensional lattice TQFT with boundary.
We now use the fact that the obtained $1+1$-dimensional TQFTs above correspond to Frobenius algebras and their representations, and can be block-diagonalized via Eq.~\eqref{eq:block_diagonal_algebra} and Eq.~\eqref{eq:representation_blockdiagonalization}. More specifically, Eq.~\eqref{eq:fusionvertex_reduction_mapping} defines a representation of $x$ times the Frobenius algebra from Eq.~\eqref{eq:321_210_mapping}. Up to isomorphism, the algebra is specified by a block dimension $I_\alpha$ for every irrep $\alpha$ and a weight vector $\omega_\alpha$, and the representation is specified by a multiplicity dimension $M_{\alpha,\beta,\gamma,\ldots}$ depending on $x$ irrep labels.

Now apply the block-diagonalizing isomorphism to all edges and vertex indices of a tensor $\calt(R)$. Using the gluing axioms in the block-diagonal basis one finds that the tensor must be obtained from a tensor $X$ that only depends on $\widetilde R$ by 1) making $n+2$ copies of the ribbon label for a ribbon with $n$ edges by contracting with a $\delta$-tensor, 2) adding a weight $\omega$ for every edge, and 3) tensoring identity matrices shared between every pair of adjacent edge or vertex indices. E.g., we have
\begin{equation}
\label{eq:umtc_blockdiagonalization}
\begin{multlined}
\begin{tikzpicture}
\atoms{vertex}{0/lab={t=$\kappa k\lambda l\nu n\sigma$,p=180:0.75}, {x/p={0:1.6},lab={t=$\tau\delta d\ldots$, p=0:0.7}}}
\atoms{regionvertex}{1/p={0:0.8}, 3/p={120:0.8}, 5/p={-120:0.8}, 2/p={0.8,1}}
\draw (0)edge[mark=arr,mark={slab=$a\alpha a'$}](1) (1)edge[mark=arr,mark={slab=$b\beta b'$}](x) (0)edge[mark=arr,mark={slab=$c\gamma c'$,r}](3) (0)edge[mark=arr,mark={slab=$e\epsilon e'$}](5);
\draw (x)edge[mark={three dots,a}]++(-60:0.2) (x)edge[mark={three dots,a}]++(60:0.2) (3)edge[mark={three dots,a}]++(120:0.2) (5)edge[mark={three dots,a}]++(-120:0.2);
\draw[mark={slab=$m\mu m'$,r},mark=arr] (2)arc(-90:270:0.3);
\end{tikzpicture}
\\
\rightarrow
\begin{tikzpicture}
\atoms{labbox=$X(\widetilde R)$}{t/}
\atoms{void}{{dots/p={0.5,0.5},lab={t=$\ldots$,p={0,0}}}}
\atoms{circ,all,tiny}{0/p={90:1.2}, 1/p={0:1.2}, 2/p={-90:1.2}, 0x/p={90:1.8}, 1x/p={1.8,0.45}, 1y/p={1.8,-0.45}, 2x/p={-90:1.8}, 3x/p={-1,0}}
\atoms{circ,small}{{a0/p={-0.3,1.8},lab={t=$\omega$,p=180:0.3}}, {a1/p={1.8,-0.75},lab={t=$\omega$,p=180:0.3}}, {a2/p={1.8,0.15},lab={t=$\omega$,p=180:0.3}}, {a3/p={-0.3,-1.8},lab={t=$\omega$,p=180:0.3}},{a4/p={-1,-0.3},lab={t=$\omega$,p=180:0.3}}}
\draw[irrep, rc] (t)--(1) (t)--(0) (t)--(2) (1x)--++(-0.3,0)--(1) (1y)--++(-0.3,0)--(1) (0)--(0x) (2)--(2x) (0x)--(a0) (1y)--(a1) (1x)--(a2) (2x)--(a3) (0x)edge[ind=\gamma]++(0,0.7) (1x)edge[ind=\beta]++(0.7,0) (1y)edge[ind=\alpha]++(0.7,0) (2x)edge[ind=\epsilon]++(0,-0.7) (0)edge[mark={three dots,a}]++(180:0.3) (2)edge[mark={three dots,a}]++(180:0.3) (t)--(3x) (3x)edge[ind=\mu](-2,0) (3x)--(a4);
\draw[irrep, rc, ind=\kappa] (1)|-++(1.3,-1.35);
\draw[irrep, rc, ind=\delta] (1)|-++(1.3,1.35);
\draw[irrep, rc, ind=\lambda] (0)-|++(0.9,1.3);
\draw[irrep, rc, ind=\nu] (2)-|++(0.9,-1.3);
\draw[rc, startind=k, ind=a] (2.5,-1.05)-|++(-0.3,0.3)--++(0.3,0);
\draw[rc, startind=a', ind=b] (2.5,-0.15)-|++(-0.3,0.3)--++(0.3,0);
\draw[rc, startind=b', ind=d] (2.5,0.75)-|++(-0.3,0.3)--++(0.3,0);
\draw[rc, startind=l, ind=c] (0.6, 2.5)|-++(-0.3,-0.3)--++(0,0.3);
\draw[rc, startind=n, ind=e] (0.6, -2.5)|-++(-0.3,0.3)--++(0,-0.3);
\draw[rc, startind=c', mark={three dots,a}] (-0.3, 2.5)|-++(-0.3,-0.3);
\draw[rc, startind=e', mark={three dots,a}] (-0.3, -2.5)|-++(-0.3,0.3);
\draw[rc, startind=m, ind=m'] (-2,-0.3)-|++(0.3,0.6)--++(-0.3,0);
\draw[] (t-tl) edge[multip]++(135:1) (t-bl) edge[multip]++(-135:1);
\path (t-tl) edge[ind=\sigma]++(135:1) (t-bl) edge[ind=\tau]++(-135:1);
% \draw[rc, mark={lab=$b$,a},mark={lab=$a'$,b}] (1.4,0.5)--++(-90:0.35)--++(0:0.4)--++(90:0.35);
% \draw[rc, mark={lab=$m$,a},mark={lab=$e'$,b}] (-0.5,-1.4)--++(0:0.35)--++(-90:0.4)--++(180:0.35);
% \draw[rc, mark={lab=$k$,a},mark={lab=$c$,b}] (-0.5,1)--++(0:0.35)--++(-90:0.4)--++(180:0.35);
% \draw[rc, mark={lab=$v$,a},mark={lab=$e$,b}] (-0.5,-1)--++(0:0.35)--++(90:0.4)--++(180:0.35);
% \draw[rc, mark={lab=$l$,a},mark={lab=$a$,b}] (1,0.5)--++(-90:0.35)--++(180:0.4)--++(90:0.35);
% \draw[rc, mark={three dots,a},mark={lab=$d'$,b}] (-0.5,2.2)--++(0:0.35)--++(90:0.2);
% \draw[rc, mark={three dots,a},mark={lab=$b'$,b}] (2.2,0.5)--++(-90:0.35)--++(0:0.2);
% \draw[rc, mark={three dots,a},mark={lab=$m'$,b}] (-0.5,-2.2)--++(0:0.35)--++(-90:0.2);
\end{tikzpicture}\;,
\end{multlined}
\end{equation}
where the right-hand side is a tensor-network diagram. $X(\widetilde R)$ has one irrep index for every ribbon of $\widetilde R$, and one multiplicity index for every vertex.

We are now ready to define $\calx(\calt)$. The ribbon labels $\calx(\calt)$ are the irreps of the block-diagonalized algebra. The fusion vector space dimensions are given by the multiplicity dimensions $M_{\alpha,\beta,\gamma,\ldots}$. The tensors are simply given by
\begin{equation}
\calx(\calt)(\widetilde R) = X(\widetilde R)\;.
\end{equation}
The quantum dimensions are given by
\begin{equation}
D= \frac{1}{\mathcal T(S_3)},\qquad d_a=D \omega_a^{-2}\;.
\end{equation}
The fusion vertex gluing axioms of $\calx(\calt)$ follow very directly from those of $\calt$. For the loop 1-surgery axiom, we find
\begin{equation}
\begin{multlined}
\begin{tikzpicture}
\draw[mark=arr] (0,0)arc(0:360:0.5);
\atoms{glueedge}{0/}
\end{tikzpicture}
=
\begin{tikzpicture}
\atoms{vertex}{0/, 1/p={0,1}}
\draw (0)edge[mark=arr,bend left=50]coordinate[midway](m0) (1) (0)edge[mark=arr,bend right=50]coordinate[midway](m1) (1);
\draw[glue] (m0)--(m1);
\end{tikzpicture}
=
\sum_{a,b} D_{a,b} D_{a,b}
\begin{tikzpicture}
\atoms{vertex}{0/, 1/p={0,1}}
\draw (0)edge[mark=arr,bend left=50,mark={slab=$a$}](1) (0)edge[mark=arr,bend right=50,mark={slab=$b$,r}](1);
\end{tikzpicture}\\
=
\begin{tikzpicture}
\atoms{labbox=$\calm(\widetilde R)$}{t/}
\atoms{void}{{dots/p={0,0.5},lab={t=$\ldots$,p={0,0}}}}
\atoms{circ,all,tiny}{d/p={0.8,0},d1/p={1.2,0.5},d2/p={1.2,-0.5},d3/p={2,0.5},d4/p={2,-0.5}}
\atoms{circ,small}{{w1/p={1.2,0.9},lab={t=$\omega$,p=180:0.25}},{w2/p={1.2,-0.9},lab={t=$\omega$,p=-180:0.25}},{w3/p={2,1.3},lab={t=$\omega^{-2}I^{-1}$,p=90:0.25}},{w4/p={2,-1.3},lab={t=$\omega^{-2}I^{-1}$,p=-90:0.25}}}
\draw[rc,irrep] (t)--(d) (d)|-(d1) (d)|-(d2) (d1)--(d3) (d2)--(d4) (d3)--++(0.3,0)|-(d4) (d1)--(w1) (d2)--(w2) (d3)--(w3) (d4)--(w4);
\draw[rc] (1.3,0.3)--++(0.8,0)--++(0,-0.6)--++(-0.8,0)--cycle (1.5,0.7)--++(1,0)--++(0,-1.4)--++(-1,0)--cycle;
\end{tikzpicture}
=
\begin{tikzpicture}
\atoms{labbox=$\calm(\widetilde R)$}{t/}
\atoms{void}{{dots/p={0,0.5},lab={t=$\ldots$,p={0,0}}}}
\atoms{circ,small}{{w/p={1.3,0},lab={t=$\omega^{-2}$,p=0:0.5}}}
\draw[rc,irrep] (t)--(w);
\end{tikzpicture}\\
=
\sum_a \omega_a^{-2}
\begin{tikzpicture}
\draw[mark=arr,mark={slab=$a$,r}] (0,0)arc(0:360:0.5);
\end{tikzpicture}
=
\sum_a \frac{d_a}{D}
\begin{tikzpicture}
\draw[mark=arr,mark={slab=$a$,r}] (0,0)arc(0:360:0.5);
\end{tikzpicture}\;.
\end{multlined}
\end{equation}
That is, taking a loop consisting of two edges/tube segments, and gluing those edges topologically performs a 1-surgery. The weights occurring in the gluing of those two edges for $\calt$ are equal to the weights occurring in the loop 1-surgery for $\calx(\calt)$. Note that a closed loop of a block index in the tensor-network diagram evaluates to $\operatorname{Tr}(\operatorname{id}_{I_\alpha}) = I_\alpha$.

To show that UMTCs and tensorial TQFTs are equivalent, it remains to show that $\calm$ and $\overline\calx(\calx(\calm))$ are equivalent, and also $\calt$ and $\calx(\overline\calx(\calt))$. For UMTCs, an appropriate notion of equivalence is given by permutations of ribbon labels and isomorphisms of fusion vector spaces. It is indeed easy to see that $\calm$ and $\overline\calx(\calx(\calm))$ are equivalent in this sense. For tensorial TQFTs (in general), an appropriate notion of equivalence is given by \emph{invertible domain walls}, which are themselves tensorial TQFTs of a more complicated type. In our case of 3-2-1-extended TQFT, those have two copies of the internal 3-region, space 1-region, and space 0-regions, as well as an internal 2-region and space 0-region that separate the two different 3-regions and 1-regions. An extended TQFT of this type is a \emph{domain wall}, and it is \emph{invertible} if it is invariant under additional topological moves of the internal regions. Now, the tensors of $\calx(\overline\calx(\calt))$ are obtained from those of $\calt$ by removing all the disconnected identity tensors in Eq.~\eqref{eq:umtc_blockdiagonalization} and adapting the weight matrices. This removal of identity tensors indeed constitutes an invertible domain wall.

Completely analogously, a Lagrangian module as defined in Section~\ref{sec:tensorial_umtc_module} is related to a type of tensorial TQFT by dimension-reduction block-diagonalization. The extended manifold type contains the one for UMTCs described above. Additionally we have one internal 2-region with point link (the boundary), a space 1-region with interval link (the boundary ribbons), and 0-regions whose link are disks with points in the interior and at the boundary (the boundary fusion vertices). Again, by pulling back a geometric/topological mapping, we obtain a special Frobenius algebra and representation corresponding to the boundary ribbons and the boundary fusion vertices. After block-diagonalization we find that it suffices to associate a single irrep label to each boundary ribbon instead of one index for each edge. This way, we arrive at our definition in Section~\ref{sec:tensorial_umtc_module}.

Finally, consider the fermionic version of the above tensorial TQFTs. All the manifolds are equipped with a spin structure, and the assigned tensors are fermionic tensors as described in Section~\ref{sec:fermionic_umtc_module}. The fermionic lattice TQFT obtained from the dimensional reduction in Eq.~\eqref{eq:321_210_mapping} is a \emph{fermionic special Frobenius algebra}, containing a \emph{super algebra}. Such super algebras can be block-diagonalized, where each block is the product of a full matrix algebra with either 1) the trivial algebra, or 2) the \emph{Clifford algebra} $Cl_1$. After discarding the disentangled full matrix algebra part, we end up with irrep labels of two types, trivial or $Cl_1$. The trivial type corresponds to $m$-type anyons, whereas the $Cl_1$ type corresponds to $q$-type anyons mentioned in the main text.

\subsection{Other types of tensorial TQFT}
\label{sec:tqft_collection}
In this section, we summarize and extend the list of tensorial TQFT types mentioned throughout this document, their physical interpretation, and which categorical/algebraic structure they correspond to. We will list those in the following format.

\deshead{Name describing the tensorial TQFT type}
\begin{description}
\item[Max.~dim.] Maximal dimension of the extended manifold type
\item[Int.~links] List of pairs (dimension, link) for all internal regions
\item[Space links] List of pairs (dimension, link) for all space regions
\item[Extra moves] ``Gluing'' operations in addition to the three kinds of Appendix~\ref{sec:tensorial_general_definition} under which the TQFT tensors are invariant. Only listed if there are any.
\item[Cat.~struct.] Categorical or algebraic structure that the simplicial tensorial TQFTs of this type (roughly) correspond to. For categorical structures we need to apply dimension-reduction block-diagonalization. This refers to the standard simplicial type, TQFTs of vertex-liquid-type could potentially be more general.
\item[Phys.~int.] Physical interpretation with regard to topological phases of matter
\end{description}

The following list is supposed to demonstrate the versatility of tensorial TQFTs, but is by far not exhaustive. For example, we did not include extra structures such as spin or pin structures, or absence of orientation. The latter together with the Hermiticity condition is understood for all TQFTs below.

\deshead{General ordinary TQFT}
\begin{description}
\item[Max.~dim.] $n$
\item[Int.~links] ($n$, $\varnothing$)
\item[Space links] ($n-1$, any manifold)
\item[Cat.~struct.] No finite set of generators in general
\item[Phys.~int.] $k$-point correlations in an $n$-dimensional topological order
\end{description}

\deshead{Ordinary TQFT in 2 dimensions}
\begin{description}
\item[Max.~dim.] $2$
\item[Int.~links] (2, $\varnothing$)
\item[Space links] (0, $S_1$)
\item[Extra moves] (0-surgery)
\item[Cat.~struct.] (Simple) Commutative Frobenius algebra
\item[Phys.~int.] $k$-point correlations in (robust) $1+1$-dimensional topological order
\end{description}
0-surgery invariance is optional, and rules out non-robust degenerate topological order such as the GHZ state. With 0-surgery, this topological order is automatically invertible.

\deshead{Open-closed TQFT in 2 dimensions}
\begin{description}
\item[Max.~dim.] $2$
\item[Int.~links] (2,$\varnothing$), (1,$B_0$)
\item[Space links] (0,$S_1$), (0,$B_1$)
\item[Cat.~struct.] Knowledgeable Frobenius algebra
\item[Phys.~int.] $k$-point correlation functions in models with boundary
\end{description}

\deshead{Lattice TQFT in 2 dimensions}
\begin{description}
\item[Max.~dim.] 2
\item[Int.~links] (2,$\varnothing$)
\item[Space links] (1,$B_0$)
\item[Cat.~struct.] Special Frobenius algebra
\item[Phys.~int.] Microscopic fixed-point models for topological order in $1+1$ dimensions.
\end{description}
Again, we could impose 0-surgery as extra moves to impose robustness/non-degeneracy.% Vertex-liquid models correspond to the most general way of defining gluing and contraction on the 1-dimensional cellulated boundary, however all $1+1$-dimensional topological phases seem to be covered by special Frobenius algebras.

\deshead{Lattice TQFT in 2 dimensions with boundary}
\begin{description}
\item[Max.~dim.] 2
\item[Int.~links] (2,$\varnothing$), (1, $B_0$)
\item[Space links] (1,$B_0$), (0, $B_1^{ps}$)
\item[Cat.~struct.] Special Frobenius algebra together with representation
\item[Phys.~int.] Microscopic fixed-point models for topological order with boundary in $1+1$ dimensions.
\end{description}

\deshead{Lattice TQFT in $n$ dimensions}
\begin{description}
\item[Max.~dim.] $n$
\item[Int.~links] ($n$,$\varnothing$)
\item[Space links] ($n-1$,$B_0$)
\item[Cat.~struct.] Pachner-move invariant simplex tensor, fusion $n-2$-category
\item[Phys.~int.] Microscopic fixed-point models for topological order in $n$ spacetime dimensions
\end{description}
%Again, vertex-liquid models could potentially cover more general phases than Pachner-move invariant simplex tensors or equivalent fusion $n-2$-categories, but no examples are known.

\deshead{2-1-0 extended on the boundary, not extended in the bulk}
\begin{description}
\item[Max.~dim.] $3$
\item[Int.~links] (3, $\varnothing$), (2, $B_0$)
\item[Space links] (1, $B_1$), (0, $B_2$ + 3 boundary points)
\item[Extra moves] (inverse) 1-handle attachment
\item[Cat.~struct.] Spherical fusion category, Weak Hopf algebra
\item[Phys.~int.] Amplitudes for fusion histories of point (in space) defects inside the boundary of a $2+1$-dimensional model. Immediately gives rise to a lattice TQFT (Turaev-Viro-Barrett-Westbury) for the $2+1$-dimensional model and its boundary.
\end{description}

\deshead{3-2-1 extended}
\begin{description}
\item[Max.~dim.] 3
\item[Int.~links] (3, $\varnothing$)
\item[Space links] (1, $S_1$), (0, $S_2$ + 3 points)
\item[Extra moves] (inverse) 0-surgery
\item[Cat.~struct.] UMTC with $c=0$
\item[Phys.~int.] Amplitudes for anyon fusion histories in $2+1$-dimensional topological models, without chiral anomaly
\end{description}

\deshead{3-2-1 extended at the boundary of 4D bulk cobordism}
\begin{description}
\item[Max.~dim.] 4
\item[Int.~links] (4, $\varnothing$), (3, $B_0$)
\item[Space links] (1, $B_2$), (0, $B_3$ + 3 boundary points)
\item[Extra moves] (inverse) 1-handle attachment, any surgery of the 4-region
\item[Cat.~struct.] UMTC
\item[Phys.~int.] Amplitudes for anyon fusion histories in $2+1$-dimensional topological models, possibly with chiral anomaly
\end{description}

\deshead{3-2-1 extended at the boundary of 4D bulk manifold}
\begin{description}
\item[Max.~dim.] 4
\item[Int.~links] (4, $\varnothing$), (3, $B_0$)
\item[Space links] (1, $B_2$), (0, $B_3$ + 3 boundary points)
\item[Extra moves] (inverse) 1-handle attachment
\item[Cat.~struct.] Braided fusion category
\item[Phys.~int.] Amplitudes for anyon fusion histories in the $2+1$-dimensional boundary of a $3+1$-dimensional topological model
\end{description}

\deshead{3-2-1 extended in the bulk, 2-1 extended at the boundary}
\begin{description}
\item[Max.~dim.] 3
\item[Int.~links] (3, $\varnothing$), (2, $B_0$)
\item[Space links] (1, $S_1$), (0, $S_2$ + 3 points), (0, $B_2$+ 1 point)
\item[Extra moves] (inverse) 0-surgery, 1-handle attachment
\item[Cat.~struct.] Commutative Frobenius algebra object inside the ($c=0$) UMTC
\item[Phys.~int.] Amplitudes for histories of anyons fusing in the bulk and condensing at the boundary
\end{description}
We can add an extra 4D bulk in order to include algebra objects in $c\neq 0$ UMTCs (with surgery moves of the 4D bulk), or braided fusion categories (without surgery moves).

\deshead{3-2-1 extended in the bulk, 2-1-0 extended at the boundary}
\begin{description}
\item[Max.~dim.] 3
\item[Int.~links] (3, $\varnothing$), (2, $B_0$)
\item[Space links] (1, $S_1$), (1,$B_1$), (0, $S_2$ + 3 points), (0, $B_2$+ 1 bulk point + 1 boundary point),  (0, $B_2$+ 3 boundary points)
\item[Extra moves] (inverse) 0-surgery, 1-handle attachment
\item[Cat.~struct.] Lagrangian module of UMTC, Fusion category together with its Drinfeld center, weak Hopf algebra together with its quantum double
\item[Phys.~int.] Amplitudes for histories of anyons in the bulk and within the boundary
\end{description}
Note that UMTCs with Lagrangian modules automatically have $c=0$. By adding an extra 4D bulk, we can include general braided modules of braided fusion categories.

\deshead{2-1-0 extended at the boundary, with bulk defect line}
\begin{description}
\item[Max.~dim.] 3
\item[Int.~links] (3, $\varnothing$), (2, $B_0$), (1, $S_1$)
\item[Space links] (1, $B_1$), (0, $B_2$ + 3 boundary points), (0, $B_2$ with 1 boundary point and 1 bulk point)
\item[Extra moves] 1-handle attachment
\item[Cat.~struct.] Fusion category + Representation of its Drinfeld center/tube algebra
\item[Phys.~int.] Amplitudes for histories of (spatial) point defects inside the boundary, together with a fixed point defect in the bulk. Gives immediately rise to a microscopic model for a 1+1D defect line inside a 2+1D topological phase. Such a defect is the same as an anyon, or a direct sum of anyons.
\end{description}

\section{Fermion condensation and the Kitaev 16-fold way}
\label{sec:condensation_appendix}
In Section~\ref{sec:z16_witt_boundaries}, we have constructed fermionic invertible boundaries for the Witt classes generated by stacks of the Ising UMTC. In this appendix, we look at the Kitaev 16-fold way representatives of those Witt classes, and discuss the relation of our fermionic CYWW boundaries to fermion condensation. The 16-fold way UMTCs are listed as $\calm^{16}_\nu$ in the following table,
\begin{equation}
\label{eq:16_fold_way}
\begin{tabular}{l|l|l|l}
$\nu$ & $\calm^{16}_\nu$ & $\calm^{16}_\nu/\psi$ & $(\calm^{16}_\nu)^{\text{loc}}_\psi$\\
0 & TC & $\zz_2$ & $\zz_2^\alpha$ \\
1 & $I^a$ & ${C_2}^a$ & $({C_2}^a)^\alpha$ \\
2 & ${U(1)_4}^a$ & $\operatorname{FTC}^a$ & $(\operatorname{FTC}^a)^\alpha$ \\
3 & $I^b$ & ${C_2}^b$ & $({C_2}^b)^\alpha$ \\
4 & $S^2$ & $S$ & $S^\alpha$ \\
5 & $I^c$ & ${C_2}^c$ & $({C_2}^c)^\alpha$ \\
6 & ${U(1)_4}^b$ & $\operatorname{FTC}^b$ & $(\operatorname{FTC}^b)^\alpha$ \\
7 & $I^d$ & ${C_2}^d$ & $({C_2}^d)^\alpha$\\
8 & 3F & $\zz_2$ & $(\zz_2^\beta)$ \\
9 & $I^e$ & ${C_2}^a$ & $({C_2}^a)^\beta$ \\
10 & ${U(1)_4}^c$ & $\operatorname{FTC}^a$ & $(\operatorname{FTC}^a)^\beta$ \\
11 & $I^f$ & ${C_2}^b$ & $({C_2}^b)^\beta$ \\
12 & $\bar{S}^2$ & $S$ & $S^\beta$ \\
13 & $I^g$ & ${C_2}^c$ & $({C_2}^c)^\beta$ \\
14 & ${U(1)_4}^d$ & $\operatorname{FTC}^b$ & $(\operatorname{FTC}^b)^\beta$ \\
15 & $I^h$ & ${C_2}^d$ & $({C_2}^d)^\beta$
\end{tabular}
\;.
\end{equation}
Here, TC stands for the \emph{toric code UMTC}, $I^x$ stands for the $8$ different variants of the Ising UMTC, ${U(1)_4}^x$ stands for the $4$ different variants of the $U(1)_4$ UMTC with $\zz_4$ fusion rules, $S^2$ stands for the square of the abelian \emph{semion UMTC} $S$
\footnote{Note that the semion UMTC itself is not part of the $\zz_{16}$ subgroup, and is in a different Witt class from $U(1)_4$ despite having the same chiral central charge $1$.}
with non-trivial anyon $s$ with spin $\theta_s=i$, $\bar{S}$ is its complex conjugate, and $3F$ stands for the three-fermion UMTC. $\calm^{16}_\nu$ contains the trivial anyon $1$, a fermion $\psi$, and either one non-abelian anyon $\sigma$ or two abelian anyons $\sigma, \sigma^*$ with topological twist $\theta_\sigma=e^{2\pi i\frac{\nu}{16}}$. $\calm^{16}_{2k+1}$ are the different variants of Ising, and $\calm^{16}_{2k}$ is an abelian UMTC. $\calm^{16}_{4k}$ has $\zz_2\times \zz_2$ fusion rules with the fermion being the $(1,1)$ element (additively written) whereas $\calm^{16}_{4k+2}$ has $\zz_4$ fusion rules with the fermion being the $2$ element.

$(\calm_\nu^{16},\psi)$ defines a \emph{spin UMTC} (c.f.~Ref.~\cite{Bruillard2016}), i.e., a UMTC with together with a fixed fermion
\footnote{The name ``spin UMTC'' might be misleading, since spin UMTCs are bosonic and involve neither spin structure nor super-vector spaces in their definition.}.
Stacking two spin UMTCs $\calm_a$ and $\calm_b$ and condensing the composite of their fermions yields product of spin UMTCs,
\begin{equation}
\label{eq:spin_mtc_product}
\begin{multlined}
(\calm_a,\psi_a) \cdot (\calm_b,\psi_b)
\\
\coloneqq
((\calm_a \otimes \calm_b)/(\psi_a\psi_b),\psi_a\simeq \psi_b)\;.
\end{multlined}
\end{equation}
The $\calm^{16}_\nu$ form a $\zz_{16}$ group under this product. The fermionic Lagrangian module of $\calm^{16}_\nu$ is such that the fermion $\psi$ condenses at the boundary with an odd $0|1$ super-dimension. A \emph{super-modular}\footnote{Also known as \emph{slightly degenerate braided fusion category}.} category \cite{Bruillard2016} is a braided category whose \emph{M\"uger center} (roughly, the restriction to the \emph{transparent} anyons that braid trivially) is the \emph{symmetric fusion category} of super-vector spaces consisting of the trivial anyon and one fermion. A super-modular category can obtained from a spin UMTC by restricting to the anyons that braid trivially with $\psi$. For all $\calm_\nu^{16}$, this super-modular category is the trivial one, namely the category of super-vector spaces itself.
It has been shown in Refs.~\cite{Bruillard2016,JohnsonFreyd2021} that every super UMTC can be extended to a spin UMTC in exactly $16$ different ways. Those $16$ different spin UMTCs are simply related by the product in Eq.~\eqref{eq:spin_mtc_product} with one of spin UMTCs $\calm^{16}_\nu$. Furthermore, taking the tensor product with the category of super-vector spaces defines a homomorphism from the ordinary Witt group to the Witt group of super UMTCs, and it has been shown in Ref.~\cite{Davydov2011} that the kernel of this homomorphism is exactly given by the 16-fold way Witt classes.

In order to discuss fermion condensation, we first need to talk about the fermionic analog of UMTCs themselves. \emph{Fermionic UMTCs} differ from their bosonic counterparts in a few points. First, in general, the extended manifolds (c.f.~Appendix~\ref{sec:tensorial_tqft}) to which a 3-2-1 extended TQFT assigns amplitudes contain 1-dimensional worldlines with arbitrary 1-manifold as link. Since every 1-manifold is a disjoint union of circles, it suffices to consider ordinary circle-link ribbons embedded into a 3-manifold. However, in the fermionic case, all manifolds must be equipped with a spin structure, including the links. There are two spin circles, one with \emph{bounding} and one with \emph{non-bounding} spin structure, which are both needed to generate all spin 1-manifolds via disjoint union. Thus, the fermionic analog of a UMTC needs to contain two different types of worldlines, which we will refer to as \emph{anyon} (with bounding circle as link) and \emph{vortex} (with non-bounding circle as link) worldlines. In an obstruction-theoretic picture for the spin structure where $d\eta_2=\omega_2$, the vortex worldlines are simply added to $\omega_2$. The fusion of anyons and vortices is $\zz_2$-graded, e.g., two vortices fuse to an anyon.%In summary, we can define a \emph{fermionic UMTC} as a map which assigns tensors to spin 3-manifolds with an embedded network of anyon and vortex ribbons, subject to gluing axioms.

The second difference is that there is an anyon called \emph{fundamental fermion}, forming a $\zz_2$ subgroup under fusion, braiding trivially with all other anyons, but having a $-1$ braiding with all the vortices. This anyon comes from the non-trivial (but invertible) topological fermionic phase in $0+1$ dimensions, given by a single degree of freedom with $0|1$ super-dimension, or more physically, a single degree of freedom with a Hamiltonian with odd-parity ground state. In a microscopic model, the fundamental fermion is just given by non-interactingly embedding this $0+1$-dimensional phase into the $2+1$-dimensional spacetime. The fundamental fermion is however not a ribbon label of the fermionic UMTC. Third, both vortices and anyons can have $Cl_1$ automorphism algebras as mentioned in Section~\ref{sec:fermionic_umtc_module}. Fourth, the morphisms between the objects form super-vector spaces instead of vector spaces, leading to reordering signs in, e.g., the pentagon equation. Fifth, for some axioms related to the topological interpretation of UMTCs beyond, e.g., the pentagon equation, we get additional sign factors from how the spin structure interacts with the tensors through the spin-statistics relation.

Fermionic UMTCs can be obtained from (bosonic) spin UMTCs through \emph{fermion condensation} \cite{Aasen2017}, the fermionic analog to \emph{anyon condensation}. Anyon condensation is a prescription that takes as input a \emph{commutative Frobenius algebra} $A$ in a UMTC $\calm$ and yields 1) a fusion category $\calm/A$ by taking the quotient with $A$, and 2) another UMTC $\calm^{\text{loc}}_A$ where we additionally remove all the anyons in $\calm$ that braid non-trivially with $A$ and thus confine. Diagrammatically, a commutative Frobenius algebra assigns amplitudes to ribbon manifolds with boundary, and boundary vertices where bulk ribbons can end, but without boundary ribbons. Physically, the ribbon labels that have a non-zero fusion dimension at such a boundary vertex are anyons that ``condense''. Condensation yields a domain wall between $\calm$ and $\calm^{\text{loc}}_A$, i.e., a Lagrangian module for $\calm\otimes \overline{\calm^{\text{loc}}_A}$. The boundary fusion category of this Lagrangian module is exactly $\calm/A$, such that we have
\begin{equation}
\label{eq:condensation_drinfeld_center}
Z(\calm/A) = \calm\otimes \overline{\calm^{\text{loc}}_A}\;,
\end{equation}
where $Z$ denotes the Drinfeld center. Physically, $\calm/A$ describes the anyons living on the domain wall, consisting of both the anyons of $\calm$ that confine and those that are shared by $\calm$ and $\calm_A^{\text{loc}}$.

In a \emph{fermionic commutative Frobenius algebra}, we can condense fermionic anyons with twist $-1$ by letting the boundary fusion vertex carry an odd ($0|1$) super-dimension. The fermion $\psi$ of any spin UMTC, including the 16-fold way UMTCs $\calm^{16}_\nu$ in Eq.~\eqref{eq:16_fold_way}, forms such a fermionic commutative Frobenius algebra. When we condense a fermionic commutative Frobenius algebra, the resulting $\calm/A$ will be a fermionic fusion category, and $\calm^{\text{loc}}_A$ will be a fermionic UMTC. Physically speaking, the condensing fermion $\psi$ becomes the fundamental fermion (which is not a ribbon label) of the fermionic categories $\calm/A$ and $\calm^{\text{loc}}_A$. The anyons of $\calm$ that braid non-trivially with $\psi$ become the vortices of $\calm^{\text{loc}}_A$.

In the third and fourth column of the table in Eq.~\eqref{eq:16_fold_way}, we have listed the fermionic fusion category $\calm^{16}_\nu/\psi$ and the fermionic UMTC $(\calm^{16}_\nu)^{\text{loc}}_\psi$ resulting from fermion condensation in the 16-fold way UMTCs. When taking the quotient with respect to $\psi$, the cosets $\{1,\psi\}$, and $\{\sigma,\sigma^*\}$ (or only $\{\sigma\}$ for odd $\nu$) become the two ribbon labels $1$ and $\beta$ of $\calm^{16}_\nu/\psi$. More precisely, $\zz_2$ stands for the corresponding untwisted (trivially fermionic) group category, and $S$ is the (trivially fermionic) semion-without-braiding, which is the $\zz_2$ group category twisted by the non-trivial group cocycle in $H^3(\zz_2, U(1))$. $\operatorname{FTC}^x$ stands for the two different versions of the input category of the \emph{fermionic toric code} \cite{Gu2013}. The fusion rules are $\zz_2$ except that the super-dimension is $0|1$ (i.e. odd-parity) rather than $1|0$ if a 3-valent fusion vertex has two $\beta$ ribbons that are directed either both inwards or both outwards. Finally, $C_2^x$ stands for the different versions of the $C_2$ fermionic fusion category introduced in Section~\ref{sec:umtc_module_examples} which has $1|1$ fusion dimension for a 3-valent vertex with two $\beta$ ribbons. Note that we have $\calm^{16}_\nu/\psi=\calm^{16}_{\nu+8}/\psi$.

The fermionic UMTC $(\calm^{16}_\nu)^{\text{loc}}_\psi$ only contains the trivial anyon (not including the fundamental fermion), since $\sigma$ (and $\sigma^*$) braids non-trivially with $\psi$ and thus confines. However, $\sigma$ yields a non-trivial vortex $\beta$. Despite not being trivial, $(\calm^{16}_\nu)^{\text{loc}}_\psi$ is invertible, meaning that stacking it with a complex-conjugated copy yields the trivial fermionic UMTC. The fusion rules and $F$-symbol of $(\calm^{16}_\nu)^{\text{loc}}_\psi$ are the same as $\calm^{16}_\nu/\psi$, when we identify the vortex $\beta$ with the boundary ribbon $\beta$. However, $(\calm^{16}_\nu)^{\text{loc}}_\psi$ and $(\calm^{16}_{\nu+8})^{\text{loc}}_\psi$ have different braiding, which we indicated by the $\alpha$ or $\beta$ superscript in Eq.~\eqref{eq:16_fold_way}. The $(\calm^{16}_\nu)^{\text{loc}}_\psi$ directly form a $\zz_{16}$ group under stacking. Physically speaking, $(\calm^{16})^{\text{loc}}_\psi$ describes the vortices of an invertible chiral fermionic topological phase known as \emph{$p+ip$ superconductor}. We observe that Eq.~\eqref{eq:condensation_drinfeld_center} still holds as
\begin{equation}
Z(\calm^{16}_\nu/\psi) = \calm^{16}_\nu\otimes \overline{(\calm^{16}_\nu)^{\text{loc}}_\psi}\;.
\end{equation}

It is believed that every fermionic UMTC comes from a (bosonic) spin UMTC by condensing the fermion. In the same way, the vortex-free part of a fermionic UMTC would come from a super UMTC. This belief, together with the results from Ref.~\cite{Davydov2011}, suggests that the (bosonic) UMTCs of the $\zz_{16}$ Witt classes are in fact the only ones that allow for a fermionic Lagrangian module.

\section{The fermionic reordering sign}
\label{sec:reordering_selflinking}
In this appendix, we explicitly prove Eq.~\eqref{eq:reordering_refinement} and Eq.~\eqref{eq:self_linking_fermion} used in Appendix~\ref{sec:3fermion_cohomology}, partly following Ref.~\cite{Gaiotto2015}. Let us start with Eq.~\eqref{eq:reordering_refinement}, stating that $\sigma$ is a \emph{quadratic refinement} of the bilinear form $\cup_1$. To show this, we look at two path integrals with tensors $X$ and $Y$ and index configurations $x$ and $y$. Considering a fixed tetrahedron and writing $x_i$ for the index at the triangle opposite to the $i$th vertex, we have
\begin{equation}
\begin{gathered}
(X, \overline{x}_0\overline{x}_2 x_1 x_3) \otimes (Y, \overline{y}_0\overline{y}_2 y_1 y_3)
\\=(X\otimes Y, \overline{x}_0\overline{x}_2 x_1 x_3\overline{y}_0\overline{y}_2 y_1 y_3)\\
=((-1)^{|x_0||y_2|+|x_3||y_1|}X\otimes Y, \overline{y_0}\overline{x_0} \overline{y_2}\overline{x_2}x_1y_1x_3y_3)\\
=((-1)^{|x_0||y_2|+|x_3||y_1|}X\otimes Y, (\overline{x_0y_0}) (\overline{x_2y_2}) (x_1y_1) (x_3y_3))\;,
\end{gathered}
\end{equation}
using the rules for equivalence of fermionic tensors in Section~\ref{sec:fermionic_umtc_module}. Writing $a=|x|$ and $b=|y|$, we find
\begin{equation}
\sigma[a+b]_{0123}
=\sigma[a]_{0123} + \sigma[b]_{0123} + (b\cup_1 a)_{0123}\;,
\end{equation}
using the formula for $\cup_1$ in Eq.~\eqref{eq:cup1_product}. Integration yields Eq.~\eqref{eq:reordering_refinement}.

Next, let us look at Eq.~\eqref{eq:self_linking_fermion}. We find
\begin{equation}
\begin{gathered}
\smallint (\gamma_1+\gamma_2)\cup d(\gamma_1+\gamma_2) - \smallint \gamma_1\cup d\gamma_1 - \smallint \gamma_2\cup d\gamma_2\\
= \smallint (\gamma_1\cup d\gamma_2+ \gamma_2\cup d\gamma_1)\\
= \smallint( d(\gamma_1\cup \gamma_2)+d(\gamma_2\cup_1d\gamma_1) +d\gamma_1 \cup_1 d\gamma_2)\\
=\smallint d\gamma_1 \cup_1 d\gamma_2\\
= \sigma[d(\gamma_1+\gamma_2)]+(\eta,d(\gamma_1+\gamma_2))\\
- (\sigma[d\gamma_1]+(\eta,d\gamma_1)) - (\sigma[d\gamma_2]+(\eta,d\gamma_2))
\;.
\end{gathered}
\end{equation}
Thus, it suffices to show that Eq.~\eqref{eq:self_linking_fermion} is true for $\gamma$ consisting of a single edge which we will also refer to as $\gamma$. More precisely, we need to show that the equation holds for every potential \emph{star} \footnote{Here, the star means the configuration of tetrahedra containing the edge, including the directions of all their edges.} of the edge $\gamma$. This part of the proof has not been explicitly carried out in Ref.~\cite{Gaiotto2015}. $\gamma\cup d\gamma$ is always trivial since the edge $\gamma$ is contained in all the triangles of $d\gamma$, and the cup product in Eq.~\eqref{eq:cup_product} is only non-zero if the edge and triangle span a tetrahedron. So it remains to show that $\sigma[d\gamma] = \omega_2(\gamma)$.

Let us start with a star consisting of $n$ tetrahedra where $\gamma$ is the $03$ edge of all tetrahedra in the star, and all the $12$ edges point clockwise. For such a star, $\omega_2(\gamma)=1+n+n=1$, since $\gamma$ is the $03$ edge of all adjacent tetrahedra and the $02$ edge of all adjacent triangles. Accordingly, we indeed have $\sigma[d\gamma]=1$ for the corresponding cyclic index ordering,
\begin{equation}
\begin{multlined}
(X,a_1\bar a_2\ldots a_{n-1}\bar a_{n} a_n\bar a_1)\\
= (-X, \bar a_1 a_1\bar a_2\ldots a_{n-1}\bar a_{n} a_n)\;.
\end{multlined}
\end{equation}
Now consider flipping the direction of an outer edge in an arbitrary star, i.e., an edge of the star spanning a tetrahedron together with $\gamma$. This does not change the vertex numbers if $\gamma$ in any of the adjacent tetrahedra or triangles, so $\omega_2(\gamma)$ is not changed. $\sigma[d\gamma]$ is unchanged as well, since we exchange the two triangles but also the orientation of the tetrahedron which reverses the Grassmann ordering. Finally, consider flipping an edge $e$, such that $\gamma$ and $e$ span a triangle $t$. For both $\sigma[d\gamma]$ and $\omega_2(\gamma)$, we get three contributions to their change, one for $t$ and two for the adjacent tetrahedra. The $t$ contribution to the change of $\sigma[d\gamma]$ comes from exchanging the Grassmann variables $\theta_t$ and $\overline{\theta_t}$, and thus is always $1$. The $t$ contribution to the change of $\omega_2(\gamma)$ comes from $\gamma$ changing between being the $01$ and $02$ edge, or between the $12$ and $02$ edge of $t$, and thus is also $1$ always. The change of $\sigma$ or $\omega_2$ due to one of the adjacent tetrahedra depends on the branching structure of this tetrahedron, which we will denote by giving the list of vertex numbers of the four vertices. In the list, we will order the four vertices according to 1) vertex of $\gamma$ but not of $e$, 2) vertex of both $\gamma$ and $e$, 3) vertex of $e$ but not $\gamma$, 4) remaining vertex neither part of $e$ nor $\gamma$. In configurations where $e$ is allowed to flip, the vertex numbers of the vertices in $e$, i.e, the second and third vertex number, have to be consecutive. So the allowed configurations before/after flipping $e$ are
\begin{equation}
\begin{multlined}
0123/0213, 3120/3210, 2013/2103,\\ 1230/1320,0231/0321, 3012/3102\;.
\end{multlined}
\end{equation}
The contribution to $\omega_2(\gamma)$ is due to $\gamma$ being the $03$ edge of the tetrahedron, thus we have a change if the first two vertex numbers are $03$ or $30$ on either side,
\begin{equation}
0321/0231, 3012/3102\;.
\end{equation}
The contribution to $\sigma[d\gamma]$ is due to the ordering of Grassmann variables of $t$ and the other triangle of both $d\gamma$ and the tetrahedron, according to Eq.~\eqref{eq:tetrahedron_grassmann_ordering}. This ordering changes for the configurations
\begin{equation}
0123/0213, 3120/3210, 2013/2103, 1230/1320\;.
\end{equation}
We see that $\sigma[d\gamma]$ gets a contribution exactly if $\omega_2(\gamma)$ does not. Since there is a contribution from \emph{two} tetrahedra, overall, the change of $\omega_2(\gamma)$ and $\sigma[d\gamma]$ when flipping $e$ is equal.

Since any potential star of $\gamma$ with $n$ tetrahedra can be obtained from a star with ``standard'' branching structure by flipping edges, this implies that $\sigma[d\gamma] = \omega_2(\gamma)$ holds for any star.

\section{\texorpdfstring{$U(1)_4$}{U(1)4} CYWW invertible boundary}
\label{sec:u1level4}
% The $\zz_8$ subgroup of 16-fold way UMTCs $\calm^{16}_{2k}$ are abelian, and hence the corresponding CYWW models allow for a description as a 2-cocycle gauge theory based on the \emph{Pontryagin square}. We can also find a cohomology description of the invertible boundary in those cases. This cohomology description of the three-fermion CYWW boundary can be found in Section~\ref{sec:3fermion_cohomology}, and for the generating abelian $U(1)_4$ CYWW models, the invertible boundary is discussed in both the cohomology and the categorical language in Appendix~\ref{sec:u1level4} \andi{still need to finish the full gauge invariance proof for that}.

In this appendix, we look at another instructive example for a fermionic invertible boundary of a CYWW model, namely that of the $U(1)_4$ UMTC, representing the $2$-element of the $\zz_{16}$ Witt subgroup spanned by Ising. We will here give the fermionic Lagrangian module of the $U(1)_4$ UMTC, the CYWW boundary is that of Section~\ref{sec:fermionic_boundaries}.
%It generates the abelian UMTCs for which the CYWW models and the invertible boundary can be formulated in terms of cohomology theory.
%\subsection{As Lagrangian module}

The $U(1)_4$ category (listed as ``$\zz_4$ MTC'' in Section 5.3 of Ref.~\cite{Rowell2007}) is an abelian UMTC with $4$ anyons $1$, $\epsilon$, $\sigma$, $\sigma^*$, with $\zz_4$ fusion rules and quantum dimensions all equal to $1$. As in Section~\ref{sec:umtc_module_examples}, the amplitude for a ribbon manifold is obtained by reducing it to the empty manifold, and any ribbon manifold can be reduced to a ribbon sphere by 1-surgery along additional ribbon loops.

% and topological spins $\theta_1=1$, $\theta_\epsilon=-1$, $\theta_\sigma=e^{2\pi i/8}$,  $\theta_{\sigma^*}=e^{2\pi i/8}$. The chiral central charge is therefore $1$. The $R$-symbols are given by $R^{\epsilon \epsilon}_1=-1$, $R^{\sigma \sigma}_\epsilon=R^{\sigma^* \sigma^*}_\epsilon=e^{2\pi i/8}$, $R^{\sigma^* \sigma}_1=R^{\sigma \sigma^*}_1=e^{-2\pi i/8}$, $R^{\sigma \epsilon}_{\sigma^*}=R^{\epsilon \sigma}_{\sigma^*}=R^{\sigma^* \epsilon}_\sigma=R^{\epsilon \sigma^*}_\sigma=-i$. The $F$-symbols are $-1$ if there are two adjacent $\epsilon$s and four $\sigma$ or $\sigma^*$s, otherwise $1$.

We first remove all $1$-labeled ribbon segments, and resolve non-$\sigma$-labeled ribbons as
\begin{equation}
\begin{gathered}
\begin{tikzpicture}
\draw (0,0)edge[mark={slab=$\sigma^*$},mark=arr](1,0);
\end{tikzpicture}
\rightarrow
\begin{tikzpicture}
\draw (0,0)edge[mark={slab=$\sigma$},mark={arr,-}](1,0);
\end{tikzpicture}\;,\\
\begin{tikzpicture}
\draw (0,0)edge[mark={slab=$\psi$},mark=arr](1,0);
\end{tikzpicture}
\rightarrow
\begin{tikzpicture}
\draw (0,0)edge[mark={slab=$\sigma$},mark=arr]++(1,0) (0,-0.3)edge[mark={slab=$\sigma$,r},mark=arr]++(1,0);
\end{tikzpicture}\;.
\end{gathered}
\end{equation}
Further, we resolve every fusion vertex into uninterrupted ribbons, e.g.,
\begin{equation}
\begin{gathered}
\begin{tikzpicture}
\atoms{vertex}{0/}
\draw (0)edge[mark={arr,-},mark={slab=$\sigma$}]++(135:0.7) (0)edge[mark={arr,-},mark={slab=$\sigma$,r}]++(45:0.7) (0)edge[mark={arr},mark={slab=$\psi$,r}]++(-90:0.7);
\end{tikzpicture}
\quad\rightarrow\quad
(2^{1/2})
\begin{tikzpicture}
\draw[mark={arr,-,p=0.7},mark={slab=$\sigma$,p=0.3,r},rc] (0.15,0)--++(0,0.7)--++(45:1);
\draw[mark={arr,-,p=0.7},mark={slab=$\sigma$,p=0.3},rc,front] (-0.15,0)--++(0,0.7)--++(135:1);
\end{tikzpicture}\;,\\
\begin{tikzpicture}
\atoms{vertex}{0/}
\draw (0)edge[mark={arr,-},mark={slab=$\sigma$}]++(135:0.7) (0)edge[mark={arr,-},mark={slab=$\sigma^*$,r}]++(45:0.7) (0)edge[mark={arr},mark={slab=$1$,r}]++(-90:0.7);
\end{tikzpicture}
\quad\rightarrow\quad
(2^{1/2})
\begin{tikzpicture}
\draw[mark={arr},mark={slab=$\sigma$,p=0.3,r},rc] (0,0)--++(-45:1)--++(45:1);
\end{tikzpicture}\;.
\end{gathered}
\end{equation}
Different choices of resolution yield different but gauge equivalent UMTCs. We obtain a collection of $\sigma$ ribbon loops which are disjoint but might be linked with another and twisted. We unlink the loops and remove the twists by
\begin{equation}
\begin{gathered}
\begin{tikzpicture}
\draw (0,0)edge[mark={arr,p=0.4},mark={slab=$\sigma$,p=0.7,r,sideoff=-0.1}](1,1.5);
\draw (1,0)edge[mark={arr,p=0.4},front, mark={slab=$\sigma$,p=0.7,sideoff=-0.1}](0,1.5);
\end{tikzpicture}
=
e^{2\pi i\frac{8-2\nu}{16}}
\begin{tikzpicture}
\draw (1,0)edge[mark={arr,p=0.4},mark={slab=$\sigma$,p=0.7,sideoff=-0.1}](0,1.5);
\draw (0,0)edge[mark={arr,p=0.4},front,mark={slab=$\sigma$,p=0.7,r,sideoff=-0.1}](1,1.5);
\end{tikzpicture}
\;,
\\
\begin{tikzpicture}
\draw[looseness=2] (0,0)edge[out=0,in=0,mark={arr,e}](0.8,1) (0.8,1)edge[out=180,in=180,mark={slab=$\sigma$,s},front](1.6,0);
\end{tikzpicture}
=
e^{2\pi i\frac{\nu}{16}}
\begin{tikzpicture}
\draw (0,0)edge[mark={slab=$\sigma$},mark=arr](1,0);
\end{tikzpicture}\;.
\end{gathered}
\end{equation}
In the end, we can remove unlinked and untwisted loops, and the empty 3-sphere has amplitude $1/D$.

The Lagrangian module is given as follows. There are two boundary ribbon labels, $1$ and $\beta$ with quantum dimension $1$, so $D^\calb=\sqrt2$. The super-dimension of the boundary fusion vertices is $0|0$ if the number of adjacent $\beta$ ribbons is not even, $0|1$ if there are two ingoing or two outgoing $\beta$ ribbons, and $1|0$ otherwise. The corresponding fusion category is the input category of the fermionic toric code \cite{Gu2013}. The non-zero super-dimensions of the boundary fusion vertices with one bulk and one boundary ribbon are given by
\begin{equation}
\begin{multlined}
\begin{tikzpicture}
\atoms{bdvertex}{0/lab={t=$1|0$,p=90:0.25}}
\draw (0)edge[bdribbon,mark={arr},mark={slab=$1$}]++(0.8,0) (0)edge[mark={arr,-},mark={slab=$1$,r}]++(-0.8,0);
\end{tikzpicture}
\;,
\begin{tikzpicture}
\atoms{bdvertex}{0/lab={t=$1|0$,p=90:0.25}}
\draw (0)edge[bdribbon,mark={arr},mark={slab=$\beta$}]++(0.8,0) (0)edge[mark={arr,-},mark={slab=$\sigma$,r}]++(-0.8,0);
\end{tikzpicture}
\;,\\
\begin{tikzpicture}
\atoms{bdvertex}{0/lab={t=$0|1$,p=90:0.25}}
\draw (0)edge[bdribbon,mark={arr},mark={slab=$1$}]++(0.8,0) (0)edge[mark={arr,-},mark={slab=$\psi$,r}]++(-0.8,0);
\end{tikzpicture}
\;,
\begin{tikzpicture}
\atoms{bdvertex}{0/lab={t=$0|1$,p=90:0.25}}
\draw (0)edge[bdribbon,mark={arr},mark={slab=$\beta$}]++(0.8,0) (0)edge[mark={arr,-},mark={slab=$\sigma^*$,r}]++(-0.8,0);
\end{tikzpicture}\;.
\end{multlined}
\end{equation}

The amplitude for a ribbon network inside a 3-ball is again obtained by reducing it to a trivial network. To this end, we neglect $1$ ribbons, push $\beta$ ribbons slightly into the bulk,
\begin{equation}
\begin{tikzpicture}
\draw[bdribbon] (0,0)edge[mark=arr,mark={slab=$\beta$}](1,0);
\end{tikzpicture}
\rightarrow
\begin{tikzpicture}
\draw (0,0)edge[mark=arr,mark={slab=$\sigma$}](1,0);
\end{tikzpicture}\;,
\end{equation}
and transform condensation- and direction-changing vertices as
\begin{equation}
\begin{gathered}
\begin{tikzpicture}
\atoms{bdvertex}{0/}
\draw (0)edge[mark={arr},bdribbon,mark={slab=$\beta$}]++(0.8,0) (0)edge[mark={arr,-},mark={slab=$\sigma$,r}]++(-0.8,0);
\end{tikzpicture}
\rightarrow
(2^{1/2})
\begin{tikzpicture}
\draw (0,0)edge[mark=arr,mark={slab=$\sigma$}]++(1,0);
\end{tikzpicture}\;,\\
\begin{tikzpicture}
\atoms{bdvertex}{0/}
\draw (0)edge[mark={arr},bdribbon,mark={slab=$1$}]++(0.8,0) (0)edge[mark={arr,-},mark={slab=$\psi$,r}]++(-0.8,0);
\end{tikzpicture}
\rightarrow
(2^{1/2})
\begin{tikzpicture}
\atoms{bdvertex}{0/}
\draw[rc,mark={slab=$\sigma$,p=0.6},mark={arr,-}] (0)--++(135:0.2)--++(-1,0);
\draw[rc,mark={slab=$\sigma$,p=0.6},mark={arr,-}] (0)--++(-135:0.2)--++(-1,0);
\end{tikzpicture}\;,\\
\begin{tikzpicture}
\atoms{bdvertex}{0/}
\draw[bdribbon] (0)edge[mark={arr},mark={slab=$\beta$}]++(0.8,0) (0)edge[mark={arr},mark={slab=$\beta$,r}]++(-0.8,0);
\end{tikzpicture}
\rightarrow
\begin{tikzpicture}
\atoms{bdvertex}{0/}
\draw (0)edge[mark={arr},mark={slab=$\sigma$}]++(0.8,0) (0)edge[mark={arr},mark={slab=$\sigma$,r}]++(-0.8,0);
\end{tikzpicture}\;.
\end{gathered}
\end{equation}
Finally, we can remove pairs of boundary fusion vertices via
\begin{equation}
\begin{tikzpicture}
\atoms{bdvertex}{0/lab={t=$\textcolor{\spincol}{0}$,p=90:0.25}, {1/p={1,0},lab={t=$\textcolor{\spincol}{1}$,p=90:0.25}}}
\atoms{bdcellvertex}{a/p={0.5,-0.6}, b/p={0.5,0.6}}
\draw (0)edge[mark={slab=$\sigma$,p=0.3},mark={arr,p=0.7}](1) (0)edge[mark={slab=$\sigma$,p=0.7,r},mark={arr,-}]++(-0.8,0) (1)edge[mark={slab=$\sigma$,p=0.7},mark=arr]++(0.8,0);
\draw[bdcellulation] (a)edge[bend left=80,mark={arr,p=0.3},looseness=2](b) (a)edge[bend right=90,mark={arr,p=0.3},looseness=2](b) (a)edge[mark={arr,p=0.3}](b);
\end{tikzpicture}
=
\begin{tikzpicture}
\draw (0,0)edge[mark=arr,mark={slab=$\sigma$}](1,0);
\end{tikzpicture}\;.
\end{equation}
Like in Eq.~\eqref{eq:fermion_segment_removal}, we consider cellulations where the boundary vertices are contained inside 2-gons and allow for the removal of two 2-gons if they are adjacent, the separating edge has $\eta_2=0$, and the Grassmann ordering is as depicted. Combined with the reduction procedure in the bulk, we obtain the amplitude for an arbitrary boundary ribbon manifold.

\section{Gauge theory formulation for abelian UMTCs}
\label{sec:cohomology_boundaries}
In this appendix we reformulate some of the invertible CYWW boundaries described in Section~\ref{sec:drinfeld_center_boundary} and Section~\ref{sec:fermionic_boundaries} in the language of gauge theories whose action is a cohomology operation. We focus on two simple examples, namely the invertible boundaries for the toric-code CYWW model, and the fermionic invertible boundary for the three-fermion CYWW model.% A simplicial gauge theory treatment of the $U(1)_4$ CYWW boundaries can be found in Appendix~\ref{sec:general_cohomology}. \andi{need to finish gauge invariance}

\subsection{(Simplicial) cohomology}
\label{sec:simplicial cohomology}
In this section, we will give a brief introduction to the basic notions of simplicial cohomology on branching-structure triangulations necessary for the following two sections. The central objects are \emph{$i$-chains}, which are functions from the set of $i$-simplices to some coefficient group, which for this section will be $\zz_2$. The \emph{coboundary} $dx$ of an $i$-chain $x$ is the $i+1$-chain whose value on a $i+1$-simplex is the sum over the values of $x$ on the contained $i$-simplices. $i$-chains with $dx=0$ are called \emph{$i$-cocycles}. Dually, the \emph{boundary} $\delta x$ is the $i-1$-chain whose value on an $i-1$-simplex is the sum over the values of $x$ on the $i$-simplices containing it, and $i$-chains with $\delta x=0$ are called \emph{$i$-cycles}. Intuitively, $1$-cycles and dually $n-1$-cocycles are networks of closed loops, $2$-cycles and dually $n-2$-cocycles are networks of membranes, etc. Central to cohomology is the relation $d^2=0$ and dually $\delta^2=0$. $i$-chains $x$ with $x=dy$ are themselves called \emph{coboundaries}, and are a subset of the cocycles. Dually, \emph{boundaries} $x=\delta y$ are a subset of the cycles. Chains, (co-)cycles and (co-)boundaries form groups under simplex-wise multiplication, and the quotient of the $i$-(co)cycles by the $i$-(co)boundaries is known as the $i$th \emph{(co)homology group}. The \emph{cup product} maps an $i$-cochain $x$ and a $j$-cochain $y$ to a $i+j$-cochain whose value on a $i+j$-simplex $S$ is given by \cite{Steenrod1947}
\begin{equation}
\label{eq:cup_product}
(x\cup y)(S)= x(S|_{01\ldots i}) \cdot y(S|_{i\ldots i+j})\;.
\end{equation}
Intuitively, the cup product corresponds the intersection of networks of loops/membranes/etc. Furthermore, one can define \emph{higher-order cup products} \cite{Steenrod1947}, of which we will need the first-order product $\cup_1$ mapping a $i$-cochain and $j$-cochain to an $i+j-1$-cochain. The value of $\cup_1$ on a $i+j-1$-simplex $S$ is given by
\begin{equation}
\label{eq:cup1_product}
\begin{multlined}
(x\cup_1 y)(S)\\ = \sum_{0\leq k< i} x(S|_{0\ldots k(k+j)\ldots i+j-1})\cdot y(S|_{k\ldots (k+j)})\;.
\end{multlined}
\end{equation}

We will need a small number of identities that can be easily checked for the simplicial formulas, and extracted from Ref.~\cite{Steenrod1947}. First, the cup product obeys a \emph{Leibniz rule},
\begin{equation}
\label{eq:leibnitz_rule}
d(A\cup B)=dA\cup B+A\cup dB\;.
\end{equation}
Second, the failure of commutativity of $\cup$ equals the failure of the Leibniz rule for $\cup_1$,
\begin{equation}
\label{eq:cup_product_property}
A\cup B+B\cup A = d(A\cup_1B)+dA\cup_1 B+A\cup_1dB\;.
\end{equation}

For a simplicial complex with boundary, a $i$-cochain ($i$-chain) $x$ restricts to another $i$-cochain ($i-1$-chain) $\partial x$ within the boundary. All of the above operations commute with the restriction to the boundary,
\begin{equation}
\begin{gathered}
d\partial x=\partial dx\;,\\
\partial(x\cup y)=(\partial x)\cup (\partial y)\;,\\
\partial(x\cup_1 y)=(\partial x)\cup_1 (\partial y)\;.
\end{gathered}
\end{equation}
Intuitively, cocycles (cycles) with $dx=0$ ($\delta x=0$) in the presence of a physical boundary are \emph{open}, meaning that they are networks of loops/membranes/etc. that are allowed to freely terminate at the boundary. \emph{Closed} $i$-cocycles ($i$-cycles) that must not terminate at a boundary can be modeled by introducing an $i-1$-cochain ($i$-chain) $\bar x$ within the boundary such that $\partial x=d\bar x$ ($\partial x=\delta \bar x$). The open cycle $x$ together with its closure $\bar x$ within the boundary then form a closed (co-)cycle.

\subsection{The invertible boundary for the toric-code CYWW}
\label{sec:gauge_toric_code}
\paragraph{The bulk model}
The toric-code CYWW model is a gauge theory based on a $\zz_2\times \zz_2$-valued 2-cocycle, or equivalently two $\zz_2$-valued 2-cocycles $A$ and $B$. The path integral for a fixed configuration is given by $(-1)^{\int \lagr}$ (like any path integral in this and the subsequent section), where
\begin{equation}
\lagr=A\cup B
\end{equation}
is a $\zz_2$-valued $4$-cocycle which is integrated/summed over.
For the lattice model branching-structure triangulation, $A$ and $B$ are labelings of the triangles by elements of $\zz_2$. By Eq.~\eqref{eq:cup_product}, on every bulk 4-simplex $01234$, we get a weight
\begin{equation}
\label{eq:toric_code_cyww_action}
(-1)^{A(012)B(234)}\;.
\end{equation}
Recalling that the toric code anyons are generated by $e$ and $m$, we see that $A$ and $B$ are just the $e$ and $m$ components of the labels at the ribbon dual to the triangles. Also the $F$-tensor is trivial and looking at Eq.~\eqref{eq:4simplex_ribbon_manifold} we see that all we need to do to evaluate the ribbon manifold corresponding to a $4$-simplex is to exchange the ribbons corresponding to the $012$ and the $234$ triangle. This is done exactly by the $R$-tensor which is $-1$ for exchanging $e$ and $m$, yielding Eq.~\eqref{eq:toric_code_cyww_action}.

The central property of the action is its invariance under gauge transformations for $1$-cochains $\Gamma$ and $\Lambda$,
\begin{equation}
\label{eq:toric_code_gauge}
\begin{gathered}
A'= A+d\Gamma\;,\\
B'= B+d\Lambda\;.
\end{gathered}
\end{equation}
The topological invariance of the partition function is a direct consequence of such a gauge invariance as follows. Consider a ball-like patch of triangulation with a fixed configuration of gauge fields (here $A$ and $B$) on its space boundary. Due to the gauge invariance, the amplitude for such a configuration does not depend on how the gauge fields extend into the bulk. Now consider a Pachner move acting on a few simplices inside the patch of triangulation. We can always find bulk gauge fields that are $0$ around where the Pachner move happens. Since the action is trivial at simplices with $0$ gauge field, the action is invariant under Pachner moves.

Intuitively, it is easy to accept that the ($\operatorname{mod} 2$) number of intersections of the 2-cocycles $A$ and $B$ does not change if we add a 2-coboundary. More precisely, at an $A$ gauge transformation, the action changes by
\begin{equation}
\begin{multlined}
\lagr(A',B)-\lagr(A,B)=(A+d\Gamma) \cup B - A\cup B\\
=d\Gamma\cup B=d(\Gamma\cup B)\eqqcolon d(\lagr^\Gamma_{\text{triv}})\;,
\end{multlined}
\end{equation}
which is a total derivative/coboundary and thus its integral vanishes. Here we used the Leibniz rule and the fact that $dB=0$. If we add a physical boundary to the spacetime, the partition function changes by $\partial \lagr_{\text{triv}}^\Gamma$. An analogous calculation yields $\lagr^\Lambda_{\text{triv}}$.

Finally, we should add normalizations such that the partition function becomes independent of the lattice size. It is given by factors of $|G|^{-1/2}$ at every $4$-cell, $2$-cell, and $1$-cell, and $|G|^{1/2}$ at every $3$-cell and $0$-cell, where $G$ is the gauge group which is $\zz_2\times \zz_2$ here. Note that we have
\begin{equation}
\label{eq:cyww_gauge_normalization}
\begin{multlined}
|G|^{-\frac12 |C_4|+\frac12 |C_3|-\frac12 |C_2|-\frac12 |C_1|+\frac12|C_0|}\\
= |G|^{-\frac12 \chi} |G|^{-|C_1|+|C_0|}\;,
\end{multlined}
\end{equation}
where $C_i$ is the set of $i$-cells and $\chi$ is the Euler characteristic. Neglecting the $|G|^{-\frac12 \chi}$ term would yield a simpler normalization that is still recellulation invariant, but not invariant under the invertibility moves described in Section~\ref{sec:state_sums_boundaries}. The normalization in Eq.~\eqref{eq:cyww_gauge_normalization} also coincides with the normalization of the CYWW state sum in Section~\ref{sec:CYWW_path_integral}.

\paragraph{The boundary}
The invertible boundary for the toric-code CYWW model is given as follows. $B$ is a closed 2-cocycle, and $A$ is an open 2-cocycle. Intuitively, the number of intersections between a closed and an open cocycle is invariant under moving them around locally. Formally, we define the closed $B$ as an open $B$ together with a 1-cocycle $\bar B$ within the boundary such that $d\bar B=\partial B$. To make the action gauge invariant at the boundary we have to add a boundary action
\begin{equation}
\lagr_\partial = \partial A \cup \bar B\;.
\end{equation}
In the lattice model, $\bar B$ is a $\zz_2$-labeling of the boundary edges, and at every boundary tetrahedron $0123$ we have a weight
\begin{equation}
(-1)^{A(012)\bar B(23)}\;.
\end{equation}

A gauge transformation in the presence of the boundary is given by
\begin{equation}
\label{eq:toriccode_boundary_gauge}
\begin{gathered}
B' = B+ d\Lambda\;,\\
\bar B'=\bar B+\partial \Lambda\;,\\
A' = A+ d\Gamma\;.
\end{gathered}
\end{equation}
Under a $\Lambda$ gauge transformations, the action changes by
\begin{equation}
\begin{multlined}
\partial \lagr^\Lambda_{\text{triv}} + \lagr_\partial(A,B')-\lagr_\partial(A,B)\\=
\partial(A\cup\Lambda) + \partial A\cup (\bar B+\partial\Lambda) - \partial A\cup \bar B\\
=\partial A\cup\partial \Lambda + \partial A\cup \partial \Lambda =0\;,
\end{multlined}
\end{equation}
which is not only a boundary but also directly $0$. A $\Gamma$ gauge transformation yields
\begin{equation}
\label{eq:toric_code_gauge_gamma}
\begin{multlined}
\partial \lagr^\Gamma_{\text{triv}} + \lagr_\partial(A',B)-\lagr_\partial(A,B)\\=
\partial(\Gamma\cup B)+\partial(A+d\Gamma)\cup\bar B-\partial A\cup \bar B\\= \partial\Gamma\cup\partial B+d\partial\Gamma \cup \bar B\\
= \partial\Gamma\cup\partial B+\partial\Gamma \cup d\bar B+d(\partial\Gamma \cup \bar B)\\
=d(\partial\Gamma \cup \bar B)\;.
\end{multlined}
\end{equation}

Analogous to Eq.~\eqref{eq:cyww_gauge_normalization}, the normalization is given by
\begin{equation}
|G|^{-\frac12 \chi} |G|^{-|C_1|+|C_0|} |G^\calb|^{-|C_0^\calb|}\;,
\end{equation}
where $G^\calb$ is the gauge group inside the boundary, in our case $\zz_2$, and $C^\calb_i$ are the $i$-cells within the boundary (which are a subset of $C_i$).

\paragraph{Invertibility of the boundary}
So far we have constructed a gauge-invariant and thus topological boundary. What remains to show is that this boundary is indeed invertible. To this end, we explicitly check the invertibility moves in Eq.~\eqref{eq:boundary_invertibility_moves} by evaluating the partition function on both sides for different space-boundary configurations. The key observation is that the weight for a bulk configuration $\beta$ is only non-zero if $\beta$ is a 2-(co-)cycle, and only depends on the (co-)homology class of $\beta$. Thus, the overall partition function for a fixed space-boundary configuration $\alpha$ is only non-zero if $\alpha$ is a 2-cocycle/1-cycle, and it only depends on the 2-cohomology/1-homology class of $\alpha$. The partition function for a fixed $\alpha$ is the sum over the weights of the different $\beta$ restricting to $\alpha$ at their space boundary, $\partial \beta=\alpha$.

The additional weight of such a boundary configuration $\alpha$ for a fixed bulk 2-cohomology class is the number of bulk 2-cocycles $\beta$ in this cohomology class (which is the same as the number of $2$-coboundaries), times the normalization in Eq.~\eqref{eq:cyww_gauge_normalization}. Combined, we get
\begin{equation}
\label{eq:cohomology_class_weight}
\begin{multlined}
|B^2| |G|^{-\frac12 \chi} |G|^{-|C_1|+|C_0|} =  |G|^{-\frac12 \chi} |B^2| |C^1|^{-1} |C^0|\\ =  |G|^{-\frac12 \chi} |Z^1|^{-1} |C^0|
=  |G|^{-\frac12 \chi} |H^1|^{-1} |B^1|^{-1} |C^0|\\ =  |G|^{-\frac12 \chi} |H^1|^{-1} |H^0| =  |G|^{-\frac12 \chi} |H_3|^{-1} |H_4|\;,
\end{multlined}
\end{equation}
where $B^i$ denotes the set of $i$-coboundaries, $C^i$ the set of $i$-cochains, $Z^i$ the set of $i$-cocycles, and $H^i$ the set of $i$-cohomology classes, and $H_i$ the set of $i$-homology classes. We used the basic identities
\begin{equation}
\begin{gathered}
|B^i|=\frac{|C^{i-1}|}{|Z^{i-1}|}\;,\\ |H^i|=\frac{|Z^i|}{|B^i|}\;,\\ |H_i|=|H^{n-i}|\;.
\end{gathered}
\end{equation}
Note that the weight in Eq.~\eqref{eq:cohomology_class_weight} also holds with a physical boundary, in which case $|H_i|$ is a homology class with open $A$-cycles and closed $B$-cycles. Also note that with a space boundary, $|B^2|$ in Eq.~\eqref{eq:cohomology_class_weight} is the number of closed-space 2-cocycles, and consequently, all the $H_i$ are closed-space homology classes.

In the following, we show invariance under the invertibility moves $M_2$ and $M_3$ in Eq.~\eqref{eq:boundary_invertibility_moves}. The moves $M_0$, $M_1$, and $M_4$ are trivial since the space 2-cohomology/1-homology is trivial. For $M_2$ and $M_3$, the partition function turns out to be $0$ for the non-trivial space cohomology class $\alpha_0$ on both sides of the equation. On one side, there is no bulk cohomology class $\beta$ restricting to $\alpha_0$ on the boundary. On the other side there are two such $\beta$ whose weights cancel.

Concretely, generating (co-)cycles of a product space $X\times Y$ can be obtained from those of $X$ and $Y$ via the \emph{Kuenneth formula}. In our simple case, the Kuenneth formula becomes
\begin{equation}
H_k(X\times Y) = \sum_i H_i(X) H_{k-i}(Y)\;.
\end{equation}
If we imagine $i$-cycles as $i$-dimensional submanifolds, then the generating $k$-cycles of $X\times Y$ can be simply obtained by taking the cartesian product of the $i$-cycles of $X$ with the $k-i$-cycles of $Y$.

Let us start with $M_2$,
\begin{equation}
\label{eq:toric_code_cohomology_m2}
\begin{aligned}
M_2: S_1\times B^{ps}_3 &= B^s_2\times B^p_2\;,\\
\begin{tikzpicture}
\draw (0,0)circle(0.5);
\end{tikzpicture}
\times
\begin{tikzpicture}
\path[manifoldbdfull] (0.5,0)arc(0:180:0.5cm and 0.2cm) arc(180:0:0.5);
\path[manifold] (0.5,0)arc(0:180:0.5cm and 0.2cm) arc(-180:0:0.5);
\path[pattern={Dots[radius=0.025cm,xshift=0.05cm,yshift=0.05cm]},pattern color=cyan] (0.5,0)arc(0:-180:0.5cm and 0.2cm) arc(180:0:0.5);
\path[manifold] (0.5,0)arc(0:-180:0.5cm and 0.2cm) arc(-180:0:0.5);
\end{tikzpicture}
&=
\begin{tikzpicture}
\path[manifold] (0,0)circle(0.5);
\end{tikzpicture}
\times
\begin{tikzpicture}
\draw[manifoldboundary,manifold] (0,0)circle(0.5);
\end{tikzpicture}\;,
\\
Z_0&=Z_1\;.
\end{aligned}
\end{equation}
The space boundary is
\begin{equation}
\label{eq:m2_space_boundary}
S_1\times B^p_2=
\begin{tikzpicture}
\draw (0,0)circle(0.5);
\end{tikzpicture}
\times
\begin{tikzpicture}
\draw[manifoldboundary,manifold] (0,0)circle(0.5);
\end{tikzpicture}
\;.
\end{equation}
This space has only one generating $B$ 1-cycle $\alpha_0$, the product of all of $S_1$ and one point of $B^p_2$, in the following marked in red,
\begin{equation}
\alpha_0:
\begin{tikzpicture}
\draw[cohomologyB] (0,0)circle(0.5);
\end{tikzpicture}
\times
\begin{tikzpicture}
\draw[manifoldboundary,manifold] (0,0)circle(0.5);
\atoms{vertex,style=\cohomologyBcol}{0/}
\end{tikzpicture}
\;.
\end{equation}
It has no non-trivial $A$ 1-cycle since $A$ is open and thus $B^p_2$ has trivial 0-homology.

The left-hand bulk $S_1\times B_3^{ps}$ has no non-trivial 2-cycle as follows. For $A$, $B_3^{ps}$ has trivial open $0$-, $1$- and $2$-homology. For $B$, only the closed $0$-homology of $B_2^{ps}$ is generated by a single point, but the $2$-homology of $S_1$ is obviously trivial. $|H_4|$ and $|H_3|$ are both trivial, and $\chi=0$
\footnote{$\chi$ can be computed by choosing small cellulations for the left-hand side of Eq.~\eqref{eq:toric_code_cohomology_m2}. $\chi$ might depend on whether we include cells of the physical or space boundary, however, different conventions change the left- and right-hand side equally since $\chi$ on a 3-manifold is always $0$. Here we choose the convention to neither include cells of space nor physical boundary.}
, so the normalization is $1$. Thus, evaluating the CYWW partition functions with the two different space boundary configurations $0$ and $\alpha_0$ yields
\begin{equation}
\begin{aligned}
Z_0(0) &= (-1)^{S(0)}=1\;,\\
Z_0(\alpha_0) &=0\;.
\end{aligned}
\end{equation}

The right-hand bulk $B_2^s\times B_2^p$ has two generating 2-cycles $\beta_0$ and $\beta_1$,
\begin{equation}
\begin{aligned}
\beta_0&:
\begin{tikzpicture}
\path[manifold] (0,0)circle(0.5);
\path[cohomologyBfull] (0,0)circle(0.5);
\end{tikzpicture}
\times
\begin{tikzpicture}
\draw[manifoldboundary,manifold] (0,0)circle(0.5);
\atoms{vertex,style=\cohomologyBcol}{0/}
\end{tikzpicture}
\;,
&\partial \beta_0&=\alpha_0\;,\\
\beta_1&:
\begin{tikzpicture}
\path[manifold] (0,0)circle(0.5);
\atoms{vertex,style=\cohomologyAcol}{0/}
\end{tikzpicture}
\times
\begin{tikzpicture}
\path[cohomologyAfull] (0,0)circle(0.5);
\draw[manifoldboundary,manifold] (0,0)circle(0.5);
\end{tikzpicture}
\;,
&\partial \beta_1&=0
\;,
\end{aligned}
\end{equation}
where $A$-cocycles are in green. It is easy to see that those are all generating 2-cycles since the homology of $B_2^p$ alone only has one non-trivial cycle for $B$ (a closed 0-cycle) as well as for $A$ (an open 2-cycle). $H_4$ and $H_3$ are both trivial, and $\chi=1$, so the overall normalization in Eq.~\eqref{eq:cohomology_class_weight} is $\frac12$. Further, we can see that
\begin{equation}
\beta_0\cup\beta_1=1\;.
\end{equation}
The partition function is thus given by
\begin{equation}
\begin{gathered}
\begin{multlined}
Z_1(0) = \frac12 ((-1)^{S(0)} + (-1)^{S(\beta_1)}) = \frac12(1+1)=1\;,
\end{multlined}\\
\begin{multlined}
Z_1(\alpha_0) = \frac12((-1)^{S(\beta_0)} + (-1)^{S(\beta_0+\beta_1)})\\= \frac12(1+(-1)^{(\beta_0)_B\cup (\beta_1)_A})=\frac12(1-1)=0\;.
\end{multlined}
\end{gathered}
\end{equation}
So we indeed find that $Z_0=Z_1$.

Let us now consider the move $M_3$,
\begin{equation}
\begin{aligned}
M_3: S_2\times B^{ps}_2 &= B^s_3\times B^p_1\;,\\
\begin{tikzpicture}
\path[doublemanifold] (0,0)circle(0.5);
\end{tikzpicture}
\times
\begin{tikzpicture}
\path[manifold] (0,0)circle(0.5);
\atoms{bdvertex}{0/p={0.5,0}, 1/p={-0.5,0}}
\draw[manifoldboundary] (0)arc(0:180:0.5);
\end{tikzpicture}
&=
\begin{tikzpicture}
\path[doublemanifold] (0,0)circle(0.5);
\end{tikzpicture}
\times
\begin{tikzpicture}
\atoms{bdvertex}{0/, 1/p={1,0}}
\draw (0)--(1);
\end{tikzpicture}\;,
\\
Z_0&=Z_1\;.
\end{aligned}
\end{equation}
The space boundary is
\begin{equation}
S_2\times B^p_1
=
\begin{tikzpicture}
\path[doublemanifold] (0,0)circle(0.5);
\end{tikzpicture}
\times
\begin{tikzpicture}
\atoms{bdvertex}{0/, 1/p={1,0}}
\draw (0)--(1);
\end{tikzpicture}\;.
\end{equation}
It has one generating $A$ 1-cycle, the product of one point of $S_2$ and all of $B^p_1$,
\begin{equation}
\alpha_0:
\begin{tikzpicture}
\path[doublemanifold] (0,0)circle(0.5);
\atoms{vertex,style=\cohomologyAcol}{0/};
\end{tikzpicture}
\times
\begin{tikzpicture}
\atoms{bdvertex}{0/, 1/p={1,0}}
\draw[cohomologyA] (0)--(1);
\end{tikzpicture}\;.
\end{equation}
There is no non-trivial $B$ 1-cycle since $B$ is closed at the physical boundary and $B^p_1$ has no non-trivial closed 1-cycle.

The left-hand bulk $S_2\times B_2^{ps}$ has two generating 2-cycles,
\begin{equation}
\begin{aligned}
\gamma_0&:
\begin{tikzpicture}
\path[doublemanifold] (0,0)circle(0.5);
\atoms{vertex,style=\cohomologyAcol}{0/};
\end{tikzpicture}
\times
\begin{tikzpicture}
\path[manifold,cohomologyAfull] (0,0)circle(0.5);
\atoms{bdvertex}{0/p={0.5,0}, 1/p={-0.5,0}}
\draw[manifoldboundary] (0)arc(0:180:0.5);
\end{tikzpicture}
\;,
&\partial\gamma_0&=\alpha_0\;,\\
\gamma_1&:
\begin{tikzpicture}
\path[doublemanifold] (0,0)circle(0.5);
\path[cohomologyBfull] (0,0)circle(0.5);
\end{tikzpicture}
\times
\begin{tikzpicture}
\path[manifold] (0,0)circle(0.5);
\atoms{bdvertex}{0/p={0.5,0}, 1/p={-0.5,0}}
\draw[manifoldboundary] (0)arc(0:180:0.5);
\atoms{vertex,style=\cohomologyBcol}{0/};
\end{tikzpicture}
\;,
&\partial\gamma_1&=0\;.
\end{aligned}
\end{equation}
Those are all generating 2-cycles since the homology of $B_2^{ps}$ alone only has one non-trivial cycle for $B$ (a closed 0-cycle) as well as for $A$ (an open 2-cycle). $|H_4|$ and $|H_3|$ are trivial, $\chi=2$, and again $\gamma_0\cup\gamma_1=1$. So we have
\begin{equation}
\begin{gathered}
Z_1(0) = \frac14((-1)^{S(0)} + (-1)^{S(\gamma_1)}) = \frac14(1+1)=\frac12\;,\\
Z_1(\alpha_0) = \frac14((-1)^{S(\gamma_0)} + (-1)^{S(\gamma_0+\gamma_1)}) = \frac14(1-1)=0\;.
\end{gathered}
\end{equation}

The left-hand bulk $B_3^s\times B_1^p$ has no non-trivial 2-cocycles: No matter whether we are open or closed at the physical boundary, $B^s_3$ has trivial 1- and 2-homology, and $B^p_1$ has trivial 2-homology. Finally, $|H_4|$ and $|H_3|$ are trivial and $\chi=1$. Thus, evaluating the CYWW model yields
\begin{equation}
\begin{aligned}
Z_0(0) &= \frac12 (-1)^{S(0)} = \frac12 \;,\\
Z_0(\alpha_0) &=0\;.
\end{aligned}
\end{equation}
Again we find $Z_0=Z_1$.

\subsection{The fermionic invertible boundary of the 3-fermion CYWW}
\label{sec:3fermion_cohomology}
In this section, we present the lattice gauge theory formulation of the fermionic invertible boundary of the three-fermion CYWW model from Section~\ref{sec:umtc_module_examples}.
\paragraph{The bulk model}
Like the toric-code CYWW model, the three-fermion CYWW model is a gauge theory with two $\zz_2$-valued 2-cocycles $A$ and $B$. This time, the action is
\begin{equation}
\lagr=A\cup A+B\cup B+A\cup B\;.
\end{equation}
$A$ and $B$ represent the configuration of the generating anyons $f_1$ and $f_2$, and the three terms in the action correspond to the $-1$ entries $R^{f_1,f_2}$, $R^{f_1,f_3}$, and $R^{f_2,f_3}$ in the braiding through Eq.~\eqref{eq:4simplex_ribbon_manifold}. The difference in action after a $\Gamma$ gauge transformation as in Eq.~\eqref{eq:toric_code_gauge} is
\begin{equation}
\begin{multlined}
\lagr(A',B)-\lagr(A,B)\\
=
(A+d\Gamma) \cup B + (A+d\Gamma)\cup(A+d\Gamma)\\- A\cup B-A\cup A\\
d\Gamma\cup B + A\cup d\Gamma + d\Gamma\cup A + d\Gamma\cup d\Gamma\\
d(\Gamma\cup B+A\cup \Gamma+\Gamma\cup A+\Gamma\cup d\Gamma)=d(\lagr^\Gamma_{\text{triv}})\;,
\end{multlined}
\end{equation}
and an analogous calculation yields $\lagr^\Lambda_{\text{triv}}$.

\paragraph{The boundary}
The invertible boundary is closed for $B$ and open for $A$ just as for the toric-code CYWW model. Intuitively, $A\cup B$ and $B\cup B$ are intersections involving at least one closed cocycle and are therefore well-defined in the presence of a boundary. Formally, the corresponding terms in $\lagr_{\text{triv}}$ can be canceled by a boundary action
\begin{equation}
\lagr_\partial=\partial B\cup \bar B+\partial A\cup \bar B\;.
\end{equation}
However, there is no suitable boundary term for $A\cup A$, which intuitively is due to the fact that $A$ is open and intersections of $A$ with itself can be pushed through the boundary.
This can be fixed by letting $\partial A$ carry an odd fermion parity. This means that $S_\partial=\int \lagr_\partial$ implicitly contains a term
\begin{equation}
\sigma[\partial A] + (\eta, \partial A)
\end{equation}
where $\sigma[\partial A]$ is the fermionic reordering sign (as an element of $\zz_2$) along $\partial A$, and $(\eta,\partial A)$ is the intersection of $\partial A$ with the spin structure $\eta$ required by the spin-statistics relation as discussed in Section~\ref{sec:fermionic_umtc_module}. Note that $\eta$ is a 2-chain in the $3$-dimensional boundary instead of a 1-cochain, and thus we compute the intersection with the 2-cocycle $\partial A$ not by the cup product but by the simplex-wise overlap $(\eta,\partial A)$. In the lattice model, we have a weight
\begin{equation}
(-1)^{\eta(012)A(012)}
\end{equation}
at every boundary triangle $012$. The fermionic reordering sign $\sigma[\partial A]$ is defined by choosing a fixed ordering of the triangles of the branching-structure tetrahedron. We will choose
\begin{equation}
\label{eq:tetrahedron_grassmann_ordering}
123-013-023-012
\end{equation}
in accordance with Ref.~\cite{Gaiotto2015}. Then, we associate two Grassmann variables (c.f.~Section~\ref{sec:fermionic_umtc_module}) $\theta_T$ and $\overline{\theta_T}$ to every triangle $T$, and evaluate
\begin{equation}
\begin{multlined}
(-1)^{\sigma[A]} = (\prod_{T\in S_2}{\smallint d\theta_T})\\
\cdot(\prod_{X\in S_3} \overline{\theta}_{X_{123}}^{A(X_{123})} \theta_{X_{013}}^{A(X_{013})} \overline{\theta}_{X_{023}}^{A(X_{023})} \theta_{X_{012}}^{A(X_{012})})^{\underline{\eta_1(X)}}\;,
\end{multlined}
\end{equation}
where $X_{abc}$ denotes the $abc$-triangle of a tetrahedron $X$, $\eta_1(X)$ is $0$ or $1$ depending on whether $X$ has right-handed or left-handed orientation, and we used the notation
\begin{equation}
(A)^{\underline B}=
\begin{cases}
A & \text{if} \; B=0\\
\overline A & \text{if} \; B=1
\end{cases}\;.
\end{equation}
Below, we will use two identities for the reordering sign that can be found in Ref.~\cite{Gaiotto2015}, and that we revisit in Appendix~\ref{sec:reordering_selflinking}. First, the reordering sign $\sigma$ is a quadratic refinement of $\cup_1$,
\begin{equation}
\label{eq:reordering_refinement}
\sigma[a+b] = \sigma[a]+\sigma[b]+\smallint a\cup_1 b\;.
\end{equation}
Second, we have
\begin{equation}
\label{eq:self_linking_fermion}
\sigma[da]+(\eta,da)=\smallint a\cup da\;.
\end{equation}
This can be understood intuitively by observing that by definition, $(-1)^{\sigma[da]+(\eta,da)}$ is the amplitude of a fundamental fermion traveling along the worldline $da$. Since fermions have spin $-1$, this should be equal to ($-1$ to the power of) the number of twists in the loop $da$ with respect to some canonical framing, i.e., the linking number of $da$ with a canonically shifted copy of $da$. As $da$ is the boundary of $a$, and $\cup$ provides a canonical shift, this number of twists is exactly given by $\smallint a\cup da$.

Using those identities, we can assert the invariance of the partition function under gauge transformations. The action is pretty directly invariant under $\Lambda$ gauge transformations as in Eq.~\eqref{eq:toriccode_boundary_gauge}, which does not involve the fermionic aspect of the boundary,
\begin{equation}
\label{eq:3fermion_cyww_gauge}
\begin{multlined}
\partial \lagr^\Lambda_{\text{triv}} + \lagr_\partial(A,B')-\lagr_\partial(A,B)\\
=\partial(A\cup \Lambda+B\cup \Lambda+\Lambda\cup B+d\Lambda\cup \Lambda)\\+\partial B\cup\partial \Lambda + \partial d\Lambda\cup \bar B+d\partial\Lambda\cup \partial\Lambda +\partial A\cup \partial \Lambda\\
=\partial\Lambda\cup\partial B + \partial \Lambda\cup d \bar B + d(\partial\Lambda\cup \bar B)=d(\partial\Lambda\cup \bar B)\;.
\end{multlined}
\end{equation}
The invariance under a $\Gamma$ gauge transformations is more involved. Let us separate $\partial \lagr^\Gamma_{\text{triv}}$ into
\begin{equation}
\begin{multlined}
\partial \lagr^\Gamma_{\text{triv} 0}= \partial\Gamma\cup \partial B\;,\\
\partial \lagr^\Gamma_{\text{triv} 1}= \partial A\cup \partial \Gamma+\partial\Gamma\cup \partial A+\partial\Gamma\cup d\partial\Gamma\\
 = \partial A\cup_1 d\partial \Gamma+\partial\Gamma\cup d\partial\Gamma + d(\partial A\cup_1 \partial \Gamma)\;.
\end{multlined}
\end{equation}
Then we have
\begin{equation}
\partial \lagr_{\text{triv0}} + \lagr_\partial(A',B)-\lagr_\partial(A,B) = d(\partial\Gamma \cup \bar B)
\end{equation}
by a calculation similar to Eq.~\eqref{eq:toric_code_gauge_gamma}. For the change of the fermion-related weights, we find
\begin{equation}
\begin{multlined}
\sigma[\partial(A+d\Gamma)] + (\eta,\partial(A+d\Gamma))- \sigma[\partial A] - (\eta, \partial A)\\
=\sigma[d\partial\Gamma]+\smallint \partial A\cup_1 d\partial\Gamma + (\eta,d\partial\Gamma)\\
=\smallint ( \partial A\cup_1 d\partial\Gamma + \partial\Gamma \cup d\partial\Gamma)\eqqcolon \smallint \Delta_{\sigma,\eta}
\end{multlined}
\end{equation}
using precisely Eq.~\eqref{eq:reordering_refinement} and Eq.~\eqref{eq:self_linking_fermion}. So the partition function is gauge invariant, as the total change
\begin{equation}
\partial \lagr_{triv} + \lagr_\partial(A',B)-\lagr_\partial(A,B)+\Delta_{\sigma,\eta}
\end{equation}
is a coboundary.

\paragraph{Invertibility}
The argument for the invertibility of the boundary is the same as for the toric-code CYWW model in the last paragraph of Section~\ref{sec:gauge_toric_code}. All that matters for those considerations is that the boundary is $A$-open and $B$-closed. As argued in Section~\ref{sec:fermionic_state_sum}, the physical boundary of the space boundary of $M_2$ in Eq.~\eqref{eq:m2_space_boundary} is a torus, on which the spin structure has to be bounding-bounding. For this bounding-bounding spin structure, the term
\begin{equation}
\smallint B\cup B+\smallint A\cup A+\sigma[\partial A] + (\eta,\partial A)
\end{equation}
evaluates trivially for all the 2-cycles spanned by $\beta_0$ and $\beta_1$ on the left-hand side, such that only the toric-code CYWW term $\int A\cup B$ remains. The same holds for $M_3$, for which there is only one spin structure for the physical boundary of the space boundary.

\end{document}